\documentclass[preprint2]{aastex} 
\pdfoutput=1

\usepackage{graphics}  
\usepackage{natbib}
\usepackage{apjfonts}
\usepackage{caption}
\usepackage{graphicx}
\usepackage{sidecap}

\newcommand{\kms}       {km~s$^{-1}$}

\newcommand{\etal}      {{et~al.}}

\newcommand{\apl}       {^{<}_{\sim}}

\newcommand{\flux}      {ergs~cm$^{-2}$~s$^{-1}$~\AA$^{-1}$}

\newcommand{\pcm}       {cm$^{-2}$}
\newcommand{\pcmV}      {cm$^{-3}$}

\newcommand{\mol}       {H$_2$}
\newcommand{\sixa}      {O~VI~$\lambda 1032$}
\newcommand{\sixb}      {O~VI~$\lambda 1037$}
\newcommand{\hdline}    {HD~6$-$0~R(0)}
\newcommand{\Hipparcos} {{\it Hipparcos\/}}
\newcommand{\Tycho}     {{\it Tycho\/}}
\newcommand{\Copernicus} {{\it Copernicus\/}}
\newcommand{\FUSE}      {{\it FUSE\/}}
\newcommand{\ROSAT}      {{\it ROSAT\/}}
\newcommand{\HST}      {{\it HST\/}}  
\newcommand{\nup}       {$N$(O~VI)$\,\sin|b|$}

\shorttitle{O~VI in the disk of the Milky Way}
\shortauthors{D. V. Bowen et al.}  

\received{xxx}
\accepted{xxx}  

\begin{document} 

\title{The \emph{Far Ultraviolet Spectroscopic Explorer} Survey of O~VI
  Absorption in the Disk of the Milky Way}         

\author{David V.~Bowen$^{1}$, 
Edward B.~Jenkins$^{1}$, 
Todd~M.~Tripp$^{1,2}$, 
Kenneth~R.~Sembach$^{3}$,
Blair~D.~Savage$^{4}$, 
H.~Warren~Moos$^{5}$, 
William~R.~Oegerle$^{6}$,
Scott~D.~Friedman$^3$,
Cecile~Gry$^{7}$,
Jeffrey~W.~Kruk$^5$,
Edward~Murphy$^8$,
Ravi Sankrit$^9$,
J.~Michael~Shull$^{10}$,
George~Sonneborn$^6$,
Donald~G.~York$^{11}$
}

\affil{ $\:$}  

\affil{$^1$ \footnotesize{Princeton University Observatory, Peyton Hall, Ivy Lane, 
Princeton, NJ 08544}} 
\affil{$^2$  \footnotesize{Dept.\ of Astronomy, University of Massachusetts, Amherst, 
  MA~01003}}
\affil{$^3$  \footnotesize{Space Telescope Science Institute, 3700 San Martin Dr.,
  Baltimore, MD~21218}} 
\affil{$^4$  \footnotesize{Dept.~of Astronomy, University of Wisconsin, 475
N.~Charter Street, Madison, WI 53706}}
\affil{$^5$ \footnotesize{ Dept.~of Physics and Astronomy, Johns Hopkins University,
Baltimore, MD~21218}}
\affil{$^6$  \footnotesize{Astrophysics Science Division,
Code 660, NASA/Goddard Space Flight Center, Greenbelt, MD 20771}}
\affil{$^7$ \footnotesize{Laboratoire d'Astrophysique de Marseille, 
    13376 Marseille, France}}
\affil{$^8$ \footnotesize{Dept.~of Astronomy, University of Virginia, Box
   400325, Charlottesville, VA 22904-4325}}
 \affil{$^9$ \footnotesize{Space Science Laboratory, University of California, Berkeley, Berkeley, CA 94270-7450}}
\affil{$^{10}$ \footnotesize{Dept. of Astrophysical and Planetary Sciences,
  University of Colorado, Boulder, CO 80309-0389}}
\affil{$^{11}$ \footnotesize{Dept.\ of Astronomy and Astrophysics, University of
  Chicago,  Enrico Fermi Institute, 5640 South Ellis Avenue, Chicago,
  IL~60637.}}

\affil{ $\:$}

\affil{ {\it{Received:}} 11-July-2007; {\it{Accepted:}} 19-October-2007}

\affil{ $\:$}
\affil{ $\:$}

\begin{abstract}
    To probe the distribution and physical characteristics of
  interstellar gas at temperatures $T\:\approx \:3\times 10^{5}$~K in
  the disk of the Milky Way, we have used the {\it Far Ultraviolet
    Spectroscopic Explorer (FUSE)\/} to observe absorption lines of
  O~VI~$\lambda$1032 toward 148 early-type stars situated at distances
  $>\:1$ kpc.  After subtracting off a mild excess of O~VI arising
  from the Local Bubble, combining our new results with earlier
  surveys of O~VI, and eliminating stars that show conspicuous
  localized X-ray emission, we find an average O~VI mid-plane density
  $n_0\:=\:1.3\times 10^{-8}{\rm\: cm}^{-3}$.  The density decreases
  away from the plane of the Galaxy in a way that is consistent with
  an exponential scale height of 3.2~kpc at negative latitudes or
  4.6~kpc at positive latitudes.  Average volume densities of O~VI
  along different sight lines exhibit a dispersion of about 0.26~dex,
  irrespective of the distances to the target stars. This indicates
  that O~VI does not arise in randomly situated clouds of a fixed size
  and density, but instead is distributed in regions that have a very
  broad range of column densities, with the more strongly
  absorbing clouds having a lower space density.  Line widths and
  centroid velocities are much larger than those expected from
  differential Galactic rotation, but they are nevertheless correlated
  with distance and $N$(O~VI), which reinforces our picture of a
  diverse population of hot plasma regions that are ubiquitous over
  the entire Galactic disk.  The velocity extremes of the O~VI
  profiles show a loose correlation with those of very strong lines of
  less ionized species, supporting a picture of a turbulent,
  multiphase medium churned by shock-heated gas from multiple
  supernova explosions.

\end{abstract}



\section{INTRODUCTION \label{sect_intro}}

Very hot plasmas are recognized as important constituents in a broad
range of astrophysical contexts, ranging from the solar and stellar
transition layers and coronae up to the largest arrangements of matter
in the universe that form web-like structures that bridge groups and
clusters of galaxies.  In all cases, the processes responsible for
heating the gas give important insights on evolutionary and dynamical
phenomena in the systems under study.  On the largest scales, there is
a recognition that hot gas in the intracluster media and even the
intergalactic media at low redshifts probably dominate the baryonic
content of the universe \citep[e.g.][]{CenOstr, tripp00a,danforth05,bregman07}.
In this paper, we focus on the presence of hot gas within the
interstellar medium (ISM) of the Milky Way, a system intermediate in
size between the extremes just mentioned.  This is a topic that has
been studied over the last 3 decades, and shortly
(\S\ref{sect_intro_early}) we will look back upon some of the key
developments in this field to help define the motivations for our
present study.

\subsection{Basic Principles of Detecting Hot Gas \label{sect_intro_basic} }

Atoms within plasmas at temperatures $T>10^5\,$K have many of their
outer electrons stripped by collisions.  As a consequence, there is an
upward shift in the energy of their resonance transitions, pushing
important spectroscopic information into the ultraviolet and X-ray
bands.  It is for this reason that progress in understanding the
distribution and properties of hot gases in space have had to await
the development of space-borne instruments that could observe
radiation at energies above those in the visible
band.\footnote{Forbidden transitions in the visible from highly
  ionized atoms are generally too weak to observe \citep{graney90},
  except as seen in emission for exceptionally bright sources such as
  supernova remnants
  \citep{woodgate74,woodgate75,teske90,sauv95,sauv99,szen00}.}  Of all
the transitions from highly ionized atoms at wavelengths longward of
the Lyman limit, the $\lambda\lambda 1032,~1037$ doublet of O~VI is
the most suitable for study.  The advantages of observing O~VI are
threefold: (1) the ionization potential of O$^{+4}$ is high enough
(114$\,$eV) to insure that there is no appreciable contamination by
cooler material photoionized by starlight,\footnote{Except for small
  contributions from very hot, He-poor white dwarfs \citep{dupree83}.}
as is the case with other partly stripped, Li-like atoms such as Si~IV
and C~IV \citep{cowie81}, (2) the transitions are strong [$\log
f\lambda = 2.138$ for the 1032~\AA\ transition \citep{morton03}], and
(3) the cosmic abundance of oxygen is high.  One mildly negative
factor is that the peak abundance of the five-times ionized form of
oxygen (22\% at $T=3\times 10^5\,$K under collisional ionization
equilibrium conditions) is lower than representative fractions that
can be reached by other ionization stages just below or above that of
O~VI \citep{shapiro76b, shull82,sutherland93}. Also, the temperature
range over which O~VI is the dominant species is smaller than that of
many other ions, such as O~VII~$\lambda$~21.6 seen in the X-ray band.

The easiest way to observe the O~VI transitions is to record them in
absorption in the spectrum of a UV-bright source.  In the disk of the
Galaxy, the most suitable stars are those hotter than spectral type B3
that have large projected rotation velocities $v \sin i$.
Difficulties with poorly behaved stellar continua sometimes arise to
create ambiguities in the absorption strengths; this issue is one that
we address later in \S\ref{sect_cont} for our current study.  For
probing the Galactic halo, AGNs and quasars make excellent background
sources.

Energetic collisions in a hot gas not only ionize the atoms, but they
also excite them to electronic states at energies well above the
ground state.  As a consequence, the atoms emit radiation in their
strong resonance lines.  O~VI can be seen in emission \citep[see,
e.g., Table~1 of][]{dixon06,otte06} although the flux levels are very
faint ($\sim1-4\times 10^{-18}$ ergs~cm$^{-2}$~s$^{-1}$~arcsec$^{-2}$)
and require long integration times to achieve a satisfactory
detection.  A principal difference between the absorption and emission
measurements is that the former is proportional to the line integral
of the electron density, while the latter senses the square of the
electron density [but with some sensitivity to temperature beyond just
the variation from the ion fraction curve \citep{shull94}].  This
difference can be exploited to measure the filling factor of the gas,
$\langle n_e\rangle ^2/\langle n_e^2\rangle$.  One drawback in
measuring the O~VI emission is that self absorption and resonant
scattering can complicate the interpretation if the optical depths are
of order or greater than unity \citep{shelton01}.  Also, the emission
can be attenuated by the absorption caused by dust in the foreground.

Line emission from atoms other than O~VI also radiate from hot gases.
As a rule, the characteristic energy bands where the emissions are
strongest and most plentiful increase toward higher energies as the
temperature rises.  For gas at temperatures of 282,000$\,$K where O~VI
has its maximum ion fraction in collisional equilibrium, the
characteristic energy of the strongest emission-line features is below
300$\,$eV \citep{kato76}.  Unfortunately, there are a number of
drawbacks that limit our ability to make good use of the X-ray
background measurements as we interpret the O~VI absorption data.
First, at energies below 300$\,$eV the absorption by foreground
neutral hydrogen and helium is strong \citep{morrison83}.  Second, it
is often difficult to know how much emission arises from beyond the
target used for the O~VI measurement.  Third, over recent years there
has been an increasing awareness that the soft X-ray background is
contaminated by emission from charge exchange between heavy solar wind
ions and the local interstellar gas \citep{cravens00, cravens01}, a
process identical to that which produces X-ray emission from comets
\citep{kharchenko00, kras02, kharchenko03}.  This contamination may be
as much as half of the observed diffuse background level
\citep{lallement04}.  Finally, with the exception of the lowest energy
X-ray measurements, such as those that operate below the 111$\,$eV
absorption cutoff of a beryllium filter \citep{bloch86}, the
observations have much stronger responses to gases that are too hot
($T\gtrsim 2\times 10^6\,$K) to yield much O~VI.  All but the second
reason listed above may help to explain why \citet{savage03} found a
poor correlation between soft X-ray emission and the strength of O~VI
absorption toward extragalactic targets at high Galactic latitudes.

\subsection{Early Developments \label{sect_intro_early}}

Prior to the advent of space astronomy, an ingenious insight on the
probable existence of a large volume of space that contains hot gases
was provided by \citet{spitzer56}, who noted that cold clouds found in
the Galactic halo had to have some means of confinement.  He
speculated that such clouds were stabilized by the external thermal
pressure of a hot Galactic corona.  It was not until some 18 years
later that firm evidence on the pervasiveness of this type of material
within the Galactic disk was demonstrated by direct observations of
absorption by the O~VI doublet at 1032 and 1038~\AA\ in the spectra of
32 stars \citep{ebj74, york74}, accompanied by a recognition by
\citet{will74} that the diffuse soft X-ray emission observed by
instruments on sounding rockets \citep{bowyer68, bunner71, davidson72,
  yentis72} probably came from the same type of material.

Following the initial discoveries, \citet{ebj78a} (hereafter J78)
conducted a survey of O~VI absorption which increased the accumulation
of cases to 72 Milky Way stars.  From this body of data, he offered
some conclusions on the nature of the hot gas and how it is
distributed in space \citep{ebj78b, ebj78c}.  In particular, the
statistical fluctuations of absorption strengths and velocities from
one sight line to the next indicated that approximately 6 hot gas
regions, each with a column density $N({\rm O~VI})\approx 10^{13}~{\rm
  cm}^{-2}$, are distributed across a representative distance of
1$\,$kpc.  Later, inspired by evidence from the soft X-ray surveys
that the Sun is surrounded by a volume of hot gas out to a distance of
around 100$\,$pc \citep{sanders77, mccammon90, snowden90, snowden98},
a structure now called the Local Bubble (hereafter, `LB'), \citet{shelton94}
conceived of a slightly more elaborate model that accounted for a
possible constant foreground contribution of O~VI, with $N({\rm
  O~VI})$ averaging about $1.6\times 10^{13}{\rm cm}^{-2}$, which
reduced the apparent fluctuations of a much more patchy distribution
at greater distances.  Their re-interpretation of the O~VI data
suggested that outside the Local Bubble the Galactic disk is populated
by regions that have column densities in the range $2-7\times
10^{13}{\rm cm}^{-2}$ with separations 450 $-$ 1300$\,$pc.  With the
advent of the {\it Far Ultraviolet Spectroscopic Explorer} (FUSE),
\citet{bill_local} looked at weak O~VI lines towards nearby white
dwarfs (WDs), and determined the average volume density to be
$2.4\times10^{-8}$~cm$^{-3}$.  They concluded that the local O~VI
contribution is much less than that estimated by \citet{shelton94}
unless it is confined to a thin shell at the edge of the LB, i.e.,
just beyond the white dwarfs. A much larger survey of nearby WDs was
conducted by \citet{savage06}, who measured the average O~VI density
to be $3.6\times10^{-8}$~cm$^{-3}$.

A challenge that was presented to the early O~VI investigators was a
proposal by \citet{castor75} that all of the O~VI arose from bubbles
that were created by stellar mass-loss winds from the target stars.
\citet{ebj78c} addressed this issue and concluded that some O~VI could
arise from the bubbles, but that most of the hot gas was truly
interstellar in nature.  We return to this topic once again in this
paper (\S\ref{sect_bubbles}).

After the end of the \Copernicus\ era, there was a long period where
only the transitions of N~V, Si~IV and C~IV could be seen in
absorption using the {\it International Ultraviolet Explorer\/} ({\it
  IUE\/}) and the {\it Hubble Space Telescope\/} ({\it HST\/}).
Because these species can be created at lower energies (ionization
potentials of the next lower stages are only 33, 48 and 77$\,$eV,
respectively), there was a persistent concern that photoionized gases
could be responsible for these species, especially in low density
environments.  Even so, important conclusions emerged about the
distributions of highly ionized material in the Galactic halo
\citep{savage87, ken91, sembach92}, the Magellanic Clouds
\citep{deboer80, fitzpatrick85}, and within the Galactic disk
\citep{cowie81, walborn84, edgar92, spitzer92, ken93, ken94, tripp93,
  savage94, ken94_zetoph, huang95}, often using distant UV sources
fainter than the sensitivity limit of {\it Copernicus}.  Further
studies of O~VI, however, were confined to results that were more
limited in scope \citep{ferguson95, hurwitz95, dixon96, hurwitz96,
  widmann98, ken99} because they came from short-lived missions, such
as the {\it Hopkins Ultraviolet Telescope\/} ({\it HUT\/})
\citep{davidsen93} and the {\it Orbiting Retrievable Far and Extreme
  Ultraviolet Spectrograph\/} ({\it ORFEUS-SPAS\/}) \citep{hurwitz98}.
With the availibility of \FUSE , we can now return to large-scale
surveys of O~VI absorption.  Already, interstellar O~VI has been
studied in the Galactic halo \citep{bill00, richter01, howk02a, fox04,
  savage03, sembach_hvcO6, wakker03}, Large Magellanic Cloud
\citep{howk02b}, Small Magellanic Cloud \citep{hoopes02}, and the
Local Bubble \citep{bill_local,savage06}.  Likewise, intergalactic
O~VI has been found for absorption systems at redshifts that are too
low for coverage with {\it HST\/} \citep{bill00, sembach01_3c273, 
tripp01, savage02, phl1811_1, prochaska04,
  richter04, sembach04, danforth05}.  With the current paper, we cover yet another
important realm, the Galactic disk.

Through the years, theoretical investigations kept pace with the
observations, starting with initial assessments by \citet{shapiro76a}
on the nature of a possible Galactic fountain, and an early proposal
by \citet{cox74} on the origin of a pervasive network (``tunnels'') of
hot gas that is kept alive by the channeling of shock waves (from
supernovae) within it.  This picture was investigated in more detail
by \citet{smith77}.  \citet{mckee77b} presented a seminal treatise on
the different phases of the medium, accounting for mass and energy
balance between a hot interstellar medium created by supernovae and
the more conventional phases at lower temperatures.  This theory was a
very important advance over an earlier model for ISM phases developed
by \citet{field69} that overlooked the hot phase and its secondary
effects.  Some noteworthy consequences that might arise from the hot
gas production include mechanical energy transport to the embedded
cool clouds \citep{cox79} and a feedback mechanism that may regulate
star formation in disk galaxies \citep{cox81}.  Enhanced accumulations
of O~VI-bearing gas will undoubtedly arise from hot gas at the edges
of, and within, the interiors of supernova remnants, which is
discussed in the models presented by \citet{chevalier74},
\citet{cowie81_snr}, \citet{slavin92}, \citet{edgar93},
\citet{smith01}, and \citet{shelton06}.

More recently, exciting realizations of how the ISM may be configured
from supernova explosions have been constucted by \citet{avillez00},
\citet{avillez04} and \citet{avillez05_mag} using hydrodynamic and
magnetohydrodynamic simulations. Their models predict global
equilibrium in the disk-halo circulation when the explosions in the
ISM are traced over several hundred Myr, and demonstrate how effective
turbulent diffusion can be in mixing hot and cold gas.
\citet{avillez05} predict specifically the expected O~VI column
densities along random lines of sight over distances of up to 1~kpc,
and find good agreement with the \FUSE\ WD and \Copernicus\ data.
Their models predict a clumpy turbulent ISM in which O~VI vortices and
filaments are interecepted every $\sim 100$~pc, each with
$N$(O~VI)$\:\simeq\:1-2\times 10^{12}$~\pcm .

\subsection{Current Topics \label{sect_current} }

A persistently debated unknown quantity is the volume filling factor
of the hot gas in the Galactic disk \citep{ferriere95}.  Equally
important is the topology of the gas: is it confined to isolated,
compact bubbles or does it comprise a matrix of multiply connected
regions within which the rest of the ISM is confined?  For observers, it
is easy to measure the average amount of O~VI in the Galactic disk,
and with reasonable assumptions we may relate this quantity to the
total amount of gas in the temperature regime $10^5-10^6\,$K.
However, a much more difficult task is to understand the structure of
the gas along the viewing direction, i.e., information that is lost by
the projection of everything along the sight line, which is compounded
by the restriction that not all directions in the sky can be sensed.
\citet{ebj78b} placed an approximate upper limit of 20\% for the
filling factor from the apparent lack of an anticorrelation between
the amount of O~VI and denser (mostly neutral) hydrogen, as revealed
by the color excesses of the target stars (however, this argument can
equally well be used to say the filling factor is greater than 80\%).
Models that propose that the hot gas has a very large filling factor
[e.g. that of \citet{mckee77b}] must be restrained by observations by
\citet{heiles80} that indicate that the neutral, warm medium (gas that
emits 21-cm radiation but has a high spin temperature and thus
negligible absorption) has its own significant filling factor.  One
feature of the McKee \& Ostriker model is that it assumes that
supernovae occur at random locations and time.  Actually, supernovae
are highly correlated \citep{mccray79, heiles90}, and this should
influence theories on the pervasiveness of O~VI-bearing material,
leading to fewer, but larger volumes whose interiors hold the hot gas.
The existence of such large, coherent structures seems to be validated
by observations of enormous shells of gas seen in 21-cm line emission
\citep{heiles79, heiles84, mcclure02}.

The kinematics of O~VI and its relationship with other forms of gas,
those that are highly ionized and those that are not, can provide
important clues on the nature of the hot gas on a more microscopic
level.  One of the first conclusions to emerge from the velocity
profiles of the O~VI features is that they were broader than those
produced by less ionized species, such as Si~II and Si~III
\citep{york74}, indicating that the O~VI was not simply a more ionized
counterpart of the lower excitation gas.  The absorptions exhibited by
York were of very good quality, which enabled him to place upper
limits to the Doppler motions that corresponded to a range $T=4\times
10^5$ to $2\times 10^6$ for different cases.  The most striking
conclusion from the survey by \citet{ebj78b} was that the scatter of
velocity centroids, amounting to only 26~\kms , is far below the shock
speeds needed to create the post-shock temperatures that can strip 5
electrons from the oxygen atoms, i.e., at least $130-160$~\kms .
\citep{ray79, shull79} A similarly small dispersion of $21\,{\rm
  km~s}^{-1}$ was reported recently by \citet{savage03} for sight
lines through the Galactic halo.  The assertion that O~VI does not
have extraordinarily large velocities must be viewed with caution
however.  We can not rule out the presence of some additional O~VI
that is moving very rapidly, for if the absorption profiles of such
high velocity material were very broad and shallow, they would be very
difficult to detect against the undulating stellar continua.

Simply put, O~VI is not created in zones that immediately follow the
shocks that heated the gas to high temperatures.  Instead, the low
velocities indicate that what we see arises either from some
interaction with the normal, low-ionization material or,
alternatively, perhaps ions whose velocities are regulated by magnetic
fields in the disk \citep{cox88} that might have been compressed and
amplified by the motions of expanding supernova remnants
\citep{spitzer90}. For the former possibility, we may consider two
alternatives.  One is that the O~VI originates in transition layers
where heat flows by conduction across interfaces established between
regions that hold very hot ($T>10^6\,$K) gas and denser, cooler
material \citep{cowie77, mckee77a, begelman90_2, bertoldi90, mckee90}.
Material is transported between the regions: at early times cool gas
evaporates into the hot region, whereas later, after radiative cooling
becomes important, hot gas condenses onto the cool gas \citep{bork90}.
If there is a substantial difference in velocity between the cool and
hot material, an additional means of transporting heat and matter
operates: instabilities in the transition zone create turbulence which
acts as a source of mechanical mixing of the cool and hot phases
\citep{begelman90, slavin93}.

Observational support for links between O~VI-bearing gas and the
cooler material was provided by \citet{cowie_ebj79}, who noted
correlations in velocity between O~VI and the ions Si~III and N~II.
Additional support was provided by the good velocity correspondence
between O~VI and C~II absorption along short path lengths through the
local ISM \citep{savage06}.  One way to gain insights on the relative
importance of the different processes discussed above, as well as to
see if the gas is simply radiatively cooling \citep{edgar86} instead
of being cooled by contact with low temperature material, is to test
whether or not various other highly ionized atoms (N~V, Si~IV and
C~IV) agree with theoretical predictions for the different regimes
\citep{spitzer96, fox03}.

In addition to the possibilities listed above, we must be watchful for
another physical condition, that, under certain circumstances, may
apply to the regions that are responsible for O~VI absorption.  In a
discussion of hot gas within the Local Bubble, \citet{breit94}
considered the possibility that an explosion in a dense region
followed by expansion in a lower density medium would lead to
adiabatic cooling of the hot gas.  The lag in recombination as the gas
cools would be even more severe than for a radiatively cooling gas.
They envision the possibility that we are surrounded by gas at $T\sim
50,000\,$K that still has an appreciable concentration of ions that
normally appear in equilibrium at much higher temperatures.  It is
important to note that we cannot depend on identifying this gas on the
basis that the lines have a velocity dispersion characteristic of the
Doppler broadening at the lower temperatures, since the expansion
process itself creates a broadening that replaces the thermal
broadening.

In the discussion presented up to now, we have considered a number of
contemporary problems in the study of gases responsible for O~VI
absorption in the disk of the Galaxy.  More comprehensive discussions
are provided by review papers written by \citet{cox87_araa},
\citet{savage87_tetons}, \citet{ebj87_iue}, and \citet{spitzer90}.

\subsection{Outline of This Paper---A Summary}

The capability of \FUSE\ to record spectra of stars much fainter than
the magnitude limit of \Copernicus\ ($V\leq 7.5$), allows us to probe
regions well beyond the original O~VI survey of J78.  In this paper,
we present a new \FUSE\ sample of disk stars which has about twice as
many stars as the older survey.  These two considerations allow us to
overcome some shortcomings of the earlier work and address such issues
as the possible presence of very large scale structures containing
either more or less than the usual amount of O~VI, and whether or not
the apparent decrease of O~VI at moderate distances from the Galactic
plane agrees with an exponential scale height of $\sim 2.3\,$kpc as
determined from sources well outside the Galaxy \citep{savage03}.
Also, with the larger column densities that arise from the greater
lengths of the sight lines, we should be less sensitive to distortions
of the results arising from {\it i)} the presence of bubbles around
the target stars and {\it ii)} an over-representation of the region
surrounding the Sun that arises from the constraint that all sight
lines must emanate from a single point.

This paper focuses on three principal topics. First, we describe in
detail the data collection and reduction of the \FUSE\ spectra, and
the analysis of the O~VI absorption line profiles. Second, we present
a re-evaluation of the distances to the targeted stars, which is an
important parameter for understanding how O~VI is distributed in the
Galactic disk. Third, we discuss the global properties of O~VI
absorption in the disk, which, of course, is the main science goal of
this survey.

As far as the first of these topics is concerned, the data reduction
of \FUSE\ spectra is not unlike that of any other data taken with
modern photon-counting devices. We were concerned, however, that we
should be able to precisely calibrate the wavelength scale of the
spectra, since \FUSE\ does not take calibration spectra
contemporaneously with an object. This led us to compare the data from
many of our sight lines with high quality, high resolution {\it HST}
data, from which we found that the standard \FUSE\ wavelength scale
for the pipeline calibration (CalFUSE 2.0.5) required a correction of
10~\kms. This analysis is introduced in \S\ref{sect_wavecal} and
described in detail in Appendix~\ref{sect_cl1}.

In analyzing the O~VI lines, we faced two main obstacles. First, J78
realized that any \sixa\ line is contaminated to some degree by the HD
$6-0$ R(0) line at 1031.9~\AA . Before measuring the physical
parameters of an O~VI line, the contribution from this HD line must be
removed.  Fortunately, in the \FUSE\ data, there exist other HD lines
that are free of the dense H$_2$ `forest' and which can be used to
model the contaminating HD line.  We found that removing the model
using Voigt profile line-fitting techniques was highly successful and
actually presented few problems for the subsequent analysis of the
\sixa\ line. Details are given in \S\ref{sect_contaminants} and
\ref{sect_hdremoval}; \S\ref{sect_epoch} also discusses a consistency
check on the efficacy of the HD removal using multi-epoch data.

A more problematic obstacle in analyzing the O~VI lines arose in
determining the shape of a star's continuum. With the data reduced, we
quickly discovered that the gradient of the continuum was probably
varying on scales less than the width of an O~VI line. The most
extreme variations were prevelant among the O9-B1 type stars.  It
became clear that we needed to track three possible continuum fits:
the adopted best fit, and two possible variations above and below this
fit --- `upper' and `lower' continuum fits, as we will refer to them
throughout this paper.  Although measuring continuum errors in
absorption-line studies is far from unusual, in many cases, the
interstellar lines under study are quite narrow, and continuum errors
are largely irrelvant. For our \FUSE\ data, however, we found that
continuum errors actually dominated the O~VI column density errors,
and not the usual Poisson counting statistical errors.  We discuss how
we chose our continua in \S\ref{sect_cont}.  With the spectra
normalized, we were able to derive column densities, Doppler
parameters and line velocities using both Voigt profile fitting
(\S\ref{sect_voigt}) and the Apparent Optical Depth (AOD) method
(\S\ref{sect_colsAOD}), both of which gave very similar results, as
expected. For the subsequent analysis, we assumed that the absorbing
gas had no small-scale velocity structures hidden within the broad
profile that might invalidate either analysis method
(\S\ref{sect_saturation}). We favored the results from the Voigt
profile fitting procedure, but used AOD measurements when no line was
clearly present and an upper limit to the O~VI column density was
required (\S\ref{sect_colsAOD} and Appendix~\ref{sect_upper_limits}).
The spectra themselves are presented in \S\ref{sect_explain_spectra}.

The second principal goal of this paper is to re-evaluate the
distances to the stars used in our survey.  There are several
databases now available (the \Hipparcos , \Tycho, and 2MASS catalogs)
which provide a more reliable estimate of a distance than was possible
a decade ago.  We introduce the problem first in
\S\ref{sect_the_disk}, where we show the distribution of stars in the
Galactic plane. We expound a full discussion in
Appendix~\ref{sect_distances}, since the methods used to derive
distances are technical and outside the main scope of our discussion
on O~VI in the Galactic disk.  A reader disinterested in the
mechanisms of deriving stellar distances can certainly skip this
Appendix, and instead simply note that the most egregious error in
determing a distance arises from the difficulty in determing the
absolute magnitude of a star, $M_V$. This is not because of a problem
in knowing $M_V$ for the particular spectral type or luminosity class
of a star, but because of the problems inherent in defining the type
and class correctly in the first place. The error in $M_V$ is not only
a function of the type and class, but depends on the quality of the
stellar spectra used to define these parameters, and on the skill and
experience of the observer making the measurements.  Despite the
significant improvements in measuring various stellar parameters with
modern techniques, we believe that errors in stellar distances range
from $10-30$\%, and that even larger deviations are possible. A full
explanantion is given in Appendix~\ref{sect_absmags}.

The final topic of this paper is, of course, the global properties of
O~VI absorption in the disk of the Milky Way. Our motivation for the
study has already been introduced in the previous sections, but we
note the following. With the huge amounts of data available from
satellites that cover many different regions of the electromagnetic
spectrum, it is now possible to compare the results from
absorption-line surveys with results from other types of studies. In
this paper, we concentrate on X-ray emission data taken with the {\it
  R\"{o}ntgen Satellite\/} (\ROSAT ). Although X-ray emitting gas is
nominally at a higher temperature than we would expect for O~VI
absorbing gas, one can imagine a natural confluence of hot gas in
structures that might be detected in both O~VI absorption and X-ray
emission. In \S\ref{sect_bubbles} and \ref{sect_SNR}, we describe the
types of X-ray emission in \ROSAT\ maps seen towards our \FUSE\ stars.
In many cases, these maps indicate regions of hot gas along the line
of sight from bubbles of gas around stellar associations or supernova
remnants. These observations lead us to classify the morphology of
X-ray emitting patches (or absence thereof) around or near our target
stars.

Using these classifications, we show in \S\ref{sect_rho_rosat} that
the average line of sight density $N$(O~VI)/$d$ is somewhat higher for
sight lines which pass through X-ray enhanced regions, leading us to
conclude that a small fraction of $N$(O~VI) towards some stars does
indeed arise in circumstellar material or wind-blown bubbles.  This
information is important, because by selecting sight lines with no
clear X-ray emission, and by correcting O~VI column densities for
absorption by the Local Bubble using recently published data, we can
derive a less biased value of the mid-plane density of O~VI, $n_0$. We
discuss this value at some length in \S\ref{sect_density}, concluding
with our best determination of $n_0$ and its likely errors in
\S\ref{sect_finally}. In calculating these values, we quantify the
significance in the difference in $N$(O~VI) towards stars in the north
and south Galactic hemispheres originally found by \citet{savage03}
(\S\ref{sect_NS}). The difference has no effect on the value of $n_0$
we derive, but does produce two different values of the scale height
of the O~VI gas layer in the plane of the Milky Way.

In \S\ref{sect_Nwithde} we show that $N$(O~VI) is directly correlated
with the distance to a star in the plane of the disk. This correlation
can most easily be understood as the effect of longer sight lines
passing through more absorbing clouds.  This shows that O~VI
absorption cannot arise primarily from the circumstellar environment
of the observed stars, but must be a truly interstellar phenomenon.
The data also reveal a discontinuity in $N$(O~VI) with distance beyond
a few hundred pc, in that column densities appear to be too high for a
given distance. As we show, however, this arises from excess
contributions to the measured $N$(O~VI) from the Local Bubble.

However, the data also suggest that the hot O~VI-absorbing ISM is
neither a smoothly distributed homogeneous gas layer, nor a population
of randomly distributed clouds with similar properties. First, there
is no acceptable fit to the decrease in O~VI density with height above
the plane of the Galaxy unless an additional `error' is introduced, a
term which represents an intrinsic `clumpiness' in the interstellar
medium (\S\ref{sect_clump}). Second, this clumpiness does not decrease
with distance, as would be expected if a sight line simply passed
through more a greater number of generic clouds
(\S\ref{sect_Nwithde}). To explain this, we propose a picture of the
O~VI absorbing ISM that contains a mixture of cloud sizes, with small,
randomly distributed small clouds and sparsely distributed large
clouds. We also note an alternative explanation, that we may be seeing
the clustering of many small clouds.

There are other lines of evidence that suggest that the O~VI absorbing
`clouds' are actually one component in a complex, evolving, turbulent
ISM. The velocities of the bulk of the O~VI absorption do not follow
the general pattern of differential co-rotation in the disk of the
Galaxy (\S\ref{sect_vel_with_long}). Moreover, if we compare the
extremes of the O~VI absorption line profiles with the extremes of
other lower ionization species (O~I, C~II, C~III and Si~III) we find
that they match each other quite closely (\S\ref{sect_highVlow}). This
apparent `coupling' of at least some of the components in the
absorption line profiles (at least those with the highest peculiar
velocities along a sight line) suggests that the structures giving
rise to absorption from hot gas are multi-ionized.

A natural explanation for all these phenomena has been discussed
above. The numerical simulations of \citet[ and references\/
therein]{avillez05} vividly realize how the ISM may look when driven
by supernovae explosions. Their two dimensional density maps (for
example) in the plane of the Galaxy show how hot gas arises in bubbles
around supernovae, which is then sheared through turbulent diffusion,
destroying the bubble and stretching the hot absorbing gas into
filaments and vortices that dissipate with time.  In
\S\ref{sect_Nwithb} we confirm the relationship between $N$(O~VI) and
Doppler parameter found by other authors, and determine that the O~VI
lines are too wide to arise from differential Galactic rotation alone.
Clearly, the velocity fields that govern the O~VI components along a
sight line are much larger than can be reproduced by the simple
motions of the Galactic disk, and of course, expanding supernova
shells naturally explain these large velocities.  When considering an
ensemble of such regions along long sight lines, it seems plausible
that the correlation of $b$ with $N$(O~VI) could arise from the
existence of exceptionally large internal motions inside large clouds,
or that smaller clouds within clusters are moving more rapidly than
the gases within individual clouds.

\citet{avillez05} found good apparent agreement between their models
and the \FUSE\ WD data and the results from \Copernicus\ over a path
length of $\sim 1$~kpc.  Our results are also likely to be consistent
with these models, although a detailed comparison is beyond the scope
of this paper.  The data presented herein should provide sufficient
information to test these, and any other detailed simulations, over
larger scales.

\section{\emph{FUSE} OBSERVATIONS, DATA REDUCTION, AND ANALYSIS \label{sect_the_data}}

\subsection{Sample Selection \label{sect_select}}

The core of our sample of sight lines comprises stars observed under
the auspices of the \FUSE\ PI Team-time programs P102 and P122.  Stars
for these programs were originally selected based on a variety of
criteria: they were more than 1~kpc distant, in order to probe regions
of the Galaxy beyond those studied by \Copernicus ; they had low
values of reddening [$E(B-V)\:<\:0.3$] to minimize contamination by
H$_2$ lines; they lay within $\pm 10$ degrees of the Galactic plane to
confine the study to the Milky Way disk; they had spectral types of O9
or earlier, or O9 to B2 if their projected rotational velocities
($v\sin\:i$) were known to be $>\: 100$~\kms ; they had luminosity
classes of III to I for O9 spectral types (or earlier), or class V if
$v\sin\: i$ was high; their spectra were not known to show emission
line characteristics; and there existed ancillary data of the star
from previous UV satellites such as {\it IUE} or {\it ORFEUS-SPAS}, or
from ground-based optical observations.

In addition to the stars selected for the PI Team-time programs, we
added data from the \FUSE\ {\it Multimission Archive at the Space
  Telescope Science Institute\/} (MAST). To investigate a range of ISM
environments, we took data from PI Team programs P116 and P216
(``Molecular Hydrogen in Translucent Clouds''), P117 (``Hot Stars''),
and P101 (``The Properties of Hot Gas in the Milky Way and Magellanic
Clouds'').  In all cases, we selected only sight lines within $|b| <
10\degr$ of the Galactic plane, again, in order to focus primarily on
absorbing gas in the disk of the Milky Way. We only included data that
had signal-to-noise (S/N) ratios at least as high as those obtained
for stars in the core program and we rejected stars that had highly
irregular continua (see \S\ref{sect_cont}).

We also searched the \FUSE\ MAST archive to find stars from Guest
Investigator (GI) programs that might also provide suitable targets.
We adopted criteria very similar to those used for selecting targets
for the P102 and P122 programs; we did, however, exclude regions
inside the Vela supernova remnant region, at Galactic longitudes
$260\degr\: \leq\: l\: \leq\: 270\degr$, and the Carina Nebula,
$280\degr\:\leq\: l\: \leq\: 300\degr$. We also avoided any \FUSE\
programs whose science goals were obviously oriented towards studying
O~VI absorption along especially selected sight lines\footnote{As we
  shall see, our Team-time programs P102 and P122 did include stars
  toward both the Vela and Carina regions, and those data are
  presented in this paper; however, by the time we searched the \FUSE\
  MAST archive it had become clear that the absorption towards these
  regions is far more complex than that seen in the rest of the
  Galactic disk (see \S\ref{sect_rho_lb}) and requires a separate
  analysis. }.  We eventually collected all suitable data available
prior to April 2003. Our sample comprises 148 stars in total, whose
parameters are listed in Table~\ref{tab_journal}.  To introduce this
table, we note the following: column 1 is a unique identifier
(``unique ID'') assigned solely to help cross-reference stars in
different figures and tables in this paper.  Column 2 gives the name
of the star, while column 3, labelled `Prog ID' is the combination of
\FUSE\ program ID (the first three numerals) and a number assigned
within that program (the last two numerals) for a particular star.
Columns 4 and 5 list Galactic coordinates $l$ and $b$; for many of the
tables in this paper, the sight lines are sorted by Galactic
longitude, in order to map any variations in the physical
characteristics of the O~VI absorption along the plane of the Milky
Way. Table~\ref{tab_lookup} lists all the stars sorted by HD number
instead, and includes the unique ID number (as well as $l$ and $b$)
which can be cross-referenced with entries in Table~\ref{tab_journal}
and any other tables or figures. The remaining entries for
Table~\ref{tab_journal} are discussed in later sections below.

\subsection{\emph{FUSE} Observations}

Of the 148 stars in our sample, 111 were observed as part of PI-Team
programs P102 and P122.  A journal of observations is given in
Table~\ref{tab_lookup}, which includes the observation date and
exposure time. For some stars, more than one day passed between return
visits to a star, in which case we list the second (or third) date of
the exposure in column 6. There are also examples where a PI Team
program star was reobserved by a Guest Observer (`GO'; the first
letter of these programs run from 'A' to 'D' for Cycles 1$-$4); these
are also listed after the PI-Team program entry.  In a few cases, some
stars were observed with very large time intervals between satellite
visits.  These data are useful for determining how strongly our
results are influenced by possible time variations in the underlying
stellar spectra.  This in turn gives us an indication of how
accurately we are able to normalize a star's continuum around an O~VI
absorption line.  We discuss this topic more fully in
\S\ref{sect_epoch}.

For each star, the data obtained usually consisted of several
individual spectra (hereafter ``sub-exposures'') which needed to be
co-added.  The exposure time $T_{\rm{exp}}$ in column 7 represents the
total time actually spent on-target after the co-addition of useable
sub-exposures; so, if individual sub-exposures were discarded (due,
e.g. to loss of signal in a channel) the sub-exposure exposure time
was not included in column 7. In total, just over 754 ksec was
expended on programs P102 and P122. All stars were observed through
the 30$''$x30$''$ LWRS aperture, except HD~175754, HD~041161 and
HD~069106, which were observed with the 4$''$x20$''$ MDRS aperture.

Details of the \FUSE\ satellite and instrumentation are given by
\citet{moos02} and \citet{sahnow02}.  The wavelength regions of
interest for detecting O~VI are covered by four separate
mirror/grating and detector combinations, namely the LiF1A/2B and the
SiC1A/2B channels. However, we made measurements mainly using only
LiF1A data, because that channel provides the highest effective area,
the astigmatism correction point was designed to be at 1030~\AA\ in
the LiF channels, and the LiF1A data were initially the best
calibrated.  There are two exceptions: the HD~041161 and HD~060196
spectra came from the SiC1A and LiF2B channels, respectively, as no
Lif1A data were available.

\subsection{Sub-Exposure Corrections and Co-Addition \label{sect_subexposures}}

The raw data for all the stars in our survey were reduced using
version 2.0.5 of CalFUSE, which was the most advanced version
available during the data reduction.  Since
the stars observed in this study are bright, the S/N ratios of
individual sub-exposures were usually high enough for shifts between
sub-exposures (caused by thermal motions in the spectrograph) to be
seen. To determine possible shifts, we used the sub-exposure with the
highest S/N ratio near the \sixa\ line as the template spectrum
against which other sub-exposures were compared.  An initial shift of
a sub-exposure was determined by cross-correlating a 10~\AA\ region of
the data (centered at 1032~\AA) with the template.  A more refined
shift was then made visually, in order to determine a shift to within
$\simeq\: \pm 0.5$ pixels ($\simeq 1$~\kms ). Both flux and error
arrays were shifted using sinc interpolation if a shift of more than
$\pm\: 0.5$ pixels was measured.  Ideally, we would like to have moved
spectra only by integer numbers of pixels to properly preserve the
noise characteristics of a spectrum. However, shifts of less than a
pixel were clearly evident in data with good S/N, so fractional shifts
were required. (As we will discuss later, continuum uncertainties
dominate the errors in our data analysis, so noise errors are only of
secondary interest.)

With individual sub-exposures mapped to the wavelength scale of the
sub-exposure with the highest S/N, the wavelength and flux arrays of
all the data were rebinned by a factor of 3, in order to increase the
signal-to-noise of the spectra but still properly sample the data.
The error arrays were rebinned by the same amount and reduced by a
factor of $\sqrt{3}$.  The initial dispersion of the data (from the
CalFUSE calibration) was $\simeq 0.0067$~\AA~pix$^{-1}$,
\footnote{Note that this dispersion was less than the dispersion of
  0.013~\AA~pix$^{-1}$ used by later versions ($>$ 3.0) of CalFUSE.
  This new number samples approximately one-quater of a \FUSE\
  spectral resolution element \citep{vandixon07}.}  but the resolution
of the LiF1A channel is between 15 and 20~\kms\ ($0.052-0.069$~\AA\ at
1030~\AA ) FWHM.  After rebinning, the final dispersion of the data
was $0.020$~\AA~pix$^{-1}$, close to that required for optimally
sampling the data.  Individual sub-exposures were
then co-added, with the flux weighted by the inverse of the variance formed
from error arrays which were usually smoothed by 7 pixels.

\subsection{Wavelength-scale Accuracy and Zero-Point Correction \label{sect_wavecal}}

An important quantity to measure in the \FUSE\ spectra is the velocity
of the detected \sixa\ lines, the precision of which depends in large
part on the accuracy of the wavelength calibration. Unfortunately,
\FUSE\ has no on-board facilities to provide such calibration in situ.
Instead, CalFUSE assigns a wavelength scale to the data, accounting
for distortions in the dispersion and spatial directions, and
correcting for instrumental distortions.

Even with these corrections, errors arise in the zero-point of the
wavelength calibrations, because, e.g., the source is not centered in
the aperture, or, for channels other than LiF1, from misalignment of
the primary mirrors.  The LiF1 channel has the most stable wavelength
calibration, since for the observations in this paper, 
the Fine Error Sensor used light from this channel
to acquire an object.  However, since we found small offsets between
sub-exposures taken during the observations of a single star, we
expected to find small errors in the absolute wavelength scale.
(Indeed, although we used the sub-exposure with the highest S/N as a
template against which to shift all the other sub-exposures, there was
no reason to believe that the zero-point for this spectrum would be
any more reliable than that of any other exposure.)

Fortunately, CalFUSE adequately corrects for extreme non-linearities
when mapping pixels to wavelength, particularly over the region of
interest around the \sixa\ line.  The measured velocities of the two
molecular hydrogen (H$_2$) lines which flank the \sixa\ line always
agreed with each other (to within a 1$\sigma$ dispersion of
$\simeq\:2-3$~\kms), which validates this assumption. We therefore
needed only to find and apply a zero-point offset to the spectrum to
correct the wavelength scale.

To derive a zero-point for a subset of the \FUSE\ stars, we compared
the velocities of H$_2$ lines near the \sixa\ line, with that of a
Cl~I~$\lambda 1347$ line in Archival {\it HST} data of the same stars,
where the wavelength calibration is more precise than
that for \FUSE . In all, we were able to correct 55 spectra out of the
total 148 spectra using this method. A comparison between the CalFUSE
wavelength scale and the corrected value showed that for the remaining
spectra, a shift of 10~\kms\ would correct the zero-point to within
$\pm\:5$~\kms\ of the true value. Full details of this analysis are
given in Appendix~\ref{sect_cl1}.

\subsection{The O~VI Wavelength Region and Contaminating Lines \label{sect_contaminants}}

An important step in the processing of the \FUSE\ data was to
normalize the spectrum of each star. Before describing this process,
however, it is helpful to first discuss the identification of
contaminating lines in the wavelength region around the \sixa\
absorption line itself. As we show in \S\ref{sect_cont}, understanding
the expected strengths of these lines was important for correctly
determining the true shape of the continuum.

\begin{figure}[t!]
\hspace*{-0.5cm}\includegraphics[width=1.25\hsize,angle=0]{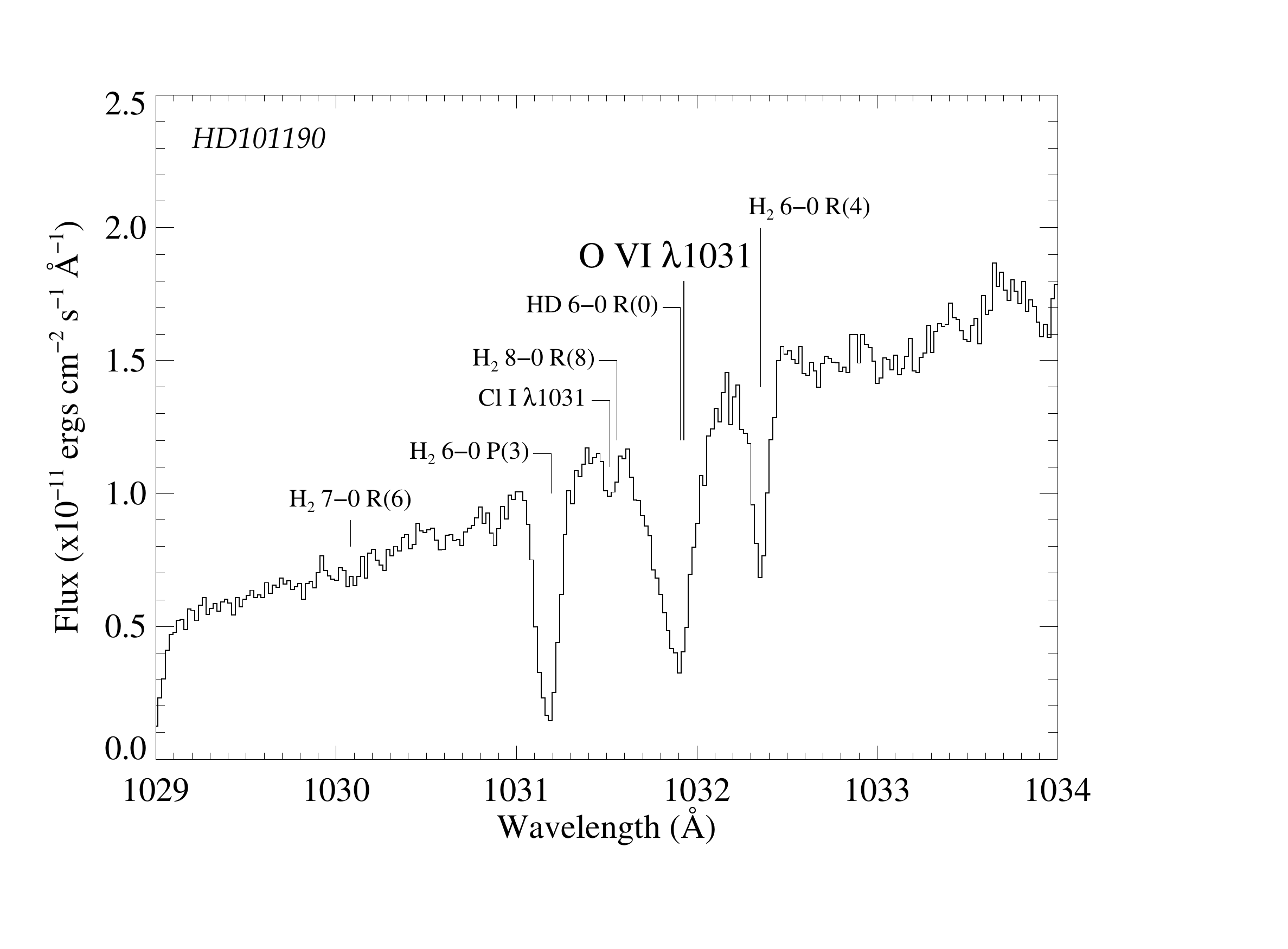}
\caption{\label{fig_where_r_lines}
The spectrum of HD~101190, showing  wavelengths of
possible contaminating absorption lines around and within the \sixa\
line.}
\end{figure}

The positions of possible contaminating lines are shown in
Figure~\ref{fig_where_r_lines}, where we plot the the un-normalized
spectrum of HD~101190 over the wavelength range 1029$-$1034~\AA . The
wavelengths of the lines shown in the figure are given in
Table~\ref{tab_lines}. This table also lists other lines that were
used in our study (often at wavelengths far from the O~VI wavelength
region discussed in this section) and which will be introduced in
later sections of this paper.

The most serious contaminant is the Lyman HD $6-0$ R(0) line at
1031.909~\AA, which is coincident in wavelength with the O~VI~$\lambda
1031.927$ line if the velocity differences are only moderate. For
strong O~VI lines, the contamination is largely negligible, but for
weaker O~VI lines, the contribution of the HD line can be significant.
We discuss in \S\ref{sect_hdremoval} how we removed the contribution
from this HD line.

Cl~I~$\lambda 1031$ is expected to arise at 1031.507~\AA. This line
was often seen in the \FUSE\ spectra, always narrow and weak. We
attempted to use the Cl~I~$\lambda 1004.678$ line to estimate the Cl~I
column density and thereby model the line at 1031~\AA, since the
1004~\AA\ line is relatively free of other contaminating lines. Our
attempts were unsuccessful, for two principal reasons.  First,
defining the rapidly changing continuum around both the 1031 and
1004~\AA\ lines proved quite difficult, a problem that made the depths
of these intrinsically weak lines highly uncertain.  Second, even in
the few (rare) cases where the continua were well behaved, we could
not fit both lines consistently with a single cloud model of column
density and Doppler parameter, leading us to suspect that problems may
exist with published oscillator strengths. Recent work by
\citet{sonnentrucker06} confirms these suspicions, at least for the
1004~\AA\ line.

Hence, in our analysis of the \sixa\ line, we note where Cl~I~$\lambda
1031$ is expected, but no attempt has been made to model it.
Fortunately, the position of the Cl~I line is usually sufficiently
blueward of the O~VI feature that its contamination is unimportant.
There are sight lines where the O~VI feature can become blended with
the Cl~I line; however, as we discuss more fully in
\S\ref{sect_voigt}, the fitting of Voigt profiles to the O~VI
line is unaffected by the presence of Cl~I, and
the contribution to the computed O~VI column density is negligible.

Close to the Cl~I line is the H$_2$ 8$-$0 R(8) line at 1031.557~\AA .
To predict if we should see this line, we looked at the 7$-$0, 6$-$0,
and 4$-$0 R(8) lines at 1042.745, 1054.520 and 1079.932~\AA,
respectively.  These lines are relatively free of contamination by
other interstellar species, and have similar oscillator strengths (see
Table~\ref{tab_lines}) compared to the 8$-$0 line, thereby providing
suitable comparisons. We found absorption from these three R(8)
transitions only along the line of sight to HDE~303308, which lies in
the Carina Nebula.  For those data we modelled the H$_2$ column
density for the 8$-$0 line from the other three transitions and
removed the profile from the spectrum. Weak R(8) lines may also be
present in the spectrum of HD~187282, but there is significant
blending with other H$_2$ features.  Fortunately, the velocity of the
8$-$0 line is well enough removed from the O~VI that it poses no
threat of contamination to the O~VI absorption.

As shown in Figure~\ref{fig_where_r_lines}, \sixa\ is flanked by the
two strong H$_2$ lines at 1031.192 and 1032.354~\AA, the 6$-$0 P(3)
and 6$-$0 R(4) Lyman lines.  These often blend with the O~VI feature,
but can be modeled and removed quite accurately, as we discuss in the
following section.

\subsection{Continuum Fitting \label{sect_cont}}

Even with a complete inventory of the lines expected near the \sixa\
line, normalization of the stellar spectra proved challenging. We
adopted the method of fitting continua as described by
\citet{sembach92}, whereby continuum levels are defined by
least-square fits of Legendre polynomials to intensities either side
of a line.  Unfortunately, for many stars, we found that the stellar
continuum varied in shape over widths comparable to that of the O~VI
features.

No interstellar absorption lines are expected between the H$_2$ 7$-$0
R(6) line at 1030.08~\AA\ and the first of the H$_2$ lines flanking
the O~VI feature, the 6$-$0 P(3) line at 1031.192~\AA. This gap of
1.1~\AA\ should provide $\sim 50$ (rebinned) pixels from which the
blue side of the continuum can be well defined. Except for O~VI with
high negative velocities, there was indeed a small amount of continuum
available between the P(3) feature and the O~VI line in our spectra,
again helping to tie down the continuum fit on the blue side of the
O~VI feature. There was, however, often contamination from the
Cl~I~$\lambda 1031$ line, potentially leading us to underestimate
where the flux should lie between the P(3) line and the O~VI.
Fortunately, the Cl~I line was always weak and narrow, and its effect
on causing continuum misplacement was small.

As we shall see, the \sixa\ lines themselves are wide, and there was
often good reason to believe that the continuum was changing shape at
1032~\AA\ in many of our spectra.  Further, at the red edge of the
O~VI line, absorption from the second flanking H$_2$ feature --- the
6$-$0 R(4) line --- is expected to set in. In many cases, we found
that the O~VI and the R(4) line were clearly beginning to blend
together. By the time the continuum had recovered redward of the H$_2$
absorption, it was often a very different flux than that predicted
simply by extrapolating from the (well fitted) continuum blueward of
the P(3) line.

We attempted to find some consistency checks for the continua that we
adopted.  As a first step, we compared the P(3) and R(4) lines with
their profiles from modeling other H$_2$ lines in the \FUSE\ spectrum.
For the R(4) line, we used the 5$-$0 and 4$-$0 R(4) lines at 1044.542
and 1057.376~\AA, respectively.  These lines are relatively free of
contaminating features, and have similar oscillator strengths to the
6$-$0 lines. For the P(3) line, we used only the 4$-$0 P(3) line at
1056.472~\AA. In both cases, we fitted Voigt profiles (convolved with
the instrumental line spread function --- see \S\ref{sect_voigt}) to
the data to derive column densities and Doppler parameters for the
R(4) and P(3) transitions, calculated the expected profiles for the
6$-$0 P(3) and R(4) lines, and compared those with the observed lines.
\footnote{In fact, the range of oscillator strengths for all the R(4)
  lines available in the \FUSE\ channels are too small to allow a
  unique determination of the R(4) H$_2$ column densities, given the
  \FUSE\ resolution.  We base our use of the 6$-$0 line on the fact
  that we can correctly recreate the equivalent width of the line
  given the equivalent width of the other R(4) lines.}

In Appendix~\ref{sect_explain_spectra} we show all the spectra in our
survey (see Fig.~\ref{fig_spectra}; a full description of this Figure is
given in Appendix~\ref{sect_explain_spectra}).  \notetoeditor{We would like to keep
  all the material in this appendix at the very {\it end} of the
  paper, which would make it Appendix~\ref{sect_explain_spectra}, and
  not the one following Appendix~\ref{sect_cl1}. Similarly,
  Fig.~\ref{fig_spectra} is out of order when mentioned here, but it
  should still be left as part of Appendix~\ref{sect_explain_spectra}}
The top panel (the first of three) for each star shows the adopted
continuum, along with the $\pm 1\sigma$ error `envelopes' which
express the positive and negative deviations possible in the
polynomial coefficients used to define the continuum.  (We will refer
to these as the `upper' and `lower' continuum fits throughout this
paper.)  The middle panel shows the {\it predicted} profile fits to
the H$_2$ lines, blended with O~VI line fits (\S\ref{sect_voigt}), for
the normalized spectra.  Given the inherent uncertainties in producing
the H$_2$ models from the other H$_2$ lines in the FUSE spectra, these
fits appear to predict the observed features quite well.  We notice a
tendency for the theoretical profiles of the flanking R(4) and P(3)
lines to be slightly stronger than the data in many cases.  This
probably arises from an incomplete knowledge of the \FUSE\ Line Spread
Function (LSF) at the wavelengths used for the H$_2$ models as well as
the lines near O~VI, since the LSF is largely uncharacterized for the
spectrograph and is known to change with wavelength.  We should say,
however, that these models did not provide particularly rigorous
constraints on the continuum fits. Since the H$_2$ lines were always
quite strong, even fairly large continuum deviations around the lines
produced very small differences in the depths of the line profiles.

\begin{figure}[t!]
\hspace*{-0.5cm}\includegraphics[width=1.05\hsize]{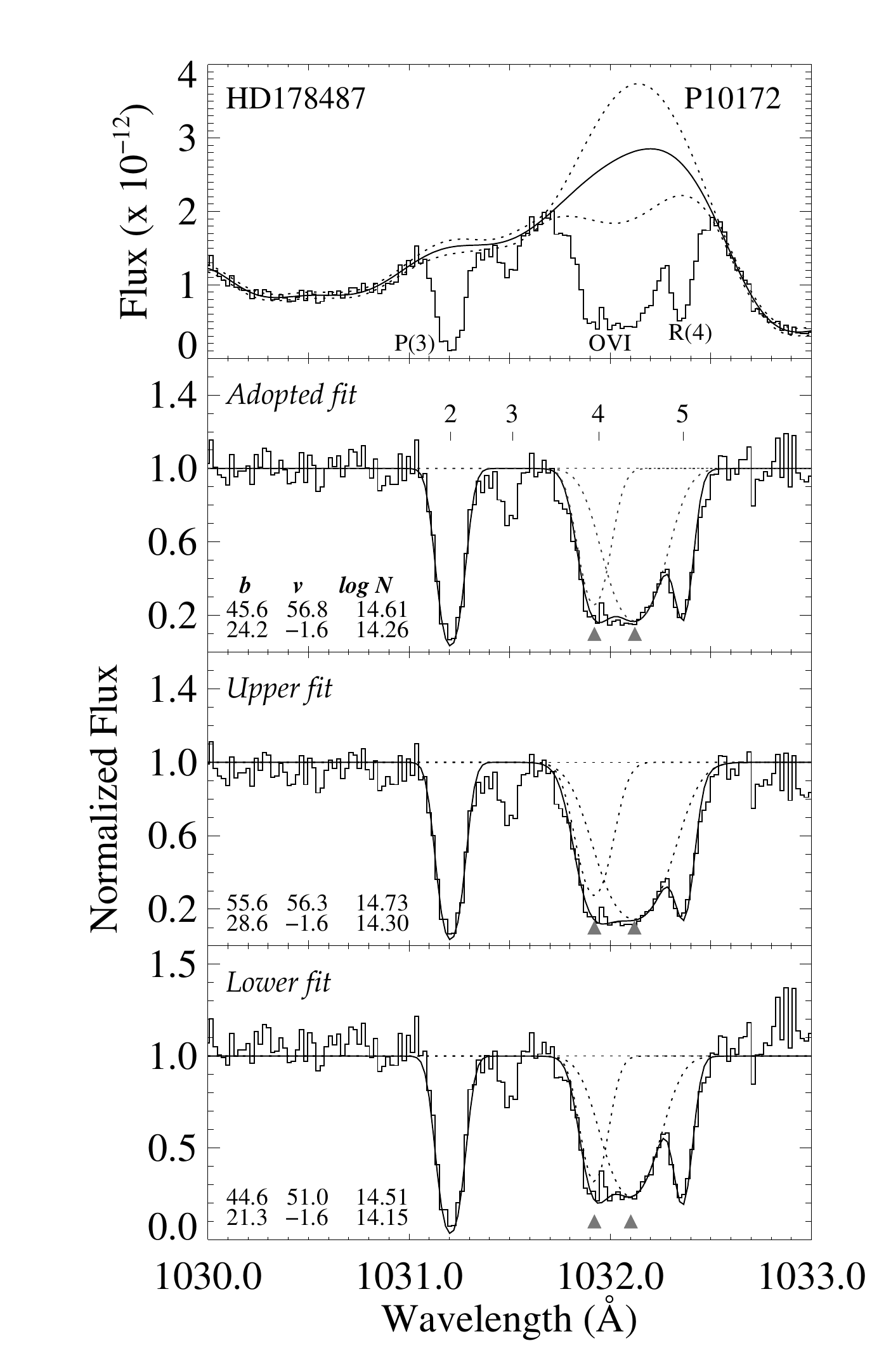}
\caption{\label{fig_wildcont} An example of the worst case for fitting
  a continuum to a star when the O~VI and H$_2$ 6$-$0 R(4) lines are
  strong and placed in a wavelength region where the continuum reaches
  a local maximum. The solid curve in the top panel shows the adopted
  fit to the observed flux, while the dotted curves mark two extremes
  arising from the $\pm 1\sigma$ errors on the Legendre polynomial
  coefficients. All three are plausible fits to the data. The second
  panel shows the profile fits for the O~VI and the molecular lines,
  given the adopted fit. Two components are needed to fit the O~VI
  profile in this example, and the centroid of each is shown with a
  gray triangle. The lower panels show the fits for the upper and
  lower continua.  However, because the absorption is strong, the
  errors in the total O~VI column density are only 0.1 dex. (See
  Appendix~\ref{sect_explain_spectra} for full details of how this
  plot is marked.)}
\end{figure}

The most difficult continua to fit occurred when the O~VI absorption
arose in a region where the stellar flux is reaching some local
maximum. For example, the top panel of Figure~\ref{fig_wildcont} shows
the flux of HD~178487, which has one of the most difficult continua to
characterize of all the stars in our sample. The solid line in the
figure shows a plausible fit, but it is clear that the dotted lines
could also represent the true underlying continuum.  As we describe
below, we can calculate column densities from the data normalized by
our adopted fit (second panel of Figure~\ref{fig_wildcont}) but we can
also re-calculate the values assuming that the upper or lower fits are
correct (third and fourth panels).  For HD~178487, the O~VI lines are
strong, and even a continuum as uncertain as the one indicated in the
figure introduces errors of only $\simeq\:\pm\: 0.1$ dex in the total
O~VI column density. Fortunately, for all the stars with these type of
continua, the \sixa\ line was indeed strong, and we were able to
characterize the likely range of column densities through the errors
adopted for the continuum fits.

\begin{figure}[t!]
\hspace*{-0.5cm}\includegraphics[width=\hsize]{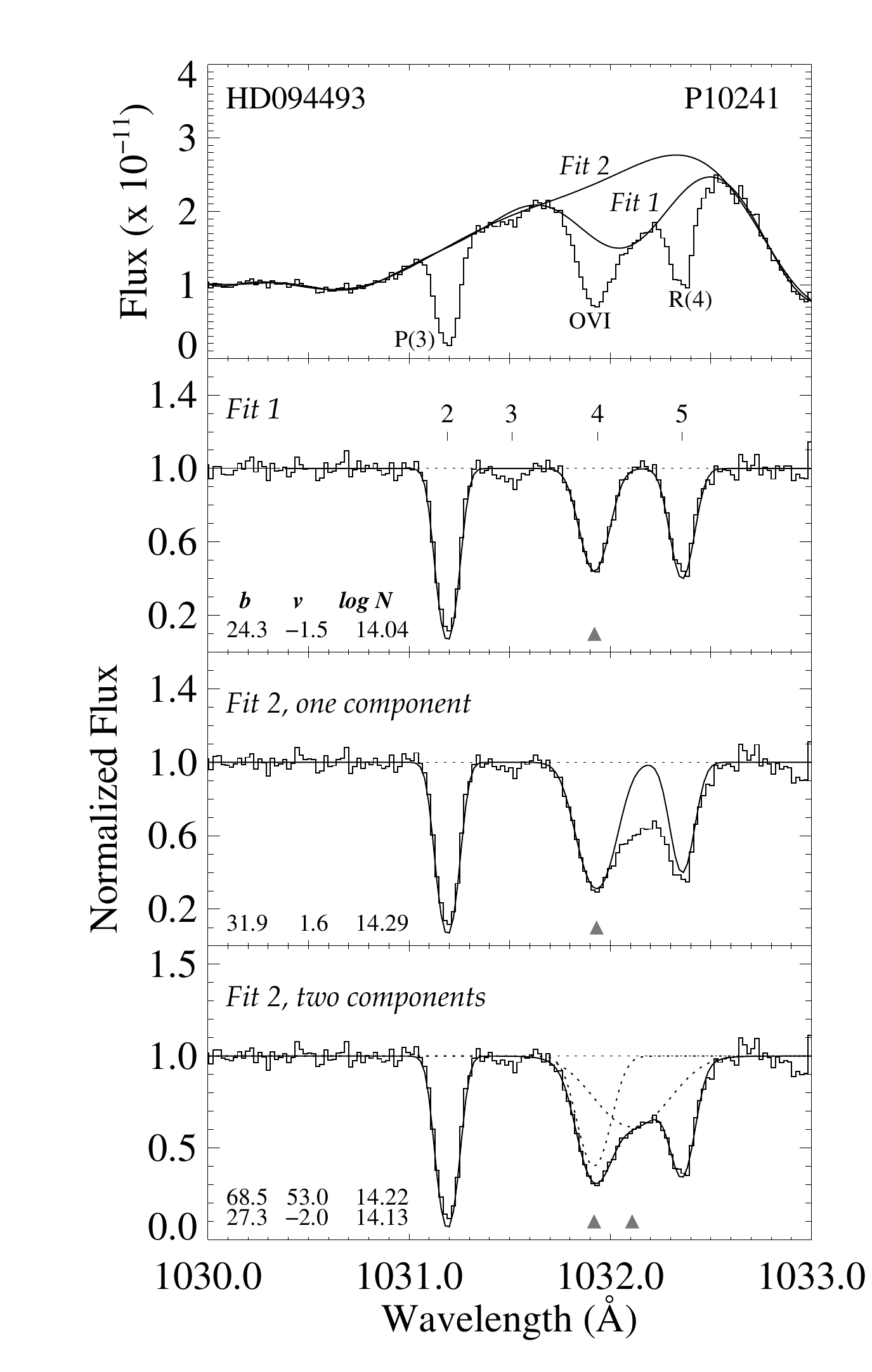}
\caption{\label{fig_diffcont}
A small subset of stars in our survey have  continua whose shape can
be fit with one of two possible choices. Here we show one such star, HD~094493. 
In the top panel, Fit-2 is based on a simple extrapolation of the continuum
from the blue side of the spectrum, while Fit-1 assumes that the flux between
the O~VI and R(4) line is real continuum. Adopting Fit-1 means that the O~VI
line is easily fit with a single component (panel 2; the centroid of
the component is indicated by a gray triangle). Using Fit-2 results in additional 
O~VI which cannot be fit with a single component (panel
3) but requires a second, broad, higher-velocity component (panel 4;
the centroids of both components are indicated with gray triangles).
Many of the stars in our sample have 
this ambiguous continuum shape, but
for reasons discussed in \S\ref{sect_cont}, we adopt continua with shapes
similar to those of 'Fit-1' above. (See Appendix~\ref{sect_explain_spectra} 
for full details of how this plot is marked.) }
\end{figure}

A far more common problem in fitting continua can be seen for one
particular star, HD~094493, in Figure~\ref{fig_diffcont}. There are
two possible continuum fits: one which continues upward from an
extrapolation of the fit blueward of the H$_2$ P(3) line, before
having to turnover beyond the R(4) (labelled `Fit-2' in the Figure);
and one which quickly drops downwards to meet the small amount of
continuum flux left between the O~VI line and the H$_2$ feature
(`Fit-1').

The difference depends on whether Fit-1 is the correct continuum,
despite the unsettling sudden deviation around the \sixa\ line, or if
the stellar flux is instead depressed by a high velocity extension to
the main O~VI absorption.  The second panel of
Figure~\ref{fig_diffcont} shows the data normalized by Fit-1, and the
resulting O~VI absorption and H$_2$ R(4) are easily reproduced with a
single component fit.  If Fit-2 is adopted instead, then a single
component fit is no longer satisfactory.  The third panel of
Figure~\ref{fig_diffcont} shows the data normalized by Fit-2, with a
single component fit to the O~VI line (along with the predicted line
profile for the H$_2$ R(4) line). There is clearly excess absorption
between the O~VI and H$_2$ line.  The blue side of the H$_2$ line is
also poorly matched.  To account for this absorption, a second
component is required at higher velocity.  Such a fit is shown in the
bottom panel of Figure~\ref{fig_diffcont}.

This type of problem occurs for 26 stars in our sample.  Perhaps
significantly, the difficulties are confined to stars of a particular
stellar type. Of the 26 problem stars, 21 (81~\%) are of type O9 to B1
(inclusive). (In contrast, for the remaining stars, only 52\% are of type
O9 to B1.)  In many cases, it seems that O~VI absorption arises in the
blue wing of a narrow O~VI emission line feature which predominates in
these types of stars.  There is no evidence of excess high (positive)
velocity O~VI absorption in the majority of earlier spectral types.
Nor are these problem stars confined to a particular region of the
sky, where, e.g.  Galactic rotation (or some other coherent
kinematical process) might be expected to produce additional O~VI
absorption.  We also note that when we compare the expected profile of
the R(4) line with the data normalized by Fit-1 type continua (marked
as line '5' in the second panel of Figure~\ref{fig_diffcont}), we find
that the predicted profile agrees well with the data every time.

For these reasons, we adopt continuum levels of the type shown by
Fit-1 for the 26 problem stars.  Can we rule
out the possibility that we are excluding real additional, high
velocity O~VI components? If such components exist, then the absorbing
clouds must have a rather unique set of properties: they must always
be of lower column density than the lower velocity component; they
must always be about twice as broad; and they must always be $\sim
50$~\kms\ redward of the lower velocity component. Such a fixed set of
characteristics from sight lines scattered across the plane of the
Milky Way seems contrived, and we believe that continua fits similar
to Fit-2 in Figure~\ref{fig_diffcont} are simply incorrect. In many
cases, the difficult continua are likely to be the
result of irregularly shaped P-Cygni profiles which arise from the
winds around the stars themselves. 

Table~\ref{tab_cols} details the results of our measurements of O~VI
lines towards all the stars in our survey except those for which we
could not detect O~VI absorption (see \S\ref{sect_colsAOD} for a
discussion of these stars, which are listed in
Table~\ref{tab_limits}).  Of particular relevance to the problem of
continuum fitting is the information given in column 15 (labelled
`Cont'), a flag that simply warns of possible problems in the
continuum fits.  For the stars discussed above, where we could have
higher continua and a second, broad, positive velocity O~VI component,
we assign a value of `2' for the continuum flag. There are also a few
examples where the continuum could actually be drawn in such a way
that additional O~VI components might exist both redward {\it and}
blueward of the O~VI line. These are assigned a value of `1'. We
consider it unlikely that these stars really have continua so very
different from those we chose for them, but given the complex nature
of stellar continua, we cannot rule out the possibility. In both
cases, it is important to note that values for the continuum flag
greater than zero are only assigned if a different observer could have
adopted a continuum different from the one we adopted, by an amount
which would {\it not} be covered by the continuum error envelopes
introduced above.

\subsection{Removal of the HD 6$-$0 R(0) Line \label{sect_hdremoval}}

As indicated in \S\ref{sect_contaminants},
Figure~\ref{fig_where_r_lines}, and Table~\ref{tab_lines}, O~VI
absorption occurs at nearly the same wavelength as the HD 6$-$0 R(0)
line at 1031.909~\AA. The line is a serious contaminant for weaker
O~VI lines, and must be removed before the properties of an O~VI line
can be measured.

To remove the HD line, we first constructed a profile of the 6$-$0
absorption line based on a fit (see \S\ref{sect_voigt}) to another HD
line, the 7$-$0 R(0) feature at 1021.453~\AA .  This line has a
similar oscillator strength to the $6-0$ line (see
Table~\ref{tab_lines}), and is usually well isolated from other
interstellar features. The resulting profile created for the $6-0$
line should therefore be, at the very least, an accurate reflection of
the amount of equivalent width needed to be removed from the O~VI
profile.

We also checked to see if our fits to the 7$-$0 line were consistent
with the 4$-$0 HD line at 1054.286~\AA , which also has an $f$-value
close to that of the 6$-$0 line and is relatively free of
contaminating interstellar lines.  We often found that although the
equivalent width of the 4$-$0 line (as predicted from the 7$-$0 line)
had roughly the correct equivalent width, the profile of the 4$-$0
feature sometimes departed from a single component Voigt profile.  We
did not therefore use the 4$-$0 line to constrain the HD model.

Having derived a plausible model for the HD absorption, we created a
synthetic HD 6$-$0 line profile and removed it from the O~VI
absorption.  The results of this removal can be seen in the top panels
of Figure~\ref{fig_spectra}, \notetoeditor{Please see previous
  note-to-editor regarding the position of
  Appendix~\ref{sect_explain_spectra} and Fig.~\ref{fig_spectra}.}
where the difference in the flux before and after the subtraction is
shown as a gray shaded region.  Towards several stars (e.g. HD~210839,
HD~005005A, HD~012323, etc.) there is no apparent O~VI absorption
after the subtraction of the HD line.  Significantly, for these cases,
our HD subtraction never leads to the production of extra flux {\it
  above} the predicted continuum (i.e., an `emission' feature), at
least beyond that expected from Poisson fluctuations.  It seems likely
therefore that our HD models are sufficiently accurate to properly
remove the HD line.

\subsection{Measurement of O~VI Absorption Lines \label{sect_no1037}}

\begin{figure}[t!]
\hspace*{-1.5cm}\includegraphics[width=7cm,angle=90]{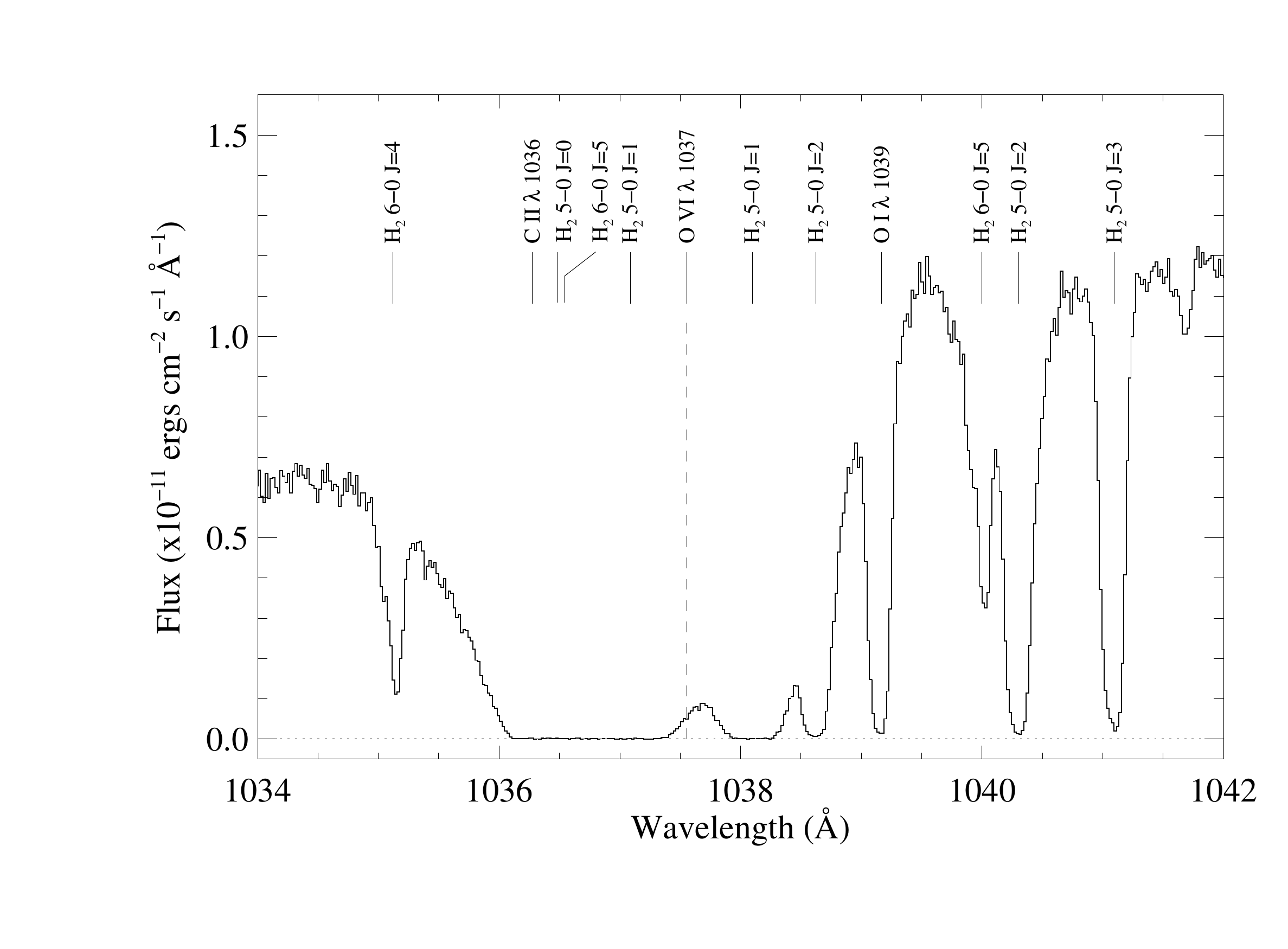}
\caption{\label{fig_bad1037}
Example of a stellar spectrum around the expected position of the
O~VI~$\lambda 1037$ line. The star is HD~67659, and shows the typical
strengths of the H$_2$ features
which make it impossible to measure the  1037 line.}
\end{figure}

To measure the column density of the O~VI absorbing clouds, we would,
ideally, like to use both members of the O~VI doublet.  Both lines
were recorded for most of the stars observed with \Copernicus , since
the H$_2$ absorption along the sight lines was usually weak. For the
\FUSE\ stars, however, H$_2$ absorption was always too strong for us
to use the 1037.6~\AA\ line. An example is shown in
Figure~\ref{fig_bad1037}, which shows the region around 1037~\AA\ for
the spectrum of HD~167659. These data are quite typical for the stars
in our survey .

The O~VI~$\lambda 1037$ could sometimes be seen in the wing of the
5$-$0 R(1) line when the H$_2$ column density was low.  In principle,
it is possible to fit $J\:=\:1$ lines in other parts of a spectrum,
then calculate and remove the H$_2$ line profiles around 1037~\AA\ and
thereby recover the O~VI~$\lambda 1037$ line. We performed such an
exercise for several sight lines where the H$_2$ column density was
low, and the H$_2$ lines weak. The results were disappointing.
Although reasonable fits could be made for the set of H$_2$ lines at,
e.g, 1078~\AA\ and 1063~\AA, it was difficult to reproduce the H$_2$
complex at 1037~\AA\ with sufficient accuracy (particularly without
knowing the continuum a priori) to recover the O~VI~$\lambda 1037$
line with anywhere near the precision needed to make the resulting
absorption line reliable enough to validate our measurements of the
1032~\AA\ line. Indeed, lines of sight with a low H$_2$ column density
along the sight line tended to be at relatively short distances; the
O~VI was itself relatively weak in these cases, making an accurate
determination of the 1037~\AA\ line even more difficult.

Consequently, all the analysis of O~VI in this paper uses the
O~VI~$\lambda 1032$ line alone.  In the following sections we
summarize the two methods used to analyze the \sixa\ lines toward our
sample stars: the fitting of theoretical (Voigt) line profiles, and
the Apparent Optical Depth (AOD) method.

\subsubsection{Column Densities: Profile Fitting \label{sect_voigt}}

As described above, a significant impediment to measuring O~VI column
densities towards some stars arose when the line was blended with
contaminating Cl~I~$\lambda 1031$ and H$_2$ lines.  The most
straightforward way to circumvent this problem was to fit theoretical
Voigt profiles to the regions of an O~VI line which were clearly not
blended with a contaminant.

This method has been used by researchers for many years. Details of
our specific line-fitting procedures are given in \citet{bowen_95}.
For the \FUSE\ spectra, we generated (one or more) theoretical Voigt
line profiles from an initial guess of the Doppler parameter $b$, line
velocity $v$, and column density $N$. These were convolved with the
instrumental LSF.  A definitive LSF is not available for the \FUSE\
channels, so we adopted a Gaussian of width 20~\kms\ FWHM.
Fortunately, the widths of most O~VI lines are wider than this, so as
long as the shape and width of the true LSF is close to the adopted
value, the errors introduced are negligible. 

For example, using an LSF with a width of 20~\kms\ for the data of
HD~134411 (\#131), we find that the line has a Doppler parameter of
20.0~\kms\ and column density $3.63\times 10^{13}$~\pcm . This line is
representative of lines with the smallest Doppler parameters in our
survey (see \S\ref{sect_bees}) and would be most affected by changes
in the LSF. In fact, if we had instead used an LSF of width 15~\kms,
we would have measured a Doppler parameter of 21.5~\kms ---an increase
of 7\% --- and a column density of $3.59\times10^{13}$~\pcm --- a
decrease of 1\%. For O~VI lines which are wider than that seen towards
HD~134411, the relative changes in column density and Doppler
paramater are even smaller.

Our fitting procedure allowed the O~VI $N$, $b$ and $v$ to vary until
a minimum was reached in the $\chi^2$ fit between profile and data
using the {\tt POWELL} minimization routine \citep{powell}.  Data
points lying within the wavelength region predicted for the Cl~I line
were excluded from the fit.  The O~VI profile was often blended with
the {\it predicted} profiles of the two flanking H$_2$ lines ---
predicted in the sense that $N$ and $b$ of the H$_2$ lines were fixed
at the values derived from fitting the relevant lines in other parts
of the spectrum (see \S\ref{sect_cont}), although $v$ was allowed to
vary for the H$_2$ lines in order to take into account any small
non-linearities in the wavelength scale of the data.

For each star, we first tried to fit only a single O~VI component to
the absorption. Additional components were added only if single
components were clearly inadequate.  It is important to note that this
does {\it not} presuppose that absorption arises from a single cloud.
Assuming such a model is likely to be erroneous, given the possible
explanations for the origin of O~VI absorption discussed in
\S\ref{sect_intro} and the results discussed later in this paper.
However, in most cases the O~VI line profiles appeared quite simple
and shaped like a single component.

Line fits were made three times for each star, and the results are
given in Table~\ref{tab_cols}: once for the data normalized by the
best continuum fit (giving column densities $N$ and Doppler parameters
$b$ --- columns 10 and 5, respectively), and twice more for the upper
($N_u$, $b_u$ --- columns 11 and 6, respectively) and lower ($N_l$,
$b_l$ --- column 9 and 4, respectively) error envelopes discussed in
\S\ref{sect_cont}.  For all sight lines, changes in $v$ for each of
the three continua fits are negligible ($<\:0.1$~\kms\ in most cases).
This procedure quantifies how the fit to the O~VI changes from errors
in the continuum, but does not address errors arising from Poisson
statistics. To estimate these values, we performed a Monte-Carlo
simulation \citep{bowen_95}; for each sight line, we used the best fit
values $N$, $b$ and $v$ to generate a theoretical line profile and
added the amount of Poisson noise given by the \FUSE\ error arrays
(wherein each pixel is assigned a deviation $\sigma_i$). This new
`synthetic' spectrum was then refitted, and new values of $N$, $b$ and
$v$ produced.  This procedure was repeated 300 times, resulting in 300
different values of $N$, $b$ and $v$. In most cases, the distributions
of these values were Gaussian in shape and the values of $\sigma$ for
each were taken to represent the errors from noise, $\sigma(N)$,
$\sigma(b)$ and $\sigma(v)$.  The values for the first of these two
quantities are given in Table~\ref{tab_cols} (columns 12 and 7,
respectively). We do not list values of $\sigma(v)$ since, along
nearly all sight lines, they are only $\sim\:1$~\kms, less than the
error in the wavelength scale.

In principle, these Monte-Carlo runs should have been repeated twice
more for each star using the best-fit profiles derived from fits to
the data normalized by the upper and lower continuum errors (i.e.
[$N_l$, $b_l$] and [$N_u$, $b_u$]).  In practice, however, since the
values of $\sigma_i$ at each pixel were the same, and the three
profiles for each continuum fit were quite similar for most stars, the
differences in $\sigma(N)$, $\sigma(b)$ and $\sigma(v)$ were
negligible.

For the O~VI lines, which are much wider than the LSF, $\sigma(b)$ and
$\sigma(N)$ are uncorrelated, and can be treated independently.  The
total errors in the Doppler parameters, $\sigma(b)_T$ are given by two
values [$b_u - b,b-b_l$] added in quadrature to $\sigma(b)$. These
values are given in column~8 of Table~\ref{tab_cols}.  In a few cases,
the positive and negative values of $\sigma(b)_T$ in column 8 appear
quite asymmetric. This usually occurred if the upper and lower
continuum fits for a star were very different in shape from each
other.  The final errors in $N$, $\sigma(N)_T$, are formed in the same
way, since the two sources of error in $N$ are independent of each
other, with [$N_u - N,N-N_l$] added in quadrature to $\sigma(N)$. 
Column~13 of Table~\ref{tab_cols} shows these errors in the form
$N$[$+\sigma(N)_T$,$-\sigma(N)_T$].

Towards some stars, more than a single component was needed to fit the
O~VI line. For convenience, column 13 of Table~\ref{tab_cols} lists
the sum of the column densities, $N_{\rm{TOTAL}}$, along with their
errors. In the case of a multicomponent fit, the final errors in
$N_{\rm{TOTAL}}$ are formed in a similar way to those of single
component fits.  Since the continuum errors affect all components in
the same way, the error in $N_{\rm{TOTAL}}$ from the continuum fit is
[($\sum_{i} N_{i,u} - \sum_{i} N_i$), ($\sum_{i} N_i - \sum_{i}
N_{i,l}$)], where $i$ is the component number. For the noise error,
the column densities for all the components are added for each
Monte-Carlo run, and $\sigma$ is calculated from the distribution of
300 values of $\sum_{i} (N_i)$. Again, the continuum error and noise
error are added in quadrature to give the errors listed in column 13
of Table~\ref{tab_cols}.

Some stars from our survey are not listed in Table~\ref{tab_cols}.
They have no appreciable O~VI absorption that can be fitted with a
theoretical line profile. The limits to their column densities are
tabulated in Table~\ref{tab_limits} and discussed below.

\subsubsection{Column Densities: Apparent Optical Depth (AOD) and Upper Limits\label{sect_colsAOD}}

Clearly, the Voigt profile fitting method described above is only
applicable when an absorption line is clearly detected. In some of our
spectra, there was good evidence that an O~VI line was not detected.
In order to rigorously define either the presence of very weak O~VI
lines, or meaningful upper limits, we used the AOD method of analysis,
which yields a straightforward formal measurement of $N$(O~VI) and its
associated error. A desire to accurately characterize marginal
detections, or the lack of any detections, with a well-posed
quantitative calculation, was the motivation for carrying out AOD
measurements.

The AOD method is described in detail by e.g.  \citet{aod91} and
\citet{ebj_96}.  To summarize: the optical depth, $\tau_i$, at a given
pixel $i$ (usually in velocity space, where the width of a pixel is
$\delta v$) is given by $\tau_i \: =\: - \ln \: F_i$ where $F_i$ is
the normalized flux.  If $\sigma_i$ is the error in the normalized
data at the $i$'th pixel, then the corresponding error in $\tau$ is
$\sigma_\tau = {\sigma_i}/{F_i}$ provided that $\sigma_i\: \ll\: F_i$.
The AOD column density associated with this single pixel, $N_i$, in
units of cm$^{-2}$, is then given by

\begin{equation}
N_i = \frac{m_e c}{\pi e^2} \frac{\tau_i}{f\lambda_0}\:\delta v =\: 
 3.768\times 10^{14} \:\: \frac{\tau_i}{f\lambda_0} \delta v
\end{equation}

\noindent
where $f$ is the oscillator strength of the absorption line, 
$\lambda_0$ the rest wavelength of the line in \AA, $\delta v$ is the
pixel velocity width in \kms , and the other symbols
have their usual meaning. Similarly, in the absence of continuum errors, the
error at the i'th pixel, $\sigma(N_i)$, is simply

\begin{equation}
\sigma(N_i) =3.768\times 10^{14}\:\:
\frac{\sigma(\tau_i)}{f\lambda_0}\:\delta v
\end{equation}

\noindent
The final column density and its variance,  integrated over
the whole line ($n$ pixels) are then

\begin{equation}
N = \sum_{i}^n \: N_i \: ; \hspace*{1cm} [\sigma(N)]^2 =  \sum_{i}^n \:[\sigma(N_i)]^2 \label{eqn_var}
\end{equation}

For the \FUSE\ data, $\sigma_i$ was taken from the error arrays
supplied by CalFUSE, which measures fluctuations from Poisson counting
statistics, and includes a contribution from the (modeled) background
subtraction. The error arrays do not take into account fixed-pattern
noise (FPN) since no stable flat-field is available. Fortunately, much
of the FPN in the one dimensional spectra is averaged out when
individual sub-exposures are co-added (\S\ref{sect_subexposures}),
provided that some drift was found in the spectra along the dispersion
direction between sub-exposures. Our re-sampling of the data by a
factor of three also ensured that the FPN was smeared out while
preserving the shape and strength of an absorption line, at least when
the FPN features were narrow.

As we discussed, the O~VI profiles towards our sample stars were often
contaminated by Cl~I or H$_2$ lines. Our approach to reconstructing
the $N(v)$ profiles without these interfering lines was to fit
theoretical Voigt profiles to the contaminants and remove them. In
most cases, there was sufficient information (i.e., a large enough
number of pixels) to define the shape of a line quite precisely. For
this procedure, we do not need to know the true column density or
width of a line to obtain a good fit; we need only assume that the LSF
is a Gaussian. If this assumption is correct, we need only fit the
profile and remove it to recover the true O~VI $N(v)$ profile.

With the contaminating lines removed, we measured the AOD column
density as follows.  For each sight line, we measured $N$ using the
number of pixels that corresponds to a velocity interval of $-120\:<\:
\Delta v \:<\:+120$~\kms. A measurement was also made of a column
density limit over the same velocity range, defined as

\begin{equation}
[\sigma(N)_T]^2\:=\: \left(\frac{N_u - N_l}{2}\right)^2 + \sigma(N)^2 \:. \label{eqn_sigtot}
\end{equation}

If the measured value of $N$ was greater than
$\sigma(N)_T$, we considered the line to be detected (see
below). Although $ \Delta v$ was the same for all stars, in some cases
the $N(v)$ profile clearly extended beyond this range, so $\Delta v$
was increased accordingly to produce the correct value of $N$. We
again measured $N$ three times, once for the data normalized by the
best continuum fit, and twice more for the upper and lower error
envelopes to give $N$, $N_u$ and $N_l$.

In principle, we could list these AOD column densities when O~VI was
detected in Table~\ref{tab_cols}. However, these quantities turned out
to have very similar values to those measured using the line profile
fitting procedure described in \S\ref{sect_voigt}.  We noted, however,
a slight systematic difference in $N$(O~VI), in that column densities
measured with the AOD method were slightly over-estimated at the
lowest column densities. This effect, caused by the difference in the
behavior of $\tau$ for positive and negative fluxes, is well
understood, and discussed, e.g., by \citet{fox05_nv}.  For this
reason, we do not list the AOD values in Table~\ref{tab_cols}. The AOD
values are given, however, in the bottom panel of the figures which
show the \FUSE\ spectra, as described in
Appendix~\ref{sect_explain_spectra}.

In some cases, the values of $N$(O~VI) measured from the AOD method
can be far less than the errors, and sometimes even negative values
can arise. For these, one has some freedom in how the numbers should
be treated, depending on the problem at hand. For example, in a study
that combines results from many sight lines, it is perhaps best to
combine the formal numbers and their errors in some optimum way to
arrive at some improved global value.  In this case, noise
uncertainties in the individual cases are allowed to cancel one
another, which allows one to achieve an answer that is more meaningful
than a large collection of upper limits. However, it may be more
useful instead to give a more precise statistical interpretation of
the outcome for an individual sight line, such as stating to a given
confidence level an upper limit for $N$(O VI).  For the latter, we
propose that one could make use of a limit calculation for
nonsignificant measures in the presence of Gaussian noise defined by
\citet{marshall92}. Full details of this procedure are given in
Appendix~\ref{sect_upper_limits}.

As mentioned above, in deciding whether only an upper limit exists for
an O~VI line, we adopted the following procedure.  If the measured
value of $N$ was less than $\sigma(N)_T$, we considered
O~VI undetected, and instead listed the star with the measurements of
$N$(O~VI) in Table~\ref{tab_limits}.  Obviously, adopting a $1\sigma$
cutoff between detection and non-detection may seem somewhat
arbitrary. Another way we could have decided whether data yielded only
an upper limit might have been to argue that a limit should be adopted
when no Voigt profile could be fit to the data.  In fact, these two
procedures are equivalent. There exists only one sight line where we
could have tried to fit a profile, but where the measured value of $N$
was less than $\sigma(N)_T$. [That star was HD~052463
(unique ID 70), and the depression at the expected position of O~VI
has a reality that indeed appears ambiguous.]

Hence Table~\ref{tab_limits} contains {\it all} the pertinent
information available for the O~VI column densities towards the stars
(except, perhaps, for HD~052463). Columns $3-5$ give the measured
values of $N$(O~VI) and the values measured using the upper and lower
continuum fits, $N_u$ and $N_l$. Column 6 lists the error from the
noise, $\sigma(N)$, while $\sigma(N)_T$ is the quadratic
sum of all the errors, as defined in equation~\ref{eqn_sigtot}.
Column 8 gives the $2\sigma$ limit as defined by \citet{marshall92}
and discussed in detail in Appendix~\ref{sect_upper_limits}.
Ultimately, whether one uses the AOD measurement of $N$(O~VI) or
$N_{\rm{limit}}$ depends on the situation, and we discuss these issues
later in the paper.

\subsection{Misleading Effects from Unrecognized Saturation? \label{sect_saturation}}

A final consideration centers on the question of whether or not the
column densities we measure for the \FUSE\ stars might be
underestimated due to saturation effects that are not evident in the
apparent optical depths.  Since the \sixb\ line is unobservable
(\S\ref{sect_no1037}), it cannot be used to check for line saturation.
If the lines are resolved, then the Voigt profile fitting (when a
suitable LSF is convolved with the theoretical profile) accounts for
any possible saturation. The more problematic situation occurs when
the line is actually composed of many components, some of which are
unresolved but blended (and perhaps hidden by broader, resolved
components). Profiles that have a mixture of clouds with high and low
optical depths instead appear as if they are composed of only a few
clouds with moderate optical depths after being smeared by the
instrumental LSF. [For more details, see \citet{ebj86}.]

There are several observations that suggest that narrow unresolved
O~VI lines do not exist. O~VI lines can arise from clouds that are
either photoionized or collisionally ionized.  If the absorbing gas
is in collional ionization equilibrium, the thermal width of the line
is given by $b^2 = 2kT/m_{\rm{O}}$, (where $m_{\rm{O}}$ is the atomic
mass of oxygen) which for oxygen gives $T=969b^2$ or, equivalently,
$b=0.0321T^{1/2}$. The ionization fraction of O~VI peaks at $\log
T_{\rm{max}}\:=\:5.45$, which would give $b\:=\:16.6$~\kms , and
decreases sharply \footnote{For temperatures $T$ which are different
  from $T_{\rm{max}}$, $b\:=\:16.6(T/T_{\rm{max}})^\frac{1}{2}$~\kms
  .}  as the temperature falls, declining to a fraction $10^{-3}$ of
the total oxygen at $\log T=5.2$ \citep[e.g.,][]{shapiro76,
  sutherland93, nahar03}.  This lower value equates to a Doppler
parameter of 13~\kms . Clearly, the Doppler parameters corresponding
to either the peak temperature or this lower value are of the same
magnitude as the \FUSE\ LSF at 1030~\AA.

It remains possible that O~VI arises in collisionally ionized gas that
is not in equilibrium (due to rapid cooling, for example), or in
photoionized gas.  In these instances, the line may be narrower.  We
consider the existence of such narrow O~VI components unlikely, for
the following reasons: {\it i)} in high resolution ($\simeq
3$~\kms\ FWHM) {\it Interstellar Medium Absorption Profile
  Spectrograph} (IMAPS) data, there is no evidence for narrow multiple
sub-components in O~VI profiles \citep{jenkins_o6_imaps}; {\it ii)}
along path lengths through the entire Galactic halo, where $N$(O~VI)
ranges from $\simeq\:1-4 \times 10^{14}$~\pcm , comparisons of the AOD
profiles for {\it both} the O~VI~$\lambda 1032$ and the $\lambda 1037$
lines show that unresolved saturation effects are relatively small
\citep{wakker03, savage03}; {\it iii)} no narrow components have been
found in High Velocity Clouds, where the high velocity of the
absorbing gas might make such features distinct from (broader)
absorption by the Galactic halo \citep{sembach_hvcO6, collins05}; {\it
  iv)} similarly, in cases where an O~VI component has a high enough
velocity to be separated from the bulk of the absorption, such as
Carina and Vela, no very narrow components are seen. Although the
environment in which these clouds exist (circumstellar or SNR
outflows) are likely different from the rest of the ISM, none have
widths $b<20$~\kms ; {\it v)} the widths of the O~VI lines seen
towards stars within only 100~pc of the sun --- although hard to
measure due to the weakness of the lines --- are rarely $< 20$~\kms\
\citep{bill_local, savage06}. We also detect few lines with $b\leq
20$~\kms\ in the \FUSE\ sample (see \S\ref{sect_bees}).  So, even when
the number of absorbing clouds along a sight line is a minimum, no
particularly narrow lines are seen. For these reasons, we conclude
that although the number of clouds intercepted in the disk may be
large, no significant column density errors arise from narrow (perhaps
saturated) O~VI components.

It is interesting to note that in a recent study of {\it
  extragalactic} O~VI absorbers, \citet{tripp07} found evidence
that approximately one-third of the extragalactic O~VI lines in
low-redshift QSO spectra arise in relatively cool gas with $\log T\:
<\: 5$.  However, those O~VI lines are frequently located in truly
intergalactic regions that have much lower densities and larger sizes
than the gas clouds in the Milky Way disk.  The cool, photoionized
clouds found in the IGM cannot exist in the higher density,
shorter-path regions that we are probing in the Galactic plane.

\subsection{The Final \emph{FUSE} Sample  \label{sect_crapstars}}

All the spectra belonging to the \FUSE\ survey described in this paper
are presented in Appendix~\ref{sect_explain_spectra} and
Figure~\ref{fig_spectra}. A detailed explanation of the material shown
for each sight line is also given.

\begin{figure*}[t!]
\hspace*{-1cm}\includegraphics[width=9cm]{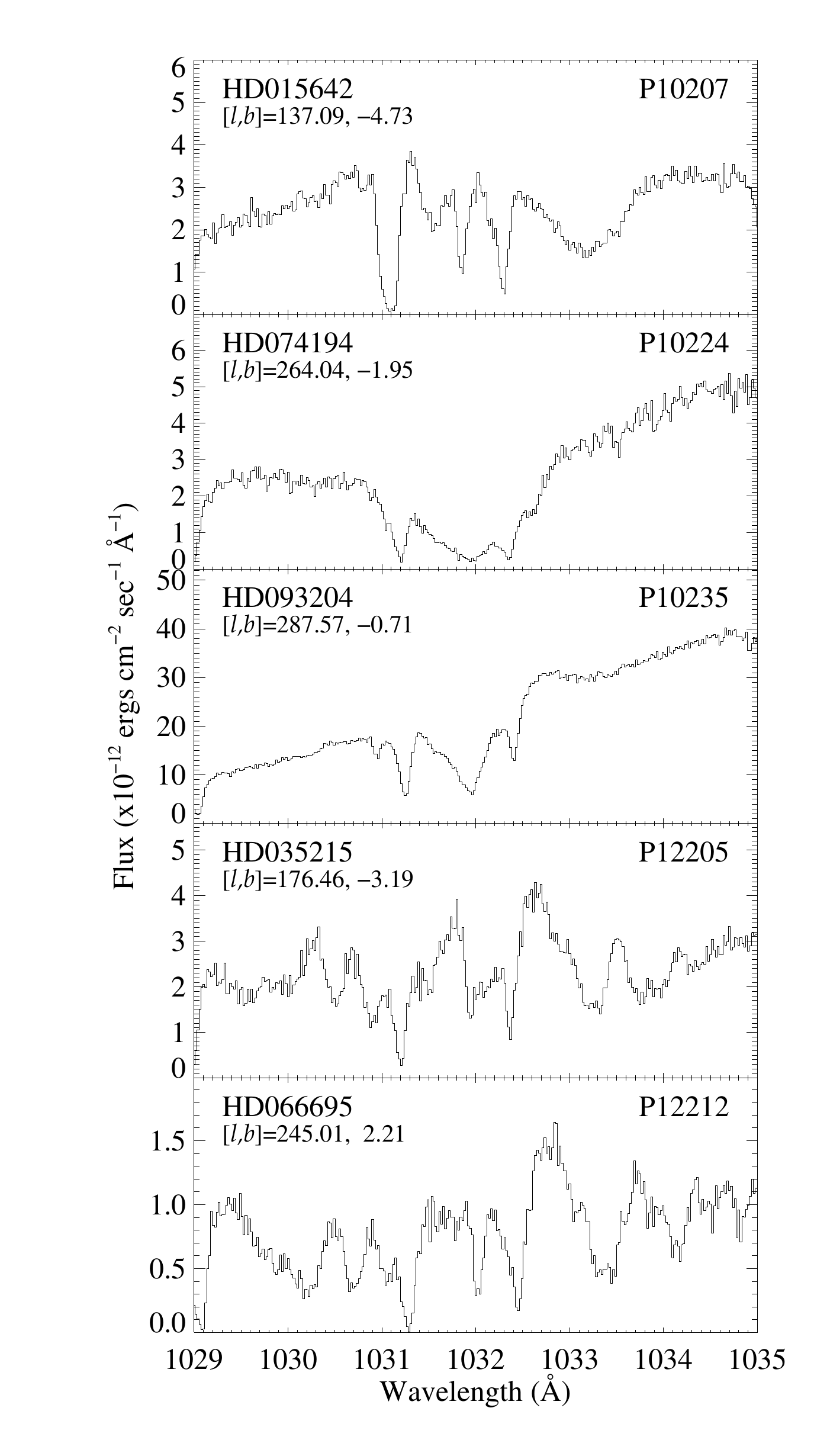}%
\hspace*{-1cm}\includegraphics[width=9cm]{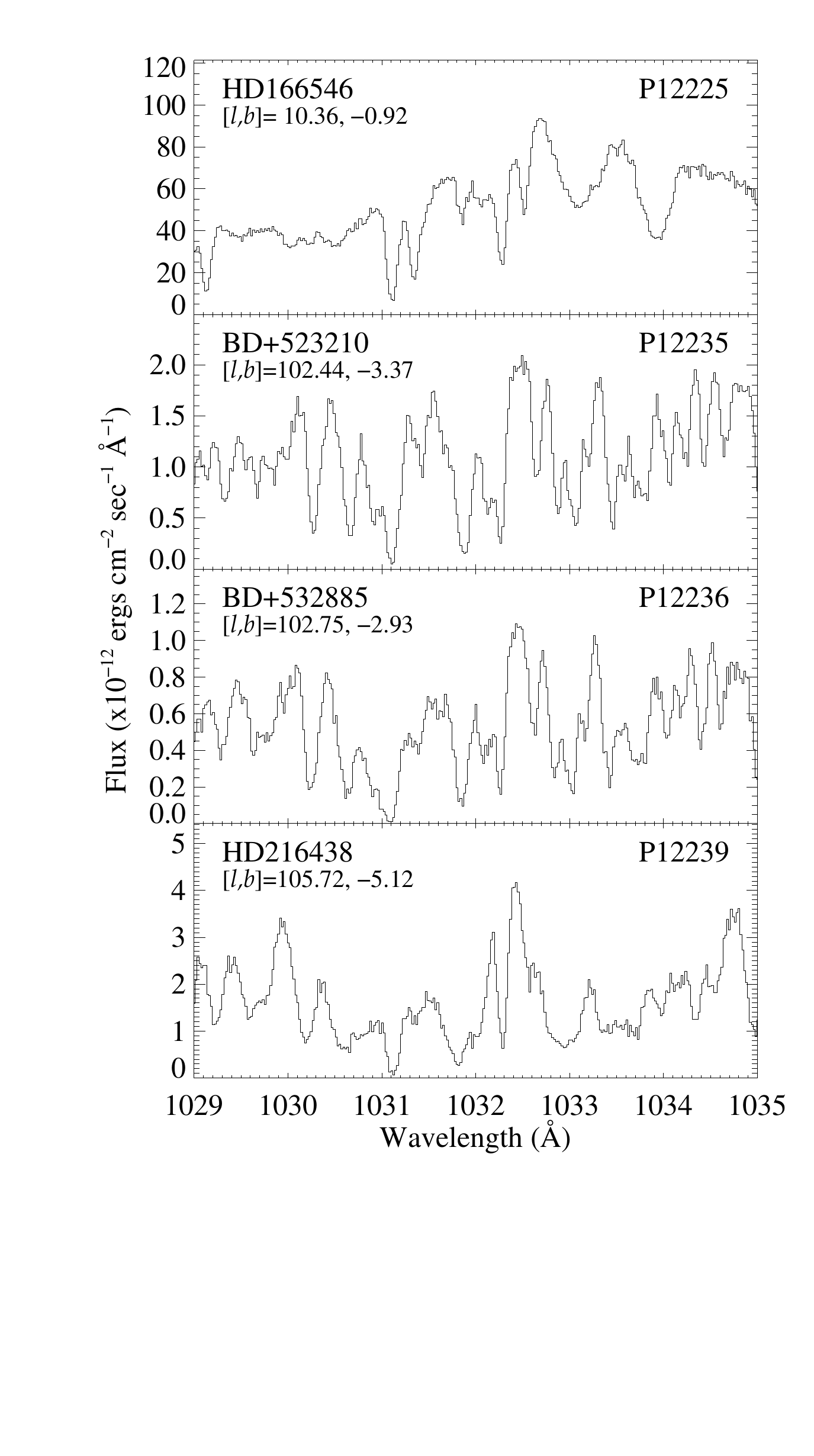}
\caption{\label{fig_rejects}
Spectra from programs P102 and P122 for which a reliable normalized \sixa\ profile could
not be analyzed.}
\end{figure*}

Although observed as part of programs P102 and P122, several stars
were excluded from the final dataset. For completeness, these are
listed in Table~\ref{tab_rejects} and plotted in
Figure~\ref{fig_rejects}. Most of these sight lines were excluded
because we could not choose reliable continua.  HD093204 (P10235) was
rejected because no reliable fit to the HD line could be obtained.

\section{INTERPRETATION}

\subsection{Inclusion of Other Data Sets}

The sight lines in our \FUSE\ survey were selected to cover distances
beyond $\sim 1$ kpc in order to study the distribution of O~VI
absorbing gas over a significant fraction of the Galactic disk.
However, there are several other sets of O~VI absorption line data
that probe gas over different scales, from a few tens of parsecs in
the local bubble (`LB'), to halo gas in the outer regions
of the Milky Way.  As we will show, we need to include these in our
analysis to fully understand the nature of the O~VI absorption. In
this section, we briefly discuss each of the datasets used and note
any modifications we made to the published data in order to avoid
systematic differences between quantities used for the \FUSE\ sample
and those used by other authors.

\subsubsection{White Dwarf Sight Lines}

Few O- and B-type stars can be found within a few tens of parsecs of
the Sun.  The nearest OB associations lie in the Sco-Cen association,
and are 118$-$145~pc away \citep{zeeuw99}.  Investigations with \FUSE\
therefore used nearby white dwarf stars (WDs) to study O~VI absorption
over very short path lengths. The first \FUSE\ survey to study LB O~VI
absorption was completed by \citet{bill_local}. This was superseded by
the larger survey of \citet[ hereafter SL06]{savage06}, whose data we
use in this paper.  We adopt most of the values given in Tables~1 and
3 of SL06, who claim that distances to the WDs are known to $\pm
20-30$\%. In our analysis we therefore use a distance range defined by
$\pm\:25$\% of the distance. O~VI column densities and Doppler
parameters are taken from measurements made using line profile fitting
procedures similar to our own.  SL06 define O~VI to be absent when the
equivalent width $W_\lambda\: <\: 2\sigma (W_\lambda)$, as measured
over an interval $\approx\:100$~\kms\ (see column 11 of their
Table~1). This is somewhat different than the method we adopted in
\S\ref{sect_colsAOD}, but since SL06 do not measure $N(v)$ in these
cases we simply use column density upper limits derived from the
$2\sigma(W_\lambda$) limits.  We do not include a WD sight line if
SL06 suggest that detected O~VI may be photospheric in origin.

\subsubsection{\emph{Copernicus} Sight Lines \label{sect_copernicus}}

The {\it Copernicus} satellite was able to obtain O~VI measurements
towards some of the nearest (brightest) stars lying a few hundred
parsecs away, most of which are too bright to be re-observed with
\FUSE , unfortunately.  The data were originally presented in
\citet{ebj74}, \citet{york74}, and \citet{ebj78a} (J78).  In this
paper, we have re-derived many of the stellar parameters for the
\Copernicus\ stars using the same procedures used for the \FUSE\
sample stars, so that the two samples can be properly combined. The
changes made are discussed more fully in
Appendix~\ref{sect_copernotes}. Two stars observed by \Copernicus\
(HDs 041161 and 186994) {\it were} re-observed with \FUSE ; these
latter data supersede the \Copernicus\ results and the information for
the stars can be found in Table~\ref{tab_journal} (unique IDs 52 and
27, respectively).

O~VI column densities for the \Copernicus\ sight lines are taken
directly from Table~1 of J78. That table also lists a value $\Delta^2
= \langle (v-\langle v \rangle)^2 \rangle$ for the \sixa\ line, which
is the second moment of the line profile after the second moment of
the \Copernicus\ LSF has been removed.  We convert this to a Doppler
parameter in the usual way, $b = \sqrt{2} \Delta$. No errors were
given for the derived column densities due to the difficulties in
fitting the stellar continua over the short wavelength regions scanned
by \Copernicus . We have, however, made some simple estimates of the
likely errors in $N$(O~VI); these can also be found in
Appendix~\ref{sect_copernotes}.

\subsubsection{Sight Lines to the Vela SNR}

Our sample contains three stars which lie behind the Vela supernova
remnant (SNR), HD~074920 (unique ID \#77), HD~074711 (\#78) and
HD~075309 (\#79). These are discussed below in \S\ref{sect_SNR} and
\S\ref{sect_rho_lb}. We added to these the stars discussed by
\citet{ebj_vela_76} and \citet{slavin04}. Distances to these stars
were recalculated according to the prescription laid out in
Appendix~\ref{sect_distances}.

\subsubsection{Distant Halo Stars and Extragalactic Sight Lines \label{sect_other_savage}} 

For some of the analysis presented later, it is useful to compare the
O~VI absorption seen in the Galactic disk with that observed by \FUSE\
over very long path lengths towards distant halo stars and
extragalactic sight lines.  \citet[ hereafter Z03]{zsargo03} observed
a small sample of stars in the Galactic halo.  Three of these were
retrieved by us from the \FUSE\ MAST archive (HD~088115, HD~148422,
and HDE~225757) before Z03's results were available, because they lie
at latitudes $|b|\:<\:10\degr$. The O~VI towards these sight lines is
strong, and we would expect errors in HD subtraction and continuum
placement to be largely insignificant (particularly since Z03 rejected
stars with difficult continua). In fact, we find that $\log N$(O~VI)
from our measurements agree well with Z03's analysis. The difference
in $\log N$(O~VI) is $\leq0.08$ dex for all three stars, and identical
within the $1\sigma$ errors.

For the rest of the stars in Z03, we have used the O~VI column
densities, Doppler parameters, and absorption line velocities given in
their Tables~4 \& 5. We have, however, re-derived some of the
distances to the stars using the same procedures used for all the
other stars described in this paper. Details are given in
Appendix~\ref{sect_zsargonotes}.

Finally, we have included the extragalactic sight lines observed with
\FUSE\ by \citet{savage03} and \citet{wakker03}.  We only use data
that have S/N ratios $> 9$ per 20~\kms\ bin [i.e., those with codes
'Q=3' or '4' in Table~2A of \citet{savage03}].  The distances to these
objects can be regarded as infinite for the purpose of studying
Galactic O~VI, and no new distances need to be computed except for one
halo star: PG~0832+675, for which we use $d\:=\:8.1$~kpc.

\subsection{Distribution of Stars in the Galactic Plane \label{sect_the_disk}}

An important parameter in the study of the O~VI distribution in the
Galaxy is the distance to any particular background star in our
sample.  Unfortunately, for stars beyond a few hundred parsecs from
the Sun, the only way to measure stellar distances is through the use
of ``spectroscopic parallaxes'', which convert the observed magnitude
of a star to its distance assuming that the spectral type and
luminosity class of a star are known, along with a calibration of the
true absolute magnitude of a star of that particular type and class.
Knowledge of the interstellar extinction along a given sightline,
$E(B-V)$, is also required.  Although distances for many of our stars
have been calculated before with this procedure, in this paper we have
re-derived stellar distances using more modern databases. Since the
details are complicated and not directly relevant to the analysis of
interstellar O~VI absorption, we document our procedures at the end of
this paper, in Appendix~\ref{sect_distances}.

\begin{figure*}[t!]
  \vspace*{-1cm}\hspace*{-1cm}\includegraphics[width=11.5cm,angle=90]{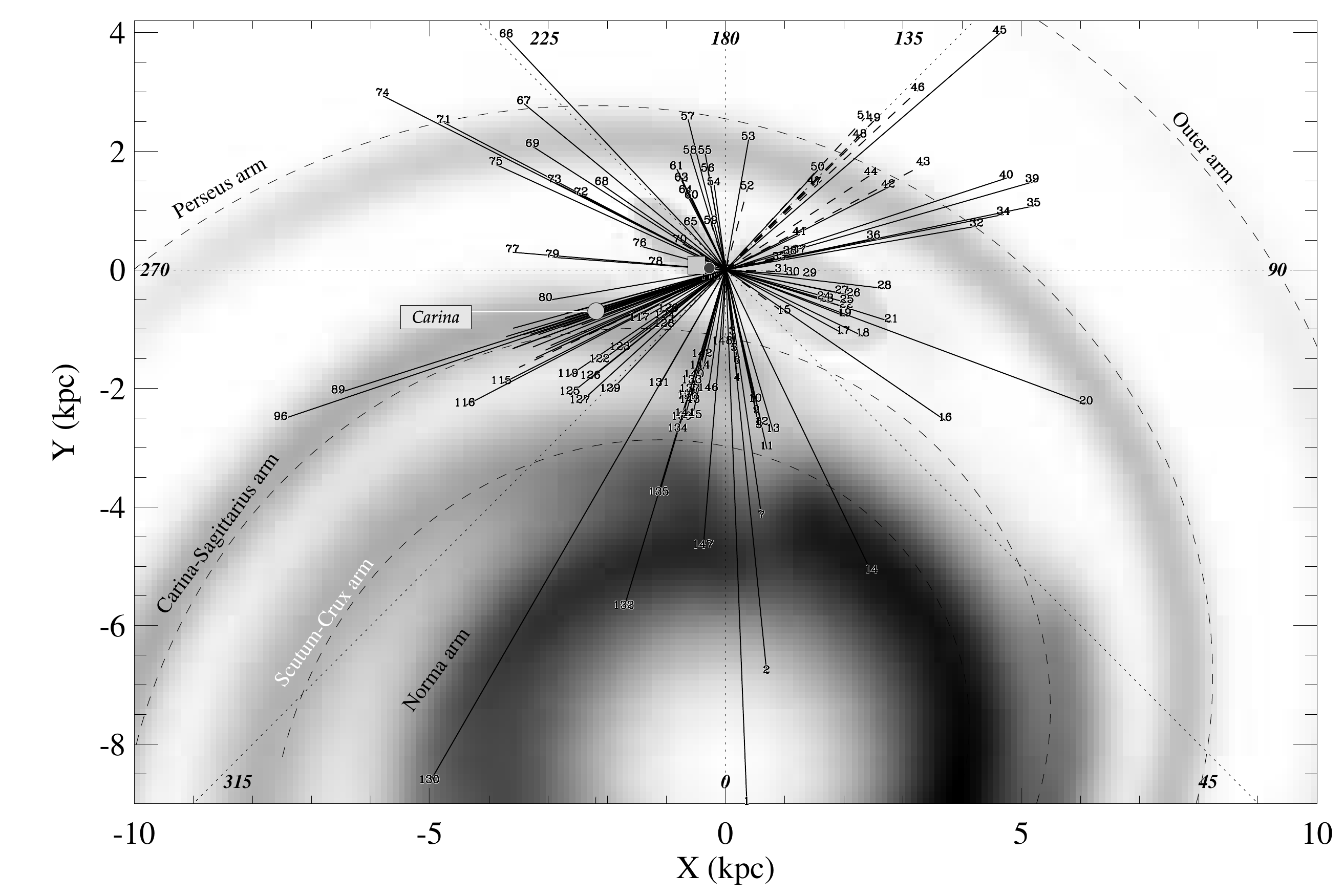}
  \caption{\label{fig_disk} Distribution of sight lines
  in our \FUSE\ survey, looking down into the Galactic plane. For
  illustrative purposes, the grayscale regions reproduce the electron
  density distribution modeled by \citet{cordes04}, while the dashed
  lines show the positions of the spiral arms from models by
  \citet{vallee02}. [The electron densities shown are the sum of
  components measured for the inner thin disk, the spiral arms, and
  the local ISM; the grayscale for the spiral arms represent electron
  densities of $\approx\: 0.01-0.1$~cm$^{-3}$---see \citet{cordes04}
  for more details.]  There is insufficient room to label all the
  stars observed towards Carina ($l=287.7$, $d=2.5$ kpc), so we
  indicate the region in the Sagittarius arm where the majority of the
  observed stars lie with a light-gray circle.  The gray box close to
  the Sun represents the Gum nebula; the small dark-gray circle next
  to it shows the position of the Vela supernova remnant ($l=263.9$,
  $d=0.25$~kpc).  The number for each star refers to the unique ID
  number given in Table~\ref{tab_lookup}. Sight lines for which only
  upper limits to the O~VI column density are found are shown as
  straight dashed lines.  }
  \end{figure*}

The distribution of our \FUSE\ survey sight lines in the plane of the
Galactic disk is shown in Figure~\ref{fig_disk}. To illustrate the
regions of the Galaxy in which the stars lie, we have plotted in gray
the distribution of the electron density in the Milky Way, as modeled
by \citet{cordes02}, using their {\tt NE2001}
program\footnote{http://astrosun2.astro.cornell.edu/\~{
  }cordes/NE2001. See also \citet{cordes04} for a summary of NE2001.}.
Overplotted are the spiral arm models from the meta-analysis of
\citet{vallee02}.  Vall\'{e}e~\etal\ used a value of 7.2~kpc for the
distance of the Sun to the center of the Galaxy; for consistency with
the rest of the analysis in this paper, we have changed this distance
to 8.5~kpc, and altered the pitch angle of the arms to be $11.3^o$.
The structures shown are {\it only meant to be illustrative} of where
in the Galactic plane the stars lie. For more precise details about
the likely structure of gas in the Milky Way, see \citet{levine06} and
references therein.

In Figure~\ref{xray_ovi_ipane} we show the distribution of stars in
the plane of the Galaxy superimposed on a map of the diffuse X-ray
emission constructed from the \ROSAT\ {\it All Sky Survey} (RASS).
This map includes all the \FUSE\ stars shown in Figure~\ref{fig_disk},
as well as stars from the other datasets discussed above.  The map is
a composite of data from three \ROSAT\ bands, the R1 ($0.11-0.28$~keV)
and R2 ($0.14-0.28$~keV) bands, and the sum of the R4$-$R6
($0.44-1.56$~keV) bands.  In the figure, the R1, R2 and R4$-$R6 bands
are assigned the colors red, green and blue, respectively. The precise
interpretation of the emissivity at different energies is somewhat
complicated, and is discussed in detail by \citet{snowden97}. However,
in \S\ref{sect_bubbles} and \$\ref{sect_SNR} below, we use the X-ray data to
examine in more detail the circumstellar environments around the stars
observed for this paper.

\begin{figure*}[t!]
  \vspace*{-1cm}\begin{minipage}[c]{0.45\linewidth}
    \caption{\label{xray_ovi_ipane} O~VI results superposed on a depiction
  of the diffuse X-ray emission recorded by the \ROSAT\ All Sky Survey
  (RASS) over the longitude range {\it a)} $\ell = 310\arcdeg$ to
  $50\arcdeg$, {\it b)} $\ell = 40\arcdeg$ to $140\arcdeg$, {\it c)}
  $\ell = 130\arcdeg$ to $230\arcdeg$, and {\it d)} $\ell =
  220\arcdeg$ to $320\arcdeg$.  Ellipses are centered on the positions
  of target probes; their vertical axes are proportional to $N$(O~VI)
  (totaled over all velocity components), and their horizontal axes
  signify the distance to the targets, both of which can be gauged by
  the scales in the lower left corner of the figure.  A circle
  corresponds to an average sight-line density of $1.62\times
  10^{-8}\:{\rm cm}^{-3}$ [i.e., $N({\rm O~VI})=5\times 10^{13}\:{\rm
    cm}^{-2}$ over a distance of 1$\,$kpc], which is a representative
  average for the whole survey.  It follows that tall, skinny ellipses
  indicate more O~VI than usual for a particular distance, while the
  converse is true for short, fat ellipses.  O~VI absorption towards
  stars at distances $<100$~pc are not plotted since the size of their
  ellipses in both the horizontal and vertical direction would be too
  small to be seen using these scaling rules.  The direction of a
  small arrow on the top of each ellipse signifies the average
  velocity of all of O~VI absorption, with an angular deflection to
  the left (negative velocities) or right (positive velocities) whose
  magnitude in degrees corresponds to the radial LSR velocity in ${\rm
    km~s}^{-1}$.  Cases with only upper limits are shown as dotted
  ellipses, where the $N$(O~VI) upper limit defines the height of the
  ellipse in the same manner as described above.  White ellipses mark
  stars in the \FUSE\ survey reported in this paper, red ellipses
  denote {\it Copernicus\/} measurements \citep{ebj74,york74,ebj78a}
  and light blue ellipses identify targets in a \FUSE\ survey of stars
  in the Galactic halo \citep{zsargo03}.  Yellow ellipses show the
  data from \citet{ebj_vela_76} and \citet{slavin04} for the Vela
  supernova remnant. Green arrows represent O~VI absorption towards
  nearby WDs from the data of \citet{savage06}.  Vertical dark blue
  bars at high Galactic latitude show the results from the \FUSE\
  survey of extragalactic objects by \citet{savage03} and
  \citet{wakker03}, and have no width because they lie outside the
  Galaxy.  Colors in the X-ray sky are keyed to the \ROSAT\ R1
  ($0.11-0.28$~keV) band (red), R2 ($0.14-0.28$~keV) band (green), and
  the sum of the R4$-$R6 ($0.44-1.56$~keV) bands (blue), with scale
  factors that yielded a good differentiation of colors across
  different regions.  The RASS results were obtained from an ftp
  download site
  http://www.xray.mpe.mpg.de/rosat/ survey/sxrb/12/fits.html maintained
  by the Max-Planck-Institut f\"{u}r Extraterrestrische Physik.}
\end{minipage}
\hspace{0.5cm}\begin{minipage}[c]{0.45\linewidth}
  \includegraphics[width=10.1cm]{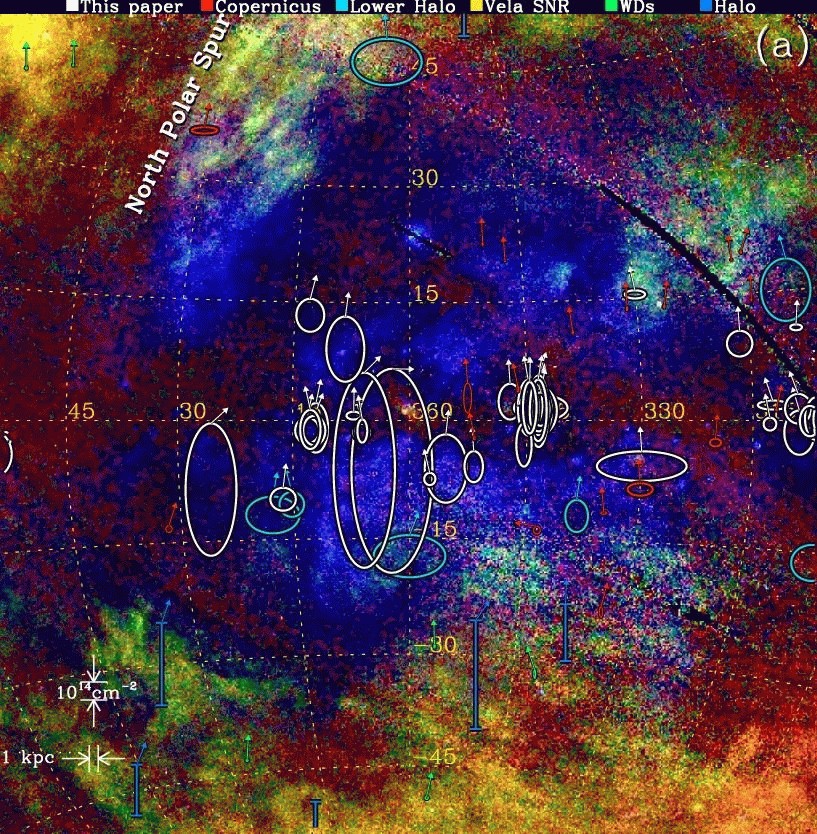}%
  \vspace{0.5cm}
  \includegraphics[width=10.1cm]{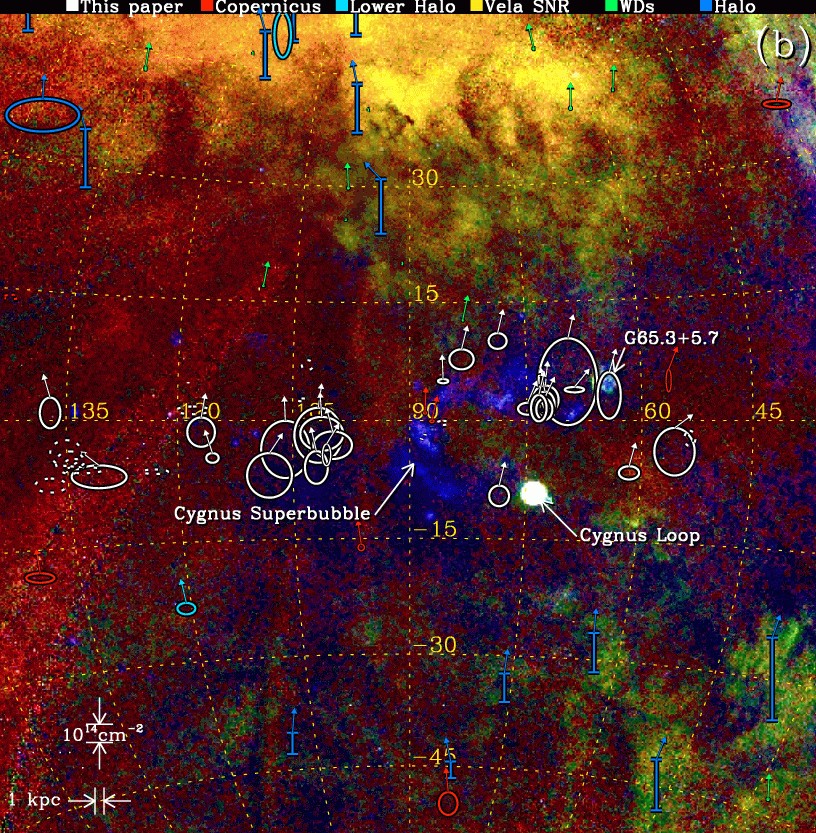}
\end{minipage}
\end{figure*}

\addtocounter{figure}{-1}
\begin{figure*}[t!]
\vspace*{-1.5cm}\hspace{4cm}\includegraphics[width=10.1cm]{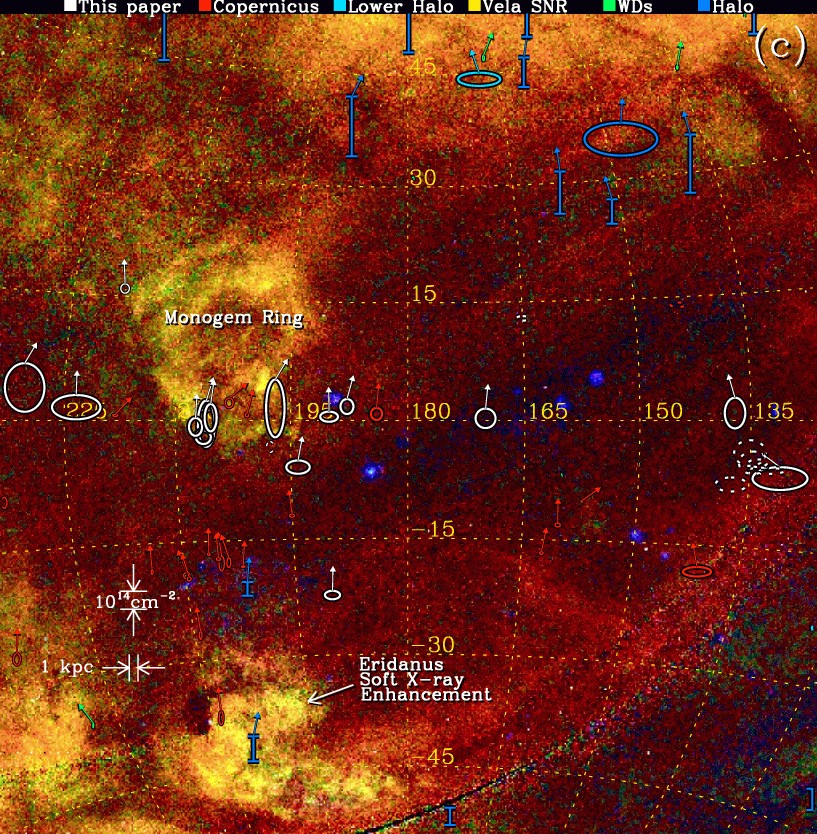}

\vspace{1cm}\hspace*{4cm}\includegraphics[width=10.1cm]{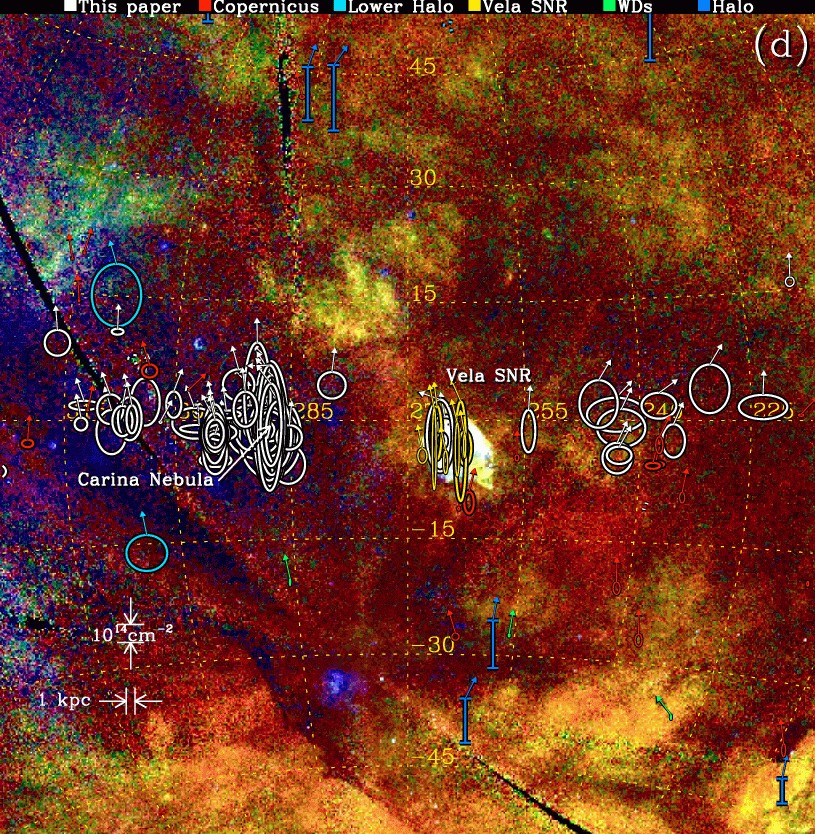}
\caption{continued}
\end{figure*}

\subsection{Variability of O~VI Absorption Lines Determined from Multi-epoch Data \label{sect_epoch}}

O~VI absorption from the Galactic ISM is superimposed on stellar
continua which may show broad O~VI P-Cygni profiles from stellar
winds.  Although the widths of these profiles can span several
thousand \kms , their shapes are often highly irregular --- a fact
which likely leads to the problems we encountered in \S\ref{sect_cont}
in fitting many difficult continua. Moreover, the shapes of these wind
profiles are known to vary on many different timescales, and can show
`discrete absorption components' (DACs) which vary in velocity as well
as width and strength \citep{york77,snow80,massa00,lehner03}.  In
principle, O~VI~$\lambda 1037$ DACs from outflowing gas at velocities
$\sim\:-1650$~\kms\ could become blended with the interstellar
O~VI~$\lambda 1032$ absorption studied in this paper. A DAC which
contaminated the interstellar O~VI line at one epoch might only be
shown to exist by temporal variations in its properties, as
recorded in multi-epoch data.

To understand whether these complicated continua variations might
adversely affect accurate measurements of interstellar O~VI
absorption, \citet{lehner01} studied a small sample of stars which had
been observed repeatedly over time intervals of either a few days or
several months. Lehner~\etal\ concluded that, at least on these
timescales, stellar wind variabily had little influence on the
measurement of O~VI column densities.

We can perform a similar experiment for our \FUSE\ sample of stars,
over slightly larger timescales. Given the assumptions we made in
\S\ref{sect_cont} for fitting the continua of stars, a comparison of
O~VI column densities measured along the same sight line at different
epochs --- particularly when the continuum is a different shape at
those epochs --- is a useful indicator of the reliability of our data
analysis. We can also search for contamination from DACs, if only
along a few specific sight lines.

\begin{figure*}[t!]
\hspace*{-0.75cm}\includegraphics[width=18cm,angle=0]{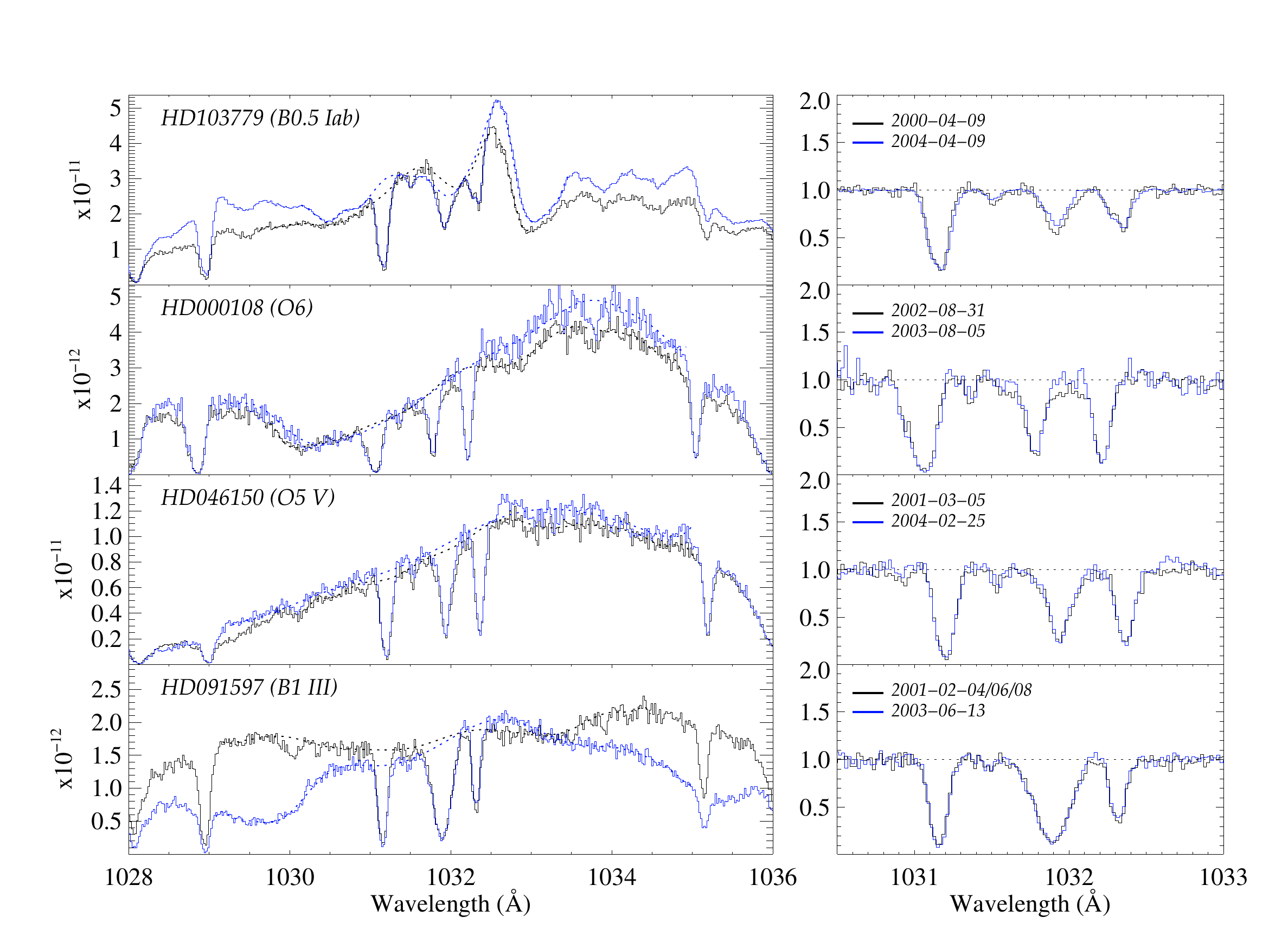}
\caption{\label{fig_time} Four stars (of six) for which multi-epoch
  data are available. Black/blue spectra represent the earlier/later
  epoch data.  The left hand panel shows the variation in absolute
  flux at two different epochs, along with the continua used to
  normalize the data.  Fluxes are in the usual units of \flux .  The
  right-hand panel shows the normalized flux {\it before} subtraction
  of the HD 6$-$0 R(0) line, while the dates of the two epochs used
  are given in the top left of each panel. Small shifts ($\leq
  0.06$~\AA ) in the wavelength scale have been added where necessary
  to align the features. }
\end{figure*}

Six stars in our \FUSE\ sample were observed at times more than a few
months apart and for which spectra with good S/N ratios were obtained
at all epochs; these are listed in Table~\ref{tab_time}.  The four
spectra with the highest S/N are shown in Figure~\ref{fig_time}. As
the left-hand panels of Figure~\ref{fig_time} show, the continua
changed to some degree for most of the stars.  To search for
variability in the O~VI line, we compare the normalized data at each
epoch in the right-hand panels of Figure~\ref{fig_time}. By accident,
several of the stars in this small sample had low O~VI column
densities compared to the sample as a whole. To avoid detecting
differences in absorption line profiles arising from difficulties in
subtracting the HD R(0) line, we show only the O~VI profile before the
removal of the HD line. (To see the strength of O~VI after the HD
removal, see Fig.~\ref{fig_spectra}.)

Figure~\ref{fig_time} shows that there is little evidence for any
variation in the O~VI(+HD) absorption. Any differences seen can be
attributed to problems in fitting the continuum. The absorption
towards HD~103779 is a particularly interesting example: the continuum
is very hard to fit at both observed epochs, yet the absorption at
1032~\AA\ is remarkably similar in observations made four
years apart. Note that the line towards HD~000108 is entirely HD
absorption (again, see Fig.~\ref{fig_spectra}, ID \#43).

Although we find no evidence for variations in absorption over
differences of $3-4$ years, the spectra do allow us to test how robust
our measurements of $N$(O~VI) really are. Although the O~VI line in
Figure~\ref{fig_time} is blended with HD absorption, we can proceed
with the analysis described in the previous sections by removing the
HD line and measuring the resulting $N$(O~VI).  The results are given
in the last column of Table~\ref{tab_time}. Despite having to contend
with difficult continua (e.g.  HD~103779), and removing HD lines that
are a significant fraction of the O~VI column density [because
$N$(O~VI) is relatively weak], the column densities measured for each
epoch agree to within their errors. (The upper limits for the spectra
of HD~000108 observed a year apart are obviously more a reflection of
the S/N of the data, which were similar due to the similar exposure
times. For both spectra, however, the HD could be modeled and removed
to show that no O~VI was obviously present.) This implies that our
measurements of $N$(O~VI) are quite insensitive to temporal
variations, continuum changes, and HD removal.

\subsection{Accounting for Circumstellar Bubbles \label{sect_bubbles}}

The target stars and their association members produce strong stellar
winds \citep{garmany81, larmers81, garmany84} which inject substantial
amounts of energy into their surrounding gaseous media
\citep{abbott82,mckee86}.  From an initial theoretical investigation
of the probable consequences of such wind-driven shells,
\citet{castor75} proposed that nearly all of the low-velocity O~VI
features seen in the spectra of early-type stars recorded by
\Copernicus\ and reported by \citet{ebj74} probably arose from a thin
conduction zone situated between an interior region of hot, shocked
wind material at $T\sim\:10^6\,$K and the surrounding medium, rather
than from the general interstellar medium.  In a more refined
investigation of these circumstellar structures, \citet{weaver77}
predicted that a typical bubble interface could produce $N({\rm
  O~VI})\approx 2\times 10^{13}~{\rm cm}^{-2}$.  While this amount of
O~VI is small compared to the typical column densities registered in
our \FUSE\ survey, we must recognize that our inventory of truly
interstellar O~VI could be elevated by the contributions from
circumstellar bubbles.

Many factors can influence the structure of a bubble, such as the
star's age, its wind energy, its motion through the medium, and the
density of the surrounding gas.  These elements, together with the
added effects from neighboring stars, make it difficult to make
trustworthy predictions about the O~VI contributions for specific
cases.  Nevertheless, we know that the internal hot gas within a
bubble should emit soft X-rays.  Thus, stars with well established
bubbles should be detectable in the RASS, although target regions for
which there is strong foreground absorption by neutral hydrogen may
show up in only the highest energy bands.  While we do not expect a
simple one-to-one correspondence between the X-ray emitting properties
of a bubble and its ability to contribute O~VI absorption, we can
still try to establish an empirical relation between the two and thus
gain some insight on the probable effects from bubbles in general.

\begin{figure*}[t!]
\vspace*{-1cm}
\hspace{1.5cm}\includegraphics[width=6.0cm]{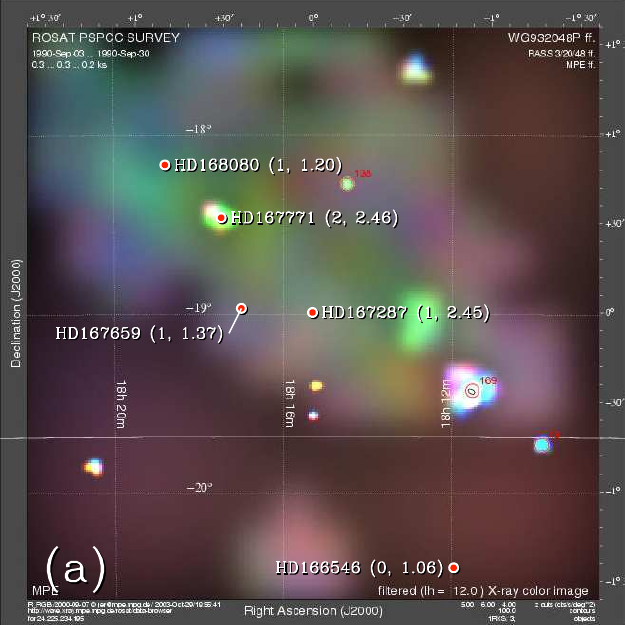}
\hspace{1cm}
\includegraphics[width=6.0cm]{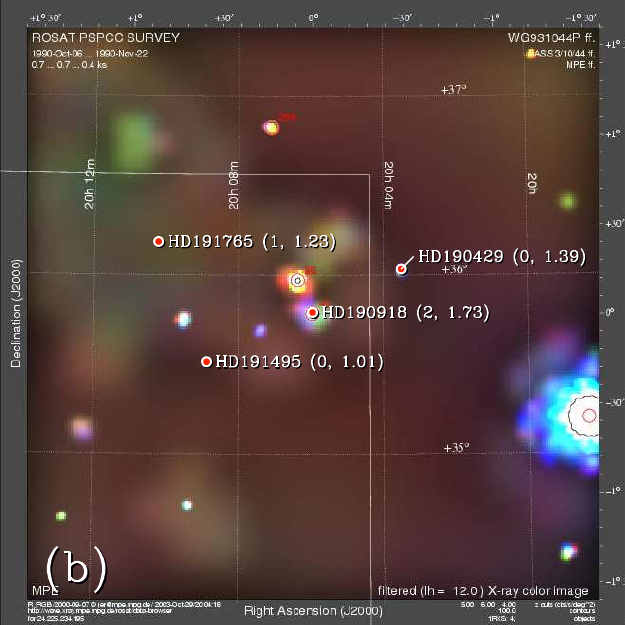}

\vspace{0.5cm}
\hspace{1.5cm}\includegraphics[width=6.0cm]{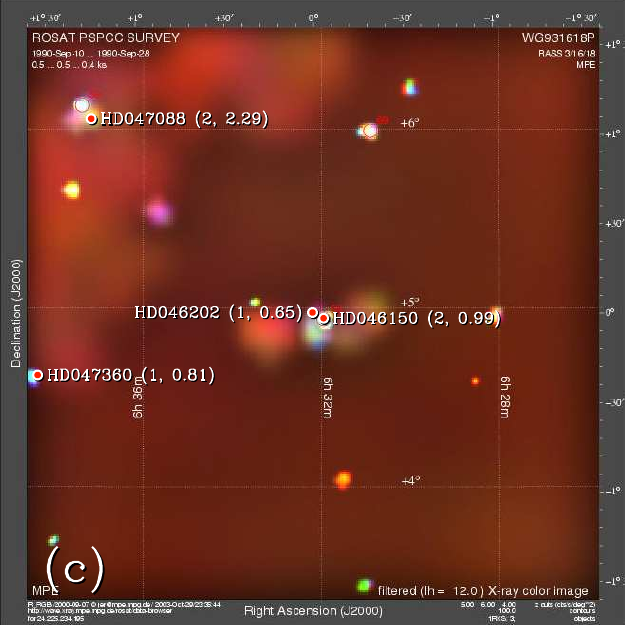}
\hspace{1cm}
\includegraphics[width=6.0cm]{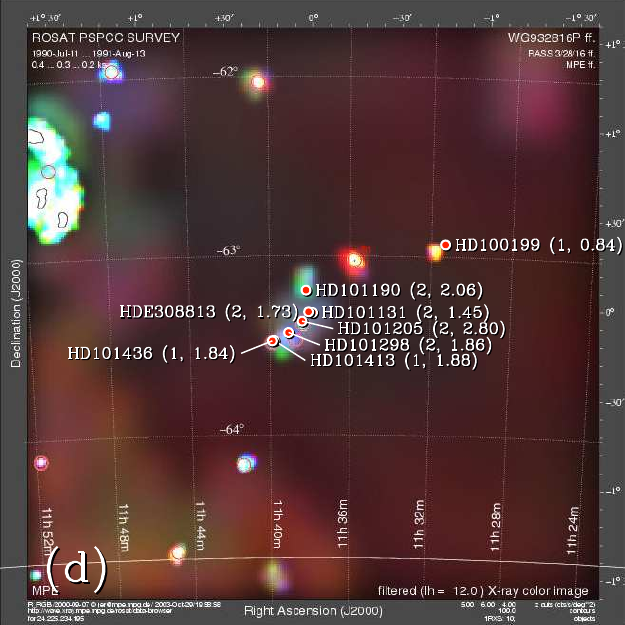}

\vspace{0.5cm}
\hspace{1.5cm}\includegraphics[width=6.0cm]{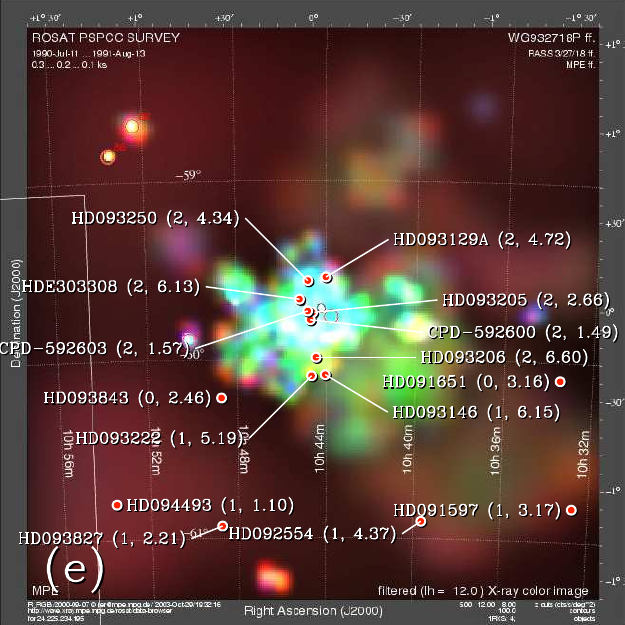}
\hspace{1cm}
\includegraphics[width=6.0cm]{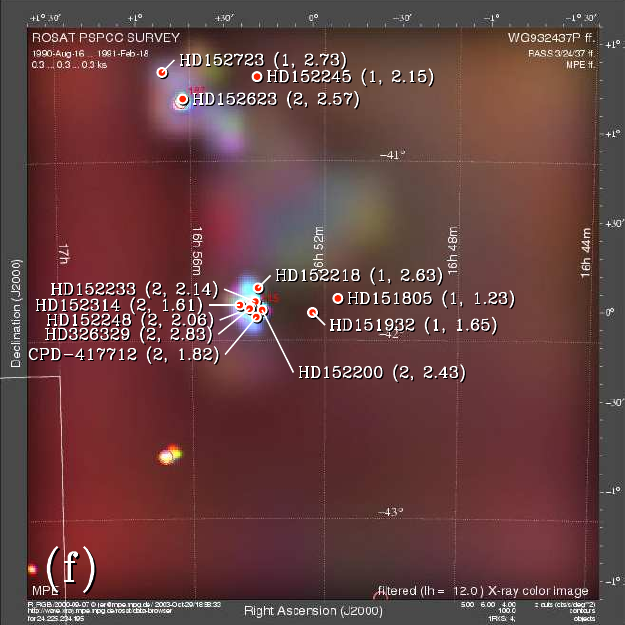}
\caption{\label{RASS_panels} 
X-ray emission observed in the \ROSAT\ All Sky Survey (RASS) within fields
centered on close groups of stars selected as targets for the O~VI survey. 
Intensities for the low, middle and high energy bands are depicted with varying
brightness levels of red, green and blue, respectively.  The positions of O~VI
survey targets are identified; following each star's name are parentheses that
contain our designation of \ROSAT\ category followed by the value for $N({\rm
O~VI})/10^{14}~{\rm cm}^{-2}$.}
\end{figure*}

Figure~\ref{RASS_panels} shows selected RASS image fields\footnote{The
  \ROSAT\ images were obtained from the {\it ROSAT Data Browser} at
  the web site http://www.xray.mpe.mpg.de/cgi-bin/rosat/data-browser,
  using the default settings except for the full-scale limits being
  set to Band~1 = 5, Band~2 = 6, Band~3 = 4 counts~${\rm
    deg}^{-2}\:{\rm s}^{-1}$ for Figs.~\ref{RASS_panels}a-d and f,
  while Band~1 = 5, Band~2 = 12, Band~3 = 8 was implemented for
  Fig.~\ref{RASS_panels}e.}  that include groups of stars in our
survey.  These examples illustrate the fact that some stars are
surrounded by clearly defined X-ray emitting zones, while others are
not.  Sometimes there are several stars that reside within a common
envelope that is bright in X-rays.

Ultimately, we need to determine the approximate magnitude of the
bubble contributions to the O~VI column densities that we measured, so
that we can arrive at a clearer understanding of how much O~VI is
distributed in the general regions of the Galactic disk.  However the
distinction between ``bubble'' and ``non-bubble'' O~VI-bearing gas is
somewhat arbitrary, since there is probably a continuum of bubble
properties that extend from well formed, identifiable bubbles with
distinct boundaries to highly disrupted structures that begin to blend
into the general medium.  Despite this limitation, we felt it was
important to make a distinction between stars that had visible
bubbles, according to some specific criterion, and those that did not.
To accomplish this, we assigned each target to one of three categories
according to appearances in the RASS fields, based on the concepts of
having no evidence of surrounding emission, having an intermediate or
doubtful level of emission, or having a clear, surrounding bubble. For
convenience, we assign the symbol $R$ to represent a category. We then
have:

\begin{itemize}
\item $R=0$: the target is in an X-ray-dark portion of the sky, well free of
any emission;
\item $R=1$: the target is in or near a general diffuse enhancement of X-ray
emission or is on the edge of a bright spot;
\item $R=2$: the target is clearly situated in projection in a bright X-ray
spot that appears to have been created by the star and/or its association
neighbors
\end{itemize}
(An additional category will be assigned in \S\ref{sect_SNR}.)

The immediate surroundings of virtually all early-type stars exhibit
some X-ray emission arising from shocked stellar winds
\citep{hillier93, bergh94, cassinelli94,bergh96,bergh97}.  Absorption
features from O~VI should arise from such material, but the large wind
speeds allow these contributions to be separated from the interstellar
features \citep{lehner01,lehner03}. For this reason, we refrained from
classifying stars as $R=2$ if the emission was not clearly extended in
the sky beyond the smearing by \ROSAT\ point-spread function.  As an
aid to determining whether or not the emission was extended, we
examined entries in the \ROSAT\ bright and faint source catalogs
\citep{voges99}.  Sources coincident with the star positions that were
not clearly extended (extent likelihood $\leq 1$) were considered to
arise from fast winds, which we considered grounds for rejecting the
assignment of $R=2$.

\subsection{Accounting for Supernova Remnants \label{sect_SNR}}

In addition to bubbles around the target stars, extraordinary
enhancements of O~VI can in principle arise if any of the lines of
sight penetrate a supernova remnant.  This is clearly the case for 3
of our stars, HDs 074920, 074711 and 075309, which lie in the
direction of the conspicuous Vela SNR [see
Fig.~\ref{xray_ovi_ipane}{\it{d)}}].  The near edge of this remnant is at a
distance of only about 250~pc \citep{cha99}, a value that is
consistent with the slightly greater distance ($\sim\: 280$~pc) to the
nebula's center as determined from the distance to the Vela pulsar
\citep{oeg89, caraveo01, dodson03} and X-ray observations of the
nebula \citep{bocchino99}.  Thus, there is no question that the
remnant is in front of these stars with distances ranging from 1.1 to
2.7~kpc.

It is important to identify less obvious cases where sight lines may
penetrate SNRs.  A comparison of our target locations with those in a
compilation of identified remnants \citep{green01} reveals a number of
coincidences, which are listed in Table~\ref{tab_SNRs}.  Once again,
in order to exclude special cases from our general assay of O~VI, we
have added one more category to the ones defined in
\S\ref{sect_bubbles} above:

\begin{itemize}
\item 
$R=3$: stars  are situated inside the
boundary of a supernova remnant and are not clearly in the foreground. 
\end{itemize}

To be conservative, when the relative distances are in doubt, we
assign a Category~3.  Unlike the situation for the Vela SNR, the
distances to the remnants presented in Table~\ref{tab_SNRs} may have
large errors.  For this reason, we sometimes experienced difficulties
in determining with much confidence whether or not the targets are
behind the SNRs.  Two stars that fall into this category are HD~168080
and HD~047417.  For HD~124314, we could find no distance estimate for
the remnant in the literature.  For other stars, such as HD~185418 and
HD~199579, the differences in distance between the stars and the
remnants are so large that we can be more certain that they are indeed
in the foreground.  Without the knowledge that HD~185418 was by chance
situated in front of G53.6~$-$2.2 (with angular dimensions of only
$33\times28\arcmin$), we would have assigned a rating of $R=2$ for
this star (i.e., by mistaking the X-ray emission as coming from a
bubble surrounding the star) instead of our adopted $R=0$.  Although
one might have said the same for HD~199579, we have retained the $R=2$
assignment because there is a bright X-ray spot centered at the
position the star that appears to be on top of the more extended
emission from the SNR.  While it may seem that HD~134411 lies far
behind its remnant, we must recognize that the distance estimate for
this SNR is probably much less accurate than the others, since it is
based on the empirical ($z$-adjusted) surface-brightness---distance
($\Sigma-D$) relation for estimating remnant distances
\citep{milne79}.

A very young remnant G266.2~$-$1.2 has a diameter of 2\arcdeg\ and is
located in projection inside the boundary of the much larger Vela SNR
\citep{asch98, iyudin98}. \citet{redman02} suggest that a nearby
nebula RCW~37 (the Pencil Nebula) is created by a collision between a
collimated flow of gas ejected from G266.2~$-$1.2 and a boundary of
the Vela SNR.  If this interpretation is correct, the distance of
G266.2~$-$1.2 should be nearly the same as that of the Vela SNR.  If
not, we can still rely on an upper limit of 0.5~kpc based on the
present-day strength of gamma ray emission from $^{44}$Ti
\citep{asch99}.  The two stars listed in Table~\ref{tab_SNRs} have
total column densities (in units of $10^{14}~{\rm cm}^{-2}$) of 1.96
(HD~074920) and 3.37 (HD~075309), along with 3.0 for HD~075821
\citep{ebj_vela_76}, which is also inside G266.2~$-$1.2.  By
comparison, stars that are still inside the main Vela remnant but
outside G266.2~$-$1.2 include HD~074711 in our survey along with HDs
074455, 072108 and 074753 in the study by \citet{ebj_vela_76}.  These
stars have $N({\rm O~VI})/10^{14}~{\rm cm}^{-2} = 0.7$, 4.5, 1.3, and
0.8, respectively.  It is thus not clear that the smaller remnant
contributes much extra O~VI.  Perhaps most of the gas inside this
young remnant is too hot to show O~VI; its X-ray spectrum indicates
that gases exist at a temperature in excess of $3\times 10^7\,$K
\citep{asch99}.

There are a number of targets in the vicinity of the Cygnus
Superbubble, a horseshoe-shaped X-ray emitting region with a diameter
of about $13\arcdeg$ \citep{cash80} (see
Fig.~\ref{xray_ovi_ipane}{\it{b)}}).  Recent investigations indicate
that this may not be a single structure, but rather a superposition of
unrelated shell and bubble-like objects at different distances
\citep{bochk85,uyan01}.  We therefore refrain from assigning $R\:=\:3$
to stars that appear near the edge of this emission.

The stars HDs 045314, 047417, 047088, and 047360 overlap with the
Monogem Ring \citep{pluc96}, but Figure~\ref{RASS_panels}{\it{c)}}
shows that HD~046202 and HD~046150 are just outside the edge of this
remnant (visible in the top left corner).  The yellow color of the
Monogem Ring depicted in Figure~\ref{xray_ovi_ipane}{\it{c)}} indicates the
presence of a strong low-energy X-ray flux that has not been
appreciably absorbed by foreground neutral hydrogen [i.e., compare
with the blue color of the Cygnus Superbubble in
Fig.~\ref{xray_ovi_ipane}{\it{b)}}], and indeed \citet{pluc96} estimate that
only $5\times 10^{19}~{\rm cm}^{-2}$ of neutral hydrogen is in front
of the nebula.  Our stars in this vicinity all have $N({\rm H~I}) >
10^{21}~{\rm cm}^{-2}$ \citep{diplas94_1}, so they are clearly at
greater distances.  As with the stars behind the Vela SNR, these stars
show more O~VI than expected for their distances.

Stars in the general vicinity of $\ell=240\arcdeg$ are behind a large
superbubble GSH~238+00+09 identified by \citet{heiles98}, who
estimated 0.8$\,$kpc for a distance to this structure.
Figure~\ref{xray_ovi_ipane}{\it{d)}} shows no appreciable departures from the
general average $N$(O~VI) per unit distance for the stars behind this
superbubble, although all of the O~VI velocities are consistently
positive.  Perhaps this velocity pattern arises from O~VI in the back
edge of an expanding bubble.

\subsection{\emph{ROSAT} Categories For Other Datasets}
 
For completeness, we also assigned \ROSAT\ categories to the stars
observed with \Copernicus , and those observed by \citet{zsargo03},
using the same system we used for the \FUSE\ stars described above.
These values are given in Tables~\ref{tab_copernicus} and
\ref{tab_zsargo} in Appendix~\ref{sect_copernotes} and
\ref{sect_zsargonotes}.

\ROSAT\ categories were also derived for the WDs of SL06. The majority
of these stars were assigned to $R=1$ or $R=2$ categories --- there
were very few WDs with no detectable ($R=0$) X-ray emission, largely
due to their small distances. The WDs give us an interesting
opportunity to look for a correlation between X-ray flux and
$N$(O~VI). Since WDs are very close by, interstellar O~VI absorption
is likely to be much less important than it is over longer path
lengths, and circumstellar O~VI absorption might in principle be the
dominant source of O~VI absorption. Using WDs also provides the
advantage of using sources with uniform properties. Although not shown
here, we searched for a correlation between X-ray flux and $N$(O~VI),
but found none. The circumstellar environment of a WD may, of course,
be different from that of a stellar association, but for the WDs, the
detected O~VI absorption is unrelated to the WD X-ray flux.

\subsection{Average O~VI Line of Sight Volume Density \label{sect_density}}

\subsubsection{Sample Definitions and Basic Statistics \label{sect_rho0} }

\citet{savage03} demonstrated that O~VI absorbing gas can be traced
several kpc above the plane of the Milky Way.  In the Galactic halo,
the volume density falls with height, but the derivation of the
mid-plane density $n_0$ requires fitting an assumed model to the
observed data (see next section, \S\ref{sect_NwithZ}).  The stars in
our sample (excluding the halo stars of Z03) have a distance above the
Galactic plane $z\:=\:d\sin b$ with a mean\footnote{The precise mean
  is $\langle z \rangle\:=\: -44$~pc, with a dispersion
  $\sigma\:=\:98$~pc.}  distributed around $z\:=\:0$ and with 98\% of
the stars below $|z|$ = 1~kpc. We would therefore expect a simple
average of the densities $n\:=\:N$(O~VI)/$d$ to be close to the
mid-plane density since the scale height is $\sim 2.3$~kpc according
to \citet{savage03}. (Again, we discuss this assumption in
\S\ref{sect_NwithZ}.) With the sight lines coded by the presence or
absence of X-ray emission, it should also be possible to see if $n$ is
different for lines of sight with and without enhanced X-ray emission.

The first problem in determining $n$ involves deciding how to treat
sight lines when only upper limits are available for $N$(O~VI). We
initially employed various survival statistics, including the well
known Kaplan-Meier (K-M) estimator, which calculates the true
distribution function of randomly censored data and yields the
distribution's mean and median \citep{feig85, isobe86}.  Given
measured values of, and upper limits to, $N$(O~VI)/$d$, we should in
principal have been able to recover $\langle n \rangle$. However, as
\citet{isobe92} makes clear, the K-M technique only works when the
upper limits are distributed randomly with respect to the overall
distribution (``Type 2 censoring''). Although the method worked well
for SL06, who had upper limits with values comparable to the
distribution of detected values of $n$ towards WDs, all of our upper
limits for $n$ were clustered below the apparent mean of $n$.

Instead, we simply defined a sample for which the column densities
were set at the upper limits given in Table~\ref{tab_limits} or as
stated in the other surveys when no line was present. We discuss the
possible errors introduced by defining the sample with these upper
limits in the next section.  Table~\ref{tab_n0} summarizes the
statistical tests employed, and the results obtained. We calculated
two statistics: (1) the average value of $n$ derived from the ensemble
of $\log(n)$ values\footnote{This is a simple average, without
  weights that account for the different magnitudes of the errors. The
  analysis presented in \S\ref{sect_clump} evaluates minima in
  $\chi^2$ which automatically recognizes differences in the size of
  the errors.}  ; and (2) the distance weighted average  $\sum
N$(O~VI)/$\sum d$.  The data were further subdivided into two subsets:
stars at distances $> 2$~kpc and those with distances between 0.2 and
2 kpc.  These distance categories partition the data into two halves
with approximately equal numbers (the median distance to the stars is
2040~pc). Of course, the two subsets contain data from different
surveys in disproportionate numbers: the $d\: = \: 0.2-2$~kpc subset
contains a mixture of \Copernicus\ and \FUSE\ data (45 and 43 sight
lines, respectively) while the $d>2$~kpc contains almost exclusively
\FUSE\ data (97 of 102 sight lines).  Although the \Copernicus\ sight
lines are essential for mapping O~VI absorption at distances less than
$\sim\:1$~kpc, the errors in $N$(O~VI)
(Appendix~\ref{sect_copernotes}) are very much larger than the \FUSE\
$N$(O~VI) errors.  The distance-selected subsets have the advantage,
however, of subdividing the data according to whether or not the sight
lines are strongly influenced by O~VI from the Local Bubble.  Finally,
within each of these categories, the samples were subdivided further
into subsets according to their \ROSAT\ classifications.

\subsubsection{The O~VI Volume Density $n$: (i) Differences for Near and Distant stars. \label{sect_rho_lb}}

\begin{figure*}[t!]
\hspace{2cm}\includegraphics[width=6cm]{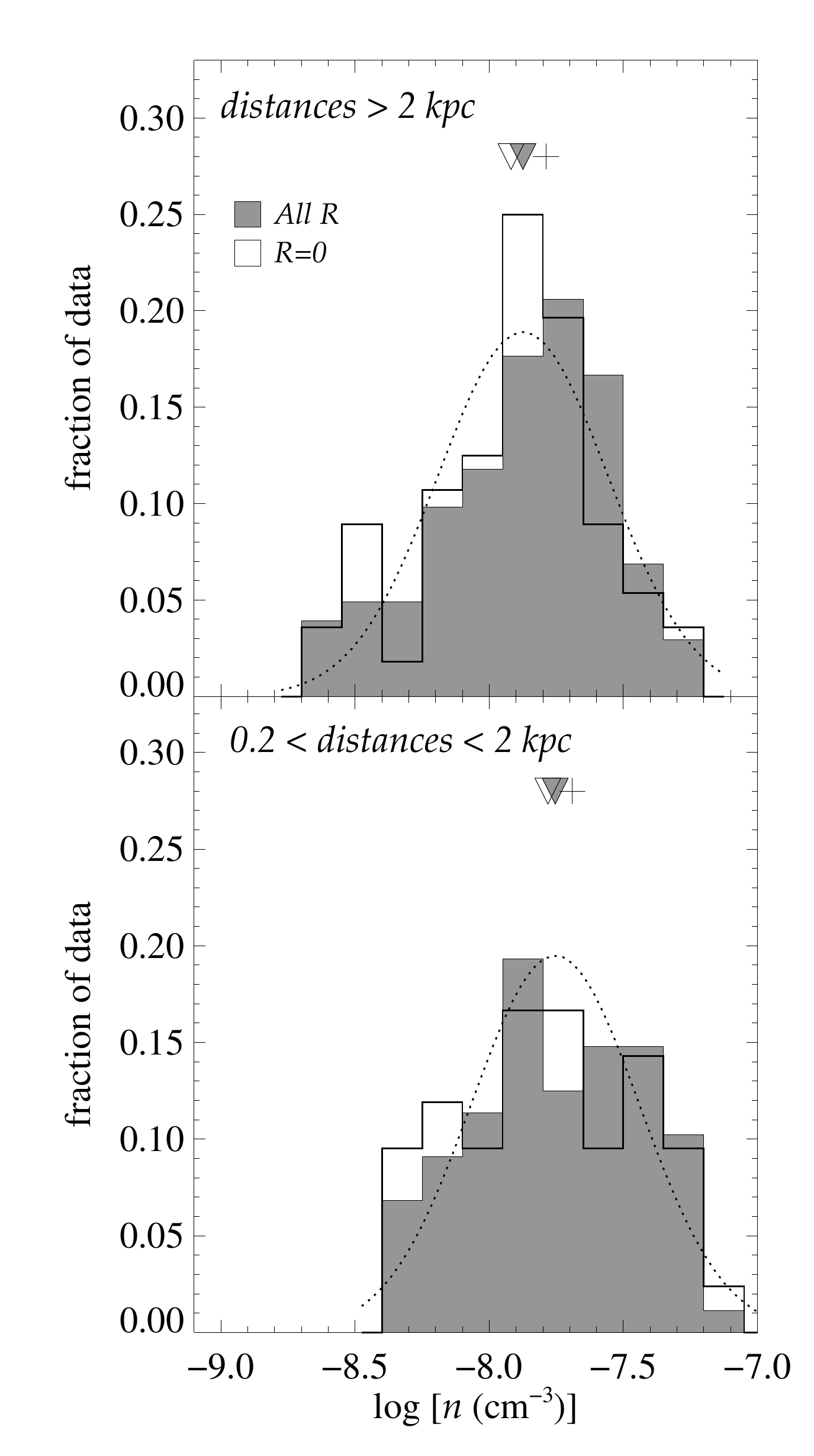}
\includegraphics[width=6cm]{fig10a}
\caption{\label{fig_rhos} {\bf Left:} Distribution of O~VI line of
  sight volume densities, $\log\:n\:=\:\log[N$(O~VI)/$d$], divided
  into two sets, one for stars beyond 2~kpc (upper panel), the other
  for stars in the interval $0.2 < d < 2$~kpc (lower panel).  The gray
  histogram shows the distribution of stars regardless of \ROSAT\
  class, while the white histogram shows only $R=0$ sight lines.  The
  dotted line shows a Gaussian with a mean and standard deviation for
  the All-$R$ samples given in Table~\ref{tab_n0}, while the inverted
  triangles show the means of each distribution. A cross marks the
  distance weighted average of $N$(O~VI). (More details are given in
  Table~\ref{tab_n0}.)  Note: distant high-latitude halo stars,
  extragalactic sight lines, and sight lines towards the Vela SNR and
  Carina Nebula are not included. Each histogram is normalized by the
  total number of objects making up the histogram. {\bf Right:} The
  same distributions, only with the Local Bubble (LB) column density
  [$N_{\rm{LB}}$(O~VI)$ = 1.11\times10^{13}$~\pcm ] removed for each
  sight line, and distances reduced by the radius of the LB, which we
  take to be 100~pc.}
\end{figure*}

Examples of the distributions of $n$ values are shown in
Figure~\ref{fig_rhos}. In each case, the distribution of $n$ for sight
lines regardless of their \ROSAT\ classification (``All-$R$'') is
given in gray, while $R\:=\:0$ sight lines are shown as white
histograms. All the histograms are normalized by the total number of
stars used to create the distribution. The top panels include stars
at ``far'' distances $d\:>\:2$~kpc; the bottom panels show stars at
``near'' distances of $ 0.2\: <\: d\: <\: 2$~kpc. The mean of a
distribution\footnote{$\bar{x}\:=\:\Sigma (\log n_i)/N$, where $N$ is
  the total number of $i$ points used. The standard deviation entries
  in Table~\ref{tab_n0} are then the usual $\sigma^2 = \Sigma(\log n_i
  - \bar{x})^2/N$; they are not the uncertainties of the listed
  averages.}  is shown by an inverted triangle, and a Gaussian profile
for the All-$R$ distribution using this mean and the standard
deviation given in Table~\ref{tab_n0} is shown with a dotted line. A
cross marks the distance-weighted average $\log[\sum N$(O~VI)/$\sum
d$].  As Figure~\ref{fig_rhos} shows, the histograms are reasonably
well approximated by log-normal distributions, enabling us to define
means and standard deviations consistent with a theoretical normal
distribution.

In parallel with these measurements, we constructed a synthetic
dataset of absorption line profiles in order to better understand the
distributions shown in Figure~\ref{fig_rhos} and the values tabulated
in Table~\ref{tab_n0}.  We were particularly interested in how well we
might reproduce the observed distributions if we based a simulation on
a single value of $n$, and hence whether the dispersion in the data
might really be some artifact of the quality of our data.  We
therefore took a set of random distances, used a fixed value of $n$ to
derive $N$(O~VI), created line profiles (adopting a fixed value of
$b\: = \:40$~\kms ), added a fixed amount of noise (similar to that
obtained for the majority of the \FUSE\ sight lines), and re-measured
$N$(O~VI) and the appropriate errors.  We also adopted the same
procedures for defining an upper limit to the column density as
described in \S\ref{sect_colsAOD} when lines were barely detected, and
included the upper limits to $n$ determined from the upper limits to
$N$(O~VI).  Finally, we calculated the `measured' values of $n$ and
determined how close they were to the original value.

The simulation provided several interesting results. How well we could
reproduce $\langle n \rangle$ (using the same methods described in the
previous section and shown in Table~\ref{tab_n0}) depended primarily
on the S/N we adopted when adding noise to the theoretical profiles.
With a high S/N, all lines were recovered and $n$ was
indistinguishable from the original value. With very low S/N data,
many lines had measured column densities which were either negative,
or were less than the upper limit determined from the noise.  As a
consequence, the mean value of $n$ became larger than the original
value (since an upper limit is larger than the true original value).

Translating these findings to our \FUSE\ and \Copernicus\ data sets
is not so straightforward, since the real data have differing S/Ns
(more precisely, the spectra all have different $N$(O~VI) upper
limits).  However, our simulations show that while the error in the
true mean of $n$ depends on the S/N of the data, it is equivalently
dependent on the relative number of lines which are lost (lower S/N
means more $n$ upper limits). Hence it is possible to simply count the
number of non-detections (upper limits) we find in our data and use
the results of the simulations as a guide to the likely
over-estimation of $\langle n \rangle$.

In the near sample ($0.2\:<\:d\:<\:2.0$~kpc) 14\% of our sight lines
show no O~VI absorption lines and have upper limits for $N$(O~VI). For
the far sample ($d\:>\:2$~kpc) only 7\% have upper limits. In our
simulation, we find that these percentages of non-detections cause an
over-estimate of $\sim\:12$\% and $\sim\:6$\% for the near and far
samples, respectively.  Hence, {\it if} the underlying O~VI column
density can be represented by a constant value, then the numbers shown
in Table~\ref{tab_n0} are likely to be over-estimated by only a small
amount after including column density upper limits.

Another result from the simulation was that the largest values of $n$
(in cm$^{-3}$) recovered arose from the weakest $N$(O~VI) lines, and
hence towards the closest stars. This was because weaker lines were
more easily lost in the noise, and were more likely to give {\it
  erroneously} large values of $n$, just because of noise
fluctuations\footnote{These inherently weaker lines were also more
  likely to give erroneously {\it small} or negative values of $n$ as
  well; however, since weak lines deemed below the noise were set to
  upper limits, the distribution of $\log n$ values discussed above
  does not include these small-$\log(n)$ outliers.}  Given this
possibility, we considered whether large values of $n$ measured in our
data were actually real, and not just a consequence of selecting lower
quality data.  In fact, values of $\log(n)\: >\: -7.5$ are found at
all distances and column densities and so are real over-densities in
the global distribution of O~VI absorption.  Moreover, seven sight
lines have $\log(n)\:>\:-7.0$, all of which arise towards the Vela
SNR.  We also noticed that most of the sight lines towards the Carina
Nebula also had large values of $n$.

Sight lines through the Carina Nebula (NGC~3372,
[$l$,$b$]$\:=\:[287.7,-0.8]$), of course, are not typical probes of
the Galactic ISM. The nebula is a giant H~II region ($1-2$ degrees in
radius in optical emission) powered by more than a dozen star
clusters, and although most of the gas has been blown away by the
clusters, low- and intermediate-mass star formation remains at the
peripheries \citep{smith03}.  The distance to Carina is somewhat
ambiguous, but appears to be in the range 2.5$-$3~kpc
\citep{feinstein95, walborn95_eta}.  Most of the stellar distances we
derive to stars towards Carina (Appendix~\ref{sect_distances} and
Table~\ref{tab_journal}) are consistent with this distance, as
expected.  The extent of the nebular in X-ray emission can be seen in
Figure~\ref{RASS_panels}e, which also shows the positions of the stars
observed by \FUSE .

It has been known for a long time that the spectra of Carina stars
show complex, multicomponent absorption line profiles, with individual
components distributed over several hundred \kms\ at both positive and
negative velocities \citep{walborn75,walborn_car82}. Absorption from
ions in both high and low ionization states is seen \citep[][ and
references therein]{walborn84,walborn02_eta,walborn07_eta}.  Most of
the stars in our \FUSE\ sample in this region are of very early
spectral type and have reasonably well determined continua, which
makes the detection of interstellar O~VI straightforward. They
generally show clear evidence for absorption at large negative
velocities away from the bulk of the absorption at $v\:\sim\:0$, high
velocity features which are not seen along other sight lines through
the ISM (except towards the Vela SNR --- see below). In some cases, a
component can be clearly resolved from the main complex; in other
cases, the additional negative velocity gas produces a discernable
asymmetry to the $v\:\sim\:0$ complex. It seems reasonable to link
this additional absorption to gas directly associated with the nebula.
Detecting high positive velocity gas in absorption is somewhat harder,
since the H$_2$ 6$-$0 R(4) line lies at a wavelength that corresponds
to any O~VI absorption offset from rest by +121~\kms , potentially
masking any high positive velocity components.  Nevertheless, it is
clear that there is no preponderance of very strong O~VI absorption
lines at $\approx\:100$~\kms , since the profiles of the 6$-$0 R(4)
lines are always close to the shape predicted by fitting other H$_2$
lines in different regions of the spectra (see \S\ref{sect_cont}).
Only towards HD~093146 (\#92) is there good evidence for excess O~VI
absorption at positive velocities.

As noted in \S\ref{sect_SNR}, the Vela SNR
([$l$,$b$]$\:=\:[263.9,-3.3]$) is much closer than Carina, at a
distance of only $250\pm30$~pc \citep{cha99}. Sight lines which pass
through Vela again show complicated multi-component absorption line
complexes \citep[see, e.g.][ and references.~therein]{slavin04}.  An
initial study of O~VI absorption was made using four stars situated
behind Vela \citep{slavin04}; our \FUSE\ sample includes an additional
three stars which lie behind the SNR: HD~074920, HD~075309, and
HD~074711 (\#77$-$79). The first two of these show evidence for
multicomponent absorption. HD~075309 in particular appears unusual,
with very weak O~VI absorption at $v_\odot\:=\:33$~\kms , and much
stronger absorption at $-74$~\kms .  Again, the kinematic structure of
the O~VI absorption seen towards HD~074920 and HD~075309 seems quite
atypical of that seen in the rest of the Galactic ISM.

A full analysis of the conditions within the Carina Nebula and Vela
SNR inferred from absorption line data is beyond the scope of
this paper.  Nevertheless, the sight lines towards these regions
present us with a dilemma. Although they present a particularly
interesting opportunity to study phenomena local to active star
formation and destruction, they also present particularly extreme
examples of O~VI absorption and probably do not represent the more
general distribution of O~VI absorbing gas in the ISM. For this
reason, we excluded Carina and Vela sight lines from all the
measurements presented in Table~\ref{tab_n0}.

We make one final note concerning the possible effect that the
non-detections of O~VI might have on the values of $\langle n \rangle$
given in Table~\ref{tab_n0}. For the distance-weighted statistic
$\Sigma N\rm{(O~VI)}$/$\Sigma d$ (column 4 of Table~\ref{tab_n0}), it
is also possible to set all upper limits to zero and see whether
$\langle n \rangle$ is very different than that calculated using the
upper limit values\footnote{Obviously, $\log[N($O~VI)]/$d$ cannot
  be calculated if $N$(O~VI) is set to zero, which is why we could not
  calculate $\langle n \rangle$ from a distribution using such a
  sample of column densities.}.  When calculated this way, we find that
$\langle n \rangle$ is only 7\% and 14\% smaller for the near and far
(``All-$R$'' --- see below) sample, respectively.

\citet{shelton94} first showed that the LB has a higher average
density than the rest of the Galactic disk. (We see this effect in
Fig.~\ref{fig_Nwithde} in \S\ref{sect_Nwithb}.)  \notetoeditor{This is
  a forward reference to a figure presented later. Please don't insert
  Fig.~\ref{fig_Nwithd} at this point} To better reflect the value of
$\langle n \rangle$ in the ISM beyond the LB, we recalculated all the
statistics after subtracting a column density representative of that
from the LB, for which we used a recent value of the LB volume density
of O~VI found by SL06, $n_{\rm{LB}}\:=\:3.6\times10^{-8}$~cm$^{-3}$.
For a bubble with a radius of 100 pc, this corresponds to a column density
$N_{\rm{LB}}$(O~VI)$\:=\:1.11\times10^{13}$~\pcm . We also subtracted
100~pc from all stellar distances. Values of $\langle n \rangle$
corrected in this way are given in Table~\ref{tab_n0}; these are
labelled as ``$-$LB'' measurements, compared to the values for which
no LB contribution is removed, ``+LB''.

The corrected distributions of $\log(n)$ are shown in the right-hand
stack-plot of Figure~\ref{fig_rhos} and tabulated in
Table~\ref{tab_n0}.  For the distant $d\:>\:2$~kpc stars, it is
unsurprising that there is little difference (0.05~dex) between the
+LB and $-$LB means. For the nearer ($0.2\:<\:d\:<\:2.0$~kpc) stars,
the difference is far more pronounced ($0.13-0.18$~dex), as would be
expected. Moreover, the near and far +LB samples (for the same \ROSAT\
classes) also show large differences ($0.09-0.14$~dex); yet with the
contribution from the LB removed, the means between the near and far
samples are nearly identical.  We conclude that {\it after} correction
for the Local Bubble, the average line of sight density of O~VI is the
same for all distances beyond 200~pc.  Note that in the bottom-right
panel, the width of the near, $-$LB sample is much wider than in the
other three panels. This is because of the increase in the relative
errors in $N$(O~VI) --- and hence $\log(n)$ --- that arise after
subtracting $N_{\rm{LB}}$(O~VI). This can also be seen in the standard
deviations listed in column 3 of Table~\ref{tab_n0}, which are much
larger for the near, $-$LB sample.

\subsubsection{The O~VI Volume Density $n$: (ii) Differences for Different \emph{ROSAT} Classes. \label{sect_rho_rosat}}

We repeated the statistical tests described above, for stars separated
by \ROSAT\ class.  The results can be seen in Table~\ref{tab_n0},
where average densities are broken into categories defined by samples
of sight lines which include ``All-$R$'', or which have $R\:=\:0$ or
$R\:>\:0$.  Table~\ref{tab_n0} shows that within each of the four
near/far, +LB/$-$LB classes, the differences in the means between the
$R\:=\:0$ and the $R\:>\:0$ sub-samples are between 0.05 and 0.1 dex.
Are such differences significant?

To test if these differences are real, we used two well established
tests on the +LB sample of stars.  We first considered the results
from a Student's $t$-test, which tests the null hypothesis that the
means of two populations are equal. For the $d\:>\:2$~kpc stars, we
found a $t$-statistic of $t\:=\:-1.6$, and a probability that we might
incorrectly reject the hypothesis that the means are equal of only
$P\:=\:0.11$. For the $0.2\:<\:d\:<\:2$~kpc stars, we found
$t\:=\:-0.8$, and $P\:=\:0.44$. We would conclude from these numbers
that there is a small difference in $n$ between $R\:=\: 0$ and
$R\:>\:0$ sight lines beyond 2~kpc, but that no significant difference
can be found nearer than 2~kpc.

Student's $t$-test is a parametric test whose accuracy relies on the
fact that the two populations being tested are normally distributed.
To avoid problems in comparing two distributions when the underlying
distributions are not known, we used the standard non-parametric
Kolmogorov-Smirnov (K$-$S) test. For the distant stars, the K$-$S $D$
parameter was measured to be $D\:=\:0.3$, with $P\:=\:0.014$, while
for the nearer stars, we found $D\:=\:0.13$ and $P\:=\:0.84$. These
numbers confirmed the conclusion that the $R\:=\: 0$ and $R\:>\:0$
have different distributions of $\log(n)$ for distant stars, but very
similar means for the nearby stars. The numbers from the K$-$S tests
and from the Student's $t$-test changed little if we used the $-$LB
instead of the $+$LB sample.

\begin{figure}[t!]
\hspace*{-1cm}\includegraphics[height=6.5cm]{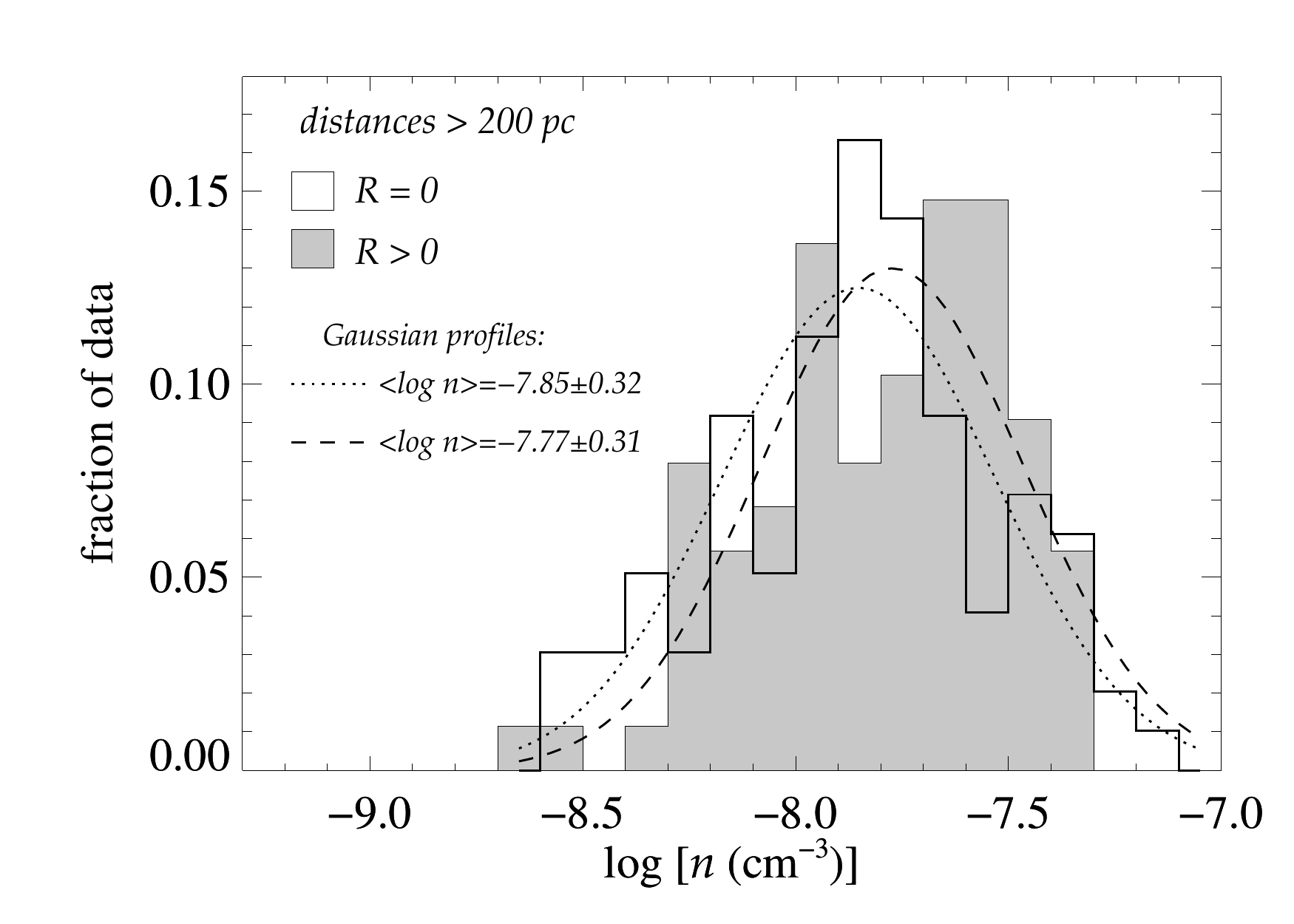}
\caption{\label{fig_diffRs} Distribution of O~VI line of sight volume
  densities for all stars in the Galactic disk (thereby excluding halo
  stars) beyond 200~pc, divided into $R\:=\:0$ (white histogram) and
  $R\:>\:0$ (gray histogram) subsets for the +LB samples. Each
  histogram is normalized by the total number of objects making up the
  histogram. Gaussian profiles for the $R\:=\:0$ (dotted line) and
  $R\:>\:0$ (dashed line) samples are drawn with means and standard
  deviations given in \S\ref{sect_rho_rosat}.}
\end{figure}

Figure~\ref{fig_diffRs} shows the distribution of $\log(n)$ values for
the two +LB \ROSAT\ classes with the two distance subsets combined,
i.e., for all stars beyond 200~pc. For the $R\:=\:0$ and the $R\:>\:0$
samples, the means and standard deviations are $\langle \log(n)
\rangle\:=\: -7.85\pm 0.32$ and $-7.77\pm0.31$, respectively.  A
$t$-test shows that the likelihood of a difference between the two
sub-classes has increased by combining the near and far samples:
$t\:=\:-1.9$, $P\:=\:0.07$. A K$-$S test gives $D\:=\:0.20$,
$P\:=\:0.03$. As we mentioned above, these tests were performed with
Vela and Carina sight lines excluded. If we include those sight lines,
most of which have $R\:>\:0$, the two distributions become very
different ($t\:=\:-3.6$, $P\:=\:3\times10^{-4}$; $D\:=\:0.26$,
$P\:=\:1\times10^{-3}$) because the lines of sight clearly intercept
regions with higher densities. Again, these values change little if we
use the $-$LB instead of the $+$LB sample.

We discussed the likely origins for higher (on average) values of $n$
towards $R\:>\:0$ sight lines in \S\ref{sect_bubbles} and
\S\ref{sect_SNR}.  Bubbles visible in X-rays and the volumes within
SNRs probably contain some gas in the temperature range that favors
the production of O~VI, and although the effect is small, sight lines
towards these structures show elevated O~VI column densities.

We note a few interesting exceptions, however.  HDs 199579 and 110432
have $R=2$, and yet they both show virtually no O~VI absorption at our
level of sensitivity.  A more spectacular example of this effect is
$\gamma$~Cas observed by J78.  The upper limit of $N({\rm O~VI}) <
10^{12}\:{\rm cm}^{-2}$ is considerably below the \FUSE\ detection
limit, and yet this star is centered on an extended source that is
19\arcsec\ in diameter, according to the \ROSAT\ {\it Bright Source
  Catalog} \citep{voges99}.  However $\gamma$~Cas is only at a
distance of 0.2~kpc, which means that the detected X-rays are emitted
within 0.02~pc of the star.  This dimension is far smaller than a
classical, well developed bubble.  Thus, the emission could arise from
the more extended parts of the hypersonic wind from the star, rather
than an established interface region. It is also likely that if this
star was ten times further away, at the typical \FUSE\ sample
distances, the X-ray emission would probably be undetected.

The reason for being able to see the differences between the $R\:=\:0$
and the $R\:>\:0$ sight lines to distant stars and not nearby ones is
likely due to selection effects. X-ray fluxes fall with distance as
$1/d^2$, and neutral gas absorbs X-rays, so for the more distant stars
we probably only see the very largest, brightest interstellar bubbles
which have the largest enhancements of $N$(O~VI).  Nearby stars are
less likely to be members of dense star-forming regions, and their
X-ray fluxes are probably easier to detect; so, for these, we assign a
$R\:>\:0$ category even if the circumstellar bubbles are not
particularly bright and there is little enhancement in $N$(O~VI) from
the bubble.

Finally, we note one other experiment we performed with the data.  We
attempted to subtract a column density that might be representative of
circumstellar O~VI absorption, $N_{\rm{circ}}$(O~VI).  Our aim was to
find a value of $N_{\rm{circ}}$(O~VI) which could be subtracted from
the $R\:=\:2$ sample, leaving a sample with a mean which was the same
as the $R\:=\:0$ sample.  Unfortunately, we were unable to find a
plausible value of $N_{\rm{circ}}$(O~VI); the distribution of $n$ for
the $d\:>\:2$~kpc sample was largely unaffected by subtracting a small
$N_{\rm{circ}}$(O~VI) from each sight line, but it greatly changed the
$0.2\: <\: d\: <\: 2$~kpc sample, sending the majority of the $n$
values to values less than zero. This probably reflects the selection
effect mentioned above: the value of $N_{\rm{circ}}$(O~VI) towards the
more distant stars is probably higher than $N_{\rm{circ}}$(O~VI)
towards nearby stars, because our $R\:>\:0$ categories include very
large, very X-ray-luminous bubbles at large distances, and smaller,
less luminous (but still detectable by \ROSAT ) bubbles at smaller
distances.

\subsubsection{Bias Toward Stars with Small Reddening}

In large part, our selection of targets is governed by the limitation
that the stars must not be so faint at 1032$\,$\AA\ that a
prohibitively large amount of observing time is needed to obtain
spectra with respectable values for the S/N.  One consequence of this
is that more distant stars tend to have much lower than normal amounts
of reddening per unit distance.  For closer stars, this selection is
not as strong.  It is fair to ask whether or not this bias in the
sampling of the Galactic disk volume distorts our conclusions on the
general distribution and properties of the regions that hold O~VI, and
in particular, on the volume densities we have derived above. This
could be an important consideration if O~VI-bearing gas is strongly
correlated or anticorrelated with the cold gas that causes reddening.
For instance, surveys of the soft X-ray background radiation exhibit
an anticorrelation, which probably arises from the effect that the
X-ray emitting gas displaces the cold material \citep{mccammon90}.

\begin{figure}[t!]
\hspace*{-1cm}\includegraphics[width=9cm]{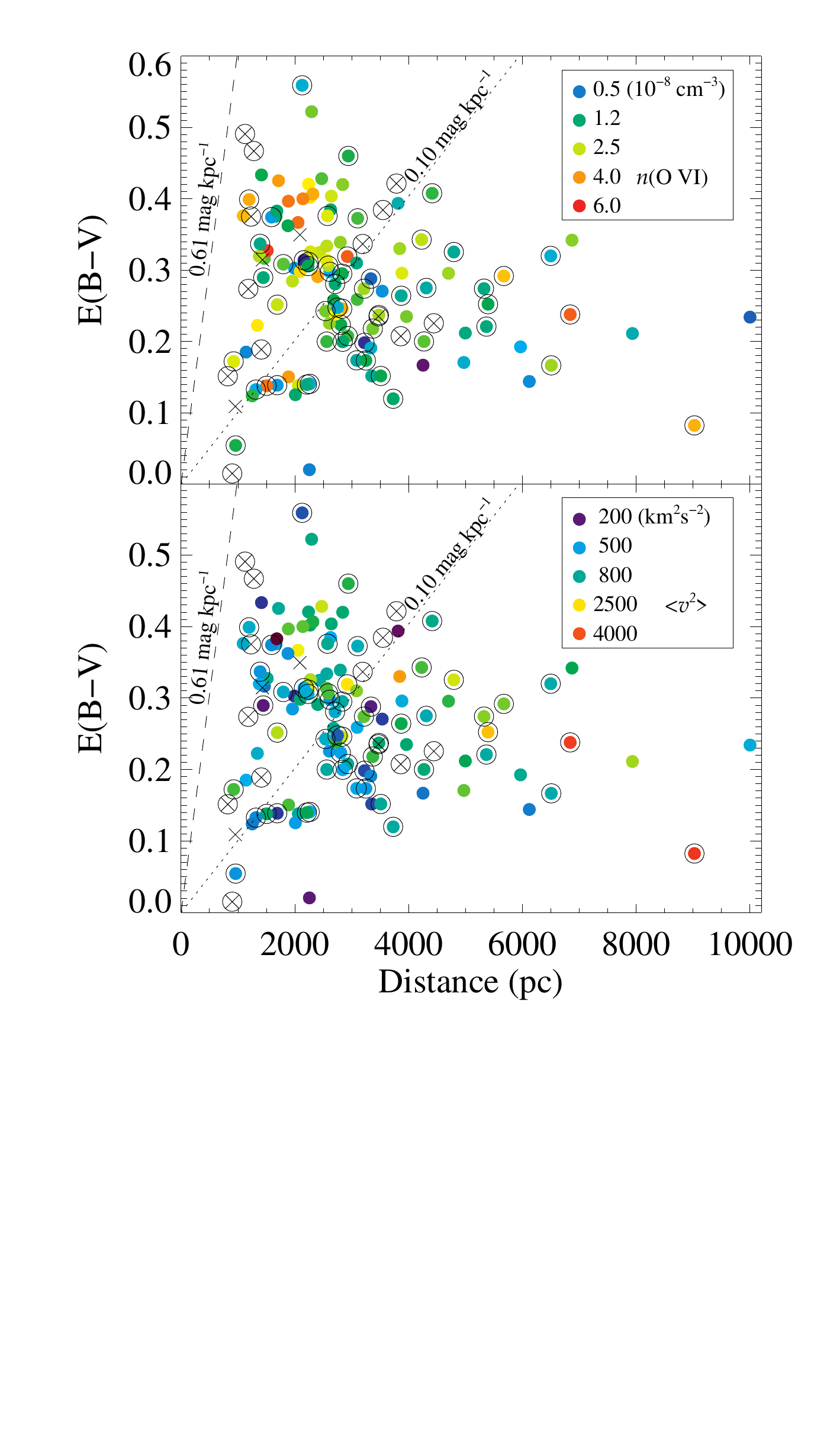}
\vspace{-5cm}\caption{\label{fig_ebj_bias} The distribution of targets with regard
  to their color excesses $E(B-V)$ and distances $d$, excluding
  targets towards the Carina Nebula or the Vela supernova remnant
  (which are atypical).  In each panel, the dotted line indicates the
  median value of $E(B-V)/d\:=\:0.10$ mag~kpc$^{-1}$ in our survey,
  and the dashed line represents the general average of 0.61
  mag~kpc$^{-1}$ in our part of the Galaxy.  Points with circles
  around them are targets that have \ROSAT\ classifications of
  $R\:=\:0$, and the crosses represent cases where only upper limits
  for $N$(O~VI) could be obtained.  The colors in the top panel
  indicate $n$(O~VI), while those in the bottom panel indicate the
  second moment of the absorption profile $\langle v^2 \rangle$.}
\end{figure}

To examine the issue of whether or not we must acknowledge and attempt
to correct for a bias in our survey, we must determine the likelihood
that the character of the O~VI, either its average volume density or
velocity dispersion, is affected somehow by the reddening per unit
distance.  Figure~\ref{fig_ebj_bias} shows the distribution of all
stars in our FUSE survey on a diagram of $E(B-V)$ vs. distance, with
the points having a color coding that represents either the O~VI
volume density $n$ (top panel) or the second moment of the absorption
profile $\langle v^2 \rangle$ (bottom panel).  The expected selection
effect in our survey is clearly evident from the fact that there are
no points in the upper right-hand portion of the diagrams.
Nevertheless, from the distribution of colors shown in the plots, we
gain the impression that internally within our collection of targets,
there is no tendency for the results to be driven much by the
reddening per unit distance.  We can easily imagine that if the
diagram were populated more evenly, i.e., without the exclusion of
cases where color excesses increase roughly in proportion to distance,
the outcome would not be appreciably different from what we have
already found.

Looking at the problem more quantitatively, we find a median reddening
per unit distance of 0.10 mag~kpc$^{-1}$, represented by the dotted
line in each panel of the figure\footnote{Our median reddening per
  unit distance is considerably lower than the general average of
  $0.61\,{\rm mag~kpc}^{-1}$ \citep{spitzer78_p155, fitzgerald68}.
  Whether or not this is important depends on the volume occupation
  fractions of gases at different densities.}.  For the half of our
measurements that are above this value, the median value of $n$ is
$1.64\times 10^{-8}~{\rm cm}^{-3}$, while for the half of the sample
below the median $E(B-V)/d$, the median value for $n$ is $1.30\times
10^{-8}~{\rm cm}^{-3}$.  This outcome is opposite to our expectation
that regions with less reddening than usual might be more likely to
hold more than the usual amount of O~VI.  However, an apparent
correspondence between $E(B-V)/d$ and $N({\rm O~VI})/d$ could arise
purely from the fact that errors in a common denominator, i.e, the
distances, might make the quantities appear to be mildly correlated
when, in fact, they are not.  A K-S test on whether the two separate
samples of $n$ could arise from the same population indicate that
there is an 8\% probability that differences this large or larger
could arise simply from chance, given that the two samples are drawn
from the same population.  (Such a test cannot be run for the velocity
dispersions because they are indeterminate when O~VI is not detected.)

Finally, we note that even if the selection of sight lines is biased
against highly reddened sight lines, the clouds which cause the
reddening are likely to occupy only a small volume compared to the
volume that the O~VI-absorbing gas occupies. If so, then the value of
$n$ we calculate along a line of sight will be unaffected by whatever
the extinction happens to be.

\subsection{O~VI Mid-Plane Volume Density \label{sect_NwithZ}}

The correlation between $N$(O~VI)$\sin|b|$ and the height $z$ above the plane
of the Galaxy has been determined before \citep{ebj78b,savage03}. However, the
increased size of our sample of disk stars, and a well determined column density
for the LB, makes it worthwhile  revisiting the problem.

The O~VI volume density as a function of $z$ is usually defined such
that $n = n_0 f(|z|)$. The column density for any individual sight
line at a height $|z| = d\sin |b|$ above or below the Galactic plane
is then


\begin{eqnarray}
  N{\rm{(O~VI)}} & = & \int_{0}^{d} n(r)\:dr  =  \int_{0}^{|z_{\rm{star}}|} n(|z|) \csc(|b|)\:d(|z|) \nonumber \\
                 & = & n_0\: \csc(|b|) \int_{0}^{|z_{\rm{star}}|} f(|z|) \:d(|z|)\:, 
\label{eqn_z}
\end{eqnarray}

\noindent
where $r$ is the distance along a sight line.  Conventionally, it has
been assumed that the O~VI volume density, $n$, falls off
exponentially with height above the disk, so that $f(|z|) =
e^{-|z|/h}$ which means that

\begin{equation}
n = n_0 e^{-|z|/h} \label{eqn_exp}
\end{equation}

\noindent
where $h$ is the scale height of the O~VI absorbing gas.
Equations~\ref{eqn_z} and \ref{eqn_exp} then lead us to an
expression for the column density projected on a line perpendicular to
the Galactic plane,

\begin{equation}
N{\rm{(O~VI)}}\,\sin |b | \: = n_0 h (1-e^{-|z|/h}) \label{eqn_NwithZ}
\end{equation}

Obviously, to measure $n_0$ and $h$, we can fit the function given in
equation~\ref{eqn_NwithZ} to the data. The data we include are those
from our own \FUSE\ survey, data from \Copernicus , the WD sample of
SL06, the halo stars observed by Z03, and the extragalactic sight
lines measured by \citet{savage03} and \citet{wakker03}. However,
including the latter dataset forces us to consider an important
problem uncovered by these authors, which we must discuss before we
attempt to derive $n_0$ and $h$, namely the difference between sight
lines in the northern and southern Galactic hemispheres. Further, we
must also introduce the problem that, although the density of O~VI
does decrease in the way suggested by equation~\ref{eqn_NwithZ}, the
layer of O~VI absorbing gas in the disk is not smoothly stratified,
but instead is rather clumpy. We discuss these two problems below.

\subsubsection{The North-South Divide \label{sect_NS}}

In their survey of extragalactic sight lines, \citet{savage03}
reported an excess of $0.20 - 0.30$ dex in the average value of $\log
[N$(O~VI)$\sin b]$ for the north Galactic Polar region towards sight
lines at $b\:>\:45^\circ$ compared to lower latitude directions in the
north and to the entire southern Galactic sky.  To investigate the
asymmetry in more detail, we examined whether the extragalactic O~VI
column densities measured at latitudes $b>0$ --- the `N' or `North'
sample --- were statistically different from those measured at $b<0$
--- the `S' or `South' sample.  In all cases, column densities were
reduced by a factor $\sin |b|$.  We
first considered the results from a Student's $t$-test; for the N and S
samples of $\log [N$(O~VI)$\sin b]$, we found a $t$-statistic of
$t=4.1$, and a probability that we might incorrectly reject the
hypothesis that the means are equal of only $P\:=\:1\times10^{-4}$.  A
K-S test yields $D\:=\:0.42$, $P=4\times10^{-4}$.  Such results
clearly bear out the assertion of Savage~\etal\ that a significant
difference between $N$(O~VI) exists in the northern and southern
Galactic hemispheres towards extragalactic sight lines.

As described in \S\ref{sect_other_savage}, we decided to use only the
higher-quality data from \citet{savage03} and \citet{wakker03}
(Q$\:>\:2$) in our analysis. If we repeat these two tests for this
sub-sample alone, the difference between the N \& S samples is
smaller, but still significant. Student's $t$-test gives $t=2.4$ and
$P\:=\:0.02$, while the K$-$S test gives $D\:=\:0.39$ and $P\:=\:0.05$.

Despite these variations, which occur in the Milky Way halo, we see no
difference in our disk stars for similar N and S samples.  For stars
beyond 200~pc, there is no clear difference between the two samples
when measured by a $t$-test ($t\:=\:0.5$, $P\:=\:0.62$) or a K-S test
($D\:=\:0.11$, $P\:=\:0.62$). This conclusion holds whether or not a
correction is made to the observed column densities for the Local
Bubble.

For this reason, in fitting the theoretical relationship of \nup\ with
$|z|$ given in equation~\ref{eqn_NwithZ} to the data, we use all our
sight lines regardless of whether they lie in the northern or southern
Galactic plane, except for the extragalactic sight lines: we derive
$n_0$ and $h$ twice, once each using the northern and southern
extragalactic sight lines.

\subsubsection{Clumpiness in the O~VI Absorbing Gas and Errors Used for \nup \label{sect_clump}}

To measure $n_0$ and $h$, we fitted the function given in
equation~\ref{eqn_NwithZ} to the data by varying $n_0$ and $h$ and
minimizing the $\chi^2$ statistic between the theoretical function and
the data. The first obvious result we found was that although the data
were clearly correlated in the way described by equation~\ref{eqn_NwithZ},
the reduced $\chi^{2}_\nu$ was much greater than the expected value of
unity\footnote{Here, we use the conventional definition of
  $\chi^{2}_\nu = \chi^{2}/\nu$, where $\nu$ is the number of data
  points used in the fit minus the number of variables used, in this
  case 2.}.

At face value, we might say that large values of $\chi^{2}_\nu$ arise
either because the assumed errors in the data are inadequate or that
the underlying model must be wrong. We discuss the second of these
conclusions below. As far as the first conclusion is concerned, we
have taken special care to derive realistic errors for the quantities
measured. A similar problem was found by \citet{savage90_al3} for the
distribution of Al~III in the Galactic plane. Their solution was to
introduce an additional ``error'', which takes into account the
patchiness or clumpiness of the absorbing gas that is ignored by using
equation~\ref{eqn_NwithZ}. Although the density of O~VI follows, primarily,
an exponential decrease with height above the plane, the distribution
is not smooth and planar, but is, at some level, clumpy. Adding an
additional ``error'', $\sigma_{\rm{CL}}$, allows for deviations caused
by this additional clumpiness when seeking an acceptable value of
$\chi^{2}_\nu$.  The $\sigma_{\rm{CL}}$ term is applied assuming that
the quantity is independent of distance, an assumption which we
demonstrate to be true in \S\ref{sect_Nwithde}.

Given the introduction of $\sigma_{\rm{CL}}$, we can combine the
natural deviations attributed to this term with the experimental
errors in determining \nup .  Ordinarily, this operation would be
problematic because variations of the parameter of a fit would respond
differently to the two different kinds of errors. In our case, we are
fortunate to have most of our measurements made at heights $|z|\:<\:h$. When
this is true, the slope of the dependent variable $x$ in the fitting
function in relation to the data points $y$ is fixed to unity. With
this constraint, errors in $x$ are equivalent to errors in $y$. For
each data point, therefore, we describe a total error exclusively in
terms of an error in just the $y-$direction equal to a value

\begin{equation}
\sigma_T^2 \:=\: \sigma_N^2 + \sigma_d^2 + \sigma_{\rm{CL}}^2 .\label{eqn_ysig}
\end{equation}

\noindent
These individual errors are all relative errors: $\sigma_N$ applies to
the measurement error in the column density\footnote{As discussed in
  \S\ref{sect_voigt}, the column density errors are asymmetric due to
  the different continuum errors. Here, for $\sigma_N$, we simply take
  the average of the two column density errors. As we will show,
  however, $\sigma_N\: \ll \: \sigma_d$, so the exact errors in the
  column densities have little effect on the derivation of $n_0$ and
  $h$.}, $\sigma[N$(O~VI)]/$N$(O~VI); $\sigma_d$ is the relative error
in the distance, $\sigma(d)/d$; and $\sigma_{\rm{CL}}$ is the
additional fractional error we derive by requiring $\chi^{2}_\nu$ to
be unity when minimizing the fit of \nup\ with $|z|$ to the data.  The
absolute error in each point is then simply
\begin{equation}
\sigma[N({\rm{O~VI}}) \sin |b|] \:=\: N({\rm{O~VI}}) \sin |b| \times \sigma_T\:.
\end{equation}

It should be understood that $\sigma_{\rm{CL}}$ is introduced {\it
  not} simply to justify a model whose functional representation is
wrong. In principle, we could propose some other trend of decreasing
$N$(O~VI) with distance from the Galactic plane, but we would still
require the use of $\sigma_{\rm{CL}}$ to account for an intrinsic
scatter in the observed O~VI column densities that exceeds the
measurement errors.

\subsubsection{The Mid-Plane Density and Scale Height of O~VI \label{sect_finally}}

\begin{figure*}[t!]
  \vspace{1cm}\begin{minipage}[c]{0.45\linewidth}
    \caption{\label{fig_NwithZ} {\bf Top:} Plot of O~VI
      column density reduced by the sine of the Galactic latitude $b$,
      against the height of a star above (or below) the plane of the
      Milky Way, $|z|\:=\:d\sin|b|$. Point shapes, explained by the key
      in the top left corner, refer to datasets obtained by various
      authors, and include the measurements made in this paper
      (``FUSE''), those obtained by \Copernicus\ \citep{ebj78a}, local
      WDs discussed by \citet{savage06}, sight lines towards the Vela
      SNR \citep{slavin04, ebj_vela_76}, those towards halo stars
      \citep{zsargo03}, and the extragalactic sight lines observed by
      \citet{wakker03} and \citet{savage03}.  Points colored black refer
      to stars which have \ROSAT\ class $R\:=\:0$, while gray points
      have $R\:>\:0$ (see \S\ref{sect_bubbles} and \S\ref{sect_SNR}).
      The extragalactic sight lines are shown as crosses at an
      arbitrarily large distance; they have have no \ROSAT\
      classification, and are separated into sight lines above
      (Northern, or `N' sight lines, left group of crosses) and below
      (Southern, or `S' sight lines, right group of crosses) the
      Galactic plane. Their distances are slightly offset from each
      other so that the errors in the column densities can be seen.
      Two-sigma upper limits are shown by downward-pointing arrows
      attached to open symbols. (The type of symbol again indicates
      which dataset the limits are taken from.) The size of a point
      indicates the relative uncertainty in the distance to the star: a
      key at the lower right indicates how the size is related to the
      uncertainty (large points represent smaller uncertainties so that
      they are more influential in guiding the eye). An absolute error
      of $\pm 30$~\% in any distance is shown below this key.  {\bf
        Bottom:} The same plot, only with a contribution from the Local
      Bubble subtracted from every sight line [$N_{\rm{LB}}$(O~VI)$ =
      1.11\times10^{13}$~\pcm ], and a distance of 100~pc removed from
      every sight line. Only $R\:=\: 0$ stars are plotted.  The solid
      line shows the best fit to the data assuming that the O~VI volume
      density $n$ varies as $n = n_0\: e^{(-|z|/h)}$ where $n_0$ is the
      mid-plane density and $h$ is the scale height.  Although the value
      of $n_0$ remains unchanged (because all stellar sight lines were
      used in fitting $n$ to the data), the value of $h$ is different
      depending on whether N or S extragalactic sight lines are used.
      The two curves show the fit for $h\:=\:4.6$~kpc (N) and
      $h\:=\:3.2$~kpc (S).}
  \end{minipage}
  \begin{minipage}[c]{0.45\linewidth}
    \includegraphics[width=11cm]{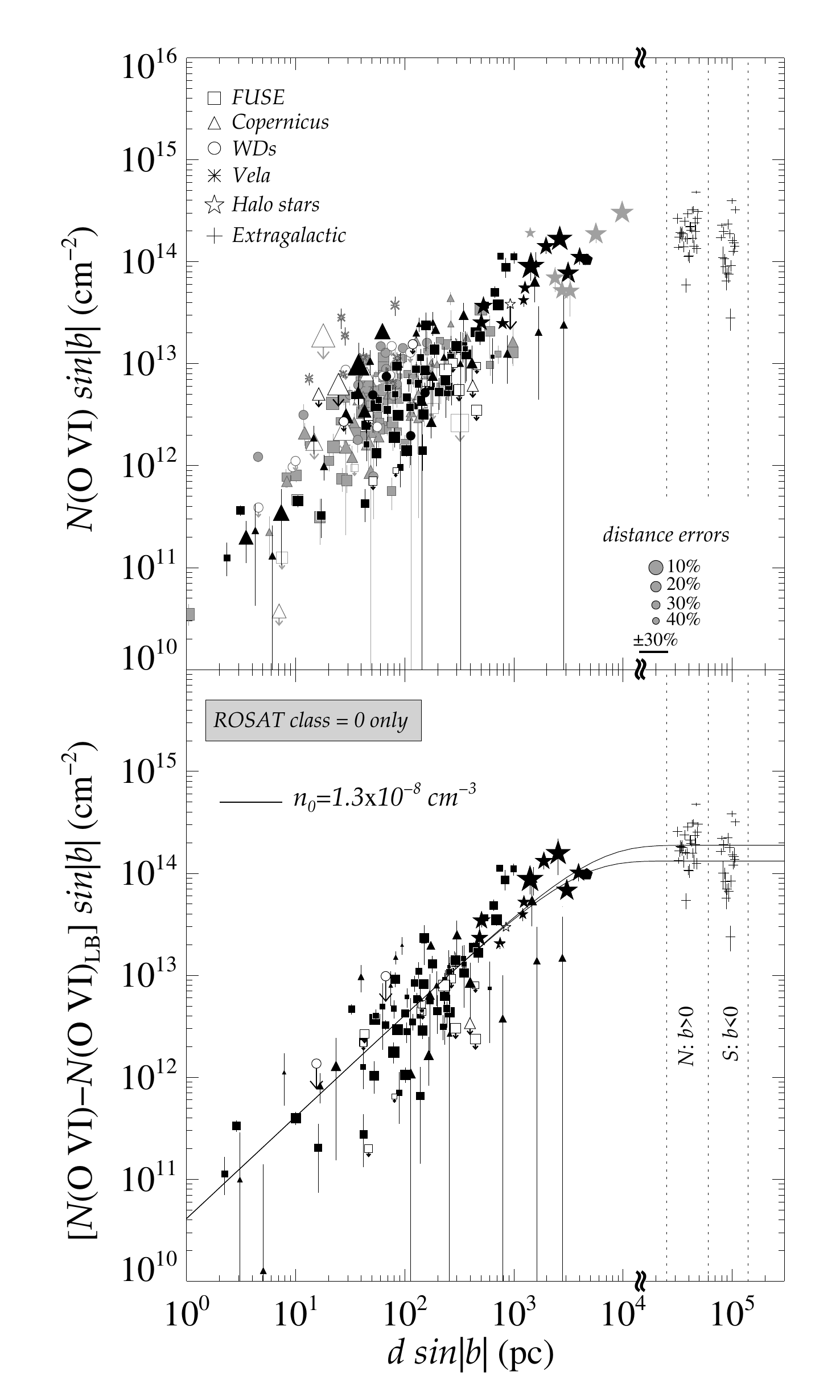}
  \end{minipage}
\end{figure*}

A plot of $\log [N{\rm{(O~VI)}}\,\sin |b|]$ with $\log(|z|)$ is shown
in the top panel of Figure~\ref{fig_NwithZ}, where we distinguish
between $R\:=\:0$ (black points) and $R\:>\:0$ stars (gray points).
In \S\ref{sect_rho_lb} we found evidence that the density of the ISM
beyond a few hundred pc was better represented with the removal of a
contribution from the LB. Hence, to derive $n_0$ and $h$, we used the
$R\:=\:0$ sample of stars, and subtracted a column density
$N_{\rm{LB}}$(O~VI)$\:=\:1.11\times10^{13}$~\pcm\ for the LB along
each sight line. Accordingly, we also subtracted the LB radius ---
100~pc --- from all distances. Sight lines with only upper limits to
$N$(O~VI) have no $\sigma_N$ errors, yet values are needed for
calculating $\chi^2$. For these points, we used the median $\sigma_N$
(0.08~dex) for the sample with detected O~VI.

The fit of \nup\ with $|z|$ must be made in log-log-space, for the
following reasons. In a $\chi^2$ fit, the errors in each point
determine the weight that that point gives to the value of $\chi^2$.
Unfortunately, the absolute error in the column density is
proportional to $N$(O~VI)\footnote{This is because strong lines are
  wider at the continuum, and have higher continuum errors. Since
  $\sigma(N)$ in the \FUSE\ data is dominated by continuum errors,
  $\sigma(N)$ and $N$ are roughly correlated, even though there is no
  correlation between distance and the quality of the data.}, and
because (as we will show in \S~\ref{sect_Nwithde}) $N$(O~VI) is
proportional to the distance to a star, the absolute error in the
column density is also proportional to distance.  Similarly,
$\sigma(d)$ is more or less a fixed constant of $\approx 20-30$~\% of
$d$ (see Appendix~\ref{sect_distances}), so the absolute error in $d$
also increases with $d$.  The result is that the final absolute error
is proportional to distance, so that more distant stars would have had
lower weights in the $\chi^2$ fit had we chosen the linear
representation.  If, instead, we fit $\log [N({\rm{O~VI}}) \sin |b|]$
against $\log |z|$ (i.e., a fit in log-log-space), then the error is
the same relative error as that given in equation~\ref{eqn_ysig}, and is
no longer proportional to the distance.

With the data divided into two samples containing separate northern
and southern extragalactic sight lines (\S\ref{sect_NS}) we found the
following:

\begin{eqnarray*}
{\bf N}:\:\: n_0 & = & 1.33\times10^{-8}\:\:{\rm{cm^{-3}}}, \\
         h   & = & 4.6\:\rm{kpc}, \:\: \sigma_{\rm{CL}}\:=\:0.25\:{\rm{dex}}\\
{\bf S}:\:\: n_0 & = & 1.34\times10^{-8}\:\:{\rm{cm^{-3}}}, \\
         h   & = & 3.2\:\rm{kpc},  \:\: \sigma_{\rm{CL}}\:=\:0.28\:{\rm{dex}}
\end{eqnarray*}

\noindent
Unsurprisingly, the value of $n_0$ does not depend on whether the N or
S samples are used, since $n_0$ depends primarily on the data below
$|z|\:\simeq 1$~kpc. Conversely, the values of $h$ do depend on which
extragalactic sight lines are used. As described in
\S\ref{sect_clump}, the values of $\sigma_{\rm{CL}}$ are set to give
$\chi^{2}_\nu\:=\:1$.  We note that these values of $\log
n_0\:=\:-7.89$ are almost identical to the centers of the Gaussian
fits made to the distribution of $n$ values discussed in
\S~\ref{sect_rho_lb} (column 5 of Table~\ref{tab_n0} for the
$R\:=\:0$, $-$LB sample).

The results from this fit are shown in the bottom panel of
Figure~\ref{fig_NwithZ} (where we only plot $-$LB, $R\:=\:0$ data
points).  Note that we only show the original column density errors,
and not the values of $\sigma_T$ used for minimizing $\chi^2$. In
fact, the derived value of $\sigma_{\rm{CL}}$ dominates the errors,
with $\sigma_N << \sigma_d < \sigma_{\rm{CL}}$. For example,
increasing the errors in the distances to the stars by as much as 1.5
times their assumed value has no effect on the values of $n_0$ and
$h$; it is the inherent clumpiness of the interstellar medium that
dominates the dispersion in \nup\ seen in Figure~\ref{fig_NwithZ}.

The new values of $h$ given above differ from the single scale height
calculated by \citet{savage03},
$h\:=\:\langle\log[N$(O~VI)$\sin|b|]\rangle/n_0\: =\:2.3$~kpc. Most of
the difference in $h$ for the southern hemisphere comes from the
reduction in $n_0$, for which Savage~\etal\ used a preliminary value
of $1.7\times 10^{-8}$~cm$^{-3}$ \citep[e.g.][]{ebj02}, a value which
included all sight lines through the Galactic disk (including those
toward Vela and Carina as well as the $R\:>\:0$ sight lines) and for
which no correction was made for the LB.

Savage~\etal\ noted that $N$(O~VI)$\sin|b|$ towards extragalactic
objects was enhanced by $\sim\:0.25$~dex in the high latitude
directions $45^\circ\:<\:b\:<\:90^\circ$ compared to lower latitudes
in the north and compared to all southern hemisphere directions.
Rather than reporting two scale heights as we have, they instead
described the distribution of O~VI in the halo with a single scale
height combined with a high latitude northern hemisphere 0.25 dex
enhancement. (Savage~\etal\ considered a number of possible
explanations for the origin of the enhancement.) With the smaller
mid-plane density found in this paper, the single scale height used in
Savage~\etal 's model would increase from 2.3 to 2.9~kpc.

\begin{figure}[t!]
\hspace*{-0.75cm}\includegraphics[width=8.5cm]{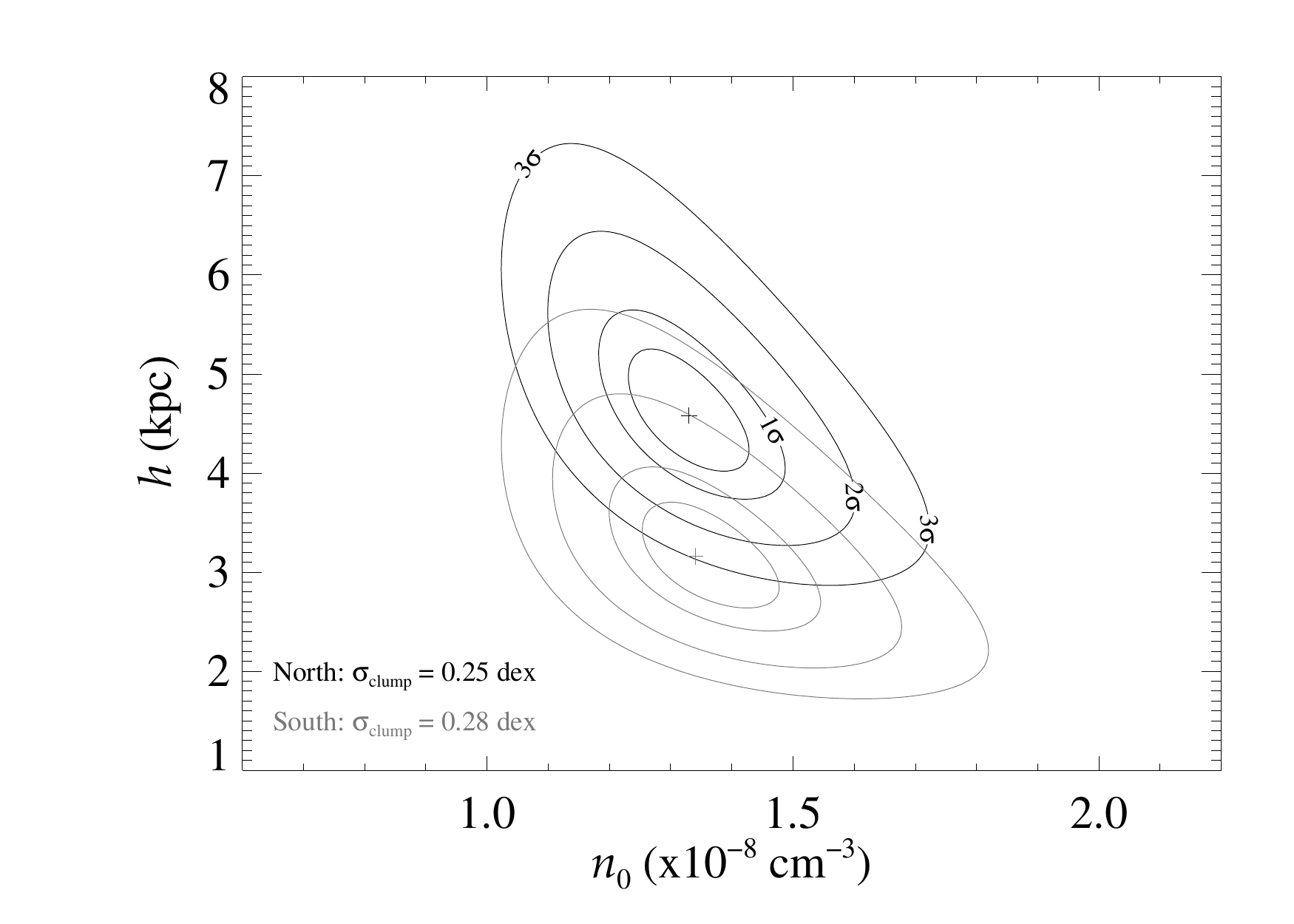}
\caption{Contour plots of $\chi^2$ about the minimum found for fitting
  $\log [N$(O~VI)$\sin|b|]$ against $\log |z|$. Contour levels are
  drawn at levels of $\chi^2$ which correspond to a significance of
  $\alpha\:=$ 0.62, 0.32, 0.046 and 0.0027, or confidence levels
  $1-\alpha\:=$ 0.38, 0.68, 0.95 and 0.997. Although the distribution
  function of $\chi^2$ is not described by a normal distribution, we
  use the analogy to describe the contours as 0.5, 1, 2 and 3 sigma
  confidence levels. \label{fig_contour}}
\end{figure}

What are the likely errors in these values of $n_0$ and $h$? It is
straightforward to calculate $\chi^{2}$ for values of $n_0$ and $h$
away from those that gave the minimum $\chi^{2}$.
Figure~\ref{fig_contour} shows contours of $\chi^{2}$ for different
values of $n_0$ and $h$, with each contour level drawn at a particular
limiting value of $\chi^{2}\:=\: \chi^2_L$. The value of $\chi^2_L$ is
calculated as $\chi^2_L\:=\:\chi^2_{\rm{min}} + \chi^2_p(\alpha)$
\citep{lampton76}.  Here, $\chi^2_p(\alpha)$ is the value of $\chi^2$
which would be exceeded a fraction $\alpha$ of the time in a set of
random trials, where the degrees of freedom is set to the number of
free parameters $p$ that we allow to vary (2 in our case). Given some
contour $\chi^2_L$ set by adopting a significance $\alpha$, we would
expect the true value of $n_0$ and $h$ to lie within the $\chi^2_L$
contour in a $1-\alpha$ fraction of similar O~VI surveys.  In
Figure~\ref{fig_contour} we adopt values of $\chi^2_L$ for confidence
levels $1-\alpha$ of 0.38, 0.68, 0.95 and 0.997, which were chosen
because they resemble the conventional interpretation of the 0.5, 1, 2
and 3$\sigma$ confidence levels that apply to a normal distribution.
It should be understood, of course, that the distribution function for
$\chi^2$ with a small number of free parameters is {\it not} normally
distributed.

The values of $\sigma_{\rm{CL}}$ given above seem to suggest that the
clumpiness of the O~VI absorbing gas may be slightly different between the
northern and southern hemispheres. Are these differences real? The
problem for estimating the errors in $\sigma_{\rm{CL}}$ are somewhat
different than deriving the confidence levels for $n_0$ and $h$
discussed above.  $\sigma_{\rm{CL}}$ is not a variable in the function
used to measure $\chi^2$, so does not enter into the minimization of
$\chi^2$. Changing $\sigma_{\rm{CL}}$ changes the numerical value of
$\chi^2$ when the minimum is found, but the values of $n_0$ and $h$ do
not change\footnote{This is not entirely true of course. With, for
  example, $\sigma_{\rm{CL}}$ set to zero, the values of $\sigma_T$
  --- and hence the weights for the $\chi^2$ fit --- are different for
  each point, which leads to small differences in the minimum value of
  $n_0$ and $h$. These differences are very small, however.}.
However, we can ask how much we can vary $\sigma_{\rm{CL}}$ before we
get `unreasonable' values of $\chi_{\rm{min}}^2$. Given the number of
degrees of freedom (number of data points - 2) for our data, we know
how $\chi_{\rm{min}}^2$ {\it should} be distributed. We can therefore
select the two values of $\chi^2$ within which the $\chi_{\rm{min}}^2$
outcomes are expected by chance 68\% of the time (i.e., a $\pm 1
\sigma$ probability interval), and then consider the limiting values
of $\sigma_{\rm{CL}}$ that make our evaluations of $\chi_{\rm{min}}^2$
conform to these two $\chi^2$ limits.

For the values of $n_0$ and $h$ given above, we find that for
values of $\chi^2$ within  $\pm 1\sigma$ of the expected value, we obtain
$\sigma_{\rm{CL}}\:=\:0.25(+0.02,-0.03)$~dex for the northern sample,
and $\sigma_{\rm{CL}}\:=\:0.28(+0.02,-0.03)$~dex for the southern
sample. Since the two values of $\sigma_{\rm{CL}}$ are within
$\pm\:1\sigma$ of each other for the N and S sight lines, it seems
that there is no significant difference in $\sigma_{\rm{CL}}$ in the
northern or southern hemispheres.  If we simply average these two
values then, we have $\sigma_{\rm{CL}}\:\simeq\:0.26\pm0.02$~dex.

We conclude this section by noting that in the simulations of
\citet{avillez05} this mid-plane density has a somewhat different
meaning than the single value which equation~\ref{eqn_exp} implies exists
at $|z|=0$.  In their models, hot gas is generated by SN explosions
throughout the plane of the Galaxy, and O~VI arises in clumps,
vortices and filaments which arise from the mixing of hot (interior)
and cold (exterior) gas associated with the SNRs.  Inside these
clumps, the O~VI volume density can vary by as much as 6 dex, although
typical clouds have densities of $\log\: n\:\sim\:-9$ to
$-6$. Although there is, therefore, no `single value' of $n$(O~VI) in
the disk of the Milky Way in these models, the regions are
sufficiently small ($\sim 100-200$~pc) that over long path lengths,
there is, obviously, an average value of $n$.  \citet{avillez05} find
that the average is constant over time, and has a value of $1.8\times
10^{-8}$~\pcmV\ (during a time period of 100 to 400~Myr and assuming
solar metallicities). This number is surprisingly close to the values
we list in Table~\ref{tab_n0} for $d\:<\:2$~kpc stars when we make no
distinction in \ROSAT\ class and subtract no contribution from the LB,
as is appropriate for comparison with the simulations.

\subsection{Distribution of O~VI Column Density with Effective Distance \label{sect_Nwithde}}

Following \cite{ebj78b} we calculate a {\it reduced} distance to a
star to take into account the decline in $n$ with $|z|$ discussed
above:

\begin{equation}
d_e = h (1-e^{-|z|/h})\,\csc |b|
\end{equation}

\noindent
In fact, these distances are nearly identical to the regular distances
$d$ for the stars studied herein---except for halo stars, whose long
path lengths and high distances above and below the plane of the
Galaxy require correcting for the change in $n$ with $|z|$.

\begin{figure*}[t!]
  \begin{minipage}[c]{0.45\linewidth}
    \caption{\label{fig_Nwithde} {\bf Top:} Plot of O~VI column density
      against the {\it effective distance} $d_e$ to a star, given by
      $d_e\:=\: h\:[1-e^{(-d\sin|b|/h)}]\:\csc|b|$, where $h$ is the scale
      height, $b$ is the Galactic latitude, and $d\:\sin |b|\:=\:|z|$ is
      the height above (or below) the plane of the Milky Way. The values
      of $h$ used are the same as those shown in Fig.~\ref{fig_NwithZ},
      but since most of the stars are in the plane of the Galaxy, $d_e$
      and $d$ are virtually indistinguishable, except for the halo stars.
      The size, shape and shading of the symbols used to plot the data are
      the same as those given in Fig.~\ref{fig_NwithZ}. The solid line
      shows the predicted value of the O~VI column density if
      $N$(O~VI)$\:=\:n_0\:d_e$, using the value of $n_0$ shown in the
      bottom plot.  (Note, this value of $n_0$ is the value derived for
      $R\:=\:0$ sight lines, which in this plot are shown by black points.
      Hence the fit should pass through the body of these data.)  A
      difference of $\pm20$~\% in $n_0$ is shown by dotted lines.  {\bf
        Bottom:} The same plot, but with a contribution from the Local
      Bubble subtracted from every sight line [$N_{\rm{LB}}$(O~VI)$ =
      1.11\times10^{13}$~\pcm ]. The two dashed lines demonstrate how the
      dispersion in $n$(O~VI) should change if $N$(O~VI) simply increases
      by intercepting more clouds of a fixed column density (in this case
      $5\times10^{12}$~\pcm ; see \S\ref{sect_Nwithde}). }
  \end{minipage}
  \begin{minipage}[c]{0.45\linewidth}
    \includegraphics[width=10cm]{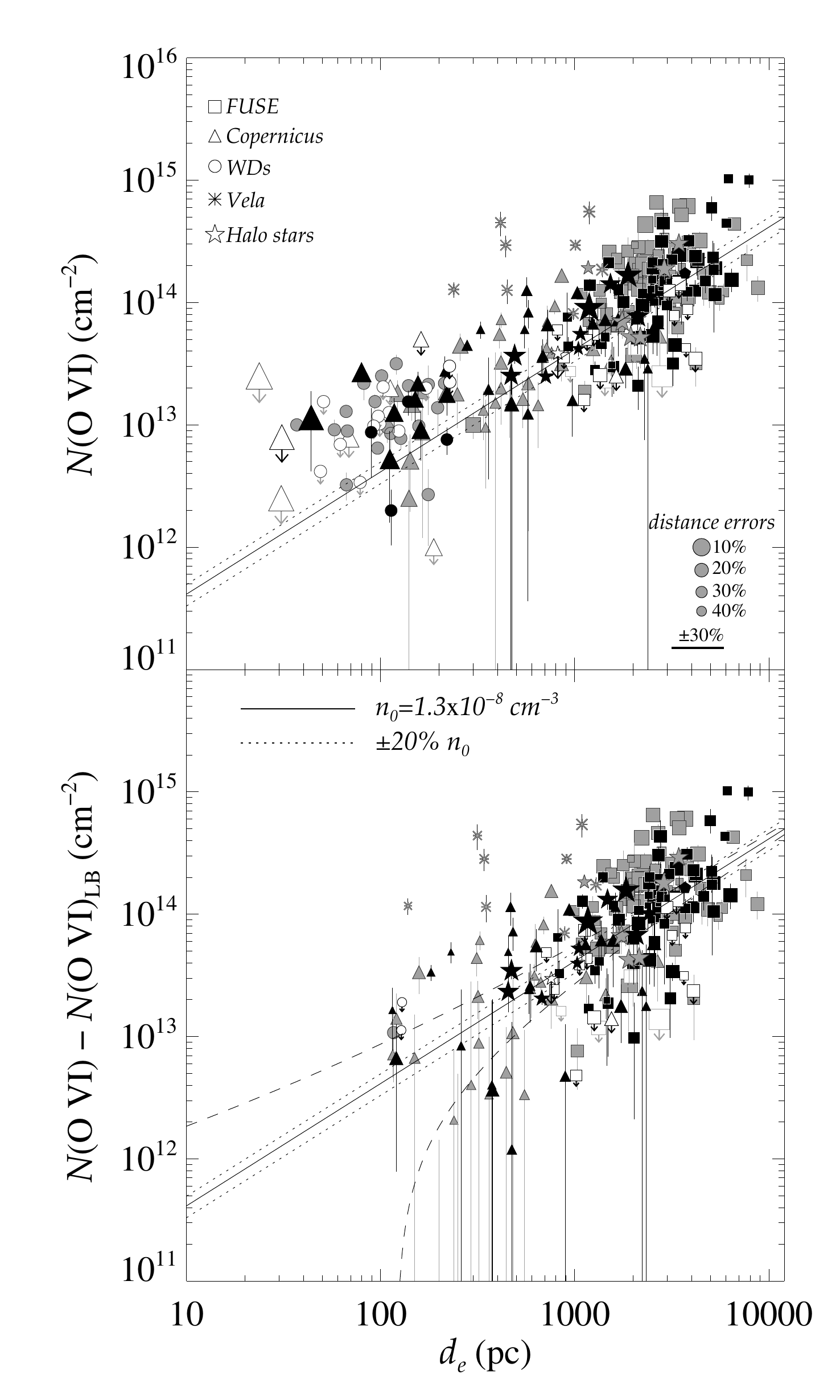}
  \end{minipage}
\end{figure*}

The top panel of Figure~\ref{fig_Nwithde} shows the correlation of
$N$(O~VI) and $d_e$, along with the expected value $N$(O~VI)$\:=\:n_0
d_e$ using the value of $n_0$ found above in \S\ref{sect_finally}. We
would expect this line to pass through the majority of the $R\:=\:0$
points, and this is the case for distances beyond a few hundred pc.
Below these distances, however, there is a clear departure from the
linear relationship. This apparent discrepancy can be removed by
plotting the column density for each point reduced by the LB
contribution. This is shown in the bottom panel of
Figure~\ref{fig_Nwithde}. The excess of points above the $n_0 d_e$
line below a few hundred pc now disappear below the $y$-axis limit
shown in the figure.

The fact that $N$(O~VI) and $d_e$ are correlated has some important
consequences. First, it means that the O~VI absorbing gas is
interstellar, and cannot be primarily circumstellar in origin. Second,
if the O~VI lines arise in discrete structures, and not in a uniform
smooth plane of gas (see below), then the lines must be composed of
many individual components which are so close in velocity to each
other that they blend together to form a `single' line with no apparent
velocity structure.  Third, because the correlation arises using sight
lines in all four quadrants of the Milky Way (Fig.~\ref{fig_disk}),
then the physical structures of the ISM which give rise to O~VI absorption
must be ubiquitous over all the Galaxy.

\begin{figure*}[t!]
\hspace{-1cm}\includegraphics[width=18cm]{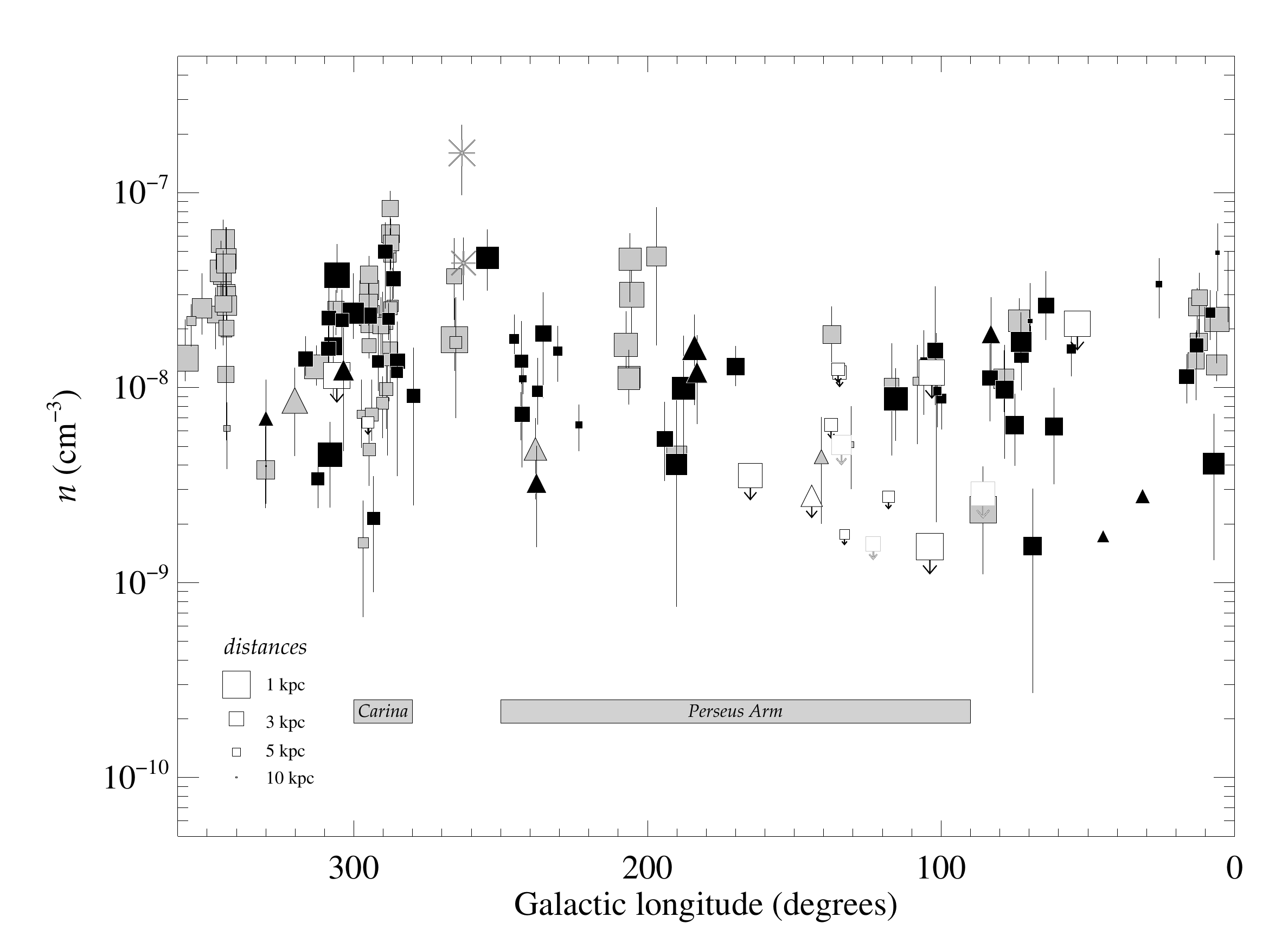}
\caption{\label{fig_Nwithl} Average O~VI volume density $n$
  [$=\:N$(O~VI)$/d$] along sight lines as a function of Galactic
  longitude. Only stars more distant than 1~kpc are included. Sight
  lines toward the Vela SNR are indicated by star symbols. More
  distant stars are shown with smaller symbols, as indicated in the
  legend.  Squares represent stars observed as part of our \FUSE\
  survey, while triangles represent stars observed with \Copernicus\
  (J78).  Squares are colored black for $R\:=\:0$ sight lines and gray
  for $R\:>\:0$ sightlines. Upper limits to $n$ are shown by open
  symbols with downward pointing arrows; again, symbols representing
  upper limits towards stars with $R\:=\:0$ have black outlines, while
  those representing upper limits towards stars with $R\:>\:0$ have
  gray outlines.  The extent of the Perseus Arm in the outer Galaxy is
  indicated (see Fig.~\ref{fig_disk}).}
\end{figure*}

Figure~\ref{fig_Nwithl} shows how $n$(O~VI) (with a contribution from
the LB removed) for stars at distances $d\:>\:1$~kpc varies with
Galactic longitude. 
Ignoring sighhtlines towards the Vela SNR and the Carina Nebular, we
find 
little difference in $n$ in
whichever direction we look, with one exception:
of the 23 \FUSE\ sight lines in the region
$l=80^\circ-140^\circ$ and with $d>1$~kpc, ten have only upper limits
to $N$(O~VI). This direction corresponds to the outer Perseus Arm (see
Fig.~\ref{fig_disk}). One could argue that there exists a genuine
deficiency of O~VI in the anti-center direction. However, more than
half the sight lines have $n$(O~VI) commensurate with the rest of the
Galactic disk. We discuss below whether the scatter in $n$ with
Galactic longitude is significant.

Figure~\ref{fig_Nwithde} shows that the dispersion of column densities
around the best-fit line at a given distance is large. This in itself
suggests that the O~VI absorbing ISM is far from being a smoothly
distributed intercloud medium --- otherwise the points would more closely
follow a $N$(O~VI)$\:=n_0\:d_e$ relationship. The figure also
indicates that absorption does not simply arise from randomly
distributed uniform clouds. If absorption arose in an ensemble of
generic clouds, each with a fixed column density $N_0$, then the total
column density measured by intercepting $p$ clouds would be $p\:N_0$.
For a random distribution of clouds the dispersion in the total column
density should go as $\pm \sqrt{p} N_0$, and the fractional error in
$N$(O~VI) would fall as $1/\sqrt{p}$. Hence the dispersion in the
observed column densities should decrease at larger distances, which
is clearly not observed.

To demonstrate this, we plot dashed lines in the bottom panel of
Figure~\ref{fig_Nwithde} to show how the error in $N$(O~VI) would
decrease with distance, assuming $N_0=5\times10^{12}$~\pcm , which is
typically the smallest column density detected towards WDs.  [Smaller
values of $N_0$ produce no dispersion in $N$(O~VI) at any distances,
while larger values exceed the smallest column densities measured.]
This suggests that the O~VI absorbing regions have deviations over a
wide range of $N_0$ (or scale sizes) compared to fluctuations caused
by randomly situated clouds with a single value of $N_0$.

It follows that the largest of these clouds have a low space density,
so that they do not appreciably increase the overall average density
$n_0$, but instead reveal themselves by creating large relative
deviations in $N$(O~VI) for long sight lines.  This is done in a
manner that compensates for the expected $1/\sqrt{p}$ decrease.
Alternatively, the maintenance of a constant level of relative
deviations in average space density at large distances could arise
from a large-scale clustering of small clouds.

Another way to measure the variation of clumpiness with distance is to
repeat our fit of \nup\ against $|z|$ for sets of stars at different
distances, and examine how the values of $\sigma_{\rm{CL}}$ change.
We can, once again, divide our stars into two sets, those at distances
between $0.2-2$~kpc and those beyond 2~kpc; we assume that there is no
difference between sight lines in the northern and southern Galactic
hemispheres, and that we can therefore use all our stellar sight
lines.  If we use column densities with a contribution from the LB
removed, and all \ROSAT\ classes, we find that at distances
$d\:>\:2$~kpc, $\sigma_{\rm{CL}}\:=\:0.28(+0.03,-0.02)$~dex, while for
$0.2<\:d\:<\:2$~kpc, $\sigma_{\rm{CL}}\:=\:0.29(+0.04,-0.03)$~dex.
These results imply that $\sigma_{\rm{CL}}$ is indeed independent of
distance.

Could these similarities simply arise from the fact that the volumes
sampled by the more distant stars have greater separations than those
toward the nearby targets? That is, could large-scale regional
differences in the Galaxy, which are not evident in the short sight
lines, create variations that almost exactly compensate for the
expected decrease in the dispersion in $\log n$ for a random
distribution of uniform clouds?  We can gain an insight on the
possible importance of this effect by performing an analysis of
variance \citep[e.g.][]{wolf74}.  We divide all 56 cases for $R=0$ and
$d\:>\:2$~kpc into $k=7$ Galactic longitude bins\footnote{The bin
  boundaries were set to the following longitudes (in degrees): 28,
  73, 104, 210, 244, 292, and 314.  The widths of the bins differ
  because of our requirement to have an equal number of samples in
  each bin. As before, we exclude sight lines to Carina and Vela.}
having indices $i$, with $m=8$ samples (with indices $j$) per bin.  We
performed a Kruskal-Wallis $H$-test to see if the deviations of the
results from one bin to the next are significant, and we found that
they are to a 96\% level of confidence ($H=13.3$).  Given that these
deviations are probably real, we now must understand quantitatively
what fraction of the overall sum of the squares of the deviations from
the mean value of $\log n$
\begin{equation}
S_0=\sum_{i=1}^k \sum_{j=1}^m (\log n_{i,j}-\langle \log n\rangle_{\rm
tot})^2
\end{equation}
arises from deviations from one longitude group to the next,
\begin{equation}
S_1=m\sum_{i=1}^k (\langle \log n\rangle_i-\langle \log n\rangle_{\rm
tot})^2
\end{equation}
as opposed to the cumulative effect of deviations within each group,
\begin{equation}
S_2=\sum_{i=1}^k\sum_{j=1}^m(\log n_{i,j}-\langle \log n\rangle_i)^2
\end{equation}
where $\langle \log n\rangle_{\rm tot}$ is the overall average $\log n$
and $\langle \log n\rangle_i$ is the average $\log n$ within each group
$i$.  It can be shown that, in general, $S_0=S_1+S_2$.  For the samples
defined above, we found the values,
\begin{equation}
S_0=5.22,~S_1=1.30,~{\rm and}~S_2=3.92
\end{equation}
which shows us that only 25\% of the variance of the entire sample
arises from deviations from one longitude set to the next. We conclude
that $\sigma_{\rm{CL}}$ is invariant with distance and that
for the most part this effect arises from the basic texture of the
volumes that hold O~VI absorbing gas, while a smaller
contribution arises from large scale regional variations in the Galactic
plane.

Finally, we note that  if the contribution from the LB is not removed,
the values of $\sigma_{\rm{CL}}$ change slightly, but there is still
no difference between nearby stars and the stars further away:
$\sigma_{\rm{CL}}\:=\:0.25(+0.03,-0.02)$~dex for $d\:>\:2$~kpc, and
$\sigma_{\rm{CL}}\:=\:0.24(+0.03,-0.02)$~dex for $0.2<\:d\:<\:2$~kpc.

\subsection{Variation of O~VI volume density with Milky Way Spiral
  Arms}

On the premise that most of the O~VI is generated by blast waves from
supernovae, we might expect to find evidence in our data that the
number density of such events and the character of their development
could be strongly influenced by whether their surrounding volumes are
located within spiral arms or interarm regions.  In order to explore
this issue, we attempted to detect an arm vs. interarm contrast in the
averages for the O~VI volume density $n$.  We defined spiral arms in
terms of their electron densities $n_e$ sensed by pulsar dispersion
measures, which ultimately were assembled into the model created by
\citet{cordes02} shown in Figure~\ref{fig_disk}.  Our test was a
simple one: we attempted to find a correlation between slight-line
values of $\langle n \rangle$ and $\langle n_e \rangle$, where the
latter was expressed in terms of the model's dispersion measure along
the sight line divided by its length.  We performed this comparison
for all stars that had a {\it ROSAT\/} class $R\: =\:0$ and distance
$d > 0.2\,$kpc, again excluding Vela and Carina stars.  The Pearson
correlation coefficient between $\log\langle n \rangle$ and
$\log\langle n_e \rangle$ was 0.08 for 98 sight lines, which means
that no significant correlation could be found.  (We recognize that
errors in distance can induce a spurious correlation, but this is not
an issue here since we are not claiming to see a correlation.)  In
short, we see no clear evidence that variations in sight-line values
of $\langle n \rangle$ are strongly influenced by the presence or lack
of spiral arms.

\subsection{Variation of O~VI Velocity with Galactic Longitude \label{sect_vel_with_long}}

Another question to consider in our quest to understand how O~VI is
distributed in the Galaxy is to ask whether the velocities of the O~VI
absorption follow that predicted from differential Galactic rotation.

To construct the LSR velocities $v$ that we would see along any
particular line of sight, assuming the absorbing gas co-rotated with
the Milky Way disk, we use the Galactic rotation curve obtained by
\citet{clemens85}, with $R_0=8.5$~kpc and $\Theta=220$~\kms. Since $v$
is a function of the distance, we can calculate a minimum and maximum
velocity where we might expect absorption.  In
Figure~\ref{fig_vel_bar1} we plot bars that show the allowed LSR
velocities as a function of Galactic longitude for each target. The
variation of permitted velocities with distance is not linear ---
velocities often approach a constant value for a given distance, for
example, and can sometimes backtrack at large distances to the same
velocities predicted at smaller distances.  To show this, each bar has
a gray-scale in which darker gray colors mark velocities where the
velocity of the gas is changing only very slowly with distance, i.e.,
darkness $\propto 1/(dv/dr)$.  [Note, each bar has grayscale coding
scaled relative to the individual sight line, so a particular gray
value for one sight line does not represent the same $1/(dv/dr)$ along
another sight line.] Bars identical to these are also shown in the
bottom panels of Figure~\ref{fig_spectra}, although the velocities
shown there are heliocentric values.

\begin{figure*}[t!]
\vspace{-2cm}\hspace{2.75cm}\includegraphics[width=11.75cm]{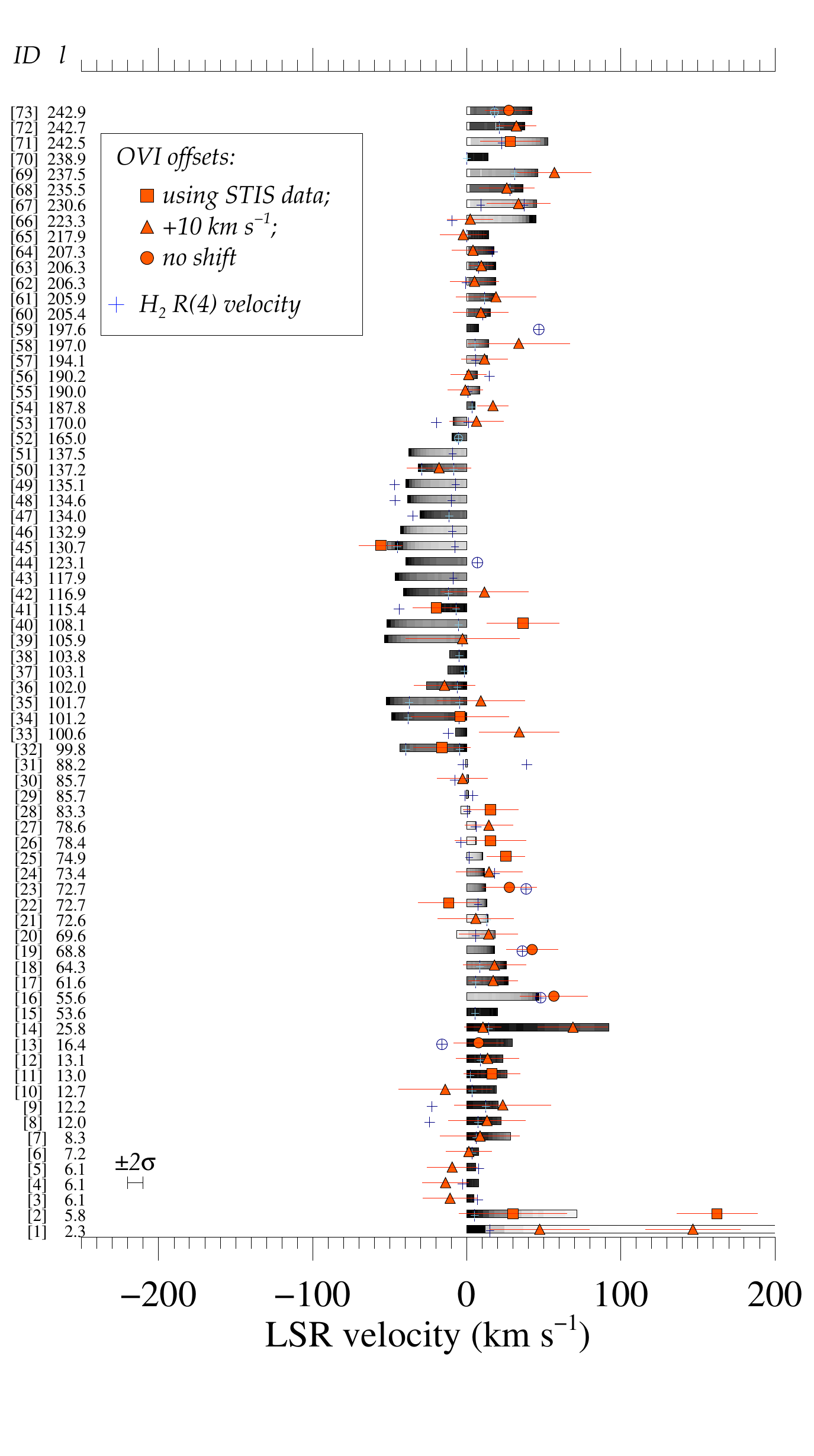}
\vspace{-1.5cm}\caption{\label{fig_vel_bar1} In this figure, permitted velocities for
  gas rotating in the plane of the Galactic disk are shown as gray
  bars, sorted by Galactic longitude. The measured velocities of
  O~VI absorption lines are overplotted with red triangles, squares,
  or circles (depending on how the offsets for the \FUSE\ wavelength
  zero-point were calibrated --- see legend box); R(4) H$_2$
  absorption lines are overplotted as blue crosses.  (Blue crosses in
  circles indicate H$_2$ measured in spectra which had no shifts
  applied.) The IDs assigned to the stars (see
  Table~\ref{tab_lookup}) are given to the left of the Galactic
  longitude of each star, in square parentheses.  To show the width of
  the O~VI absorption, a red line extends a distance of $\pm\:b_i/2$
  from the centroid velocity of each absorption component (where $b_i$
  is the Doppler parameter of component $i$).  }
\end{figure*}

\addtocounter{figure}{-1}
\begin{figure*}[t!]
\vspace{-2cm}\hspace{2.75cm}\includegraphics[width=11.75cm]{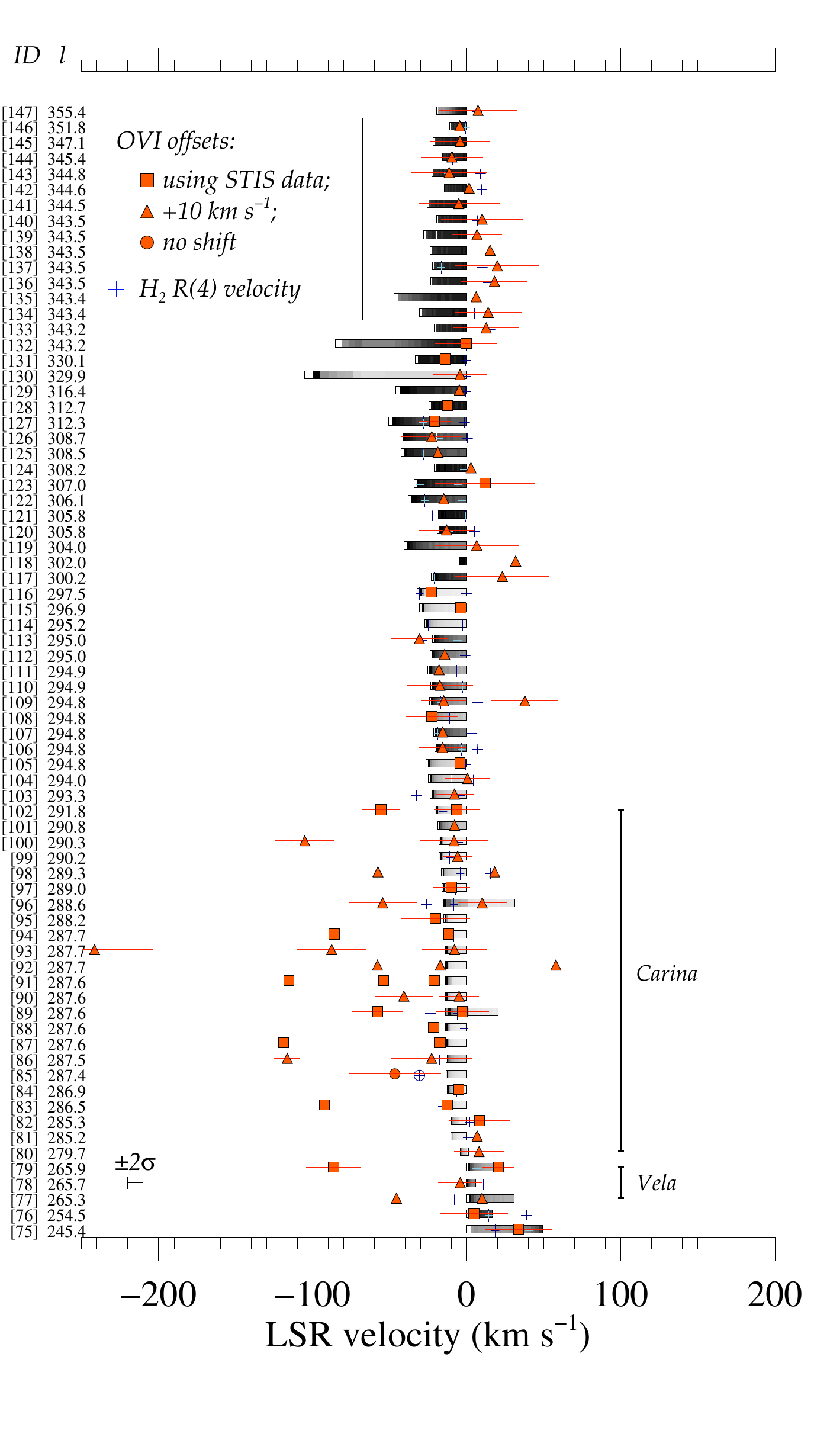}
\vspace{-1.5cm}\caption{continued.}
\end{figure*}

On top of each of the predicted-velocity bars, we plot the velocity
and width of the O~VI absorption components measured towards a star
either as a square, if the wavelength scale of the data was corrected
using available STIS data, or a triangle, if the adopted $+10$~\kms\
shift was applied (see \S\ref{sect_wavecal} and
Appendix~\ref{sect_cl1}).  If no shift was thought to be appropriate,
we plot a circle.  We also plot the velocity of the H$_2$ 6$-$0 R(4)
line as a cross.  Regardless what kind of velocity correction was
made, the location of the H$_2$ velocity with respect to that of the
O~VI was always correct, since the two were adjusted in the same
manner.  All O~VI and H$_2$ components are shown, and sight lines with
no detectable O~VI are still included (although obviously no O~VI
positions are marked).

The figure shows that there is a considerable lack of correlation
between the velocities of O~VI components and those predicted from
differential Galactic rotation, even if one ignores regions where the
gas is violently disturbed (i.e., Vela and Carina). The H$_2$ lines
often agree better with that predicted from rotation, but they too can
be found outside the permitted range. At some longitudes (e.g.
$l\:\sim\:343^\circ$), there appears to be some coherence where all
the O~VI is at the same velocity. In most cases, however, this is
simply because the stars are all at nearly the same longitude. For the
147 individual O~VI components plotted in Figure~\ref{fig_vel_bar1},
52~\% have centroids outside the predicted range.

Of course, the O~VI absorption velocities only show the centroid of
the absorption (values listed in Table~\ref{tab_cols}, but converted
to LSR velocities) and its width. As can be seen in
Figure~\ref{fig_spectra}, there is often some overlap between
velocities allowed by Galactic rotation, and some part of an O~VI
profile. Again, of the 147 individual O~VI components plotted in
Figure~\ref{fig_vel_bar1}, only 16\% have Doppler widths which do not
extend into the region predicted for differential Galactic rotation.
If the O~VI line is comprised of many components, then some
O~VI absorbing regions will be correlated with the general kinematics
of the ISM. However, Figure~\ref{fig_vel_bar1} shows that {\it the
  bulk} of the O~VI knows little about the rotation of the Galactic
disk, or that a mechanism exists to generate large peculiar motions in
clouds that do follow Galactic rotation.

\subsection{Comparison of Low and High Ion Velocity Extremes \label{sect_highVlow}}

Figure~\ref{fig_vel_bar1} shows that the velocities of the O~VI
centroids often fall outside the kinematic limits defined by
differential Galactic rotation.  Also, a direct comparison of H$_2$
and O~VI velocities show that the two correspond to each poorly.  This
implies that the bulk of the O~VI absorption does not mimic the
behavior of the cool neutral interstellar medium of the Milky Way
which makes up the Galactic disk.

Measuring the centroids of the O~VI absorption and comparing them with
either the range of velocities predicted from differential Galactic
rotation or the velocities of other interstellar features tells only
part of the story, however. If one considers {\it the outer edges} of
the O~VI absorption line profiles and compares them to these
velocities (see the bottom panels of Fig~\ref{fig_spectra}) the
differences are even larger.

To see how the extremes of the O~VI lines compare to the extremes of
other lower ionization species, we have investigated the absorption
line profiles of four other species: C~III~$\lambda 977$, O~I~$\lambda
1039$, Si~III~$\lambda 1206$ and C~II~$\lambda 1335$ lines.  All these
transitions have large $f$-values and produce extremely strong
absorption lines with sharp edges.  These features are so strongly
saturated that we can expect even relatively low column density gas
that is well removed from the core velocities to yield detectable
absorption components. The first two of these four lines are recorded
in our \FUSE\ data, while the second two arise at longer wavelengths
and were taken from the STIS data for the stars listed in
Table~\ref{tab_shifts}.  Since STIS data were not available for all
the stars in our \FUSE\ sample, the number of Si~III and C~II lines
studied is smaller than the number of C~III and O~I lines.

The C~III~$\lambda 977$ line lies in one of the SiC channels, and for
our analysis we usually used the SiC2A channel. To ensure that the
wavelength calibration of the SiC channel matched that of the LiF
channel, we compared the H$_2$ line nearest in velocity to the C~III
line [the 11$-$0 P(3) line at 978.217~\AA ] with the 6$-$0 P(3) line
next to the OVI absorption at 1031.192~\AA .  Matching these lines in
velocity provided the correct {\it relative} wavelength scales.  No
relative correction was made for the O~I~$\lambda 1039$ since the line
lies in the same channel as the O~VI line, and we assumed that the
wavelength scale was sufficiently accurate at both wavelengths. The
{\it absolute} wavelength of the C~III, O~I, and O~VI lines finally
rested on how well the LiF channel zero-point was corrected using
available STIS data or assuming a 10~\kms\ shift
(\S\ref{sect_wavecal}). The Si~III~$\lambda 1206$ and C~II~$\lambda
1335$ lines in the STIS data were assumed to have negligible
wavelength scale errors.

\begin{figure*}[t!]
\hspace{1.5cm}\includegraphics[width=13cm]{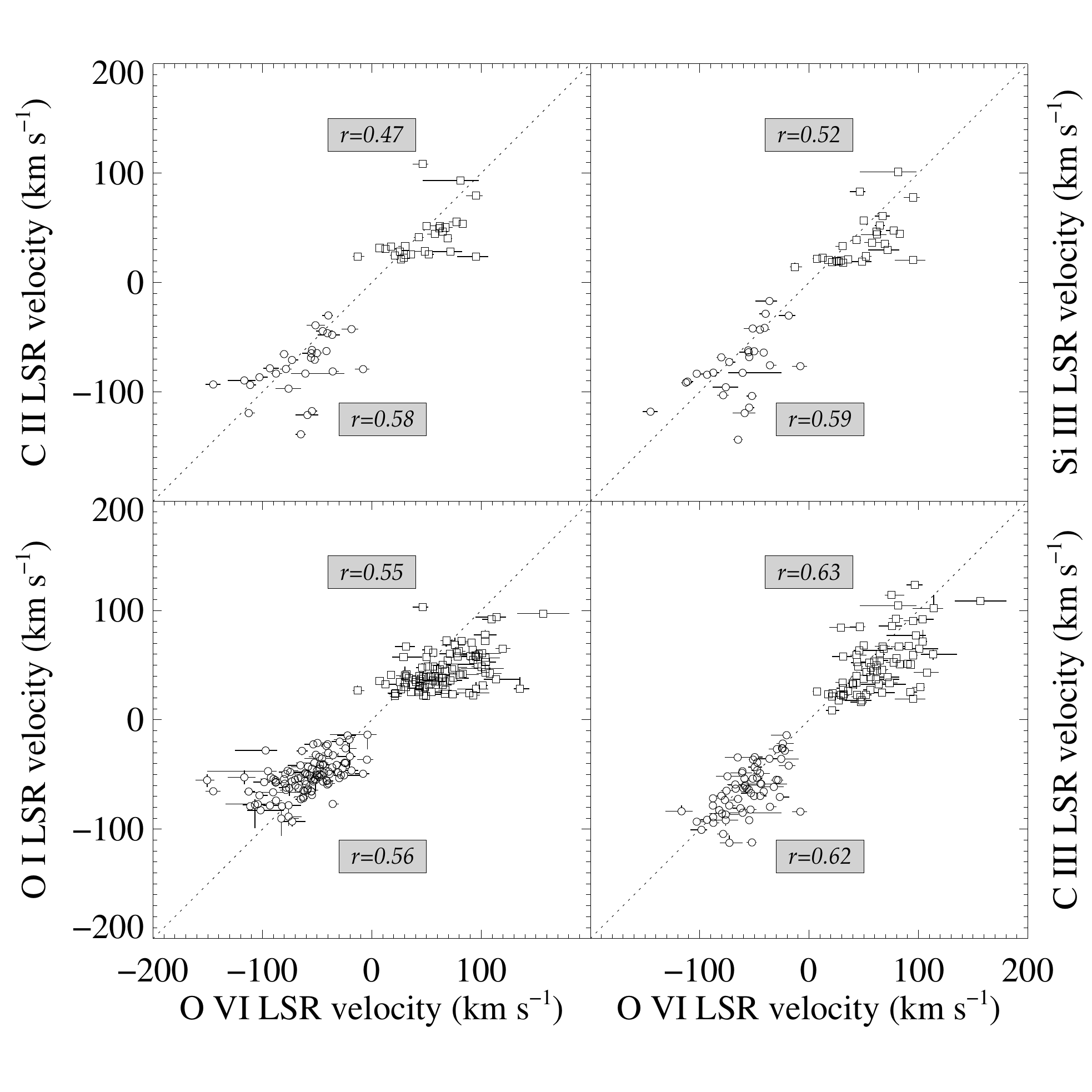}
\caption{\label{fig_highVlow} Velocity extremes for the profiles of the
  C~II~$\lambda 1335$, Si~III~$\lambda 1206$, C~III~$\lambda 977$ and
  O~I~$\lambda 1039$ lines plotted against the velocity extremes of the
  O~VI~$\lambda 1032$ lines. Circles represent negative edges, while
  squares represent positive edges of the absorption lines. For the
  positive and negative edges, the Pearson correlation coefficient $r$
  is shown at the top and bottom of each panel, respectively. The
  dotted line represents the line of equality for the velocities of
  any two ions.}
\end{figure*}

To calculate the velocity extremes of all these species, $v_e$, we
first converted normalized absorption line profiles to AOD $N(v)$
column density profiles. For O~VI, we used the $N(v)$ profiles for
which the H$_2$ lines had been removed (\S\ref{sect_colsAOD}).  We
then fitted a high-order polynomial to the wings of each $N(v)$
profile and measured the velocity extreme {\it at a given} $N(v)$. The
values of $N(v)$ used for the O~VI, O~I, C~II, C~III and Si~III lines
were: $2 \times10^{11}$, $5 \times 10^{12}$, $1 \times 10^{12}$,
$4\times10^{11}$, and $1\times 10^{11}$~\pcm\ (${\rm
  km~s}^{-1})^{-1}$, respectively. (We comment on these values below.)
In the few cases where individual absorption components could be seen
resolved from the main body of the absorption, we used the outer edge
of that component if it had $N(v)$ values that crossed the adopted
$N(v)$ limits.

The results are shown in Figure~\ref{fig_highVlow}.  The error bars
arise from re-measuring $v_e$ for profiles normalized by upper and
lower continua error envelopes (see \S\ref{sect_cont}). The figure
shows that there is a good correlation between the velocity extremes
of all the species selected and those of the O~VI absorption.  In each
panel, we show the Pearson correlation coefficients for the positive
and negative edges considered seperately.  The figure demonstrates
that there is some form of coupling between the hot medium and the
warm or cool phases.  This correlation was seen earlier by
\citet{cowie_ebj79} from the \Copernicus\ O~VI survey, and was thought
to arise from conduction/evaporation interfaces between the two
phases. The interface origin for some of the O~VI absorption has been
confirmed over short paths through the local ISM by \citet{savage06}.

This behavior is not unprecedented.  Very strong lines from
low-ionization species exhibit extraordinary dispersions of their
velocity extremes, a phenomenon studied in detail by \citet{cowie78a,
  cowie78b}.  One may imagine that very small fractions of the
low-ionization material are stirred up by the same dynamical processes
that produce the O~VI absorbing material, such as the influences from
supernova blast waves or mass-loss flows from stars, as we have
discussed previously. In this case, we would expect little correlation
between the velocity of gas in these structures and the bulk motion of
the cool ISM, as we discussed above in \S\ref{sect_vel_with_long}.

We note, however, that Figure~\ref{fig_highVlow} should be interpreted
with some degree of caution. The exact value of $v_e$ depends on the
value of $N(v)$ at which $v_e$ is evaluated, which in turn can be
influenced by the resolution of the data. The Si~III and C~II lines in
the STIS data have very sharp edges because of the high resolution at
which they were measured. In these cases, the exact value of $N(v)$
adopted is less important, since the line edges are well defined. For
the O~I and C~III lines measured at lower resolution in the \FUSE\
data, the edges are smeared by the instrumental LSF. Inspection of the
profiles used in this study show that these effects are likely to be
small; the values of $N(v)$ we adopted to measure $v_e$ were set so as
to ensure that the majority of lines would have reliable measurements
of $v_e$ as close to $N(v)=0$ as possible, given the typical S/N of
the \FUSE\ and STIS data.

\subsection{Minimum O~VI Line Widths \label{sect_bees}}

The fraction of oxygen in the form of O~VI peaks for collisionally ionized
gas at temperatures of $\log T_{\rm{max}} = 5.45$ if the gas is in
collisional ionization equilibrium, at which point 20\% of the oxygen
is in the form of O$^{+5}$ ions. This temperature corresponds to a
Doppler width of $b\:=\:0.0321\sqrt{T}\:=\:16.6$~\kms.  What is the
minimum reliable Doppler parameter in our sample? There are several
sight lines for which we have measured $b\:\apl\: 16$~\kms. For these,
however, either the O~VI line is weak, and the removal of strong
\hdline\ absorption has made the line profile uncertain (HD~110432),
or the line is actually a separate component blended with other
stronger components (HD~093205, HDE~303308).  The sight lines with the
narrowest, reliable values of $b$ are towards HD~046202, with
$b=16.4^{+4.3}_{-3.7}$~\kms\ and HD~093250 with $b=16.7\pm1.6$~\kms .
These are consistent with the smallest values of $b$ found towards WDs
(see, e.g. Fig.~10 of SL06).  There are many more profiles with $b\geq
20$~\kms , a number which provides a reliable lower limit for $b$.
Thus, our smallest value of $b$ is consistent with detecting single
thermally broadened, collisionally ionized clouds. We cannot be sure
however, that our narrowest absorption lines actually arise from
single clouds, since their column densities (0.65 and
0.83$\times10^{14}$~\pcm ) are much greater than the weak lines seen
towards local WDs and nearby \Copernicus\ stars.  These sight lines
may intercept several clouds whose relative motions along the line of
sight are at a minimum.  Alternatively, if sight lines intercept
clouds of very different sizes (\S\ref{sect_Nwithde}) the narrowest
lines we detect may indeed represent absorption from single large
clouds with moderate column densities.

\subsection{Distribution of O~VI Column Density with Doppler Parameter,  and with Distance \label{sect_Nwithb}}

\begin{figure*}[t!]
  \hspace{1cm}\includegraphics[width=14cm]{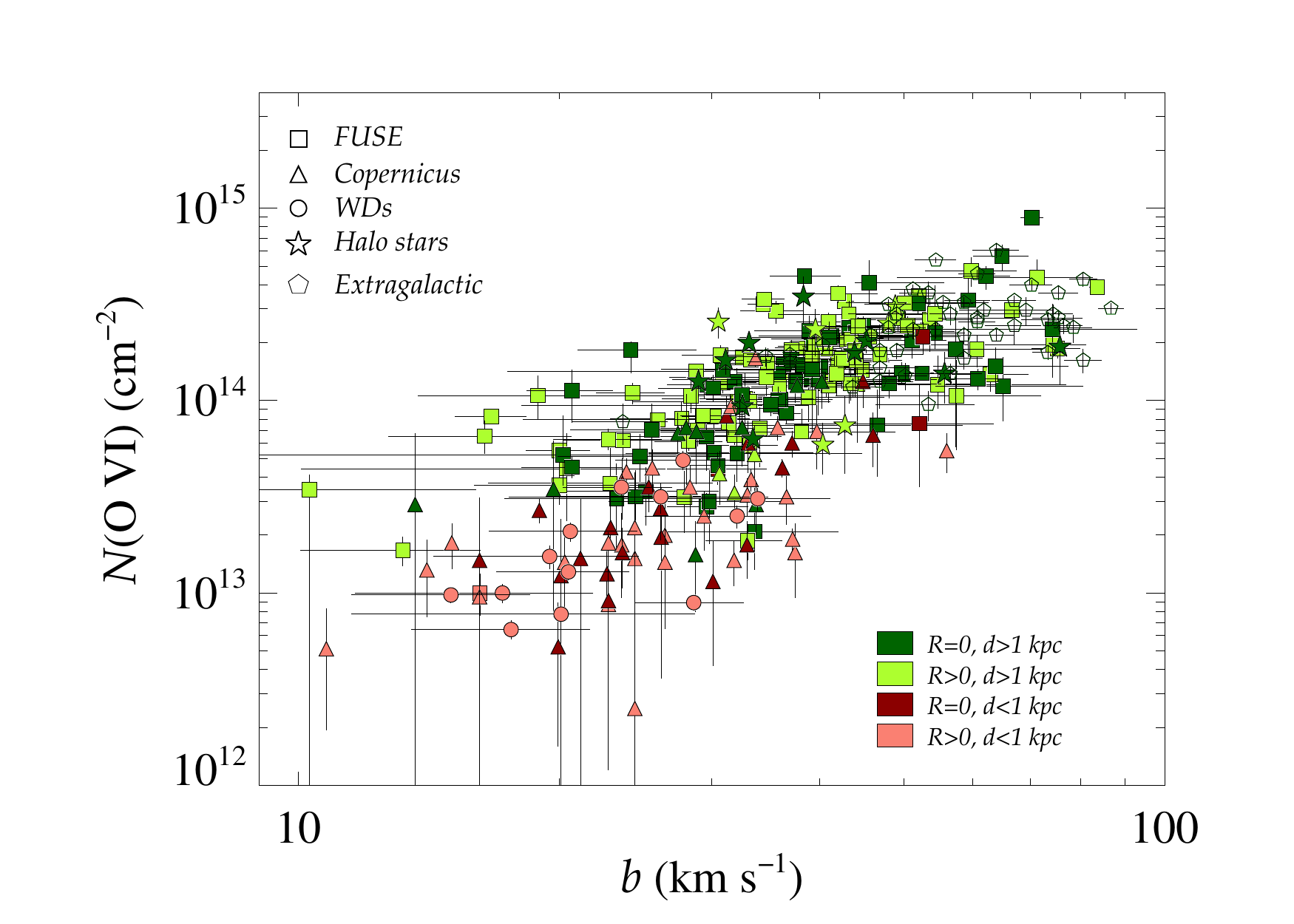}
  \caption{\label{fig_bvN} Doppler parameter $b$ against O~VI column
    density $N$(O~VI). Reliable measurements exist for $b\geq\:
    16$~\kms ; profiles of lines with $b$ less than this are either
    weak and suffer from HD contamination, or are one of multiple
    component systems, and are strongly blended with other components.
    Errors in $b$ are not available for the \Copernicus\ or
    \citet{zsargo03} datasets. Extragalactic objects have no distance
    or \ROSAT\ assignments, so are not colored.}
\end{figure*}

A Gaussian profile well fits the distribution of Doppler parameters
$b$ found for the \FUSE\ dataset. The profile is centered at $\langle
b \rangle\:=\:37$~\kms\ and has a standard deviation of 11~\kms.
However, the Doppler parameters are not entirely independent
quantaties.  A correlation between $N$(O~VI) and $b$ was first discussed by
\citet{heckman02}, who studied the trend using extragalactic sight
lines and data from early investigations of O~VI in our Galaxy.
Although our survey does not include O~VI absorption systems from
clouds outside of the Local Group of galaxies, our increased sample
size of Galactic systems permits a closer examination of the
correlation. Figure~\ref{fig_bvN} shows that the correlation exists in
the datasets adopted in this paper. To understand better the origin of
the trend, we have again indicated which sight lines have \ROSAT\
categories $R=0$ or $R>0$, and which have distances $d\:<\:1$~kpc or
$d\:>\:1$~kpc. The figure shows that there is no dependency on $R$; it
also shows that most low-$N$(O~VI), low-$b$ points are towards nearby
stars.

We suggest that the $N$ vs.~$b$ correlation has a simple
interpretation: we recall from our findings on the lack of change in
$\sigma_{\rm CL}$ with distance (\S\ref{sect_Nwithde}) that stars that
are further away intercept not only more O~VI absorbing structures,
but probably also larger, more sparsely distributed ones with higher
values of O~VI column densities.  On theoretical grounds, it is plausible to imagine
that these larger clouds have large internal motions because they have
recently evolved from explosive events.  The more numerous low-$N$(O~VI)
systems may represent pieces of much older regions that have been
partitioned by incursions of the ambient, cooler medium, but have
not yet had a chance to cool appreciably below the temperatures
representative of O~VI.  These smaller regions may have had their
internal velocities moderated because they have had a chance to couple
dynamically to the surrounding quiescent gas phases.  Alternatively,
we can imagine that what we interpret to be large $N$(O~VI) systems are
actually clusters of small clouds, which have cloud-to-cloud velocity
dispersions that are larger than that of the material inside
individual clouds.

\section{SUMMARY}

\begin{enumerate}

\item We have used \FUSE\ to observe early-type (mainly O2$-$B3) stars
  in the Galactic disk in order to characterize \sixa\ absorption
  lines arising from interstellar gas in the plane of the Milky Way. A
  total of 120 stars was observed as part of \FUSE\ PI team programs
  P102 and P122, of which nine were eventually rejected from our
  survey; the remaining 111 stars constitute a total investment of 754
  ksec of \FUSE\ time.  We supplemented these sight lines with
  archival data available April 2003 or earlier; our final sample
  consists of 148 stars, most of which are at Galactic latitudes
  $|b|<10\degr$ and distances of $>\: 1$~kpc.
  
\item In order to derive accurate spectroscopic distances to the
  stars, we have used several modern databases to improve
  upon published distances (Appendix~\ref{sect_distances}).
  \Hipparcos\ and \Tycho\ catalogs have provided precise
  optical magnitudes, as well as identification of multiple systems
  and variable stars. 2MASS infrared magnitudes have enabled us to
  estimate the interstellar $V-$band extinction along a sight line,
  and the absolute magnitudes of stars --- given a spectral type and
  luminosity class --- have been constructed by collating available
  surveys from the literature. Despite these improvements, imprecise
  knowledge of stellar types and classes continues to have the
  greatest uncertainty in deriving a spectroscopic distance, limiting
  the precision in a distance measurement to $\sim\:10-30$\%.
  
\item Where appropriate, we have added the O~VI measurements from
  several other O~VI studies to our sample, in order to probe
  distances not covered by our \FUSE\ stars. These include nearby
  ($\sim\: 20 -100$ pc) white dwarf stars (WDs) \citep{savage06},
  intermediate distance ($\sim 100-1000$~pc) stars observed by the
  \Copernicus\ satellite \citep{ebj78a}, distant halo stars
  \citep{zsargo03}, and extragalactic sight lines
  \citep{savage03,wakker03}. For
  many of the stars (but not including the WDs), we have rederived
  distances using the same set of tools used to derive distances
  towards the \FUSE\ sample of stars.
  
\item \ROSAT\ X-ray maps show that many stars lie towards regions of
  enhanced X-ray emission. This emission likely arises from hot
  bubbles blown out from the star or its stellar association. Other
  sight lines, however, are found at positions with very little X-ray
  emission, which we label as $R\:=\:0$ sight lines.  In contrast,
  lines of sight intercepting X-ray emitting regions have designations
  of $R\:>\:0$ (\S\ref{sect_bubbles} and \S\ref{sect_SNR}).  The
  average volume density towards stars in the Galactic disk,
  $n\:=\:N$(O~VI)$/d$, depends on whether we use $R\:=\:0$ or
  $R\:>\:0$ sight lines, and whether we subtract a contribution to the
  O~VI column density from the Local Bubble (LB; \S\ref{sect_rho0}).
  When we subtract a value of
  $N_{\rm{LB}}$(O~VI)$=1.11\times10^{13}$~\pcm\ arising from the LB,
  we find little difference in $n$ between stars with distances
  $0.2\:<\:d\:<\:2.0$~ kpc and those with $d\:>\:2$~kpc
  (\S\ref{sect_rho_lb}); we also find that for stars $d\:>\:2$~kpc,
  the $R\:>\:0$ sight lines have slightly higher densities than
  $R\:=\:0$ sight lines, even after excluding stars in the direction
  of the Vela SNR and Carina Nebula,
  which have the highest densities in our sample
  (\S\ref{sect_rho_rosat}).  This is probably because the
  circumstellar environment of the star contributes a small excess to
  the O~VI column density.

\item Our data well fit the relationship $N$(O~VI)$\sin |b|\:=\:n_0 h
  (1-\exp^{-|z|/h})$, which naturally arises from assuming that the
  density $n$ of O~VI absorbing gas falls with height $|z|$ above the
  Galactic plane as $n\:=\:n_0 \exp^{-|z|/h}$. Here, $n_0$ is the
  mid-plane density and $h$ is the characteristic scale height. We
  find $n_0\:=\:1.3\times 10^{-8}$~\pcmV\ [after subtracting a
  contribution to $N$(O~VI) from the LB and using the $R\:=\:0$ sight
  lines; \S\ref{sect_finally}], but that $h$ depends on whether we use
  extragalactic lines of sight in the northern (N) or southern (S)
  Galactic hemispheres \citep[as first noted by][
  ---see \S\ref{sect_NS}]{savage03}. We derive values of
  $h\:=\:[4.6,3.2]$~kpc when using the [N,S] sample.  We also find it
  neccessary to include a clumpiness factor $\sigma_{\rm{CL}}$ to
  represent the fact that the dispersion in $\log [N$(O~VI)$\,\sin
  |b|$] around the theoretical fit is much larger than can be
  accounted for by experimental errors alone (\S\ref{sect_clump}). Our
  best estimate is that $\sigma_{\rm{CL}}\:\simeq\:0.26\pm0.02$~dex.

\item $N$(O~VI) and the distance to a star are correlated
  (\S\ref{sect_Nwithde}). This shows that the processes which give
  rise to O~VI absorption are ubiquitous over the entire Galactic
  disk, and certainly do not arise primarily from circumstellar
  environments. The only inhomogeneity of the OV~I volume density $n$
  can be found in a concentration of $N$(O~VI) upper limits (10 out of
  23 sight lines) towards $d\:>\:1$~kpc \FUSE\ sight lines in the
  region $l\:=\:80^\circ-140^\circ$, the outer Perseus arm of the
  Galaxy. There is no evidence that the volume density of O~VI is
  influenced by the presence of Galactic spiral arms along the stellar
  sight lines.

\item Although O~VI is distributed smoothly enough for $N$(O~VI) to
  correlate with distance, the relative {\it dispersion} on either
  side of this correlation does not decrease with distance, as would
  be expected if more distant sight lines intercepted more clouds.
  Quantitively, we find no difference in the dispersion
  ($\sigma_{\rm{CL}}$) between nearby ($0.2-2$~kpc) and distant
  ($>\:2$~kpc) stars. This indicates that in addition to a random
  placement of small clouds, there may also be sparsely distributed
  large clouds.  Or, alternatively, some of the small clouds may
  simply be clustered over very large scales.  \label{point_Nwithde}

\item The velocity edges of the O~VI lines and those of the strong transitions
  of C~III~$\lambda 977.02$, O~I~$\lambda 1039.23$, Si~III~$\lambda 1206$ and
  C~II~$\lambda 1335$, are all correlated. This suggests that, at least to
  first order, the processes that move the hot gas to high velocity also
  affect the velocities of other species (\S\ref{sect_highVlow}).

\item The smallest reliable O~VI Doppler parameter we find in our
  sample is $b\:\simeq\:16-17$~\kms, which is consistent with that
  expected for a thermally broadened cloud at a temperature close to
  the peak in the ionization fraction curves, $\log T_{\rm{max}}\:\simeq\: 5.45$
  (\S\ref{sect_bees}).  We also confirm earlier research which found
  that $b$ and $N$(O~VI) are correlated. We show, however, that the
  width of O~VI lines cannot be accounted for by differential Galactic
  rotation; the lines are far wider than expected if absorption came
  from a smoothly distributed ISM corotating with the disk of the
  Milky Way (\S\ref{sect_Nwithb}). The correlation suggests that the
  large clouds mentioned in Point~\ref{point_Nwithde} above
  have large internal motions, or, alternatively, that the small clouds
  within clusters are moving more rapidly than the gases within
  individual clouds.

\end{enumerate}

\bigskip
\acknowledgements

This work is based on data obtained for the Guaranteed Time Team by
the NASA-CNES-CSA FUSE mission operated by Johns Hopkins University.
Financial support has been provided by NASA contract NAS5$-$32985.  We
have made extensive use of the SIMBAD database, operated at CDS,
Strasbourg, France, and the VizieR catalog service \citep{vizieR},
from which the \Hipparcos\ catalogs \citep{esa_mission} were
downloaded.  The {\it Two Micron All Sky Survey} (2MASS), is a joint
project of the University of Massachusetts and the Infrared Processing
and Analysis Center/California Institute of Technology, funded by the
NASA and the NSF. Archive data were obtained from the {\it
  Multimission Archive at the Space Telescope Science Institute}
(MAST). The Space Telescope Science Institute is operated by the
Association of Universities for Research in Astronomy, Inc., under
NASA contract NAS5$-$26555. Support for MAST for non-{\it{HST}} data
is provided by the NASA Office of Space Science via grant NAG5$-$7584
and by other grants and contracts.  \ROSAT\ data came from the \ROSAT\
Data Archive of the Max-Planck-Institut f\"{u}r extraterrestrische
Physik (MPE) at Garching, Germany. The data used for this survey can
be found on-line at {\tt www.astro.princeton.edu/$\sim$dvb/o6}.

\bibliographystyle{apj} 
\bibliography{apj-jour,bib_stars,bib2}


\clearpage
\begin{deluxetable}{
c 
l 
c 
r 
r 
c 
l 
r 
l 
c 
c 
l 
c 
c 
c 
c 
}
\rotate
\tabletypesize{\scriptsize}
\tablecolumns{16}
\tablewidth{0pc} 
\tablecaption{Stellar Parameters for \FUSE\ Target Stars \label{tab_journal}}
\tablehead{
\colhead{ }                 
& \colhead{ }                 
& \colhead{Prog}    
& \colhead{$l$ }  
& \colhead{$b$ } 
& \colhead{ }   
& \colhead{ }               
& \colhead{ }       
& \colhead{ } 
& \colhead{ } 
& \colhead{ } 
& \colhead{$d$}    
& \colhead{$d_r$}    
& \colhead{$v_c$} 
& \colhead{}
& \colhead{Sp/L} \\
\colhead{ID }                 
& \colhead{Star}         
& \colhead{ID}          
& \colhead{(deg)}  
& \colhead{(deg)} 
& \colhead{$B$} 
& \colhead{$V$}
& \colhead{$K$}
& \colhead{Sp/L}       
& \colhead{$E(B-V)$ } 
& \colhead{$E(K-V)$ } 
& \colhead{(kpc)}      
& \colhead{(kpc)}      
& \colhead{(\kms)}
& \colhead{[R]}
& \colhead{ref} \\
\colhead{(1) }                 
& \colhead{(2)}    
& \colhead{(3)}  
& \colhead{(4) } 
& \colhead{(5) }   
& \colhead{(6) }               
& \colhead{(7) } 
& \colhead{(8)  } 
& \colhead{(9)} 
& \colhead{(10)}    
& \colhead{(11)}  
& \colhead{(12)}
& \colhead{(13)} 
& \colhead{(14)} 
& \colhead{(15)}                 
& \colhead{(16)}
}
\startdata
  1 & HD167402 & P10162 &   2.27 & $ -6.39$ &  8.97 &  9.09$^{} $ &  \ldots  & B0.5Ib/B0II & $0.08\pm0.03$ &  \ldots  &  9.0 &  5.7,12.2 & $   8.6$ & 0 &        1\\
  2 & HD168941 & P10165 &   5.82 & $ -6.31$ &  9.36 &  9.37$^{} $ &  \ldots  & O9.5II-III & $0.24\pm0.04$ &  \ldots  &  6.8 &  4.4, 9.6 & $   9.3$ & 0 &        2\\
  3 & HD164906 & P10277 &   6.05 & $ -1.33$ &  7.54 &  7.40$^{g} $ &  \ldots  & B1IVpe & $0.38\pm0.03$ &  \ldots  &  1.1 &  0.7, 1.4 & $  10.0$ & 2 &            2\\
  4 & HD165052 & P10278 &   6.12 & $ -1.48$ &  6.91 &  6.84$^{} $ &  \ldots  & O6.5Vnf & $0.36\pm0.02$ &  \ldots  &  1.9$^{a}$ &  1.6, 2.3 & $  10.0$ & 2 &            2\\
  5 & HDE315021 & P12224 &   6.12 & $ -1.33$ &  8.52 &  8.41$^{} $ &  8.26 & B2IVn & $0.32\pm0.07$ & $-0.75\pm 0.07$ &  1.4 &  0.9, 1.9 & $  10.1$ & 1 &        3\\
  6 & HD163892 & P10276 &   7.15 & $  0.62$ &  7.57 &  7.48$^{} $ &  7.11 & O9IVn & $0.37\pm0.02$ & $-1.16\pm 0.04$ &  1.6 &  1.2, 2.1 & $  10.5$ & 0 &            2\\
  7 & HD158661 & P12222 &   8.29 & $  9.05$ &  8.32 &  8.18$^{} $ &  7.71 & B0.5Ib & $0.34\pm0.02$ & $-1.00\pm 0.04$ &  4.2 &  3.2, 5.4 & $  11.6$ & 0 &        4\\
  8 & HD167287 & P12226 &  11.99 & $ -0.93$ &  7.13 &  7.08$^{} $ &  6.81 & B1Ib & $0.24\pm0.02$ & $-0.78\pm 0.04$ &  2.7 &  1.8, 3.6 & $  11.2$ & 1 &            4\\
\enddata
\end{deluxetable}


\clearpage
\begin{deluxetable}{lcrcrrlr}
\tabletypesize{\small}
\tablecolumns{8}
\tablewidth{0pc} 
\tablecaption{Sight Lines Observed with \FUSE\ Sorted By Name, and Observation Log \label{tab_lookup}}
\tablehead{
\colhead{Star} 
& \colhead{\FUSE} 
& \colhead{Unique}
& & \colhead{$l$} 
& \colhead{$b$} 
& \colhead{Start}
& \colhead{$T_{\rm{exp}}$}
\\
\colhead{Name} 
& \colhead{Prog ID} 
& \colhead{ID\tablenotemark{a}}
& & \colhead{(deg)} 
& \colhead{(deg)} 
& \colhead{Date}
& \colhead{(ksec)\tablenotemark{b}}
}
\startdata
BD+532820 & P12232 & 34 &  & 101.24 & $ -1.69$ & 2000 08 06 & 11.3\\
CPD-417712 & B02506 & 134 &  & 343.41 & $  1.17$ & 2001 08 09 &  9.6\\
CPD-592600 & P12214 & 90 &  & 287.60 & $ -0.74$ & 2000 03 21 & 15.6\\
 &  &  &  &  &  & 2000 03 23 & \\
 &  &  &  &  &  & 2000 03 25 & \\
CPD-592603 & P12215 & 88 &  & 287.59 & $ -0.69$ & 2000 03 21 &  6.0\\
HD000108 & Z90101 & 43 &  & 117.93 & $  1.25$ & 2002 08 31 &  7.9\\
 & D06401 &  &  &  &  & 2003 08 05 & \\
\enddata
\tablenotetext{a}{ID number corresponding to panel numbers shown in 
Fig.~\ref{fig_spectra} as well as entries in Tables~\ref{tab_journal}, 
\ref{tab_cols}, \ref{tab_limits}, and \ref{tab_SNRs}.}
\tablenotetext{b}{Total exposure time from coadded spectra actually used,
  summed over all dates and \FUSE\ programs}
\tablecomments{All observations were made with the LWRS aperture
  except for 
HD175754,
HD041161,
and HD069106,
 which were observed using
  the MDRS aperture.}
\end{deluxetable}

\clearpage
\begin{deluxetable}{lrl}
\tabletypesize{\small}
\tablecolumns{3}
\tablewidth{0pc} 
\tablecaption{UV lines used in the analysis of \FUSE\ data  \label{tab_lines}}
\tablehead{
\colhead{    }       & \colhead{$\lambda_0$} & \colhead{}\\
\colhead{Line}       & \colhead{(\AA)} & \colhead{ $f$-value} 
} 
\startdata
HD    7$-$0 R(0)           &     1021.453  &   0.024 \\ 
\\
H$_2$ 7$-$0 R(6)           &     1030.080  &   not used   \\
H$_2$ 6$-$0 P(3)           &     1031.192  &   0.011   \\ 
Cl~I~$\lambda1031.5$        &     1031.507  &   0.151   \\ 
H$_2$ 8$-$0 R(8)           &     1031.557  &   0.019   \\
HD    6$-$0 R(0)           &     1031.909  &   0.023 \\ 
O~VI~$\lambda$1031         &     1031.927  &   0.133    \\ 
H$_2$ 6$-$0 R(4)           &     1032.354  &   0.017   \\ 
O~VI~$\lambda 1037$        &     1037.617  &   0.066  \\ 
\\
H$_2$ 7$-$0 R(8)           &     1042.745  &   0.018   \\
H$_2$ 5$-$0 R(4)           &     1044.542  &   0.015   \\ 
HD    4$-$0 R(0)           &     1054.286  &   0.016 \\ 
H$_2$ 6$-$0 R(8)           &     1054.520  &   0.017   \\
H$_2$ 4$-$0 P(3)           &     1056.472  &   0.009  \\
H$_2$ 4$-$0 R(4)           &     1057.376  &   0.013   \\
H$_2$ 4$-$0 R(8)           &     1079.932  &   0.012   \\ 
Cl~I~$\lambda1347$\tablenotemark{b} &     1347.240  &   not used   \\
\enddata 
\tablenotetext{b}{Measured in STIS data, used for calibration of zero point in \FUSE\ data.}
\end{deluxetable}


\clearpage
\thispagestyle{empty}

\begin{deluxetable}{clrccccrcccrcrcc}
\tabletypesize{\small}
\rotate
\tablecolumns{16}
\tablewidth{0pc} 
\tablecaption{Voigt Profile Fits to Detected O~VI Lines Toward Target Stars \label{tab_cols}}
\tablehead{
\colhead{}                 
& \colhead{}                 
& \colhead{$v$}  
& \colhead{$b_l$}   
& \colhead{$b$}  
& \colhead{$b_u$}    
& \colhead{$\sigma(b)$}    
& \colhead{$\sigma(b)_T$}
& & \colhead{$N_l$ } 
& \colhead{$N$ } 
& \colhead{$N_u$ }   
& \colhead{$\sigma(N)$ }   
& \colhead{$N_{\rm{TOTAL}}$}
& \colhead{} 
& \colhead{}\\
\cline{3-8}\cline{10-14}
 \colhead{ID} & \colhead{Star} 
& \multicolumn{6}{c}{(\kms )} & & \multicolumn{5}{c}{($10^{14}$~\pcm )} 
& \colhead{ZPF} 
& \colhead{Cont} \\
\colhead{(1)}                 
& \colhead{(2)}  
& \colhead{(3)}   
& \colhead{(4)}  
& \colhead{(5)}    
& \colhead{(6)}    
& \colhead{(7)} 
& \colhead{(8)} 
& & \colhead{(9)}   
& \colhead{(10)}   
& \colhead{(11)}   
& \colhead{(12)} 
& \colhead{(13)} 
& \colhead{(14)} 
& \colhead{(15)} 
}
\startdata
  1 & HD167402 & $   38$ &   60.7 &   64.9 &   69.0 &    2.3 & +4.7$,-$4.8 &  &    4.76 &    5.63 &    6.41 &    0.35 &   & F &  0 \\
 &  & $  138$ &   59.6 &   62.2 &   65.3 &    2.7 & +4.1$,-$3.7 &  &    3.88 &    4.43 &    4.90 &    0.36 &   10.05($+   1.26,-   1.42$) &  \\
  2 & HD168941 & $   20$ &   68.2 &   70.3 &   72.3 &    0.6 & +2.1$,-$2.2 &  &    8.51 &    8.94 &    9.35 &    0.07 &   & S &  0 \\
 &  & $  153$ &   51.6 &   52.5 &   53.6 &    2.3 & +2.6$,-$2.5 &  &    1.30 &    1.39 &    1.47 &    0.06 &   10.33($+   0.50,-   0.52$) &  \\
  3 & HD164906 & $  -20$ &   33.4 &   35.9 &   38.6 &    0.6 & +2.8$,-$2.6 &  &    1.07 &    1.18 &    1.29 &    0.02 &    1.18($+   0.12,-   0.11$) & F &  0 \\
  4 & HD165052 & $  -23$ &   27.8 &   30.2 &   32.7 &    0.7 & +2.6$,-$2.5 &  &    0.75 &    0.83 &    0.91 &    0.01 &    0.83($+   0.08,-   0.08$) & F &  0 \\
  5 & HDE315021 & $  -19$ &   30.3 &   33.2 &   36.3 &    1.0 & +3.2$,-$3.1 &  &    0.88 &    0.99 &    1.11 &    0.02 &    0.99($+   0.12,-   0.11$) & F &  0 \\
  6 & HD163892 & $   -9$ &   23.1 &   29.8 &   36.7 &    2.0 & +7.2$,-$7.0 &  &    0.18 &    0.30 &    0.44 &    0.01 &    0.30($+   0.14,-   0.12$) & F &  1 \\
\enddata
\end{deluxetable}


\clearpage
\begin{deluxetable}{clrrcccr}
\tabletypesize{\small}
\tablecolumns{8}
\tablewidth{0pc} 
\tablecaption{AOD Column Density Upper Limits Towards Stars With No Detected O~VI \label{tab_limits} }
\tablehead{
  \colhead{ID} 
& \colhead{Star}
& \colhead{$N_l$ }
& \colhead{$N$ }
& \colhead{$N_u$ }
& \colhead{$\sigma(N)$ }
& \colhead{$\sigma (N)_T$ }
& \colhead{$N_{\rm{limit}}$\tablenotemark{a} }\\
\cline{3-8}
  &  & \multicolumn{6}{c}{($10^{13}$~\pcm )}\\
  \colhead{(1)}
& \colhead{(2)}
& \colhead{(3)}
& \colhead{(4)}
& \colhead{(5)}
& \colhead{(6)}
& \colhead{(7)}
& \colhead{(8)}
}
\startdata
 15 & HD185418 & $  -3.11$ & $   0.58$ & $   3.95$ & $   0.45$ & $   3.56$ & $   8.52$ \\
 29 & HD199579 & $  -0.17$ & $   0.55$ & $   1.37$ & $   0.28$ & $   0.82$ & $   2.29$ \\
 31 & HD202347 & $  -0.94$ & $   0.22$ & $   1.29$ & $   0.16$ & $   1.13$ & $   2.72$ \\
 37 & HD207198 & $  -2.01$ & $   0.17$ & $   2.28$ & $   0.94$ & $   2.34$ & $   5.45$ \\
 38 & HD210839 & $  -1.50$ & $  -0.67$ & $   0.12$ & $   0.34$ & $   0.88$ & $   1.59$ \\
 43 & HD000108 & $  -2.24$ & $  -0.33$ & $   1.43$ & $   0.69$ & $   1.96$ & $   4.25$ \\
 44 & HD005005A & $  -0.74$ & $   0.23$ & $   1.11$ & $   0.44$ & $   1.02$ & $   2.49$ \\
 46 & HD012323 & $  -2.12$ & $  -0.49$ & $   1.10$ & $   0.44$ & $   1.67$ & $   3.47$ \\
\enddata
\tablenotetext{a}{$2\sigma$ confidence upper limit. See
  \S\ref{sect_colsAOD} and Appendix~\ref{sect_upper_limits} for details.}
\end{deluxetable}

\clearpage
\begin{deluxetable}{clrrlcr}
\tablecolumns{7}
\tabletypesize{\small}
\tablewidth{0pc} 
\tablecaption{Rejected Stars from the \FUSE\ Disk Survey \label{tab_rejects}}
\tablehead{
\colhead{Prog} 
& \colhead{Star} 
& \colhead{$l$} 
& \colhead{$b$} 
& \colhead{}
& \colhead{Start}
& \colhead{$T_{\rm{exp}}$ }
\\
\colhead{ID} 
& \colhead{Name} 
& \colhead{(deg)} 
& \colhead{(deg)} 
& \colhead{Sp/L\tablenotemark{a}}
& \colhead{Date}
& \colhead{(ksec)}
}
\startdata
P10207 & HD015642 & 137.09 & $ -4.73$ & O9.5III:n & 2000 09 15 &   5.1\\
P10224 & HD074194 & 264.04 & $ -1.95$ & O8.5Ibf & 1999 12 27 &   3.4\\
P10235 & HD093204 & 287.57 & $ -0.71$ & O5Vf & 2000 02 01 &   4.7\\
P12205 & HD035215 & 176.46 & $ -3.19$ & B1V & 2000 02 29 &  10.1\\
P12212 & HD066695 & 245.01 & $  2.21$ & B0.5IV & 2000 01 24 &   8.2\\
P12225 & HD166546 &  10.36 & $ -0.92$ & O9.5II-III & 2000 09 03 &   3.4\\
P12235 & BD+523210 & 102.44 & $ -3.37$ & B1V & 2000 08 06 &  15.7\\
 &  &  &  &  & 2000 08 09 & \\
P12236 & BD+532885 & 102.75 & $ -2.93$ & B2III & 2000 08 02 &  11.8\\
 &  &  &  &  & 2000 08 05 & \\
P12239 & HD216438 & 105.72 & $ -5.12$ & B1II-III & 2000 07 22 &   5.1\\
\enddata
\tablenotetext{a}{Spectral Type and Luminosity Class of star}
\end{deluxetable}

\clearpage
\begin{deluxetable}{llcclcrcl}
\tablecolumns{9}
\tabletypesize{\small}
\tablewidth{0pc} 
\tablecaption{Stars observed at more than one epoch \label{tab_time}}
\tablehead{
\colhead{Unique}   &
\colhead{}         &
\colhead{}         &
\colhead{FUSE}     &
\colhead{}         &
\colhead{}         &
\colhead{$T_{\rm{exp}}$} &
\colhead{}         &
\colhead{ $N$(O~VI)$^{a}$} \\
\colhead{ID}       &
\colhead{Star}     &
\colhead{Date}     &
\colhead{root}     &
\colhead{Aperture} &
\colhead{Channel}     &
\colhead{(ksec)}   &
\colhead{}         &
\colhead{ ($10^{14}$~\pcm )} 
}
\startdata
84  & HD091597  & $2001-02-04/06/08$ & P1023002/3/4 & LWRS$^{b}$ & LiF1A & 14.3  && $3.17\pm0.12$  \\
    &           & $2003-06-13$       & D0640301     & LWRS       & LiF1A & 13.5  && $3.14\pm0.13$  \\
76  & HD069106  & $2001-04-04$       & P1022301     & LWRS       & LiF1A & 0.06  && $1.99\pm0.28$  \\
    &           & $2002-12-31$       & P1022302     & MDRS$^{b}$ & LiF1A & 0.4   && $2.12\pm0.23$$^{c}$ \\
62  & HD046150  & $2001-03-05$       & P1021401     & LWRS$^{b}$ & LiF1A & 4.9   && $0.96\pm0.16$ \\
    &           & $2004-02-25$       & C1680201     & LWRS$^{b}$ & LiF1A & 3.7   && $0.80\pm0.13$ \\
    &           & $2004-02-24$       & P1021402/3   & MDRS       & LiF1A & 16.3  && $0.90\pm0.22$$^{c}$ \\ 
25  & HD192639  & $2000-06-12$       & P1162401     & LWRS$^{b}$ & LiF1A & 4.8   && $0.90\pm0.39$ \\
\enddata
\tablenotetext{a}{Column densities derived from profile fitting, with
  HD 6$-$0 R(0) line removed, as
  described in \S\ref{sect_voigt}.}
\tablenotetext{b}{These datasets were eventually coadded for each star 
to produce a final spectrum.}
\tablenotetext{c}{Assuming a MDRS instrumental resolution of 3~\kms\ FWHM \citep{feldman01}.}
\end{deluxetable}

\clearpage

\begin{deluxetable}{
l
l    
c    
l    
l    
l    
}
\tabletypesize{\small}
\tablecolumns{6}
\tablewidth{0pt}
\tablecaption{\FUSE\ Sight Lines Projected Through SNRs\tablenotemark{a} \label{tab_SNRs}}
\tablehead{
\colhead{Unique}
& \colhead{Target} 
& \colhead{Star's distance} 
& \colhead{SNR} 
& \colhead{SNR's distance} 
& \colhead{Ref.\tablenotemark{b}}\\
\colhead{ID}
& \colhead{star} 
& \colhead{(kpc)} 
& \colhead{Ident.} 
& \colhead{(kpc)}
}
\startdata
12  & HD168080 &2.7&G13.3 $-$1.3&2.0$-$4.7&1\\
15  & HD185418 &0.8&G53.6 +5.7\tablenotemark{c}&6.7&2\\
29  & HD199579 &1.1&G85.9 $-$0.6&5.&3\\
60  & HD047417 &1.2&G205.5 +0.5\tablenotemark{d}&0.8&4,5\\
77  & HD074920\tablenotemark{e} &1.5&G266.2 $-$1.2&0.3\tablenotemark{f}&6\\
79  & HD075309 &1.5&G266.2 $-$1.2&0.3\tablenotemark{f}&6\\
128 & HD124314 &1.1&G312.4 $-$0.4&?&\nodata\\
131 & HD134411 &2.1&G330.0 +15.0\tablenotemark{g}&0.5:&7\\
\enddata
\tablenotetext{a}{Excluding sight lines intercepting Vela}
\tablenotetext{b}{References for distance estimates: [1]  Seward et al. 1995, [2] 
Rosado 1983, [3]  Kothes et al. 2001, [4]  Kirshner, Gull, \& Parker 1978,
[5]  Lozinskaya 1981, [6]  Redman et al. 2002, [7]  Milne 1979}
\tablenotetext{c}{3C400.2}
\tablenotetext{d}{Monoceros Nebula (not to be confused with the much larger
Monogem Ring)}
\tablenotetext{e}{The star is located on the edge of the SNR, so the sight-line
interception, even if the SNR is in the foreground, is uncertain.}
\tablenotetext{f}{This SNR is presumed to be within the Vela SNR; see text for
details.}
\tablenotetext{g}{The Lupus Loop}
\end{deluxetable}


\clearpage 
\thispagestyle{empty}

\begin{deluxetable}{lccc}
\tablecolumns{4}
\tablewidth{0pt}
\tablecaption{Line of sight volume densities of O~VI in the Galactic Plane \label{tab_n0}}
\tablehead{
  \colhead{}
  &\colhead{}
  &\colhead{$<\log [N$(O~VI)$/d]>$}
  &\colhead{$\log\:\Sigma N\rm{(O~VI)}$/$\Sigma d$} \\
  \colhead{}
  &\colhead{No.}
  &\colhead{[$\log$ (cm$^{-3}$)]}
  &\colhead{[$\log$ (cm$^{-3}$)]}\\
  \colhead{(1)}
  &\colhead{(2)}
  &\colhead{(3)}
  &\colhead{(4)}
} 
\startdata
\multicolumn{4}{c}{{\it stellar distances} $>$ {\it 2000 pc}} \\
\cline{1-4} 
+LB, All $R$        & 102 &  $-7.87 \pm 0.32$        & $-7.79$              \\
+LB, $R=0$          & 56  &  $-7.92 \pm 0.31$        & $-7.79$              \\
+LB, $R>0$          & 46  &  $-7.82 \pm 0.32$        & $-7.79$             \\
&&& \\
$-$LB, All $R$      & 96  &  $-7.92 \pm 0.36$        & $-7.81$           \\
$-$LB, $R=0$        & 55  &  $-7.97 \pm 0.36$        & $-7.81$            \\
$-$LB, $R>0$        & 41  &  $-7.87 \pm 0.36$        & $-7.62$            \\
\enddata 
\tablecomments{Description of columns: 
(1)--- ``$-$LB'' indicates that a contribution from the Local Bubble,  
[$N_{\rm{LB}}$(O~VI)$ =  1.11\times10^{13}$~\pcm ] has been removed 
from all column densities, while ``+LB'' indicates that no such 
correction has been made; $R$ indicates the \ROSAT\ class assigned 
in \S\ref{sect_bubbles} and \S\ref{sect_SNR}; for the $-$LB samples,
distances have also been reduced by the radius of the LB, which we
take to be 100~pc;
(2) --- number of points used in statistical test. (Note that the
numbers in the +LB sample for a given \ROSAT\ class in each distance catagory are not
quite the same as the number for the equivalent $-$LB class; this is because subtracting
100~pc for the $-$LB stars moves a few objects either from the far to
near sample, or out of the near sample altogether.);
(3) --- mean of $\log N$(O~VI)$/d$ and the standard deviation of the
distribution of column densities (see \S\ref{sect_rho0});
(4) --- distance weighted average of O~VI column densities.
 Note: None of the statistics above include sightlines to Carina or Vela.}
\end{deluxetable}


\clearpage

\appendix

\section{THE ZERO-POINT OF THE \emph{FUSE} WAVELENGTH SCALE \label{sect_cl1}}

As discussed in \S\ref{sect_wavecal}, measuring precise velocities for
O~VI absorption lines potentially requires a correction for any errors
which might exist in the \FUSE\ wavelength-scale zero-point associated
with CalFUSE v2.0.5.  Fortunately, many of the stars in our sample
have also been observed with the {\it Space Telescope Imaging
  Spectrograph} (STIS) aboard \HST, which has a well determined
wavelength scale.  In particular, a substantial number of E140M and
E140H echelle data have been recorded which cover the Cl~I~$\lambda
1347$ line.  According to the STIS Handbook, the relative accuracy of
the STIS echelle modes is $0.25 - 0.5$ pixels across the full format
of the MAMA detectors, and the absolute accuracy is $0.5 - 1.0$ pixels
\citep{stis_ihb_v7}. Somewhat larger wavelength-scale errors have been
found in some STIS echelle spectra \citep{ebjC1, tripp05} indicating
relative wavelength scale accuracies $\approx\:1$ pixel.  To be
conservative, we assume that the uncertainty in the STIS wavelength
calibration is $\approx$1 pixel. This corresponds to 1.5 and 3.0 km
s$^{-1}$ for the E140H and E140M STIS gratings, respectively.

To derive corrections to the \FUSE\ wavelength scale, we compared
absorption lines of neutral chlorine with the molecular hydrogen lines
flanking the O~VI features, along individual stellar sight lines; H$_2$
clouds are efficient producers of Cl~I (Jura 1974; Jenkins 1995), so a
direct comparison between the \mol\ lines which flank the \sixa\
absorption in the \FUSE\ data, and a strong (but unsaturated) Cl~I
line whose velocity is known precisely, should provide a suitable
anchor to the \FUSE\ wavelength scale.  While in principle we could
have used the Cl~I~$\lambda 1031.5$~\AA\ line (see
Table~\ref{tab_lines} and Figure~\ref{fig_where_r_lines}), it was often
too weak, or absent altogether, for its wavelength to be measured.

The STIS data are recorded at higher resolution than the \FUSE\
spectra, so in order to compare the \mol\ and Cl~I lines directly, we
smoothed the STIS data, scaled the Cl~I to match the strength of the
\mol\ $6-0$ R(4) and P(3) lines, and shifted the \FUSE\ spectrum in
wavelength space until the lines overlapped. The shape of the
smoothed Cl~I line well matched the \FUSE\ H$_2$ line profiles in all
cases, and we were able to record shifts to within $\pm 0.2$ pixels,
or just over 1~\kms\ at the wavelength of the \sixa\ line.

\begin{figure}[t!]
  \hspace{0.5cm}\includegraphics[width=17cm]{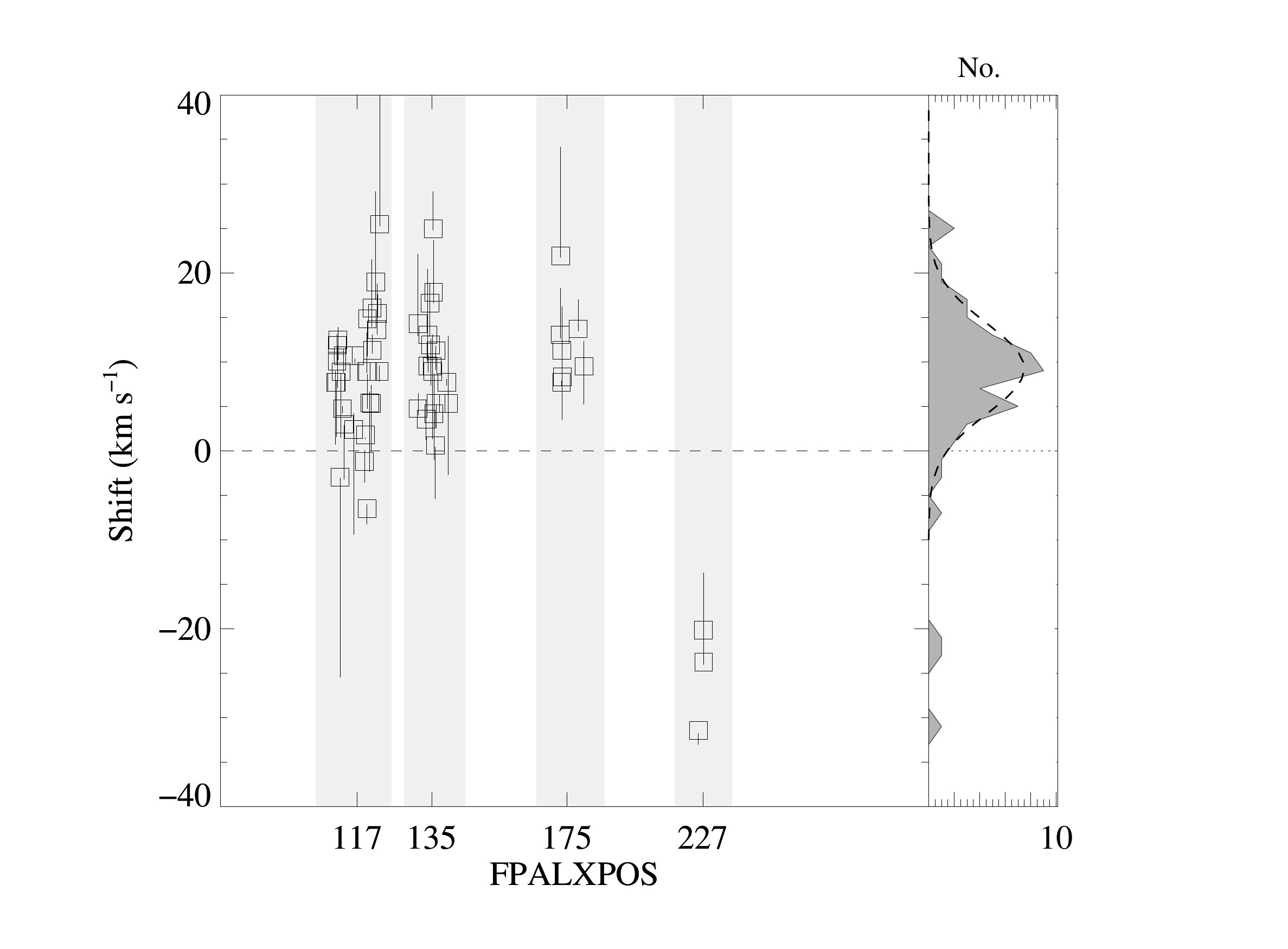}
  \caption{\label{fig_fxpos} Plot of the shifts required to accurately
    calibrate the zero-point of the \FUSE\ wavelength scale around the
    \sixa\ line, as a function of X-position of the Focal Plane
    Assembly (as determined from the FPALXPOS keyword in the LiF1A
    headers).  Values of FPALXPOS for our data fall close to one of
    four values: 117, 135, 175 and 227. To better show the error bars
    in the Y direction (see text for a full explanation of the error
    bars), we have randomized the FPALXPOS values by small amounts to
    make the points slightly offset from each other. It should be
    understood that points within a particular gray box actually have
    FPALXPOS values almost identical to the value shown on the X-axis
    at the bottom of each gray box. The distribution of the shifts is
    shown as a histogram to the right. For spectra with FPALXPOS in
    the nominal range $\leq 175$, the shift is centered at $9.2$~\kms,
    with an rms deviation of $\simeq\: \pm 5.1$~\kms . }
\end{figure}

Apart from 46 sight lines in our Galactic disk sample which had also
been observed with STIS, we were able to examine \FUSE\ and STIS
spectra of a few other non-survey stars. A total of
55 \FUSE\ and STIS spectra enabled us to
investigate whether the errors in the \FUSE\ wavelength scale were
dependent on the state of the spectrograph. We found a crude
correlation between the more extreme shifts in the wavelength
zero-point and the X-position of the Focal Plane Assembly, as recorded
in the FPALXPOS keyword of the LiF1A headers. The data used in the
study are listed in Table~\ref{tab_shifts}, while the measured shifts
and their values of FPALXPOS are shown in Figure~\ref{fig_fxpos}. We
found no correlation between measured shifts and the Z-position of the
FPA, as recorded in the data header keyword FPALZPOS.

We note the following in Figure~\ref{fig_fxpos}. The error bars in the
ordinate direction are {\it not} statistical errors. As explained in
\S\ref{sect_subexposures}, we used the sub-spectrum with the highest
S/N ratio to act as a template against which the other sub-exposures
were compared before sub-exposures were coadded. The bars in
Figure~\ref{fig_fxpos} represent the minimum and maximum excursions in
the values of the shifts measured for a set of sub-exposures.

The majority of the \FUSE\ spectra were taken with FPALXPOS in the
range 117.0$-$175.0. For these data, a consistent shift of $\sim
+10$~\kms\ is required to correct the \FUSE\ wavelength scale. (Again,
this shift was not necessarily the shift of the first sub-exposure
taken of a star, but was derived from the sub-exposure with the
highest S/N.) The histogram on the right-hand side of
Figure~\ref{fig_fxpos} shows the distribution of the shifts, along with
a Gaussian fit to the distribution, which peaks at 9.2~\kms\ and has a
width of $\sigma = 5.1$~\kms . This implies that a +10~\kms\ shift to
\FUSE\ data which have no companion STIS data 
will provide an accurate re-calibration of the wavelength scale to
within $\pm 5$~\kms.

With this information, we applied the following re-calibration to the
survey stars. For the 46 stars also observed by STIS, the \FUSE\ data
were shifted by the numbers given in Table~\ref{tab_shifts}. Nine
stars were observed either with the MDRS aperture, with a channel
other than the LiF1A, or when the X-position of the FPA was $ >\:
200$; no shifts were applied to these data.  The remaining 93 spectra
were shifted in wavelength by 0.0344~\AA , or 10~\kms\ at 1032~\AA.

Finally, we note that later versions of CalFUSE (v3.0) automatically
applied a correction of 10~\kms\ to the zero-point of the wavelength
calibration to correct for the offset described above.  Versions of
CalFUSE later than v3.2 used an improved scheme for the wavelength
calibration of the raw data as a whole, reductions which are described in detail by
\citet{vandixon07}.


\clearpage

\thispagestyle{empty}

\begin{deluxetable}{lllclccrrr}
\tablecolumns{10}
\tabletypesize{\small}
\tablewidth{0pc} 
\tablecaption{Derived shifts for \FUSE\ data \label{tab_shifts}}
\tablehead{
\colhead{FUSE}     
& \colhead{}      
& \colhead{STIS} 
& \colhead{STIS} 
& \colhead{STIS}  
& \colhead{}         
& & \multicolumn{3}{c}{Shifts\tablenotemark{a}} \\
\cline{8-10}
\colhead{Root} 
& \colhead{Star}   
& \colhead{dataset}  
& \colhead{ID\tablenotemark{b}}  
& \colhead{grating}& \colhead{FPALXPOS} 
& & \colhead{(pixs)} 
& \colhead{(\AA)} 
& \colhead{(\kms)}
}
\startdata
B07101 & HD001383 & O5C07C010 & 8241 & E140H & 117.0 &  & $  1.7$ & $  0.035$ & $  10.1$\\
B07104 & HD091983 & O5C08N010 & 8241 & E140H & 117.0 &  & $  2.1$ & $  0.043$ & $  12.5$\\
P10165 & HD168941 & O6LZ81010 & 9434 & E140M & 117.0 &  & $  0.8$ & $  0.016$ & $   4.7$\\
P10179 & BD+354258 & O6LZ89010 & 9434 & E140M & 117.0 &  & $  1.8$ & $  0.037$ & $  10.7$\\
P10188 & HD218915 & O57R05010 & 7270 & E140H & 117.0 &  & $  0.5$ & $  0.010$ & $   3.0$\\
P10285 & HD190918 & O6359J010 & 8662 & E140M & 117.0 &  & $  1.5$ & $  0.031$ & $   8.9$\\
P10286 & HD192035 & O6359K010 & 8662 & E140M & 117.0 &  & $  1.5$ & $  0.031$ & $   8.9$\\
P11631 & HD210839 & O54304010 & 8043 & E140H & 117.0 &  & $  2.7$ & $  0.055$ & $  16.0$\\
P12231 & HD210809 & O5C01V010 & 8241 & E140H & 117.0 &  & $  2.3$ & $  0.047$ & $  13.6$\\
Z90101 & HD000108 & O5LH01010-40 & 8484 & E140H & 117.0 &  & $  4.3$ & $  0.088$ & $  25.5$\\
B03001 & HD203374A & O5LH08010-30 & 8484 & E140H & 117.2 &  & $  1.3$ & $  0.027$ & $   7.7$\\
B03004 & HD208440 & O5C06M010 & 8241 & E140H & 117.2 &  & $  1.3$ & $  0.027$ & $   7.7$\\
B03005 & HD209339 & O5LH0B020 & 8484 & E140H & 117.2 &  & $  2.0$ & $  0.041$ & $  11.9$\\
B07105 & HD122879 & O5LH07020 & 8484 & E140H & 117.2 &  & $ -0.5$ & $ -0.010$ & $  -3.0$\\
B07109 & HD206773 & O5C04T010 & 8241 & E140H & 117.2 &  & $  1.5$ & $  0.031$ & $   8.9$\\
P10231 & HD091651 & O6LZ34010 & 9434 & E140M & 117.2 &  & $  0.4$ & $  0.008$ & $   2.4$\\
P10232 & HD092554 & O6LZ36010 & 9434 & E140M & 117.2 &  & $  1.8$ & $  0.037$ & $  10.7$\\
P10265 & HD134411 & O6LZ60010 & 9434 & E140M & 117.2 &  & $ -0.2$ & $ -0.004$ & $  -1.2$\\
P10266 & HD151805 & O6LZ63010 & 9434 & E140M & 117.2 &  & $  0.3$ & $  0.006$ & $   1.8$\\
P10279 & HD165955 & O63599010 & 8662 & E140M & 117.2 &  & $ -1.1$ & $ -0.022$ & $  -6.5$\\
P10289 & HD202347 & O5G301010 & 8402 & E140H & 117.2 &  & $  0.9$ & $  0.018$ & $   5.4$\\
P11624 & HD192639 & O5C08T010 & 8241 & E140H & 117.2 &  & $  0.9$ & $  0.018$ & $   5.3$\\
P11627 & HD206267 & O5LH09010-20 & 8484 & E140H & 117.2 &  & $  1.9$ & $  0.039$ & $  11.3$\\
P12216 & HDE303308 & O4QX04020 & 7301 & E140H & 117.2 &  & $  3.2$ & $  0.065$ & $  19.0$\\
P12230 & HD201345 & O5C050010 & 8241 & E140H & 117.2 &  & $  2.6$ & $  0.053$ & $  15.4$\\
P12232 & BD+532820 & O6359Q010 & 8662 & E140M & 117.2 &  & $  1.5$ & $  0.031$ & $   8.9$\\
P10275 & HD157857 & O5C04D010 & 8241 & E140H & 117.5 &  & $  2.5$ & $  0.051$ & $  14.8$\\
P10123 & HD088115 & O54305010-30 & 8043 & E140H & 134.9 &  & $  2.4$ & $  0.049$ & $  14.3$\\
P10257 & HD104705 & O57R01010 & 7270 & E140H & 134.9 &  & $  3.0$ & $  0.061$ & $  17.8$\\
P10262 & HD124314 & O54307010 & 8043 & E140H & 134.9 &  & $  1.9$ & $  0.039$ & $  11.3$\\
P10118 & HD066788 & O6LZ26010 & 9434 & E140M & 135.1 &  & $  0.8$ & $  0.016$ & $   4.7$\\
P10221 & HD063005 & O63531010 & 8662 & E140M & 135.1 &  & $  0.6$ & $  0.012$ & $   3.6$\\
P10230 & HD091597 & O6LZ33010 & 9434 & E140M & 135.1 &  & $  2.2$ & $  0.045$ & $  13.0$\\
P10236 & HD093205 & O4QX01020 & 7301 & E140H & 135.1 &  & $  2.8$ & $  0.057$ & $  16.6$\\
P10237 & HD093222 & O4QX02020 & 7301 & E140H & 135.1 &  & $  2.0$ & $  0.041$ & $  11.9$\\
P10241 & HD094493 & O54306010 & 8043 & E140H & 135.1 &  & $  1.6$ & $  0.033$ & $   9.5$\\
P10245 & HD099857 & O54301010-30 & 8043 & E140H & 135.1 &  & $  4.2$ & $  0.086$ & $  24.9$\\
P10246 & HD099890 & O6LZ45010 & 9434 & E140M & 135.1 &  & $  0.7$ & $  0.014$ & $   4.2$\\
P10258 & HD114441 & O6LZ53010 & 9434 & E140M & 135.1 &  & $  0.1$ & $  0.002$ & $   0.6$\\
P10261 & HD116781 & O5LH05010-20 & 8484 & E140H & 135.1 &  & $  0.9$ & $  0.018$ & $   5.3$\\
P12215 & CPD-592603 & O40P01D6Q & 7137 & E140H & 135.1 &  & $  1.3$ & $  0.027$ & $   7.7$\\
P12219 & HDE308813 & O63559010 & 8662 & E140M & 135.1 &  & $  0.9$ & $  0.018$ & $   5.3$\\
P10227 & HD075309 & O5C05B010 & 8241 & E140H & 135.4 &  & $  1.6$ & $  0.033$ & $   9.5$\\
P10240 & HD093843 & O5LH04010-20 & 8484 & E140H & 135.4 &  & $  1.5$ & $  0.031$ & $   8.9$\\
P10203 & HD013268 & O63506010 & 8662 & E140M & 175.1 &  & $  2.2$ & $  0.045$ & $  13.0$\\
P10205 & HD014434 & O63508010 & 8662 & E140M & 175.1 &  & $  1.3$ & $  0.027$ & $   7.7$\\
P10206 & HD015137 & O5LH02010-40 & 8484 & E140H & 175.1 &  & $  1.4$ & $  0.029$ & $   8.3$\\
P12201 & HDE232522 & O5C08J010 & 8241 & E140H & 175.1 &  & $  2.3$ & $  0.047$ & $  13.7$\\
S3040201 & HD224151 & O54308010 & 8043 & E140H & 175.1 &  & $  1.6$ & $  0.033$ & $   9.5$\\
P10202 & HD012323 & O63505010 & 8662 & E140M & 175.4 &  & $  3.7$ & $  0.075$ & $  21.9$\\
P10204 & HD013745 & O6LZ05010 & 9434 & E140M & 175.4 &  & $  1.9$ & $  0.039$ & $  11.3$\\
S3040202 & HD224151 & O54308010 & 8043 & E140H & 227.0 &  & $ -4.0$ & $ -0.082$ & $ -23.7$\\
P11623 & HD185418 & O5C01Q010 & 8241 & E140H & 227.2 &  & $ -5.3$ & $ -0.108$ & $ -31.5$\\
P12241 & HD224151 & O54308010 & 8043 & E140H & 227.2 &  & $ -3.4$ & $ -0.069$ & $ -20.2$\\
\enddata
\tablenotetext{a}{This is the shift which must be added to the \FUSE\ data to align the 
P(3) and R(4) 6$-$0 H$_2$ lines with the Cl~I~$\lambda 1347$ line in the STIS data}
\tablenotetext{b}{Data from GO programs: 8241---``A SNAPSHOT survey of
  Interstellar Absorption Lines'', Lauroesch PI; 8662, 9434---``A
  SNAPSHOT survey of the Hot Interstellar Medium'',  Lauroesch PI;
  8043, 8484---``Physical Properties of H~I and H~II Regions'', Jenkins PI.}
\end{deluxetable}

\clearpage

\section{STELLAR DISTANCES, THEIR ERRORS, AND CORRECTIONS TO AVAILABLE DATASETS \label{sect_distances}}

A key set of parameters for our \FUSE\ sample is the distance to each
star, $d$, and its uncertainty.  These quantities are important
because one of our objectives was to measure as accurately as possible
the mean and dispersion of the average density of O~VI using all the
available sight lines.  To derive $d$ in pc we need to calculate the
spectroscopic distance (often referred to as the ``spectroscopic
parallax'') given by

\begin{equation}
5\log d = (m_V - A_V - M_V +5) \label{eqn_dist}
\end{equation}

\noindent
where $m_V$ is the observed apparent magnitude in the $V$ band, $A_V$ is the
interstellar extinction in the $V$-band, and $M_V$ is the absolute $V$-band
magnitude of the star of a particular spectral type and luminosity class
(hereafter, `Sp/L'). In the following subsections, we discuss how we derived
each of these quantities. We show that it is possible to derive observed
magnitudes and interstellar extinctions quite accurately using \Hipparcos\ 
satellite data and infrared magnitudes obtained from the ground.
We also show that the absolute magnitude of a star of
known Sp/L is probably known to $\sim
0.2-0.3$~mags.  What remains poorly known, however, is how well an Sp/L can be
measured for a star, particularly for stars later in type than B0.

\subsection{Observed Magnitudes; Multiple Systems \& Intrinsic Variability \label{sect_mags}}

In order to use a set of magnitudes obtained in a uniform way and at a single epoch, 
we took stellar magnitudes from the \Tycho\ Starmapper
\citep{esa_mission} catalogs. This instrument,  
part of the ESA \Hipparcos\ mission, 
provides two-color photometry, $B_T$ and $V_T$, in a
magnitude system close to that of Johnson $B$ and $V$ bands \citep{leeuwen97}.
The two magnitude bands are related via the transformation
\begin{eqnarray}
V     & = & V_T - 0.090 (B -V)_T \label{eqn_vconvert}\\
(B-V) & = & 0.850 (B-V)_T\label{eqn_bconvert}
\end{eqnarray}

\noindent
provided $ -0.2 \: < \: (B - V )_T\: <\: 1.8$ 
\citep[Eqn.~1.3.20 of ][]{esa_mags}\footnote{The
  ``Guide to the Data'' can be found at
  http://www.rssd.esa.int/Hipparcos/CATALOGUE\_VOL1/catalog\_vol1.html}. 
The errors in the $V$ magnitudes and in the $B-V$ colors are given by
equations~1.3.21 and 1.3.22 of \citet{esa_mags} and are reproduced here
for convenience:

\begin{eqnarray}
\sigma(V)                                    &  =      & (1.09\:\sigma_{V_{T}}^2 + 0.09 \sigma_{B_{T}}^2)^{1/2} \label{eqn_aa} \\
\sigma(B-V) \: = \: G\:\times\:\sigma(B-V)_T & \simeq  & G\:\times\: (\sigma_{V_{T}}^2 + \sigma_{B_{T}}^2)^{1/2} \label{eqn_bb}
\end{eqnarray}
The factor $G$ in the above equation provides the slope of the local
($B-V$) against ($B-V$)$_T$ relation and ranges from 0.85 to 0.97. We
calculate a precise value of $G$ for each value of
($B-V$)$_T$ by extrapolating the values given in Table 1.3.4 of
\citep{esa_mags}. Using this scheme, the majority of the stars in our
sample has
$B$ and $V$ magnitude errors of $<\:0.04$ mags.

We note the following exceptions where we did not use these
relationships: HD~326329 only has an entry for $V_T$ (=8.90) in the
\Tycho\ catalog, with no $B_T$.  For this star we used $V$ and $B$
from \citet{baume99}.  HD~152314 is also not in the \Tycho\ catalog,
but does exist in the \Tycho-2\ catalog\footnote{VizieR catalog I/259}
\citep{hog00}, which lists both $B_T$, $V_T$ and associated
errors. Several stars have unusually large errors in the Tycho
catalog, and their values are replaced by values from the
literature. These stars are: CPD$-$417712 \citep{baume99}, and
CPD$-$592600 \& CPD$-$592603 \citep{massey93}.  All the adopted
magnitudes are listed in Table~\ref{tab_journal}.

\subsubsection{Photometric Binaries \label{sect_photobin}}

It is well known that stars are often members of multiple stellar systems. This
presents the simple problem that a star's measured magnitude may not be a true
measure of its distance if the measured flux really comes from two (or more)
unresolved stars. The \Hipparcos\ data can again be used --- at least for a
subset of stars --- to help identify  multiple systems.

Of the 148 stars in our sample, 106 are listed in the \Hipparcos\ 
catalog and could therefore be analyzed in more detail than was possible from
their \Tycho\ data alone. Of these, 19 are listed in the \Hipparcos\ {\it
  Double and Multiples System Annex: Component solutions} (DMSA/C) catalog
\citep{esa_doubles}, which means that these system can be modeled as individual
stars. (The \Hipparcos\ catalogs labels these as ``Component Solutions''.) For
a particular system, we adopt the magnitudes of the brightest component listed
in the DMSA/C catalog; we thereby assume that the bulk of the flux which defines the
Sp/L of the star comes from the brightest component. Six stars (HDs 099857,
101190, 101413, 101436, 190918, 152248) are cataloged in the
DMSA/C catalog, but the magnitude of only one component is given. We take this
to mean that the secondary is too faint to be modeled; its presence is
therefore not expected to alter the magnitude of the primary star. (We use the
DMSA/C catalog magnitude nonetheless.) 

In the DMSA/C catalog, only \Hipparcos\ magnitudes $H_p$ are available, and
not the $V_T$ used to convert to Johnson $V$ magnitudes for the other stars. A
comparison between $H_p$ and Johnson $V$ for stars not labelled as multiple
show that in all cases, differences in $V$ (calculated from $V_T$) and $H_p$
are small ($\leq 0.05$ mags).  The transformations suggested by
\citet{harmanec98} also show that for $B-V=\pm0.2$ --- the minimum and maximum
values for our stars --- the difference in $H_p$ and $V$ should only be
$\mp0.06$ mags.

Two of the 106 stars (HD~185418 and HDE~232522) are listed in the \Hipparcos\ 
catalog as being multiple due to their non-linear motion (``Acceleration
Solutions'' --- similar to astrometric binaries) although they have no entry
in the DMSA/C catalog. Both are actually spectroscopic binaries, and their $V$
magnitudes are corrected for in same way as the rest of the spectroscopic
binaries in our sample, discussed below.

For the 42 stars not in the \Hipparcos\ catalog, the \Tycho\ catalog provides a
flag to indicate which stars may be multiple systems\footnote{Field T36 of the
  Tycho catalog, {\it Source of photometric data}; see \citet{esa_tycho}}.
None of the 42 stars not in the \Hipparcos\ catalog were flagged as multiples
in the \Tycho\ catalog.

\subsubsection{Spectroscopic Binaries \label{sect_specbins}}

An important class of multiple systems are the spectroscopic binaries (SBs). These
stars are too close together to be resolved by  \Hipparcos\ (except for those
which might also be eclipsing binaries --- see below). Given the luminosity of
the primary and secondary stars to be $L_1$ and $L_2$, the correct distance to the
system $d'$ is 

\begin{equation}
d' = \left( 1+\frac{L_2}{L_1} \right)^{1/2} d 
\label{eqn_distcorr}
\end{equation}

\noindent
where $d$ is the distance to the system estimated from the integrated ($L_1 +
L_2$) luminosity. In the worst case, $L_1$ and $L_2$ are equal and $d$ is
incorrect by a factor $\sqrt{2}$. If the magnitudes of the primary and
secondary, $m_1$ and $m_2$ respectively, are known, the ratio of the
luminosities can be calculated in the usual way $L_2/L_1 = 10^{0.4(m_1 -
  m_2)}$. In many cases, however, the value of $\Delta
m$ is unavailable. How then can we estimate what correction to make for these
systems?

For double-lined SBs, the ratio of the radial velocities for each component
gives the mass ratio of the system $M_2/M_1$. The mass-luminosity relation for
main sequence (MS) stars with mass $>0.2 M_\odot$ is given by \citet{sk82}:

\begin{equation}
\log\frac{L}{L_\odot} = 3.8 \log\frac{M}{M_\odot} - 0.8 .  \\
\label{eqn_MtoL}
\end{equation}

To convert between the bolometric luminosity $L$, and $V$-band luminosity $L_V$,
we must apply a bolometric correction (BC) to equation~\ref{eqn_MtoL}. The BC is
a function of luminosity, but can be derived for a given Sp/L; for MS stars,
the BC can be shown to be $-0.9\:\log(L/L_\odot ) + 1.2$ \citep{sk82} which gives

\begin{equation}
\log\frac{L_1}{L_2} = 2.4 \log\frac{M_1}{M_2} 
\label{eqn_final}
\end{equation}

\noindent
where the luminosities are those measured in the $V$-band. This can then be
applied to equation~\ref{eqn_distcorr} to correct the distance to the system.

To identify spectroscopic binaries in our sample of stars, we used {\it The
  Ninth Catalogue of Spectroscopic Binary Orbits} (SB9) by \citet{pourbaix04},
and the list of binaries in NGC~6231 by \citet{garcia01}, who note that
HD~152314 (S161), HD~152200 (S266), HD~152233 (S306) and HD~152248 (S291) are
probably spectroscopic binaries. We identify 21 of our stars to be SBs. For 11
of these stars, values of $\Delta L$ or $\Delta M$ are available, and their
distances can be corrected directly from equation~\ref{eqn_distcorr} or indirectly
from equation~\ref{eqn_final}.  For the remaining ten stars, we note that
\citet{pins06} find that 50\% of binaries have mass ratios $>0.87$, and use
this as a generic correction. This mass ratio corresponds to $\Delta L\: =\:
0.72$, and hence the distance to the system is increased by a factor of 1.3.
A superscript next to the distance to a star in Table~\ref{tab_journal}
indicates that the star was treated as a SB.

As noted above, equation~\ref{eqn_MtoL} is probably only applicable for MS
stars, and evolved stars may well not obey the same $M-L$
relationship. Unfortunately, only seven of our SBs are MS stars. It
may be that equation~\ref{eqn_MtoL} does not represent a precise conversion
between mass and luminosity, and, therefore, that the correction we
make to the stellar distance is similarly imprecise.  The distances to
these systems should be treated with caution.

\subsubsection{Variable Stars \label{sect_variables}}

Stars which are not multiple systems may still exhibit variability.
These may be rotating, pulsating, eruptive or cataclysmic types of
variability. B-type emission line stars may also show variations in
luminosity. 14 stars in our sample which are not in SB9 or in the
\Hipparcos\ {\it Double and Multiples} catalog are listed in the
\Hipparcos\ {\it Variability Annex} \citep{esa_var}. For these stars,
we take the maximum $H_p$ magnitude (minimum luminosity) from the
fitted light curve derived from the \Hipparcos\ photometry. The
assumption here is that the minimum luminosity of the star represents
its `true' luminosity for its given spectral type and luminosity
class.  This assumption is based on the idea that Be stars are
brighter than B stars for a given spectral type \citep{briot97};
obviously, for stars whose luminosities are not increased by
mechanisms such as the emission from circumstellar disks which are
thought to surround Be stars, this assumption may be invalid. However,
of the 14 stars corrected, all but one have a difference between the
faintest magnitude and the mean magnitude of less than 0.06 mags.
(HD~042401 shows a difference of 0.17 mags). Although these corrections
are therefore largely irrelevant, they have been applied to our data.
Distances adjusted this way are noted with a superscript in
Table~\ref{tab_journal}.

Variable stars may, of course, be (partially) eclipsing binaries, more akin to
the SBs discussed above. However, most of the stars in the
{\it Variability Annex} which are labelled as likely (or already known) 
eclipsing binaries (field 'P5' in the {\it Annex}) are already
identified in SB9, and their distances are corrected in the way described
above. (Only HD~042401 is identified as a $\beta$ Lyrae type of eclipsing
binary and is not in SB9.) Hence the luminosities for these stars are not
corrected in the same way that the spectroscopic binaries were corrected
above.

Unfortunately, for the 42 stars of our sample not in the \Hipparcos\ catalog, the \Tycho\ 
analysis provides no clear information on which stars are intrinsically
variable. Although a variability flag does exist, the flag appears unreliable
(failing to note variability when detected by \Hipparcos , and flagging
variability when \Hipparcos\ detects none) and we have not included the
information in Table~\ref{tab_journal}. Instead, we searched for our stars in
the {\it Combined General Catalogue of Variable Stars} (GCVS) and the GCVS
{\it Suspected Variable stars and Supplement} catalogs
\citep{samus04}\footnote{vizieR catalog II/250}.  This catalog contains a very
inhomogeneous set of data, but can indicate that a star was recorded to have a
fainter magnitude than was seen during the $\simeq 3$ year \Hipparcos\ mission.

For stars which have no entries in the
\Hipparcos\ catalogs, are not SBs, or which have distances
that are not taken from the literature, we find that: a) only one has a minimum-maximum
variation listed in the GCVS catalogs greater than 0.06 mags
[\citet{lipunova85} noted that HDE~332407 fell from $V=8.51$ (close to the
\Hipparcos\ value of $V=5.82$) by $\simeq 0.1$ mags. \Hipparcos\ found no
variation, but we have made this correction in Table~\ref{tab_journal}]; and
b) six stars have variations less than 0.06 mags
\footnote{Using a value of 0.06 mags to separate 'weak' from 'strong'
  variability is purely arbitrary, but the value is used in the \Hipparcos\
  catalog to provide a course variability flag (Field H6 of the main catalog),
  which is why we use that same value here. Stars in our sample which show a
  variability $<0.06$ mags usually do not show up in the  {\it Variability
    Annex}. Hence we cannot correct the magnitudes of these stars anyway.}
, which we consider small enough to ignore. 

Given the short lifetime of the \Hipparcos\ mission, it is also interesting to
consider whether the maximum magnitudes of the variable stars that are indeed
in the {\it Variability Annex} are significantly different from the maxima
given by the GCVS catalogs. In fact, for the six stars for which there is a
difference $> 0.06$ mags (excluding SBs), the distances we use are taken from the literature,
measured by \Hipparcos\ parallax, or the GCVS magnitude is not from
$V$-band measurements. There is therefore no evidence that our distances are affected by 
not considering a long enough timeline for the variable stars\footnote{Such variations
may exist of course, but they have not been recorded in the literature.}.

\subsection{Interstellar Extinction \label{sect_av}}

The relationship between interstellar extinction $A_V$ and the color excess of
stars $E(B-V)$ is usually taken to be $A_V\:=\:R_V\:E(B-V)$, where $R_V$ is the ratio
of total to selective visual extinction. In the Milky Way, a typical value
of $R_V$ through the diffuse ISM is  $\simeq\:3.1$. In a compilation of $R_V$ values for
417 O3- to B5-type stars, \citet{valencic04} found a Gaussian distribution of
$R_V$ centered at 3.0 (largely as expected), but with a 1$\sigma$ dispersion
of 0.35. Assuming that a large fraction of this dispersion arises from real
line-of-sight differences in $R_V$, and not just from measurement errors, then
using a value of 3.1 for $R_V$ to calculate $A_V$ might be lead to errors of
at least $0.35\:E(B-V)$ mags in calculating the distance to a star in
equation~\ref{eqn_dist}, even if the errors in $E(B-V)$ were zero. For values of
$R_V$ very different from $\simeq\: 3$  the errors are even larger.  

We would therefore like to have a way to measure  $A_V$ along a sight line
directly. \citet{whittet92} describes a method
which uses infrared (IR) magnitudes: $R_V\:\simeq\:1.1\:E(V-K)/E(B-V)$. Since
$A_V\:=\:R_V\:E(B-V)$, we can directly measure the extinction from

\begin{equation}
A_V\:=\:1.1\:E(V-K)
\label{eqn_ir}
\end{equation}

To obtain $K$-band magnitudes, we used the {\it Two Micron All Sky
  Survey} (2MASS) \citep{skrutskie06}\footnote{vizieR catalog II/246}. The
2MASS magnitude was converted to Johnson $K$ using the relation\footnote{From
  the {\it Explanatory Supplement to the 2MASS All Sky Data Release} at
  http://www.ipac.caltech.edu/ 2mass/releases/allsky/doc/sec6\_4b.html by
  R.~M.~Cutri~\etal .}  $K\:=\:K_{\rm{2MASS}} +0.027 + 0.00787(J_{\rm{2MASS}}
- K_{\rm{2MASS}})$ unless the $J_{\rm{2MASS}}$ was flagged as unreliable, in
which case the color term was neglected.  Values of $K_{\rm{2MASS}}$ were only used if
an `A' photometric quality flag (Qflg) was given (see the 2MASS catalog for a
full explanation of this flag). Intrinsic IR colors for early type stars were
taken from \citet{wegner94}.

We calculated the error in the IR color excess as $\sigma(E_{V-K})^2
= \sigma(V)^2 + \sigma(K)^2 + 0.03^2$, where the last term is adopted as a
typical error in the intrinsic $V-K$ colors given by \citet{wegner94};
$(V-K)_0$ appears to vary slowly with luminosity class, but strongly with
stellar type. With this error, we were able to reject poorly determined IR
color excesses when $E(V-K) < 2\sigma (E_{V-K})$, and did not use
equation~\ref{eqn_ir}. $E(V-K)$ was also not used if more than one
stellar type or luminosity class was given for a particular star, or if the
star was in the \Hipparcos\ {\it Variability Annex} or DMSA/C catalogs, since
we have no information on how a star might vary in the IR from 2MASS.
We allowed $E(V-K)$ to be used for the SBs; here we assume that $\Delta
L$ in the optical is similar to $\Delta L$ in the IR. 

If we were unable to use $E(V-K)$ to estimate $A_V$, we fell back to the more
usual practice of calculating  $A_V\:=\:3.1\:E(B-V)$. $E(B-V)$ is given for
most of our stars in Table~\ref{tab_journal}, except when distances were taken
from the literature. The error in the Johnson $B-V$ color is given by
equation~\ref{eqn_bb} above; we again used
\citet{wegner94} for the intrinsic ($B-V$) colors of the stars, and assumed a value
of $\sim0.01$ mags as a likely error, since the intrinsic color varies
very slowly with spectral type and luminosity class for early type stars.

\subsection{Absolute Magnitudes of O-type and Early B-type Stars \label{sect_absmags}}

Beyond the problems in measuring the observed magnitudes of, and the
extinction towards, a particular star, the largest uncertainty in
measuring its distance comes from a poor knowledge of its absolute
magnitude $M_V$ in equation~\ref{eqn_dist}. There are two problems: knowing
the absolute magnitude of a star given its MK-system spectral type and
luminosity class [at least, the extension developed by
\citet{walborn71}, since the original MK classification by
\citet{mk43} did not cover stars earlier than O9]; and, more
importantly, knowing precisely the spectral type and luminosity class
of any given star.

The first of these problems is largely tractable, thanks to surveys
which use stars with well-defined Sp/Ls, carefully measured observed
magnitudes, and distances measured using methods other than
the spectroscopic parallax. Although the measurements of all these
quantities are fraught with their own uncertainties, there appears to
be consensus on how $M_V$ varies with Sp/L for early type stars. To
produce our own $M_V$-Sp/L scale, we collated data from
\citet{walborn72}, \citet{walborn73}, \citet{lesh79}, \citet{sk82},
\citet{humph84}, and \citet{howarth89}.  We also added, when an
Sp/L was available, magnitudes from models by \citet{vacca96} and
\citet{martins05} for comparison.  Not every available published scale
was included: for example, absolute magnitudes for OB stars have
been given by \citet{wegner00} based on \Hipparcos\ parallaxes, but
these values differ so significantly from the other authors just cited
that we did not include them. The final $M_V$ magnitudes for a given
Sp/L are tabulated in Table~\ref{tab_absmags}.

\begin{figure}[t!]
\hspace*{2cm}\includegraphics[width=14cm]{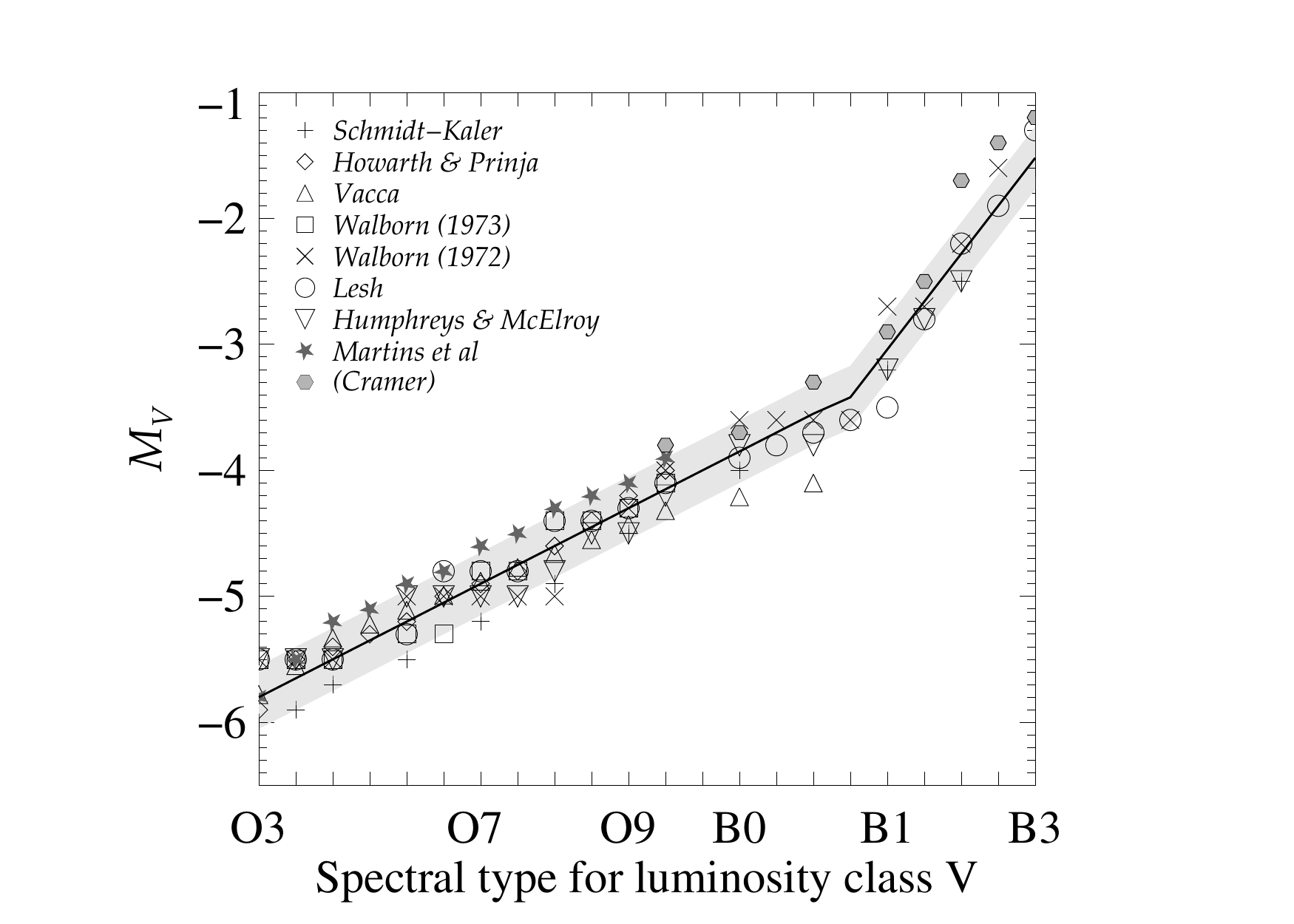}
\caption{\label{fig_classV} Plot of absolute $V$-band
  magnitude $M_V$ against spectral type (O3$-$B3) from the authors
  listed at the top left of the figure. Our adopted relationship
  between $M_V$ and spectral type is shown as a solid black line; the
  gray region either side marks a difference of $\pm 0.25$ mags which
  we take as the error in knowing $M_V$ for any given stellar type.}
\end{figure}

An example of one $M_V$-Sp/L scale is shown in
Figure~\ref{fig_classV}, which plots $M_V$ against spectral type for
luminosity Class V stars. (These, along with Class III stars, are the
most well studied of the luminosity classes.)  In general, authors
tend to agree on the relation between $M_V$ and spectral type to
within several tenths of a magnitude for this luminosity class.  In
Figure~\ref{fig_classV} we show our adopted scale as a straight line,
along with an error of $\pm 0.25$ mags (shown by gray boundaries).
Theoretical values are also in good agreement with the data: the
models by \citet{martins05} appear to be too faint by only $\approx
0.25$ mags, while those from \citet{vacca96} agree well with
observations for the early stellar types until B0.


Figure~\ref{fig_classV} also shows the values of $M_V$ derived by
\citet{cramer97} for B-type stars whose parallaxes were measured by
\Hipparcos . We did not use these data in setting our $M_V$-Sp/L scale
as no O-type stars were included, and a complete analysis remains
unpublished. The variation of $M_V$ with Sp is not entirely smooth,
presumably due to the small number of B-type stars close enough to the
Sun for an accurate parallax to be determined. Nevertheless, although
the stars as a group seem to be slightly fainter than our adopted
scale, the values of $M_V$ are largely consistent with our adopted
scale, to within an error of $\approx\:0.25$ mags. (We return to the
question of how well our spectroscopic distances compare to
\Hipparcos\ parallaxes in Appendix~\ref{sect_hippcf}.)

Plots of $M_V$ and spectral type for other luminosity classes show similar
results, although the Class I stars have no `break' around B0 but have roughly
constant absolute magnitudes for all spectral types, as expected.

One other problem which remains is that of the Be stars, those which presently, or in
the past, have shown emission lines. There are nine Be stars in our sample;
the adopted Sp/L for these should probably be used with caution, since the
classification of Be stars is known to be problematic, due to emission
lines filling-in the diagnostic absorption lines, and due to the large
rotation speeds which are usually associated with these systems.  Unpublished
studies of the difference in $M_V$ between B and Be stars using \Hipparcos\ 
parallax distances by \citet{briot97} and \citet{briot00} suggests that
although Be stars of a given Sp/L are indeed brighter than B stars of the same
Sp/L, the differences for the stars of interest in this paper, B0-B3, are
negligible. Hence we make no additional corrections to $M_V$ for Be stars.

Given the relative small dispersions in $M_V$ and the general agreement with
theoretical values (when available), we assume that $M_V$ can be known for a
star of a particular Sp/L to within $\pm 0.25$ mags.

The second problem in determining $M_V$ for a given star --- that of
knowing its Sp/L precisely --- is not so straightforward.  Stellar
classification depends not only on the quality, resolution, and
wavelength extent of the spectra used to make the classification, but
on the experience of the observer making the classification. (Many
stars were assigned their Sp/Ls several decades before the development
of automatic classification of digital spectra, for example.)
Moreover, although a particular set of stars can be used to define a
class and type \citep[e.g.][]{garrison94}, Sp/L categories are
discrete, and any given star may in reality fall between the
designated category.  For example, the absolute magnitudes in
Figure~\ref{fig_classV} decreases rapidly for stars between types B1 and
B2, though only very small changes are seen in line ratios used to
define the Sp/Ls. Hence a star may have a magnitude a few tenths
different from the $M_V$ assigned from the adopted Sp/L.  More
serious, however, is the fact that the use of a poor quality spectrum
of a star can lead to a complete mis-classification of a star, an
occurrence which appears more frequently for the B-type stars.

In most cases, we have no access to the original data used to
determine the Sp/L of the stars in our survey. We have therefore
searched through the available literature to find suitable Sp/Ls for
our stars. The references for the Sp/Ls we adopt are given in the
final column (16) of Table~\ref{tab_journal}.

Given the difficulties which exist in assigning an Sp/L to a star, we
have developed a simple procedure for measuring the likely range of
plausible distances to an individual star.  Our primary goal is to
quantify a generic error in $M_V$ for each star. Assigning only a
single number for this error is inappropriate, since uncertainties in
$M_V$ are larger for, e.g. B-type stars than for O-type stars of the
same luminosity class. Instead, we calculate an error in three parts:
we first take the difference in magnitude between the adopted value
and the magnitude of the next spectral type in the series; we then
calculate the difference in magnitude between the adopted spectral
type and that of the same spectral type in the next luminosity class.
Finally, we add these two values in quadrature with the error we adopt
for assigning $M_V$ to a given Sp/L, which we take to be
$\sigma(M_V)_C\:=\:0.25$ mags, as discussed above.  In notation form,
we calculate an error in $M_V$ for possible positive deviations from
the adopted $M_V$ thus:

\begin{equation}
\sigma[M_V, ({\rm{Sp/L}})_i]^2 \: =\: \Delta M_V ({\rm{Sp}}_{i+1} -
{\rm{Sp}}_{i})^2 + \Delta M_V ({\rm{L}}_{i+1} - {\rm{L}}_{i})^2 +
\sigma(M_V)_C^2 
\label{eqn_spl}
\end{equation}

\noindent
So, for example, the positive error in a B3 III star is formed from taking
$M_V$(B4 III) $-$ $M_V$(B3 III) in Table~\ref{tab_absmags}, $M_V$(B3 IV) $-$
$M_V$(B3 III), and adding these in quadrature, along with 0.25 mags for
$\sigma(M_V)_C$. The negative deviation is calculated in the same way, only
taking differences between Sp$_{i}$ and Sp$_{i-1}$, and L$_{i}$ and L$_{i-1}$.
Using these values of $M_V$ leads to an upper and lower bounds for the
distance, $d_u$ and $d_l$, respectively.

Even this method is imprecise however, since its is easier to
classify, e.g. an O4 star than a B1 star. In this sense, the ranges in
$M_V$ determined for O type stars may be somewhat over-estimated,
while for B type stars the errors may be too small. However, the
magnitude of the errors are likely to be roughly correct, since in the
scheme outlined in equation~\ref{eqn_spl}, O stars have smaller errors than
B stars. 

From estimating distance `errors' in this way,
we conclude that {\it the errors we derive in $M_V$ 
dominate all other errors discussed in this Appendix.}
For example, for the Class V dwarfs and Class IV subgiants, errors range from $\simeq\:
0.3-1.0$ mags for O3$-$B3 type stars, or a 14$-$46\% error in the 
distance\footnote{$\sigma(d)/d = 0.46\sigma(M_V)$}.
The largest errors in this scheme occur for the B3 Class II giants, since their
magnitudes are furthest from both the Ia supergiants as well as the Class III
giants;  errors range from $0.4-1.5$ mags for  O3$-$B3 type stars, 
$20-70$~\% in distance. Clearly, the errors associated with the extinction $A_V$
($\simeq 0.1-0.2$ mags) and those with the \Tycho\ and \Hipparcos\ catalog
magnitudes ($<<\:0.1$~mags), are insignificant compared to the errors in $M_V$.  
 
Further, in some cases, the stars in our sample have no published luminosity
class, or only a range of luminosity classes are given. In these cases we
calculate the distances for the brightest and faintest of the specified classes
(or `V' and `Ia' if no Class is given), and use the average value for
the distance. The minimum and maximum error are simply the minimum and maximum
values calculated for each class in the way described above. The value of
$E(B-V)$ in Table~\ref{tab_journal} is also an average value
(although the average was not used in the calculation of the
distance.)

\subsection{Distances from the Literature \label{sect_lit_distances}}

There are some exceptions to this scheme for calculating the distance to our
survey stars.  Our sample contains four Wolf-Rayet stars; distances and values
of $E(B-V)$ for three of these, HD~151932 (WR~78), HD~191765 (WR~134) and
HD~190918 (WR~133) are taken from the catalog of \citet{vanderhucht}, while
data for HD~187282 (WR~128) come from \citet{conti90}.  HD~93129A is an O2 If
star \citep{walborn_O2s} in Trumpler 14, one of the ionizing clusters in
Carina, and since O2 stars have no defined values of $M_V$, we adopt the
distance given by \citet{tapia03}.  The \Hipparcos\ DMSA/C catalog lists four
components to the multiple system of which HD~005005A is a member, but their
membership of the IC~1590 cluster seems unequivocal --- unlike the Sp/L of
HD~005005A --- and an accurate distance is given by \citet{guetter97}, which we
use. LS~277 has no published Sp/L, and its distance is taken from
\citet{reed93}. Finally, HD~110432 is close enough that the parallax distance
from \Hipparcos\ ($300\pm 50$~pc) should be more accurate than its distance
determined from its Sp/L (see Appendix~\ref{sect_copernotes}).

Note that in Table~\ref{tab_journal}, use of a distance from the literature is
indicated with a superscript next to the distance. If no errors for a published
distance are available, we adopt a value of $\pm 20$~\% of the distance.



\begin{deluxetable}{lcccccccc}
\tabletypesize{\small}
\tablecolumns{9}
\tablewidth{0pt}
\tablecaption{Adopted $M_V$ mags for given Spectral Types and Classes \label{tab_absmags}}
\tablehead{
\colhead{} &
\colhead{} &
\multicolumn{7}{c}{Luminosity Class} \\
\cline{3-9}
\colhead{Type} &
\colhead{} &
\colhead{V} &
\colhead{IV} &
\colhead{III} &
\colhead{II} &
\colhead{Ib} &
\colhead{Iab} &
\colhead{Ia} 
}
\startdata
    O3    &&  $-$5.80 &  $-$6.00 &  $-$6.30 &  $-$5.95 &  $-$6.30 &  $-$6.65 &  $-$6.90 \\
    O4    &&  $-$5.65 &  $-$5.88 &  $-$6.20 &  $-$5.94 &  $-$6.29 &  $-$6.64 &  $-$6.91 \\
    O5    &&  $-$5.50 &  $-$5.76 &  $-$6.10 &  $-$5.93 &  $-$6.28 &  $-$6.63 &  $-$6.92 \\
    O5.5  &&  $-$5.35 &  $-$5.64 &  $-$6.00 &  $-$5.92 &  $-$6.27 &  $-$6.62 &  $-$6.93 \\
    O6    &&  $-$5.20 &  $-$5.52 &  $-$5.90 &  $-$5.91 &  $-$6.26 &  $-$6.61 &  $-$6.94 \\
    O6.5  &&  $-$5.05 &  $-$5.40 &  $-$5.80 &  $-$5.90 &  $-$6.25 &  $-$6.60 &  $-$6.95 \\
    O7    &&  $-$4.90 &  $-$5.28 &  $-$5.70 &  $-$5.89 &  $-$6.24 &  $-$6.59 &  $-$6.96 \\
    O7.5  &&  $-$4.75 &  $-$5.16 &  $-$5.60 &  $-$5.88 &  $-$6.23 &  $-$6.58 &  $-$6.97 \\
    O8    &&  $-$4.60 &  $-$5.04 &  $-$5.50 &  $-$5.87 &  $-$6.22 &  $-$6.57 &  $-$6.98 \\
    O8.5  &&  $-$4.45 &  $-$4.92 &  $-$5.40 &  $-$5.86 &  $-$6.21 &  $-$6.56 &  $-$6.99 \\
    O9    &&  $-$4.30 &  $-$4.80 &  $-$5.30 &  $-$5.85 &  $-$6.20 &  $-$6.55 &  $-$7.00 \\
    O9.5  &&  $-$4.15 &  $-$4.68 &  $-$5.20 &  $-$5.84 &  $-$6.19 &  $-$6.54 &  $-$7.01 \\
    O9.7  &&  $-$4.00 &  $-$4.56 &  $-$5.10 &  $-$5.83 &  $-$6.18 &  $-$6.53 &  $-$7.02 \\
    B0    &&  $-$3.85 &  $-$4.44 &  $-$5.00 &  $-$5.82 &  $-$6.17 &  $-$6.52 &  $-$7.03 \\
    B0.2  &&  $-$3.70 &  $-$4.32 &  $-$4.90 &  $-$5.70 &  $-$6.10 &  $-$6.51 &  $-$7.04 \\
    B0.5  &&  $-$3.55 &  $-$4.20 &  $-$4.80 &  $-$5.50 &  $-$6.05 &  $-$6.50 &  $-$7.05 \\
    B0.7  &&  $-$3.42 &  $-$4.08 &  $-$4.60 &  $-$5.30 &  $-$6.00 &  $-$6.49 &  $-$7.06 \\
    B1    &&  $-$3.04 &  $-$3.96 &  $-$4.25 &  $-$5.10 &  $-$5.95 &  $-$6.48 &  $-$7.07 \\
    B1.5  &&  $-$2.66 &  $-$3.50 &  $-$3.90 &  $-$4.90 &  $-$5.90 &  $-$6.47 &  $-$7.08 \\
    B2    &&  $-$2.28 &  $-$3.10 &  $-$3.55 &  $-$4.70 &  $-$5.85 &  $-$6.46 &  $-$7.09 \\
    B2.5  &&  $-$1.90 &  $-$2.70 &  $-$3.20 &  $-$4.50 &  $-$5.80 &  $-$6.45 &  $-$7.10 \\
    B3    &&  $-$1.52 &  $-$2.30 &  $-$2.85 &  $-$4.30 &  $-$5.75 &  $-$6.44 &  $-$7.11 \\
\hline
Refs:\tablenotemark{a}  &&   $1-7$  & 1,$3-5$ & $1-7$ & 1,2,4,5 & 1,2,4,5 & 1,2,4,5 & 1,2,4,5,7 \\
\enddata
\tablenotetext{a}{Collation of data from: 
(1)---\citet{walborn72};
(2)---\citet{walborn73};
(3)---\citet{lesh79};
(4)---\citet{sk82};
(5)---\citet{humph84};
(6)---\citet{howarth89};
(7)---\citet{vacca96}}
\end{deluxetable}

\clearpage

\section{REVISITING EARLIER SURVEYS \label{sect_other_surveys}}

\subsection{Corrections Made to \emph{Copernicus} Data \label{sect_copernotes}}

\subsubsection{Notes on Stellar Parameters}

In \S\ref{sect_copernicus} we introduced the stars studied by J78. Since
they are both bright and comparatively nearby, there is a
considerable amount of data available in the literature for these stars. In
deriving distances to the \Copernicus\ stars, we have tried to treat them in
exactly the same way as our \FUSE\ sample. All the \Copernicus\ stars are listed
in the \Hipparcos\ main catalog, which means that they can be treated even more
uniformly than the \FUSE\ sample of stars. The stellar data for the \Copernicus\ 
stars are given in Table~\ref{tab_copernicus}; below, we outline some of the
changes we have made to the stellar data since the stars were studied by
J78.

{\bf Sp/Ls:} these are taken directly from Table~1 of J78. Many of 
these stars have been studied since the advent of stellar spectroscopy, and their
Sp/Ls are well understood (and often actually define a particular
Sp/L). Small changes have been made to the Sp/Ls of $\gamma$~Cas, $\epsilon$~Per,
$\zeta$~Ori, $\zeta$~Pup, $\beta$~Cen, $\beta$~Lup, $\delta$~Sco, and
$\sigma$~Sgr since improved classifications are available in \citet{garrison_web}.

{\bf Magnitudes:} $V$ and $B$ magnitudes are converted from $B_T$ and $V_T$
given in the \Tycho\ catalog in the same way as described above. We have again
tried to use $E(V-K)$ to calculate $A_V$, but we find that many of the stars
are saturated in the 2MASS data. 

{\bf Emission line stars:} more information is now available for whether emission
lines have been detected in the stellar spectra of the \Copernicus\ stars, and
we have appended 'e' suffixes to the Sp/Ls in Table~\ref{tab_copernicus} 
for the following stars: 
$\alpha$~Eri \citep{hanuschik96},
$\eta$~Tau \citep{tycner05},
$\alpha$~Cam \citep{morel04},
$\lambda$~Eri \citep{hanuschik96},
25~Ori \citep{banerjee00},
$\epsilon$~Ori \citep{morel04},
HD~051283 \citep{irvine90},
HD~052918 \citep{cote93},
$\delta$~Sco \citep{galazutdinov06},
$\lambda$~Pav \citep{mennickent91},
2~Vul \citep{barker83},
$\kappa$~Aql \citep{cote93},
60~Cyg \citep{koubsky00},
and 8~Lac \citep{chauville01}.

{\bf Spectroscopic binaries:} we have again used the SB9 to flag which stars
are SBs, and have adopted values of $\Delta L$ or $\Delta M$ when available
(Appendix~\ref{sect_specbins}).  For HD~057060 (29 CMa), $\Delta L$ and its
Sp/L are taken directly from \citet{bagnuolo94}. HD~037303 is cited by
\citet{morrell91} as being an SB in the Orion OB1
association, though the star is not listed in SB9. We assume this designation
is correct. For HD~047839 (15~Mon) \Hipparcos\ gives the magnitude 
of the primary component ($V=H_p=4.61$), and we take $\Delta m = 1$ mag 
between primary and secondary from \citet{gies93}.
HD~158926 ($\lambda$ Sco) is composed of two early-type B stars and a
low-mass pre-main-sequence star, according to \citet{uytterhoeven04}, but \Hipparcos\ finds
no multiplicity in the DMSA/C catalog, and the system is not listed in SB9. We
use these authors' estimate for the mass ratio of the two stars, $\Delta M \simeq
0.84$. 

{\bf Variables:} in general, stars flagged by the GCVS as variable are also found in the
\Hipparcos\ catalog. There is little published evidence that any of the
\Copernicus\ stars have been substantially fainter over the last century than
the faintest magnitudes found by \Hipparcos . We know of two clear exceptions:
the long period $V$-band light curve for HD~005394 ($\gamma$ Cas) has a
minimum luminosity of $V=2.75$ \citep{doazan83}, while HD~200120 is known to
show a minimum of at least $V=5.05$ \citep{harmanec02}. We adopt these
magnitudes as the `true' values for these two stars.

{\bf Multiple systems:} HD~36486 ($\delta$ Ori) is a triple system.
\Hipparcos\ resolves the eclipsing binary, but the primary is an SB.
\citet{harvin02} show that the flux from the SB is dominated by the O9.5II
primary, and that $\Delta L$ is small, only 0.09. We use the magnitude of component
Aa, and correct that using $\Delta L = 0.09$.  For HD~57061 ($\tau$ CMa),
\Hipparcos\ finds two components; using the magnitude of the primary, and
assuming this is composed of two stars, we obtain a distance of 1224~pc.
However, the system appears to be quite complicated [see the discussions by
\citet{leeuwen_taucma} and \citet{stickland98}]. Although we could use the
distance to NGC~2362, [e.g. 1480~pc by \citet{moitinho01}],
\citet{leeuwen_taucma} note that HD~57061 may no longer be near the center of
the cluster. Given the lack of information, we simply retain the derived
distance of 1224~pc.

\subsubsection{Errors on \emph{Copernicus} O~VI Column Densities}

The \Copernicus\ stars tend to probe distances in the Galactic disk in the
range between the WDs and the stars studied in the
present survey. As such, their
inclusion in an analysis of how, e.g., $N$(O~VI) varies with distance, is valuable.
Unfortunately, no {\it errors} on the column density $\sigma(N)$ were given by
either J78 or \citet{ebj74}. Yet $\sigma(N)$ for these stars is an
important quantity, since we need to know how much statistical weight to give
 $N$(O~VI) when deriving quantities such as the average
volume density in the Milky Way disk (as discussed in \S\ref{sect_NwithZ}).

We can estimate these errors by using the coadded scan data of the
stars from the \Copernicus\ MAST archive. These data
include a $2\sigma$ noise error array, which can be used to measure
$\sigma(N)$.

Deriving $N$(O~VI) from \Copernicus\ MAST archive data is complicated; estimating
the continuum of the star around the O~VI line is often difficult, and
the precise subtraction of the HD absorption (when other HD lines are
available to produce a satisfactory model of the absorption) is 
challenging. We believe we are unlikely to improve upon the actual values of
$N$(O~VI) given by J78, and so concentrate on how best to obtain
$\sigma(N)$. The \Copernicus\ data are of variable quality, so a fixed
percentage (for example) of the measured $N$(O~VI) would be an inappropriate
measure of the error in the column density.  Since we only require approximate
estimates of $\sigma(N)$, our approach is to produce a very crude
re-measurement of $N$(O~VI) and extract a value of $\sigma(N)$ associated with
that column density. To first order, this value of $\sigma(N)$ should be close
to the real value. The values are likely to be less precise than those derived
for the \FUSE\ sample, but they can be used to correctly weight the \Copernicus\ 
values when data from all samples are combined.

We fitted the available \Copernicus\ coadded scans made around
the O~VI~$\lambda 1032$ line in the same way as described for the
\FUSE\ sample in \S\ref{sect_cont}, and measured $N$(O~VI) with the
AOD method (\S\ref{sect_colsAOD}).  The velocity range used by J78
to measure $N$(O~VI) was variable, anywhere between
[$-50$,$+50$]~\kms\ to twice that range. For simplicity, we measured
$\sigma(N)$ between $-70$ and $+70$~\kms. We calculated a final value
of $\sigma(N)$ by adding in quadrature the errors from the continuum
uncertainties and the noise errors given by the coadded scan error
arrays.

Despite the simplicity of this approach, our measurements of
$N$(O~VI) and those published by J78 are in reasonable agreement.
More importantly, J78 plotted profiles of the optical
depth at each pixel, $\tau_i$, along with $2\sigma$ error envelopes,
$2\sigma(\tau_i)$, for about half of the \Copernicus\ sample. This means we
can directly compare the values of $\sigma(\tau_i)$ measured from the Archive
data with those shown in Figure~2 of J78 for many of the stars. In
general, the agreement was very good.  Only three stars showed apparent
problems: HD~064740, HD~106490 and HD~121263 had unusually large values of
$\sigma(\tau_i)$ compared to the envelopes shown in J78, all
because of inexplicably large noise errors $\sigma_i$ in the Archive data
error arrays. We estimated that the errors were too large by approximately a
factor of $4-5$, and subsequently reduced all values of $\sigma_i$ by a factor
of 5 before calculating $\sigma(\tau_i)$.

Compared with the dataset as a whole,  $\sigma(\tau)_i$ for another three
stars, HD~52918, HD~157246 and HD~158926, also seemed unreasonably large given
the apparently good S/N of the data. These stars were originally discussed by
\citet{ebj74} but neither they or J78 showed plots of $\tau(v)$
against which we could compare the new values. (The original analysis of
these stars is no longer available.) We therefore again reduced the
values of $\sigma_i$ by a factor of five, to give errors in $\sigma(\tau_i)$ apparently
commensurate with the S/N of the data.

Values of $\sigma(N)$ are given in Table~\ref{tab_copernicus} [along with the
original values of $N$(O~VI) given by J78]. Again, we emphasize
that {\it these values of $\sigma(N)$ are only approximations} and may only be useful
for the statistical weights they provide for the \Copernicus\ column densities.

\subsubsection{Comparison Between Sp/L and \Hipparcos\ Parallax  Distances for   \emph{Copernicus} Stars \label{sect_hippcf}}

While we derived the distances to the \Copernicus\ stars in exactly
the same way as we did for \FUSE\ stars, it is also true that reliable
\Hipparcos\ parallax distances are also available for many of the
\Copernicus\ stars.  The parallax distance $d_\pi$ can be derived
from the parallax $\pi$ given in the \Hipparcos\ catalog from the relation
$d_\pi\:=\:1000/\pi$~pc, where $\pi$ is the parallax measured in mas,
which has an error of $\sigma(\pi)$.


\begin{figure}[t!]
\hspace{3cm}\includegraphics[width=13cm]{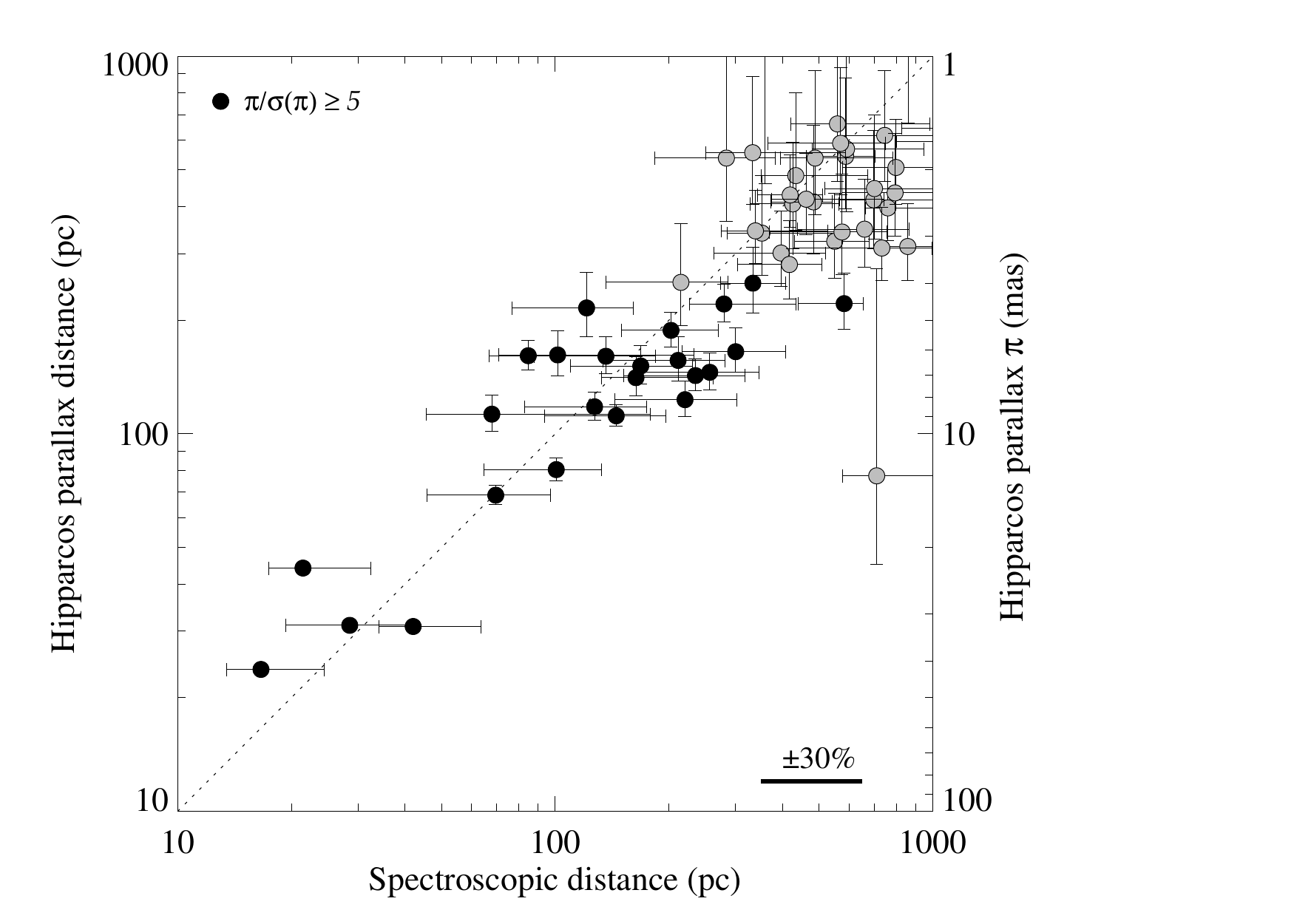}
\caption{ \label{fig_hipparcos1} 
Comparison between parallax distances estimated from {\it Hipparcos}
  data and those derived from adopted spectral type and luminosity classes, for
  stars observed with Copernicus. One star in the sample (HD~149881)
  has a negative parallax, and is not plotted.} 
\end{figure}

Apart from providing a direct measure of the distance to a star, these
data also enable us to compare distances with those derived from given
Sp/Ls, and examine how good the $M_V$ and color calibrations discussed
above actually are.  In Figure~\ref{fig_hipparcos1} we plot the
Hipparcos distance against the distance derived from spectral typing.
In the figure, we distinguish between those
stars with $\pi/\sigma(\pi) \geq 5$ (black circles) and the rest of the sample
(gray circles). Within
$100-200$~pc, the correlation between the distances derived in the two
different ways is quite good, and the spectroscopic distances are
generally equal to the \Hipparcos\ distances to within at least twice
our estimates of the error range determined by equation~\ref{eqn_spl}. This
substantiates our use of the errors from equation~\ref{eqn_spl} as a useful
`1$\sigma$' estimate. Beyond $200-300$~pc, parallaxes become
too small to be measured accurately by \Hipparcos . 
We note that the Be
stars do not have errors in spectroscopic distances much different
from B stars.

Given that the \Hipparcos\ distances for the nearest stars have much smaller
errors than the spectroscopic distances, we adopt the \Hipparcos\ distance for
any star with $\pi/\sigma(\pi) > 5$. (We note our adoption of the
\Hipparcos\ parallax distance with a superscript `{\it{c}}' against
the value of $d$
in column 11 of Table~\ref{tab_copernicus}.) For these stars,
the errors in their distances should be $<5$~\% \citep{brown97}. In keeping with
the range in errors adopted for the \FUSE\ targets, we take the range in
possible distances, $d_l$ and $d_u$, to be 1000/[$\pi + \sigma(\pi)$]
and 1000/[$\pi - \sigma(\pi)$], respectively.

\subsection{Notes on Corrections Made to Data from Z03 \label{sect_zsargonotes}}

All the stars observed by \citet{zsargo03} can be found in the
\Hipparcos\ catalog, so we have again used the cataloged values for
the stellar magnitudes.  In many cases, however, Sp/Ls are either not
known, are only poorly determined, or are irrelevant---as in the case
of the small number of post-asymptotic giant branch (PAGB) stars.  For
these stars, we have taken distances directly from the literature.
Details are given in Table~\ref{tab_zsargo}.


\clearpage
\thispagestyle{empty}

\begin{deluxetable}{c 
l 
r 
r 
c 
l 
c 
l 
c 
c 
l 
c 
r 
c 
r 
c 
r 
c } 
\rotate
\tabletypesize{\scriptsize}
\tablecolumns{18}
\tablewidth{0pc} 
\tablecaption{{\it Copernicus} Data Obtained by J78 \label{tab_copernicus}}
\tablehead{
\colhead{Star}                 
& \colhead{Alt}    
& \colhead{$l$ }  
& \colhead{$b$ } 
& \colhead{ }   
& \colhead{ }        
& \colhead{ }       
& \colhead{ } 
& \colhead{ } 
& \colhead{ } 
& \colhead{$d$}    
& \colhead{$d_r$}    
& \colhead{$v_c$} 
& \colhead{}
& \colhead{$v_{\odot}$}
& \colhead{$b$}
& \colhead{$N$}
& \colhead{$\sigma(N)$}\\
\colhead{Name}                 
& \colhead{Name}         
& \colhead{(deg)}  
& \colhead{(deg)} 
& \colhead{$B$} 
& \colhead{$V$}
& \colhead{$K$}
& \colhead{Sp/L}       
& \colhead{$E(B-V)$ } 
& \colhead{$E(K-V)$ } 
& \colhead{(kpc)}      
& \colhead{(kpc)}      
& \colhead{(\kms)}
& \colhead{[R]} 
& \colhead{(\kms )}
& \colhead{(\kms )}
& \colhead{($10^{13}$\pcm )}
& \colhead{($10^{13}$\pcm )} \\
\colhead{(1)}                 
& \colhead{(2)}    
& \colhead{(3)}  
& \colhead{(4)} 
& \colhead{(5)}   
& \colhead{(6)}
& \colhead{(7)} 
& \colhead{(8)} 
& \colhead{(9)} 
& \colhead{(10)}    
& \colhead{(11)}  
& \colhead{(12)}
& \colhead{(13)} 
& \colhead{(14)} 
& \colhead{(15)}
& \colhead{(16)}                 
& \colhead{(17)}                 
& \colhead{(18)}                 
}
\startdata
HD175191 & $\sigma$~Sgr &   9.56 & $-12.44$ &  1.93 &  2.06$^{} $ &  \ldots  & B3IV & $0.05\pm0.01$ &  \ldots  &  0.07$^{c}$ &  0.06, 0.07 & $   9.1$ & 1 &  \ldots  &  \ldots  & $<$ 0.79 & \ldots\\
HD149881 & HD~149881 &  31.37 & $ 36.23$ &  6.84 &  7.01$^{} $ &  7.57 & B0.5III & $0.05\pm0.01$ & $-0.11\pm 0.04$ &  2.85$^{a}$ &  2.10, 4.06 & $  15.3$ & 0 & $ -3.1$ & $ 19.7$ &  3.47 &   2.7\\
HD184915 & $\kappa$~Aql &  31.77 & $-13.29$ &  4.92 &  4.96$^{g} $ &  \ldots  & B0.5IIIne & $0.17\pm0.01$ &  \ldots  &  0.70 &  0.52, 1.00 & $  12.0$ & 0 & $  6.6$ & $ 23.7$ &  3.72 &   1.3\\
HD214080 & HD~214080 &  44.80 & $-56.92$ &  6.68 &  6.82$^{g} $ &  7.24 & B1Ib & $0.05\pm0.01$ & $-0.09\pm 0.04$ &  3.42 &  2.27, 4.48 & $   2.2$ & 0 & $ 20.1$ & $ 13.6$ &  2.88 &   3.9\\
HD180968 & 2~Vul &  56.36 & $  4.85$ &  5.49 &  5.50$^{d} $ &  \ldots  & B1IVe & $0.22\pm0.01$ &  \ldots  &  0.57 &  0.37, 0.75 & $  15.5$ & 0 & $  7.6$ & $ 44.9$ & 12.59 &   3.5\\
HD219188 & HD~219188 &  83.03 & $-50.17$ &  6.91 &  7.04$^{g} $ &  7.48 & B0.5IIIn & $0.09\pm0.01$ & $-0.23\pm 0.04$ &  2.08 &  1.53, 2.96 & $   2.9$ & 0 & $ -9.1$ & $ 40.2$ & 12.59 &   3.6\\
HD200310 & 60~Cyg &  87.15 & $ -0.10$ &  5.17 &  5.37$^{} $ &  \ldots  & B1Vne & $0.03\pm0.01$ &  \ldots  &  0.46 &  0.37, 0.74 & $  12.4$ & 1 & $  1.7$ & $ 20.3$ &  1.45 &   0.7\\
HD200120 & 59~Cyg &  88.03 & $  0.97$ &  4.85 &  5.05$^{} $ &  \ldots  & B1.5Ve & $0.02\pm0.01$ &  \ldots  &  0.34 &  0.27, 0.53 & $  12.4$ & 1 & $ -9.9$ & $ 14.1$ &  1.32 &   0.6\\
HD214168 & 8~Lac &  96.37 & $-16.15$ &  6.35 &  6.49$^{d} $ &  \ldots  & B1Ve & $0.09\pm0.01$ &  \ldots  &  0.71 &  0.58, 1.14 & $   8.5$ & 1 & $-17.6$ & $ 28.3$ &  3.55 &   1.0\\
HD120315 & $\eta$~UMa & 100.70 & $ 65.32$ &  1.75 &  1.85$^{g} $ &  \ldots  & B3V & $0.08\pm0.01$ &  \ldots  &  0.03$^{c}$ &  0.03, 0.03 & $  10.5$ & 1 &  \ldots  &  \ldots  & $<$ 0.25 & \ldots\\
HD005394 & $\gamma$~Cas & 123.58 & $ -2.15$ &  2.70 &  2.75$^{} $ &  \ldots  & B0IVe & $0.21\pm0.01$ &  \ldots  &  0.19$^{c}$ &  0.17, 0.21 & $   4.7$ & 2 &  \ldots  &  \ldots  & $<$ 0.10 & \ldots\\
HD014633 & HD~014633 & 140.78 & $-18.20$ &  7.22 &  7.44$^{} $ &  8.15 & O8V & $0.07\pm0.01$ & $-0.13\pm 0.04$ &  3.13$^{a}$ &  2.74, 3.99 & $  -1.6$ & 1 & $ -9.6$ & $ 33.6$ &  5.25 &   2.2\\
HD030614 & $\alpha$~Cam & 144.07 & $ 14.04$ &  4.29 &  4.30$^{} $ &  \ldots  & O9.5Iae & $0.23\pm0.01$ &  \ldots  &  1.73$^{a}$ &  1.35, 1.94 & $   1.4$ & 0 &  \ldots  &  \ldots  & $<$ 2.51 & \ldots\\
HD022928 & $\delta$~Per & 150.28 & $ -5.77$ &  2.93 &  3.03$^{d} $ &  \ldots  & B5III & $0.05\pm0.01$ &  \ldots  &  0.16$^{c}$ &  0.14, 0.19 & $  -2.6$ & 0 &  \ldots  &  \ldots  & $<$ 5.01 & \ldots\\
HD024760 & $\epsilon$~Per & 157.35 & $-10.09$ &  2.73 &  2.90$^{} $ &  \ldots  & B0.5IV & $0.07\pm0.01$ &  \ldots  &  0.17$^{c}$$^{a}$ &  0.15, 0.19 & $  -4.9$ & 2 & $ 62.5$ & $ 22.8$ &  0.87 &   0.8\\
HD024912 & $\xi$~Per & 160.37 & $-13.11$ &  4.02 &  4.04$^{g} $ &  \ldots  & O7.5IIInf & $0.26\pm0.01$ &  \ldots  &  0.59 &  0.46, 0.70 & $  -5.9$ & 2 & $  7.1$ & $ 24.5$ &  2.19 &   2.1\\
HD024398 & $\zeta$~Per & 162.29 & $-16.69$ &  2.97 &  2.88$^{g} $ &  \ldots  & B1Ib & $0.27\pm0.01$ &  \ldots  &  0.40 &  0.26, 0.52 & $  -6.7$ & 2 & $ 20.2$ & $ 24.5$ &  1.51 &   2.4\\
HD023630 & $\eta$~Tau & 166.67 & $-23.46$ &  2.81 &  2.87$^{g} $ &  \ldots  & B7IIIe & $0.05\pm0.01$ &  \ldots  &  0.11$^{c}$ &  0.10, 0.13 & $  -8.3$ & 1 &  \ldots  &  \ldots  & $<$ 2.00 & \ldots\\
HD093521 & HD~093521 & 183.14 & $ 62.15$ &  6.78 &  7.01$^{} $ &  7.72 & O9Vp & $0.05\pm0.01$ & $-0.08\pm 0.04$ &  1.76 &  1.54, 2.29 & $   1.7$ & 0 & $-15.8$ & $ 32.5$ &  7.24 &   2.7\\
HD040111 & 139~Tau & 183.97 & $  0.84$ &  4.74 &  4.82$^{} $ &  5.06 & B1IB & $0.10\pm0.01$ & $-0.28\pm 0.04$ &  1.24 &  0.82, 1.63 & $  -9.7$ & 0 & $ 15.5$ & $ 27.4$ &  6.76 &   1.9\\
HD036861 & $\lambda$~Ori & 195.05 & $-12.00$ &  3.32 &  3.50$^{d} $ &  \ldots  & O8IIIf & $0.10\pm0.01$ &  \ldots  &  0.55 &  0.43, 0.68 & $ -13.0$ & 2 & $  5.1$ & $ 37.4$ &  1.62 &   0.7\\
HD045995 & HD~045995 & 200.62 & $  0.68$ &  6.02 &  6.13$^{d} $ &  \ldots  & B2.5Ve & $0.08\pm0.01$ &  \ldots  &  0.36 &  0.29, 0.55 & $ -12.5$ & 0 & $ 28.7$ & $ 26.2$ &  1.95 &   1.6\\
HD035439 & 25~Ori & 201.96 & $-18.29$ &  4.70 &  4.87$^{g} $ &  \ldots  & B1Vne & $0.06\pm0.01$ &  \ldots  &  0.35 &  0.29, 0.57 & $ -14.4$ & 1 & $ 17.7$ & $ 16.2$ &  0.95 &   0.7\\
HD047839 & 15~Mon & 202.94 & $  2.20$ &  4.38 &  4.61$^{d} $ &  \ldots  & O7Vf & $0.07\pm0.01$ &  \ldots  &  0.86$^{a}$ &  0.75, 1.07 & $ -12.7$ & 3 & $ 34.8$ & $ 36.6$ &  3.16 &   0.9\\
HD036486 & $\delta$~Ori~A & 203.86 & $-17.74$ &  2.25 &  2.41$^{d} $ &  \ldots  & O9.5II & $0.08\pm0.01$ &  \ldots  &  0.42$^{a}$ &  0.30, 0.51 & $ -14.6$ & 2 & $ -5.9$ & $ 56.0$ &  5.50 &   1.3\\
HD036695 & VV~Ori & 204.84 & $-17.81$ &  5.18 &  5.35$^{} $ &  5.89 & B1V & $0.06\pm0.01$ & $-0.12\pm 0.04$ &  0.59$^{a}$ &  0.48, 0.95 & $ -14.7$ & 0 & $  4.6$ & $ 31.2$ &  8.32 &   2.2\\
HD037128 & $\epsilon$~Ori & 205.21 & $-17.24$ &  1.59 &  1.70$^{g} $ &  \ldots  & B0Iae & $0.10\pm0.01$ &  \ldots  &  0.48 &  0.37, 0.54 & $ -14.7$ & 0 & $  8.7$ & $ 16.2$ &  1.48 &   1.6\\
HD037742 & $\zeta$~Ori & 206.45 & $-16.59$ &  1.71 &  1.82$^{d} $ &  \ldots  & O9.5Ib & $0.13\pm0.01$ &  \ldots  &  0.25 &  0.21, 0.31 & $ -14.8$ & 1 & $ 13.2$ & $ 23.6$ &  1.78 &   0.8\\
HD028497 & HD~028497 & 208.78 & $-37.40$ &  5.39 &  5.59$^{g} $ &  \ldots  & B1.5Ve & $0.02\pm0.01$ &  \ldots  &  0.43 &  0.35, 0.67 & $ -15.1$ & 1 & $  7.2$ & $ 35.7$ &  7.24 &   1.1\\
HD033328 & $\lambda$~Eri & 209.14 & $-26.69$ &  4.07 &  4.25$^{g} $ &  \ldots  & B2IVne & $0.03\pm0.01$ &  \ldots  &  0.28 &  0.18, 0.38 & $ -15.3$ & 0 & $  4.3$ & $ 36.2$ &  4.47 &   0.5\\
HD037043 & $\iota$~Ori & 209.52 & $-19.58$ &  2.56 &  2.76$^{} $ &  \ldots  & O9III & $0.06\pm0.01$ &  \ldots  &  0.43$^{a}$ &  0.33, 0.56 & $ -15.3$ & 2 & $ -3.6$ & $ 26.5$ &  2.00 &   1.0\\
HD037303 & HD~037303 & 209.79 & $-19.15$ &  5.81 &  6.01$^{} $ &  \ldots  & B1.5V & $0.01\pm0.01$ &  \ldots  &  0.70$^{a}$ &  0.57, 1.08 & $ -15.3$ & 0 & $ -8.4$ & $ 25.4$ &  3.55 &   0.9\\
HD038771 & $\kappa$~Ori & 214.51 & $-18.50$ &  1.94 &  2.05$^{g} $ &  \ldots  & B0.5Ia & $0.09\pm0.01$ &  \ldots  &  0.22 &  0.19, 0.27 & $ -15.7$ & 2 & $  9.9$ & $ 29.4$ &  2.51 &   0.9\\
HD052918 & HD~052918 & 218.01 & $  0.61$ &  4.78 &  4.96$^{g} $ &  \ldots  & B1IVe & $0.04\pm0.01$ &  \ldots  &  0.57 &  0.37, 0.76 & $ -14.4$ & 0 & $ 61.0$ & $ 20.1$ &  1.23 &   1.2\\
HD087901 & $\alpha$~Leo & 226.43 & $ 48.93$ &  1.37 &  1.40$^{} $ &  \ldots  & B7V & $0.10\pm0.01$ &  \ldots  &  0.02$^{c}$ &  0.02, 0.02 & $  -4.5$ & 1 &  \ldots  &  \ldots  & $<$ 2.51 & \ldots\\
HD051283 & HD~051283 & 234.01 & $ -9.33$ &  5.14 &  5.29$^{} $ &  \ldots  & B2IIIe & $0.03\pm0.01$ &  \ldots  &  0.56 &  0.42, 0.98 & $ -15.9$ & 0 & $ 28.8$ & $ 33.0$ &  6.03 &   1.8\\
HD091316 & $\rho$~Leo & 234.89 & $ 52.77$ &  3.69 &  3.84$^{g} $ &  \ldots  & B1Iab & $0.04\pm0.01$ &  \ldots  &  1.09 &  0.83, 1.47 & $  -3.5$ & 0 & $-12.1$ & $ 28.7$ &  1.58 &   0.8\\
HD038666 & $\mu$~Col & 237.29 & $-27.10$ &  4.89 &  5.15$^{} $ &  6.01 & O9.5V & $0.01\pm0.01$ & $ 0.10\pm 0.04$ &  0.76 &  0.66, 1.00 & $ -16.5$ & 0 & $ 17.6$ & $ 46.1$ &  6.61 &   2.1\\
HD058978 & HD~058978 & 237.41 & $ -3.00$ &  5.51 &  5.64$^{e} $ &  \ldots  & B0.5IVnpe & $0.11\pm0.01$ &  \ldots  &  0.79 &  0.57, 1.08 & $ -15.3$ & 1 & $ 45.3$ & $ 31.6$ &  9.33 &   1.3\\
HD057060 & 29~CMa & 237.82 & $ -5.37$ &  4.74 &  4.90$^{} $ &  \ldots  & O7Iabfp & $0.15\pm0.02$ &  \ldots  &  1.87$^{a}$ &  1.53, 2.30 & $ -15.5$ & 0 & $ 32.1$ & $ 33.8$ &  2.88 &   0.9\\
HD057061 & $\tau$~CMa & 238.18 & $ -5.54$ &  4.73 &  4.89$^{d} $ &  \ldots  & O9II & $0.11\pm0.01$ &  \ldots  &  1.57$^{a}$ &  1.19, 1.92 & $ -15.5$ & 2 & $ 30.9$ & $ 31.9$ &  3.31 &   0.8\\
HD044506 & HD~044506 & 241.63 & $-20.78$ &  5.35 &  5.52$^{g} $ &  \ldots  & B1.5IIIn & $0.02\pm0.01$ &  \ldots  &  0.75 &  0.57, 1.23 & $ -16.3$ & 1 & $ 17.2$ & $ 39.7$ &  6.92 &   2.5\\
HD066811 & $\zeta$~Pup & 255.98 & $ -4.71$ &  2.01 &  2.21$^{g} $ &  \ldots  & O5Ibnf & $0.12\pm0.01$ &  \ldots  &  0.42 &  0.34, 0.51 & $ -14.3$ & 2 & $ 15.0$ & $ 33.0$ &  3.24 &   0.7\\
HD064760 & HD~064760 & 262.06 & $-10.42$ &  4.09 &  4.23$^{g} $ &  \ldots  & B0.5Ib & $0.05\pm0.01$ &  \ldots  &  1.06 &  0.80, 1.34 & $ -14.1$ & 0 & $ 28.7$ & $ 37.7$ & 12.02 &   0.7\\
HD068273 & $\gamma^2$~Vel & 262.80 & $ -2.69$ &  1.67 &  1.81$^{} $ &  \ldots  & WC8+O9I & $0.00\pm0.01$ &  \ldots  &  0.26$^{c}$$^{a}$ &  0.23, 0.30 & $ -13.3$ & 2 & $  1.7$ & $ 25.6$ &  4.47 &   1.1\\
HD042933 & $\delta$~Pic & 263.30 & $-27.68$ &  4.49 &  4.71$^{} $ &  \ldots  & B0.5IV & $0.02\pm0.01$ &  \ldots  &  0.80$^{a}$ &  0.58, 1.08 & $ -14.7$ & 2 & $ -3.7$ & $ 33.3$ &  3.89 &   0.5\\
HD064740 & HD~064740 & 263.38 & $-11.19$ &  4.40 &  4.61$^{g} $ &  \ldots  & B1.5Vp & $0.01\pm0.01$ &  \ldots  &  0.22$^{c}$ &  0.20, 0.25 & $ -14.0$ & 0 & $ 22.2$ & $ 33.0$ &  1.78 &   0.6\\
HD104337 & HD~104337 & 286.92 & $ 41.63$ &  5.08 &  5.26$^{} $ &  5.85 & B1V & $0.04\pm0.01$ & $-0.07\pm 0.04$ &  0.49$^{a}$ &  0.40, 0.78 & $  -2.0$ & 0 & $-13.6$ & $ 21.2$ &  1.51 &   1.6\\
HD010144 & $\alpha$~Eri & 290.84 & $-58.79$ &  0.47 &  0.53$^{} $ &  \ldots  & B3Vpe & $0.13\pm0.02$ &  \ldots  &  0.04$^{c}$ &  0.04, 0.05 & $ -10.1$ & 0 & $ 17.0$ & $ 30.1$ &  1.15 &   0.7\\
HD106490 & $\delta$~Cru & 298.23 & $  3.79$ &  2.59 &  2.77$^{g} $ &  \ldots  & B2IV & $0.02\pm0.01$ &  \ldots  &  0.11$^{c}$ &  0.10, 0.12 & $  -5.8$ & 0 & $ 55.7$ & $ 19.9$ &  0.52 &   0.4\\
HD112244 & HD~112244 & 303.55 & $  6.03$ &  5.38 &  5.40$^{e} $ &  \ldots  & O8.5Iabf & $0.25\pm0.01$ &  \ldots  &  1.71 &  1.41, 2.15 & $  -4.2$ & 0 & $-15.1$ & $ 28.0$ &  7.24 &   3.6\\
HD122451 & $\beta$~Cen & 311.77 & $  1.25$ &  0.44 &  0.58$^{d} $ &  \ldots  & B1III & $0.06\pm0.02$ &  \ldots  &  0.16$^{c}$ &  0.15, 0.18 & $  -2.8$ & 0 & $ 16.6$ & $ 22.8$ &  0.91 &   0.4\\
HD121263 & $\zeta$~Cen & 314.07 & $ 14.19$ &  2.34 &  2.52$^{} $ &  \ldots  & B2.5IV & $0.02\pm0.01$ &  \ldots  &  0.12$^{c}$$^{a}$ &  0.11, 0.13 & $  -0.6$ & 0 & $ -2.5$ & $ 22.7$ &  1.26 &   0.4\\
HD120307 & $\nu$~Cen & 314.41 & $ 19.89$ &  3.19 &  3.39$^{} $ &  \ldots  & B2IV & $0.01\pm0.01$ &  \ldots  &  0.15$^{c}$$^{a}$ &  0.13, 0.16 & $   0.3$ & 1 & $ 18.2$ & $ 31.8$ &  1.48 &   0.4\\
HD121743 & $\phi$~Cen & 315.98 & $ 19.07$ &  3.61 &  3.81$^{g} $ &  \ldots  & B2IV & $0.02\pm0.01$ &  \ldots  &  0.14$^{c}$ &  0.13, 0.16 & $   0.5$ & 1 & $ -9.1$ & $ 10.8$ &  0.51 &   0.3\\
HD116658 & $\alpha$~Vir & 316.11 & $ 50.84$ &  0.91 &  1.04$^{} $ &  \ldots  & B1IV & $0.10\pm0.02$ &  \ldots  &  0.08$^{c}$$^{a}$ &  0.08, 0.09 & $   4.3$ & 0 & $  1.1$ & $ 19.0$ &  2.69 &   0.4\\
HD135591 & HD~135591 & 320.13 & $ -2.64$ &  5.33 &  5.43$^{} $ &  \ldots  & O7.5IIIf & $0.18\pm0.01$ &  \ldots  &  1.25 &  0.99, 1.50 & $  -1.1$ & 2 & $  6.2$ & $ 30.6$ &  4.17 &   1.3\\
HD132058 & $\beta$~Lup & 326.25 & $ 13.91$ &  2.49 &  2.66$^{} $ &  \ldots  & B2IV & $0.03\pm0.01$ &  \ldots  &  0.16$^{c}$ &  0.14, 0.18 & $   2.5$ & 1 & $  4.0$ & $ 37.2$ &  1.91 &   0.3\\
HD150898 & HD~150898 & 329.98 & $ -8.47$ &  5.48 &  5.57$^{g} $ &  5.77 & B0.5Ia & $0.11\pm0.01$ & $-0.35\pm 0.04$ &  2.80 &  2.12, 3.14 & $   0.8$ & 0 & $ -5.6$ & $ 28.8$ &  6.92 &   3.2\\
HD136298 & $\delta$~Lup & 331.32 & $ 13.82$ &  3.01 &  3.20$^{g} $ &  \ldots  & B1.5IV & $0.02\pm0.01$ &  \ldots  &  0.16$^{c}$ &  0.14, 0.18 & $   3.8$ & 0 & $ -7.0$ & $ 22.9$ &  2.19 &   0.5\\
HD173948 & $\lambda$~Pav & 333.61 & $-23.87$ &  4.06 &  4.21$^{g} $ &  \ldots  & B2IIIe & $0.05\pm0.01$ &  \ldots  &  0.33 &  0.25, 0.59 & $  -0.3$ & 0 & $ 17.3$ & $ 37.1$ &  6.03 &   1.0\\
HD157246 & $\gamma$~Ara & 334.64 & $-11.48$ &  3.18 &  3.31$^{} $ &  \ldots  & B1IB & $0.05\pm0.01$ &  \ldots  &  0.66 &  0.44, 0.86 & $   1.6$ & 2 & $ -6.7$ & $ 26.5$ &  1.45 &   0.8\\
HD143118 & $\eta$~Lup & 338.77 & $ 11.01$ &  3.21 &  3.41$^{g} $ &  \ldots  & B2.5IV & $0.00\pm0.01$ &  \ldots  &  0.15$^{c}$ &  0.14, 0.17 & $   5.3$ & 0 & $-15.2$ & $ 23.7$ &  1.62 &   0.6\\
HD165024 & $\theta$~Ara & 343.33 & $-13.82$ &  3.56 &  3.66$^{} $ &  \ldots  & B2Ib & $0.06\pm0.01$ &  \ldots  &  0.73 &  0.43, 0.99 & $   3.4$ & 2 & $-21.2$ & $ 23.9$ &  4.27 &   0.7\\
HD151890 & $\mu^1$~Sco & 346.12 & $  3.91$ &  2.82 &  2.98$^{} $ &  \ldots  & B1.5IV & $0.05\pm0.01$ &  \ldots  &  0.22$^{a}$ &  0.14, 0.29 & $   6.3$ & 0 & $-19.3$ & $ 26.2$ &  2.75 &   0.9\\
HD143018 & $\pi$~Sco & 347.21 & $ 20.23$ &  2.73 &  2.88$^{} $ &  \ldots  & B1V & $0.08\pm0.01$ &  \ldots  &  0.14$^{c}$$^{a}$ &  0.13, 0.16 & $   8.2$ & 1 & $-12.5$ & $ 24.5$ &  0.25 &   0.5\\
HD209952 & $\alpha$~Gru & 350.00 & $-52.47$ &  1.69 &  1.76$^{g} $ &  \ldots  & B7IV & $0.06\pm0.01$ &  \ldots  &  0.03$^{c}$ &  0.03, 0.03 & $  -1.4$ & 0 &  \ldots  &  \ldots  & $<$ 0.79 & \ldots\\
HD143275 & $\delta$~Sco & 350.10 & $ 22.49$ &  2.31 &  2.39$^{d} $ &  \ldots  & B0.3IVe & $0.00\pm0.01$ &  \ldots  &  0.12$^{c}$$^{a}$ &  0.11, 0.14 & $   8.9$ & 1 & $-11.5$ & $ 15.0$ &  1.82 &   0.5\\
HD158926 & $\lambda$~Sco & 351.74 & $ -2.21$ &  1.48 &  1.62$^{d} $ &  \ldots  & B1.5IV & $0.08\pm0.01$ &  \ldots  &  0.22$^{a}$ &  0.18, 0.27 & $   6.9$ & 2 & $-19.9$ & $ 22.8$ &  1.82 &   0.2\\
HD155806 & HD~155806 & 352.59 & $  2.87$ &  5.59 &  5.64$^{e} $ &  \ldots  & O7.5Ve & $0.23\pm0.01$ &  \ldots  &  0.86 &  0.75, 1.09 & $   7.7$ & 1 & $-12.8$ & $ 33.7$ & 16.60 &   1.5\\

\enddata
\tablecomments{ Explanation of columns:  
(1)---HD number; 
(2)---Alternative star name;
(3,4)---Galactic $l$ and $b$; 
(5,6)---Johnson $B$- and $V$-band magnitudes,
  converted from \Tycho\ $B_T$ and $V_T$  values ; 
(7)---Johnson $K$-band magnitude, converted from 2Mass $K$-band magnitude. 
Value is not given if $E(K-V)$ was not derived for the sight line (see Appendix~\ref{sect_av}); 
(8)---spectral type and luminosity class; 
(9)---optical $B-V$ color excess;  
(10)---IR $K-V$ color excess;
(11) distance to star based on spectral type and luminosity class, except
  for stars with \Hipparcos\  parallaxes such that  $d_p/\sigma(d_p) > 5$;
(12)---range in possible distances, based on 
  the uncertainties discussed in Appendix~\ref{sect_absmags}, except for distances
  derived from \Hipparcos\ parallax values; 
(13)---correction required  to move heliocentric velocities
  observed along a sight line to LSR velocities ($v_{\rm{LSR}}=v_c +v_\odot$);
(14)---our \ROSAT\ classification discussed in \S\ref{sect_bubbles}
and \S\ref{sect_SNR}. 
(15)---Heliocentric velocity of O~VI absorption, converted from Table~1 of
\citet{ebj78a} via $v_\odot\:=\:v_{\rm{LSR}} - v_c$;
(16)---Doppler parameter of O~VI absorption, converted from Table~1 of
\citet{ebj78a} with $b\:=\:\sqrt{2}\Delta$;
(17)---O~VI column density from Table~1 of \citet{ebj78a};
(18)---Approximate errors in the O~VI column density, derived from Archive
coadded data scans.
Note that the superscript marks for columns 6 and 11 refer to the same
  footnotes given in Table~\ref{tab_journal}:
that is, `{\it{a}}' means that the star is identified as a spectroscopic binary (SB) either in the
  SB9 catalog \citep{pourbaix04} or by  \citet{garcia01}; and `{\it{c}}'
  means that a \Hipparcos\ parallax distance was used.
}
\end{deluxetable}


\clearpage
\thispagestyle{empty}
\begin{deluxetable}{l 
r 
r 
r 
r 
c 
l 
c 
c 
l 
c 
r 
c 
r 
c 
r 
c} 
\rotate
\tabletypesize{\scriptsize}
\tablecolumns{17}
\tablewidth{0pc} 
\tablecaption{{\it FUSE} Data Obtained by  Zsarg\'{o} et al \label{tab_zsargo}}
\tablehead{
\colhead{Star}        
& \colhead{$l$}       
& \colhead{$b$}       
& \colhead{ }         
& \colhead{ }         
& \colhead{ }         
& \colhead{ }         
& \colhead{ }         
& \colhead{ }         
& \colhead{$d$}       
& \colhead{$d_r$}     
& \colhead{$v_c$}     
& \colhead{}          
& \colhead{$v_\odot$} 
& \colhead{$b$}       
& \colhead{$\log N$}  
& \colhead{dist} \\   
\colhead{Name}                 
& \colhead{(deg)}  
& \colhead{(deg)} 
& \colhead{$B$} 
& \colhead{$V$}
& \colhead{$K$}
& \colhead{Sp/L}       
& \colhead{$E(B-V)$ } 
& \colhead{$E(K-V)$ } 
& \colhead{(kpc)}      
& \colhead{(kpc)}      
& \colhead{(\kms)}
& \colhead{[R]} 
& \colhead{(\kms )}
& \colhead{(\kms )}
& \colhead{[$\log$(\pcm )]}
& \colhead{ref\tablenotemark{a}} \\
\colhead{(1)}                 
& \colhead{(2)}    
& \colhead{(3)}  
& \colhead{(4)} 
& \colhead{(5)}   
& \colhead{(6)}
& \colhead{(7)} 
& \colhead{(8)} 
& \colhead{(9)} 
& \colhead{(10)}    
& \colhead{(11)}  
& \colhead{(12)}
& \colhead{(13)} 
& \colhead{(14)} 
& \colhead{(15)}
& \colhead{(16)}                 
& \colhead{(17)}                 
}
\startdata
NGC6723$-$III60 &   0.02 & $-17.30$ &  \ldots  &  \ldots \phantom{$^{x}$} &  \ldots  & PAGB &  \ldots  &  \ldots  &  8.0 &  6.8, 9.2 & $   6.5$ & 1 & $ 17.5$ & $ 44.3$ & 14.37$\pm$0.12 &  1\\
NGC$-$5904$-$ZNG$-$1 &   3.87 & $ 46.80$ &  \ldots  &  \ldots \phantom{$^{x}$} &  \ldots  & PAGB &  \ldots  &  \ldots  &  7.7 &  7.0, 8.5 & $  11.8$ & 1 & $-17.6$ & $ 36.5$ & 14.41$\pm$0.08 &  2\\
HD175876 &  15.28 & $-10.58$ &  6.80 &  6.92\phantom{$^{x}$} &  7.28 & O6.5III & $0.16\pm0.01$ & $-0.49\pm 0.04$ &  2.7 &  2.2, 3.1 & $  10.3$ & 0 & $-26.6$ & $ 59.2$ & 14.14$\pm$0.13 &  3\\
HD177989 &  17.81 & $-11.88$ &  9.22 &  9.34\phantom{$^{x}$} &  9.61 & B0III & $0.11\pm0.01$ & $-0.42\pm 0.05$ &  6.0 &  4.5, 8.9 & $  10.5$ & 0 & $  0.7$ & $ 49.5$ & 14.31$\pm$0.06 &  3\\
vZ$-$1128 &  42.50 & $ 78.69$ &  \ldots  &  \ldots \phantom{$^{x}$} &  \ldots  & PAGB &  \ldots  &  \ldots  & 10.0 &  9.1,10.9 & $   9.7$ & 1 & $-41.0$ & $ 52.8$ & 14.49$\pm$0.03 &  4\\
HD121800 & 113.01 & $ 49.76$ &  8.98 &  9.07$^{b} $ &  \ldots  & B1.5V & $0.13\pm0.02$ &  \ldots  &  1.8 &  1.5, 2.9 & $  10.2$ & 1 & $  1.9$ & $ 52.0$ & 14.40$\pm$0.03 &  3\\
HD003827 & 120.79 & $-23.23$ &  7.76 &  7.95\phantom{$^{x}$} &  8.74 & B0.7V & $0.05\pm0.02$ & $ 0.12\pm 0.04$ &  2.0 &  1.8, 2.9 & $   2.4$ & 0 & $-17.3$ & $ 39.0$ & 13.80$\pm$0.04 &  3\\
HDE233622 & 168.17 & $ 44.23$ &  9.82 &  9.97$^{b} $ &  \ldots  & B2V &  \ldots  &  \ldots  &  4.7 &  4.0, 5.4 & $   0.3$ & 1 & $-21.7$ & $ 47.2$ & 13.87$\pm$0.19 &  5\\
BD+382182 & 182.16 & $ 62.21$ & 10.95 & 11.18\phantom{$^{x}$} &  \ldots  & B3V &  \ldots  &  \ldots  &  4.5 &  3.8, 5.2 & $   1.8$ & 0 & $ -5.2$ & $ 35.2$ & 14.10$\pm$0.07 &  6\\
HD018100 & 217.93 & $-62.73$ &  8.23 &  8.44\phantom{$^{x}$} &  \ldots  & B5II/III &  \ldots  &  \ldots  &  3.1 &  2.6, 3.6 & $ -12.8$ & 1 & $  4.5$ & $ 45.0$ & 13.77$\pm$0.13 &  7\\
HD100340 & 258.85 & $ 61.23$ &  9.84 & 10.07\phantom{$^{x}$} &  \ldots  & B1V &  \ldots  &  \ldots  &  3.0 &  2.8, 3.2 & $  -0.4$ & 0 & $ 10.3$ & $ 78.3$ & 14.28$\pm$0.16 &  8\\
HD097991 & 262.34 & $ 51.73$ &  7.18 &  7.39\phantom{$^{x}$} &  \ldots  & B1V & $0.02\pm0.01$ &  \ldots  &  1.2 &  1.0, 1.9 & $  -2.6$ & 0 &  \ldots  &  \ldots  & $<$13.69 &  3\\
JL212 & 303.63 & $-61.03$ & 10.28 & 10.35$^{b} $ &  \ldots  & B2V &  \ldots  &  \ldots  &  2.2 &  1.9, 2.6 & $  -8.5$ & 0 & $ 22.7$ & $ 37.0$ & 14.21$\pm$0.07 &  9\\
HD116852 & 304.88 & $-16.13$ &  8.37 &  8.49\phantom{$^{x}$} &  8.80 & O9III & $0.14\pm0.01$ & $-0.46\pm 0.04$ &  4.5 &  3.5, 6.0 & $  -6.4$ & 0 & $ -6.7$ & $ 38.7$ & 14.30$\pm$0.02 &  3\\
NGC$-$5139$-$WOR$-$197 & 308.94 & $ 15.07$ &  \ldots  &  \ldots \phantom{$^{x}$} &  \ldots  & PAGB &  \ldots  &  \ldots  &  5.5 &  5.2, 5.8 & $  -1.7$ & 0 & $-15.8$ & $ 43.2$ & 14.54$\pm$0.12 & 10\\
HD121968 & 333.97 & $ 55.84$ & 10.04 & 10.16\phantom{$^{x}$} &  \ldots  & B1V &  \ldots  &  \ldots  &  3.8 &  3.4, 4.2 & $   7.4$ & 0 & $ -6.8$ & $ 38.2$ & 13.97$\pm$0.06 & 11\\
NGC$-$6397$-$162 & 338.19 & $-11.94$ &  \ldots  &  \ldots \phantom{$^{x}$} &  \ldots  & PAGB &  \ldots  &  \ldots  &  2.5 &  2.2, 2.9 & $   2.4$ & 0 & $  6.8$ & $ 48.2$ & 14.25$\pm$0.06 & 12\\
\enddata
\tablecomments{ 
See Table~\ref{tab_journal} for a full explanation of the table entries.}
\tablenotetext{a}{Distances to the stars taken from the following references:
 1---\citet{martins87};
 2---\citet{testa04};
3---this paper;
 4---\citet{rood99};
 5---\citet{ryans97};
 6---\citet{keenan95};
 7---\citet{keenan86_dist};
 8---\citet{ryans99};
 9---\citet{magee01};
10---\citet{bellazzini04};
11---\citet{little95};
12---\citet{gratton03}.}
\tablenotetext{b}{Detected by \Hipparcos\ as variable---faintest $H_p$ used.} 
\end{deluxetable}

\clearpage

\section{Upper Limits \label{sect_upper_limits}}

Astronomical observations sometimes exhibit measurement outcomes that
are not significantly above the noise.  Ideally, in such circumstances
one should report the formal measurement and the associated error, as we
have done in columns 3--7 of
Table~\ref{tab_limits}.
This procedure conveys the most information.  Nevertheless, it is a
common practice to define a minimum measurement
level above which one can claim a ``detection'' (with an associated
uncertainty), and this comfort-level threshold is usually expressed as
some factor $N$ times the standard deviation of the noise $\sigma$.  On
the occasion that a measurement is found to fall below this threshold,
it is customary to state that only an upper limit at the $N\sigma$ level
of confidence for the true signal is warranted.  In defining this upper
limit, many investigators arbitrarily discard the actual measurement
outcome and claim that the upper limit is simply equal to $N\sigma$. 
When this is done, there is a complete loss of potentially useful
information about whether the measurement came out at just barely below
the threshold, or alternatively, below zero by some one or two standard
deviations of the noise (which can happen with reasonable frequency for
null signal levels in the absence of any noise).  The former represents
a weaker case for an upper limit than the latter, since we know that it
is much more difficult for a noise deviation to drive any moderately
positive signal to a negative outcome than to cancel it just enough to
go somewhat below the defined threshold level.  This difference is not
reflected in the claimed degree of confidence for the upper limit. 
Moreover, there is an untidy discontinuity in the upper limits inferred
from the ``upper limit only'' cases for outcomes just below the
threshold, compared with ``detections plus $N\sigma$'' for outcomes just
above the threshold.  For these reasons, we adopted a more refined
procedure. 

In \S\ref{sect_colsAOD} we made use of a scheme described
by \citet{marshall92} for creating upper limits at a given level of
confidence $\alpha$ (we chose to use the ``$2\sigma$'' level of
significance, making $\alpha=0.97725$) for marginal or negative
measurement outcomes, under the condition that we have {\it a priori\/}
information that the quantity being measured must be positive.  Here, we
derive the formula for evaluating these upper limits and show a plot
that allows one to find quickly what the upper limits would be at a few
levels of significance that differ from the ``$2\sigma$'' value.

Our derivation differs somewhat from that given by \citet{marshall92}, and
makes use of an extension of Bayes' Theorem for determining the
probability of condition $A$, given conditions $B$ and $C$,
\begin{equation}\label{BT2}
{\rm Prob}(A\vert B,C)={\rm Prob}(A\vert B){\rm Prob}(C\vert A,B)/{\rm
Prob}(C\vert B)~.
\end{equation}
If we designate the true value of the quantity as $t$, our stated upper
limit as $u$, and have a measurement outcome datum $d$ with an
associated noise amplitude $n$, we can make the assignments for the
conditions $A\equiv t>u$, $B\equiv t>0$ and $C\equiv d,n$. Restating
Eq.~\ref{BT2} in terms of our explicit conditions yields
\begin{equation}
{\rm Prob}(t>u\vert t>0,d,n)={\rm Prob}(t>u\vert t>0){\rm Prob}(d,n\vert
t>u, t>0)/{\rm Prob}(d,n\vert t>0)~,
\end{equation}
We set the first term on the right-hand side of this equation equal to 1
(this clearly must be true if $u<0$ given that $t>0$; for $u>0$ there
is no constraint on the probability).   The second term represents the
small probability $(1-\alpha^\prime)$ that we could have obtained an
outcome $d$ with noise $n$ when the true value $t$ violates our declared
upper limit $u$.  This term is given by the area under the one-sided
tail of a Gaussian distribution with unit variance and zero mean
integrated from $(u-d)/n$ to infinity, i.e.,
\begin{equation}\label{CAB}
{\rm Prob}(d,n\vert t>u, t>0)=\onehalf {\rm erfc}\left( {u-d\over
\sqrt{2}n}\right)
\end{equation}
where erfc is the complementary error function, defined as
\begin{equation}\label{erfc}
{\rm erfc}(x)\equiv 1-{\rm erf}(x)={2\over \sqrt{\pi}}\int_x^\infty
\exp(-t2)dt
\end{equation}
Similarly,
\begin{equation}\label{CB}
{\rm Prob}(d,n\vert t>0)=\onehalf {\rm erfc}\left( {-d\over
\sqrt{2}n}\right)=1-\onehalf {\rm erfc}\left( {d\over \sqrt{2}n}\right)
\end{equation}
so that our final result for the probability that there will be a
violation of the upper limit is given by
\begin{equation}\label{1-alpha}
1-\alpha={\rm erfc}\left( {u-d\over \sqrt{2}n}\right)\Bigg/\left[ 2-{\rm
erfc}\left( {d\over \sqrt{2}n}\right)\right]~,
\end{equation}
for which $\alpha$ represents our level of confidence that $u$ is a
valid upper limit for $t$.

\begin{figure}[t!]
\hspace*{2cm}\includegraphics[width=12cm]{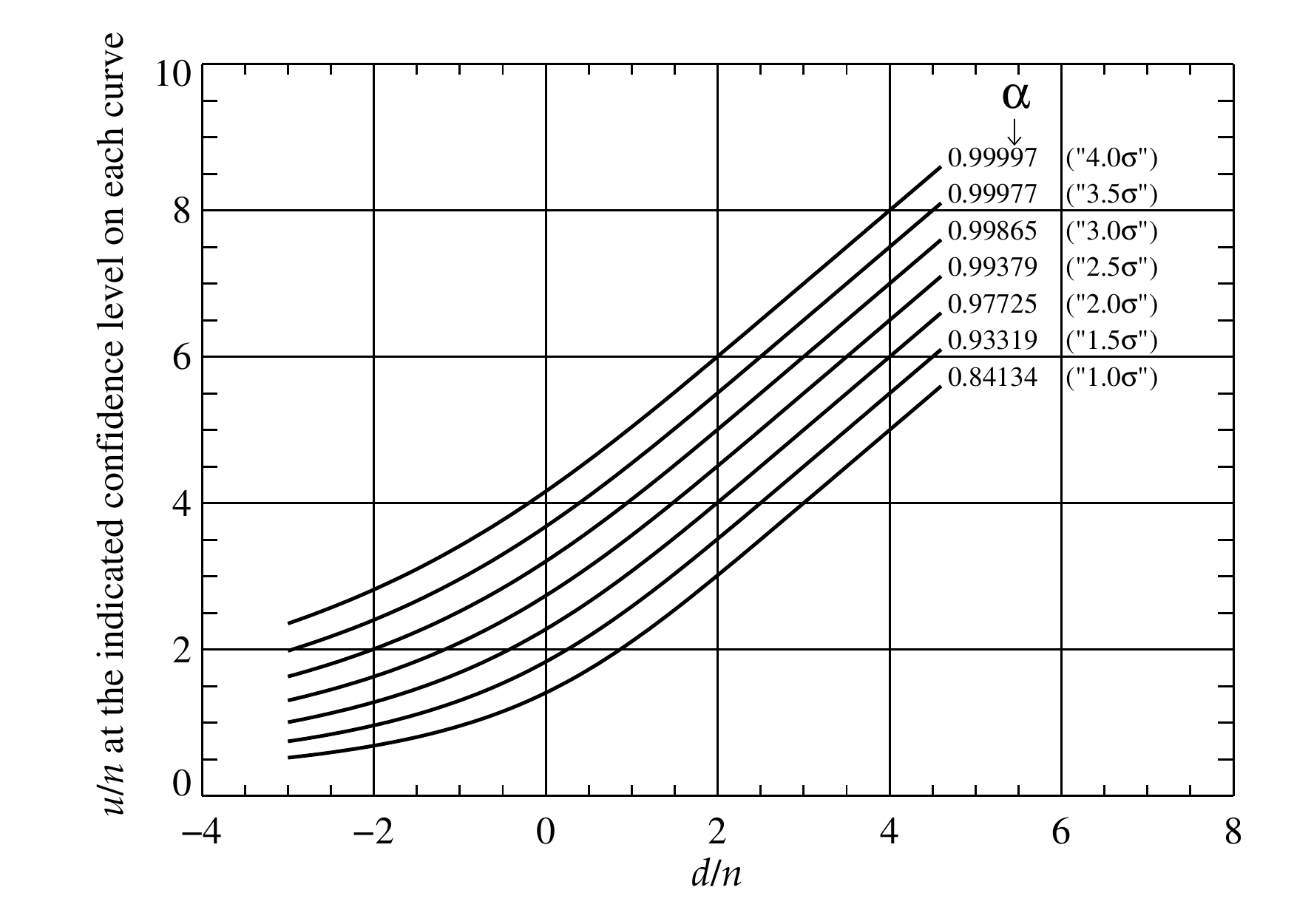}
\caption{ \label{fig_ebj_upper_lim} The family of curves for different
  levels of confidence $\alpha$, plotted in terms of the declared
  upper limits divided by the measurement uncertainties (ordinates),
  $u/n$, vs. the measurement outcomes divided by their uncertainties
  $d/n$ (abscissae).}
\end{figure}

Figure~\ref{fig_ebj_upper_lim} shows the solutions to equation~\ref{1-alpha} for
various confidence levels $\alpha$.  This figure should help one to
convert our raw measurements of barely or undetected O~VI into upper
limits of one's choice.  The values that we had listed in
Table~\ref{tab_limits} corresponded to $\alpha=0.97725$,
i.e., a ``$2\sigma$'' upper limit.

\section{Final Spectra \label{sect_explain_spectra}}

\begin{figure}
\vspace*{-10mm}\hspace*{1cm}\includegraphics[scale=0.44]{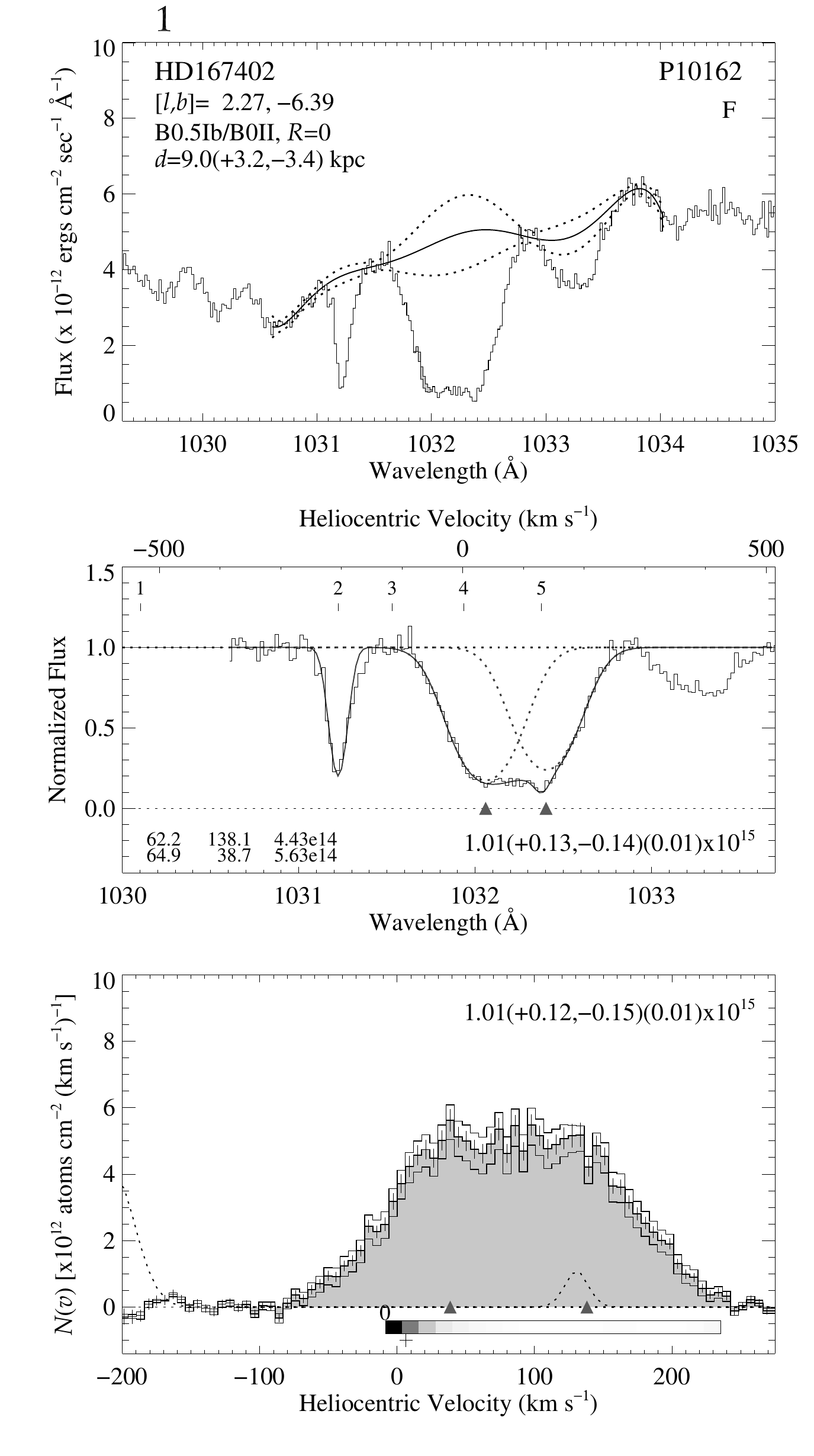}
\vspace*{-10mm}\hspace*{1cm}\includegraphics[scale=0.44]{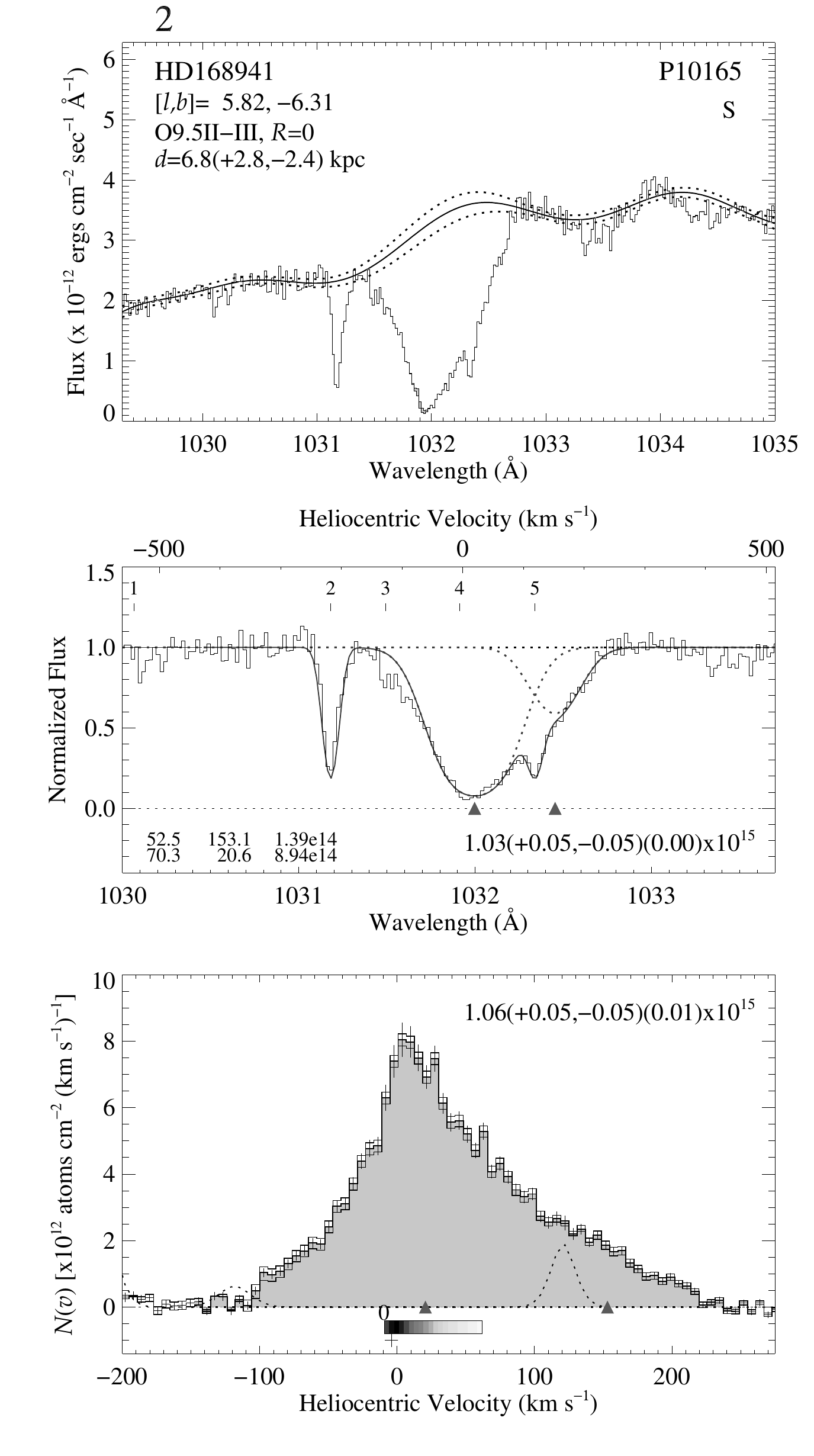}

\vspace*{1cm}\hspace*{1cm}\includegraphics[scale=0.44]{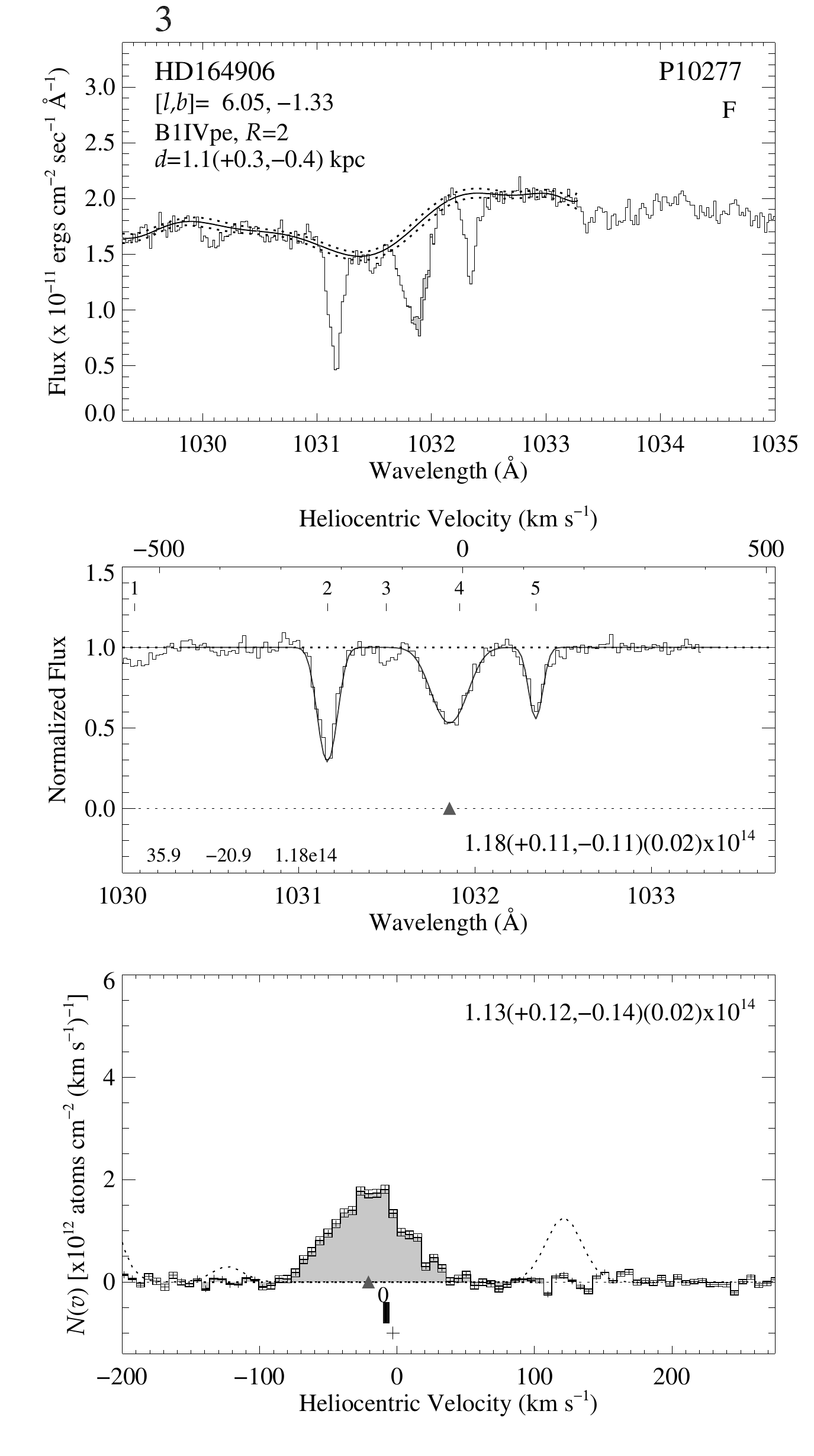}
\vspace*{1cm}\hspace*{1cm}\includegraphics[scale=0.44]{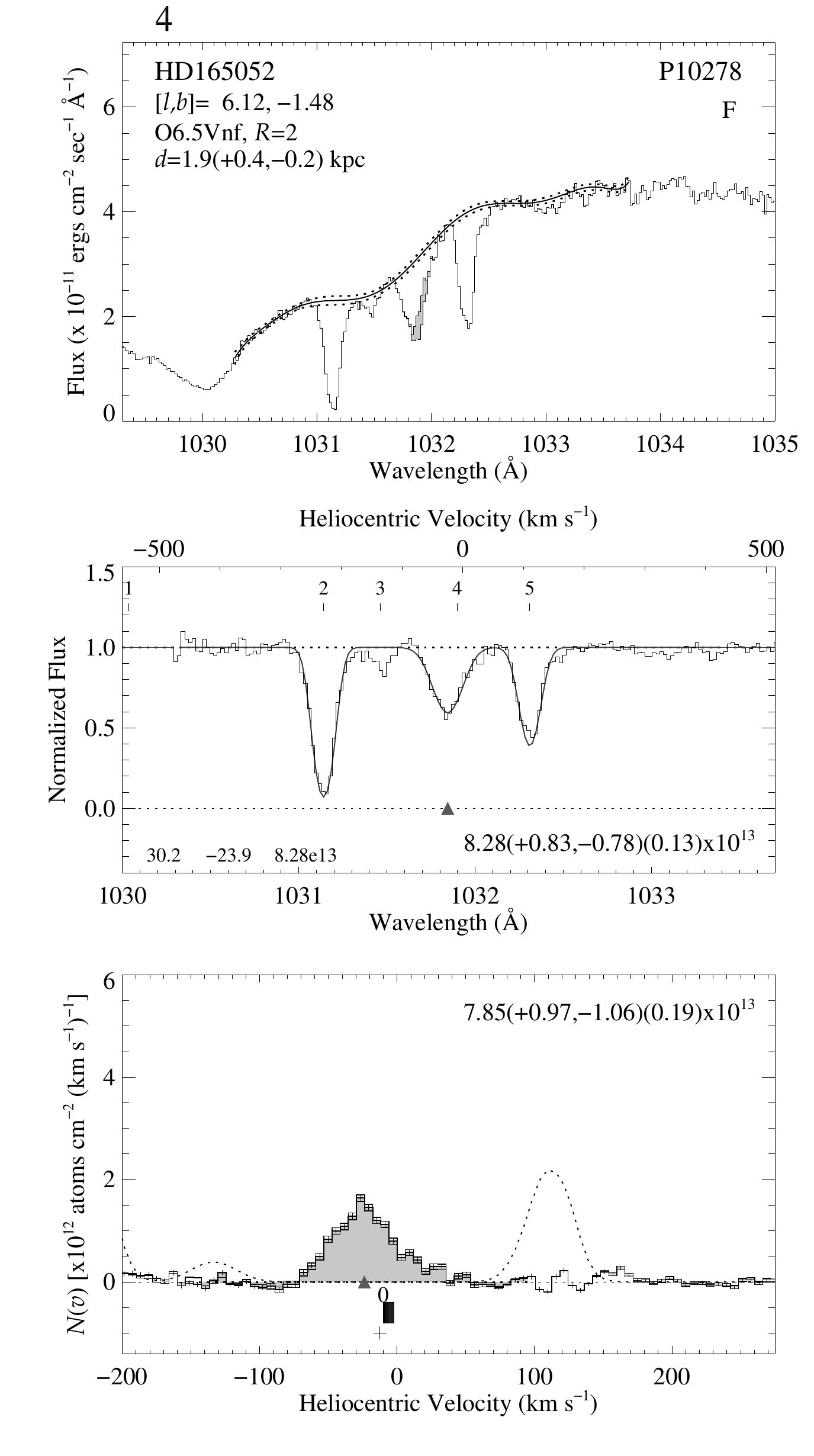}
\vspace*{-0.5cm}\caption{\label{fig_spectra}
Spectra of Milky Way disk stars observed with \FUSE\ around
the wavelength region of \sixa . See text in Appendix~E for full details.}
\end{figure}

The spectra obtained from our \FUSE\ survey are shown in
Figure~\ref{fig_spectra}.  These include all the data we deemed usable
from the PI Team programs P102 and P122 as well as sight lines taken
from the \FUSE\ Archive prior to April 2003 (see \S\ref{sect_select}).
Several stars from programs P102 and P122 were not used, and are not
shown here (see \S\ref{sect_crapstars}, and Fig.~\ref{fig_rejects}
instead).  For each spectrum in Figure~\ref{fig_spectra}, there are
three panels which depict the data analyzed in the ways discussed
throughout \S\ref{sect_the_data}.  Below, we describe the information
presented in each panel.

{\bf Top panel:} This panel shows the coadded spectra for each
sight line.  All data come from the LiF1A \FUSE\ channel, unless
indicated at the bottom right of the panel, where the channel used is
indicated.  Spectra are plotted in specific flux units, along with the
continuum used to normalize the data (solid line). The $\pm 1\sigma$
error `envelopes' to the fitted continuum are shown as dotted lines
about the adopted best-fit continuum. The star's name, Galactic
co-ordinates [$l$,$b$], spectral type, and \ROSAT\ designation $R$
(see \S\ref{sect_bubbles} and \ref{sect_SNR}) are shown at the top
left of the panel.  The distance to the star $d$, along with the
difference between the distance and the upper and lower bounds to the
distance (Appendix~\ref{sect_absmags}), are also given at the top left
of the panel.  The \FUSE\ program identification number is given at
the top right. A unique identification number is shown above the top
left corner of the top panel.  This number corresponds to the ID
numbers given in Tables~\ref{tab_journal}, \ref{tab_lookup},
\ref{tab_cols}, and \ref{tab_limits}.

If the wavelength scale has been corrected using data from \HST , then
an ``S'' is marked beneath the program number. Similarly, an ``X''
indicates that no shift has been made (for the reasons given in
Appendix~\ref{sect_cl1}), and the velocities of the absorption should
be considered uncertain.  Otherwise, a $+10$~\kms\ shift (+0.0344~\AA
) has been applied to the data, which is indicated by an ``F''.  (This
is the same nomenclature used in column 14 of Table~\ref{tab_cols}.)
Stars that lie against the X-ray emission of the inner regions of the
Carina Nebula or the Vela SNR are noted as such with a label at the
top of the panel. Note that these are included only to indicate the
region of interest, and do {\it not} imply that the star is at the
same distance as Carina or Vela.  Sight lines which likely pass
through the SNRs listed in Table~\ref{tab_SNRs} have a label `SNR'
(except for Vela).  The flux is shown with the HD 6$-$0 R(0) line
already removed; the region shaded in gray near the bottom of the O~VI
profiles represents the difference in the spectrum before and after
the removal of the HD line.

{\bf Middle panel:} The data normalized by the best-fit continuum are
plotted in wavelength (bottom $x$-axis) and observed heliocentric
velocity for the \sixa\ transition (top $x$-axis). The wavelengths of
five lines are marked as follows.  ``1'': the {\it predicted} position
of the H$_2$ 7$-$0 R(6) line at 1030.08~\AA, based on the measured
velocity of the H$_2$ 6$-$0 R(4) line; ``2'': the measured position of
the H$_2$ 6$-$0 P(3) line; ``3'': the {\it predicted} wavelength of
the Cl~I~$\lambda 1031$ line, based on the velocity of the H$_2$ 6$-$0
R(4) line. ``4'': the measured position of the HD 6$-$0 R(0) line
which is removed from the data as discussed in
\S~\ref{sect_hdremoval}; ``5'': the measured position of the H$_2$
6$-$0 R(4) line.

Line profiles corresponding to the best-fit solutions discussed in
\S\ref{sect_voigt} over-plot the data (histogram) with curved lines.
If more than one component is fitted to the O~VI line, individual
profiles are shown with dotted lines.  The best-fit values of $b$, $v$
and $N$ for the O~VI feature are listed at the bottom left of the
panel, in units of \kms, \kms, and \pcm , respectively.  The total
column density is given at the bottom right of the panel, along with
two `errors': the first two values in parentheses represent the range
in $N$(O~VI) from the upper and lower continuum fits, $N_u-N$ and
$N-N_l$, where $N$ is the O~VI column density measured from the
best-fit continuum; the second value in parentheses is the error from
Poisson statistics alone.  The central velocities/wavelengths of O~VI
absorption components are marked with filled triangles.  The $b$ and
$N$ values which are used to model the H$_2$ lines shown, are fixed by
fits to other H$_2$ lines in different regions of the \FUSE\ data.
Two components are sometimes fitted to the H$_2$ lines in these other
regions; these are indicated with two tick marks for an individual
H$_2$ line.

{\bf Bottom panel:} The variation of $N(v)$ as a function of
heliocentric velocity for the \sixa\ line derived from the AOD method
is shown in the bottom panel.  All $N(v)$ profiles are plotted at the
same scale except for HDs 167402 (star \#1), 168941 (\#2), 178487
(\#14), and HDE~225757 (\#20); the O~VI absorption towards these stars
is so strong that the y-axis is compressed to accommodate the larger
profile.

To enable calculation of an AOD column density even when the O~VI line is
contaminated by one or both of the H$_2$ lines, the P(3) and R(4) lines have
been removed from the profile (see \S\ref{sect_colsAOD} for more details). The
dotted lines show the positions and strengths of the removed H$_2$ lines,
scaled as if they had the same $f$-value as \sixa . They are included
to illustrate how the O~VI AOD profile was affected by the
H$_2$ lines before they were removed.
 
$N$($v$) profiles derived using the continuum-fitting error envelopes
(yielding values of $N_u$ and $N_l$) are shown above and below the
best-fit $N$($v$) (which forms the middle, shaded histogram).  The
1$\sigma$ Poisson noise errors in $N$($v$) are shown as normal
vertical error bars for each pixel, although they are only plotted for
the middle (adopted) $N$($v$) profile.

The total column density measured over all velocities using the AOD
method is quoted at the top right of the bottom panel. The velocity
range used was $-120$~\kms $\:<\:v_\odot\:<\: +120$~\kms, although
this was modified for profiles that clearly extended beyond this
interval. The velocity range used can be inferred from the total range
covered by gray pixels; these are the $N(v)$ values that contribute to
the calculation of the total column density.  As with the total column
densities that we listed for the profile fits (middle panel), the
differences in the column density from the upper and the best
continuum fit, and the best and the lower continuum fit, 
[$N_u -N$, $N - N_l$], are shown in
the first set of parentheses.  The value in the second set of
parentheses gives the total error from Poisson noise.

Below the $N(v)$ profile, a gray bar indicates the range of velocities
expected from differential Galactic rotation for absorbing gas along
the line of sight. (These are the same bars that are shown in
Fig.~\ref{fig_vel_bar1}.)  Darker colors mark velocities where the
velocity of the gas is changing only very slowly with distance. This
is discussed more fully in \S\ref{sect_vel_with_long}. A small `0'
above the bar marks the value of $v_c$ given in
Table~\ref{tab_journal} to indicate which end of the bar corresponds
to distances of $d=0$.  As in the middle panel, small triangles along
the baseline show the central velocities of the O~VI components
derived from profile fitting (for comparison with the AOD profile),
along with the velocities of the H$_2$ 6$-$0 R(4) components, which
are shown as crosses.


\clearpage
\thispagestyle{empty}

\clearpage

\begin{center}
\begin{huge}

\bigskip
The remaining 144 panels showing all the spectra in this
survey can be found in the full-resolution preprint, available at:

\medskip
http://www.astro.princeton.edu/$\sim$dvb/o6

\end{huge}
\end{center}

\end{document}